\documentclass[aps,prx,superscriptaddress,reprint,notitlepage]{revtex4-2}
\usepackage{amsmath}
\usepackage{graphicx}
\usepackage{bm}
\usepackage{float}
\usepackage{appendix}
\usepackage{placeins}
\newcommand{\ph}[1]{\textcolor{black}{#1}}
\usepackage{xcolor}

\renewcommand{\vec}[1]{\mathbf{#1}}

\newcommand{\mctwo}{Department of Microtechnology and Nanoscience -- MC2,
Chalmers University of Technology, SE-41296 Gothenburg, Sweden}
\newcommand{\kist}{Computational Science Research Center, 
Korea Institute of Science and Technology, Seoul 02792, Republic of Korea}
\newcommand{\mf}{Molecular Foundry, Lawrence Berkeley National 
Laboratory, Berkeley, California 94720-7300, USA}
\newcommand{\lbl}{Materials Sciences Division, Lawrence Berkeley National 
Laboratory, Berkeley, California 94720-7300, USA}
\newcommand{\berkeley}{Department of Physics, 
University of California, Berkeley, California 94720-7300, USA}
\newcommand{\kavli}{Kavli Energy NanoScience Institute at Berkeley,
Berkeley, California 94720-7300, USA}

\begin{document}

\title{Accurate non-empirical range-separated hybrid 
van der Waals density functional for complex molecular problems, solids, and surfaces}


\author{Vivekanand Shukla$^\S$}
\affiliation{\mctwo}
\thanks{$\S$ Equal contribution}
\author{Yang Jiao$^\S$}
\affiliation{\mctwo}
\thanks{$\S$ Equal contribution}

\author{Jung-Hoon Lee}
\affiliation{\kist}
\affiliation{\mf}
\author{Elsebeth Schr{\"o}der}
\affiliation{\mctwo}
\author{Jeffrey B. Neaton}
\affiliation{\lbl}
\affiliation{\berkeley}
\affiliation{\kavli}
\author{Per Hyldgaard}
\email[]{hyldgaar@chalmers.se}
\affiliation{\mctwo}

\date{\today}

\begin{abstract}
We introduce a new, general-purpose, range-separated hybrid van der Waals density \ph{functional, termed
vdW-DF-ahbr,} within the non-empirical vdW-DF method 
[JPCM 32, 393001 (2020)]. It combines correlation from vdW-DF2 with a screened Fock exchange that is fixed by \ph{a new
model of exchange effects} in the density-explicit vdW-DF2-b86r functional [PRB 89, 121103(R) (2014)]. 
The new vdW-DF2-ahbr prevents spurious exchange binding and has a small-density-gradient form
set from many-body perturbation analysis. It is accurate 
for \ph{bulk as well as layered materials} and 
it systematically and 
significantly improves the performance of present
vdW-DFs for molecular problems. Importantly, vdW-DF2-ahbr also outperforms present-standard (dispersion-corrected) range-separated hybrids on a broad collection of noncovalent-interaction 
benchmark sets, while at the same time successfully mitigating the density-driven errors that 
often affect the description of molecular transition states and isomerization calculations. vdW-DF2-ahbr furthermore improves on
state of the art density functional theory approaches by 
1) correctly predicting both the substrate 
structure and the site preference for CO adsorption
on Pt(111), 2) outperforming existing non-empirical vdW-DFs for the description of CO$_2$ adsorption 
in both a functionalized and in a simple metal-organic 
framework, and 3) being highly accurate 
\ph{for the} set of base-pair interactions 
in a model of DNA assembly.
\end{abstract}

\maketitle

\section{Introduction}

The van der Waals (vdW) density functional (vdW-DF) method 
\cite{anlalu96,dobdint96,ryluladi00,rydberg03p126402,Dion,dionerratum,thonhauser,lee10p081101,hybesc14,behy14,bearcoleluscthhy14,Berland_2015:van_waals,Thonhauser_2015:spin_signature,DFcx02017,JiScHy18b,ChBeTh20,DefineAHCX,Hard2Soft} 
for strictly nonempirical density functional theory (DFT) has been successfully applied in materials and chemistry for more than two decades. vdW-DF \cite{2001surfscience,rydberg03p606,rydberg03p126402,Dion} 
opened the door for early DFT predictions of 
adhesion among graphene sheets and in  lubricants \cite{rydberg03p126402},
weak molecular binding \cite{kleis05p192,kleis05p164902,kleis07p100201,kleis08p205422,hooper08p891,cooper08p1304,cooper08p204102},  and the weak adhesion of nuclic bases and other organics on graphene and oxides
\cite{chakarova-kack06p146107,chakarova-kack06p155402,chakarova-kack10p013017}. The functionals of the vdW-DF method have no empirical parameters and avoid double counting of correlation. 
Predating the set of also-popular dispersion-corrected DFTs \cite{grimme1,becke05p154101,becke07p154108,grimme2,silvestrelli08p53002,ts09,ts2,ts-mbd,ambrosetti12p73101,grimme4,KiKiGo20},
the accuracy and robustness have systematically improved
over time.

The success of vdW-DF motivates continued investments to design even better non-empirical versions of vdW-DF. The vdW-DF method is built from  many-body perturbation theory (MBPT) analysis of the nature of the
fully interacting electron ground state. This strategy led, for example, to a straightforward extension to include spin \cite{Thonhauser_2015:spin_signature,Hard2Soft}. We can also directly interpret the quality and performance differences in terms of the spatial variation in and hence nature of the different  contributions to the exchange correlation (XC) energy  \cite{JiScHy18a,JPCMreview,MOFdobpdc}.

The vdW-DF method provides a formally exact framework for a systematic inclusion of nonlocal-correlation effects \cite{JPCMreview}. Part of the MBPT foundation for the vdW-DF method  \cite{ra,lavo87,anlalu96,dobdint96,ryluladi00,Dion,thonhauser,hybesc14,Berland_2015:van_waals,JPCMreview} was first described in the same paper that established logic for correlations in the constraint-based generalized gradient approximation (GGA)
\cite{lavo87}. As such, it represents a third 
generation of XC-energy functionals in an electron-gas tradition that started with the 
local spin density approximation  \cite{helujpc1971,gulu76,pewa92} (LDA) and led to the highly successful PBE \cite{pebuer96} and PBEsol \cite{PBEsol} GGAs. The overall logic of this tradition
is to gradually introduce a controlled increase in flexibility so that we can reliably benefit from more of the pool of physics insight and trusted MBPT inputs \cite{lavo87,ra,Dion,thonhauser,lee10p081101,SCAN,SCANvdW,JPCMreview}.

\ph{Finding an accurate, general-purpose functional is important since theory often concerns complex materials, i.e., systems where the atomic structure is not fully established. There, DFT calculations must be used to first assert which are the most plausible of several candidate motifs, for example, as in  Ref.\ \onlinecite{MOFdobpdc}.
The consistent-exchange 
vdW-DF-cx version  \cite{behy14} (abbreviated CX) is  crafted to seek high accuracy simultaneously for molecules, bulk, and surfaces \cite{bearcoleluscthhy14} but (as
discussed elsewhere \cite{JPCMreview})
it uses a type of XC guidelines \cite{lape79ssc,lavo87,thonhauser} that favors dense-matter and noncovalent (NOC) interaction problems \cite{Tran19,JPCMreview,DefineAHCX,Hard2Soft}. The vdW-DF2-b86R (abbreviated B86R) \cite{hamada14} uses a different nonlocal correlation \cite{lee10p081101} but retains and, in fact, enhances the general-purpose character \cite{bearcoleluscthhy14,Tran19}. Other vdW-DFs 
\cite{Dion,lee10p081101,optx,vv10,cooper10p161104,vdwsolids,Sabatini2013p041108,ChBeTh20} are often found better
at some rather than other types of problems \cite{Tran19}; See supplementary information 
(SI) material 
for an illustration of molecular-performance
variations and see SI Table S.I for a list of functional abbreviations 
that we shall systematically use below. All of the vdW-DFs,
including the unscreened-hybrid forms
\cite{DFcx02017,JiScHy18b}, 
fail in some cases to correctly balance vdW attraction with the repulsion provided by the gradient correction to exchange \cite{Harris85,rydberg03p606,rydberg03p126402,mulela09},
for example, in complex metal-organic-framework systems \cite{MOFdobpdc}. Nevertheless, in DFT we seek to characterize
and predict} molecular reaction energies and transition barriers at the 1 kcal/mol (or 43 meV) limit that defines so-called chemical accuracy \cite{beckeperspective,BurkePerspective,G1set,G2RC,gmtkn55}. Higher accuracy still is needed for understanding chemical fuels \cite{HydMOF,langrethjpcm2009,kong09p081407,lee10p081101,li12p424204,cooper12p34},
CO$_2$ capture \cite{CO2MOF,CO2dobpdc1,CO2sepMOF,CO2dobpdc2,zuluaga2014,poloni_understanding_2014,CO2MOF74survey,MOFdobpdc}, 
batteries \cite{lee2012li,persson2010thermodynamic,shukla2019modelling,shukla2018borophene}, 
and biochemistry \cite{cooper08p1304,le12p424210,umrao2019anticarcinogenic,Hard2Soft}. These are cases where we must understand 
the role of NOC interactions \cite{gmtkn55} as they act in concert and competition \cite{chakarova-kack06p155402,behy13,bearcoleluscthhy14}.
\ph{To get at the complex-materials challenge, we must correctly balance the XC terms in just one general-purpose, yet highly accurate, vdW-DF design that also avoids density-driven DFT errors
\cite{BurkeSIE,DDerrorQaA,MOFdobpdc}.}

This paper reports the design and testing of a range-separated hybrid (RSH)
vdW-DF. It is termed vdW-DF2-ahbr and abbreviated AHBR because it builds on the vdW-DF2 nonlocal-correlation description and on an analytical-hole (AH) analysis \cite{HJS08,DefineAHCX} of the nature of exchange in the \ph{B86R variant \cite{hamada14}.}
We show that it stands out by having an exceptional general-purpose capability and clearly outperforms \ph{even the recent AHCX design 
of a RSH vdW-DF \cite{DefineAHCX}, for reasons
we explain.}

\begin{figure*}
\includegraphics[width=0.9\textwidth]{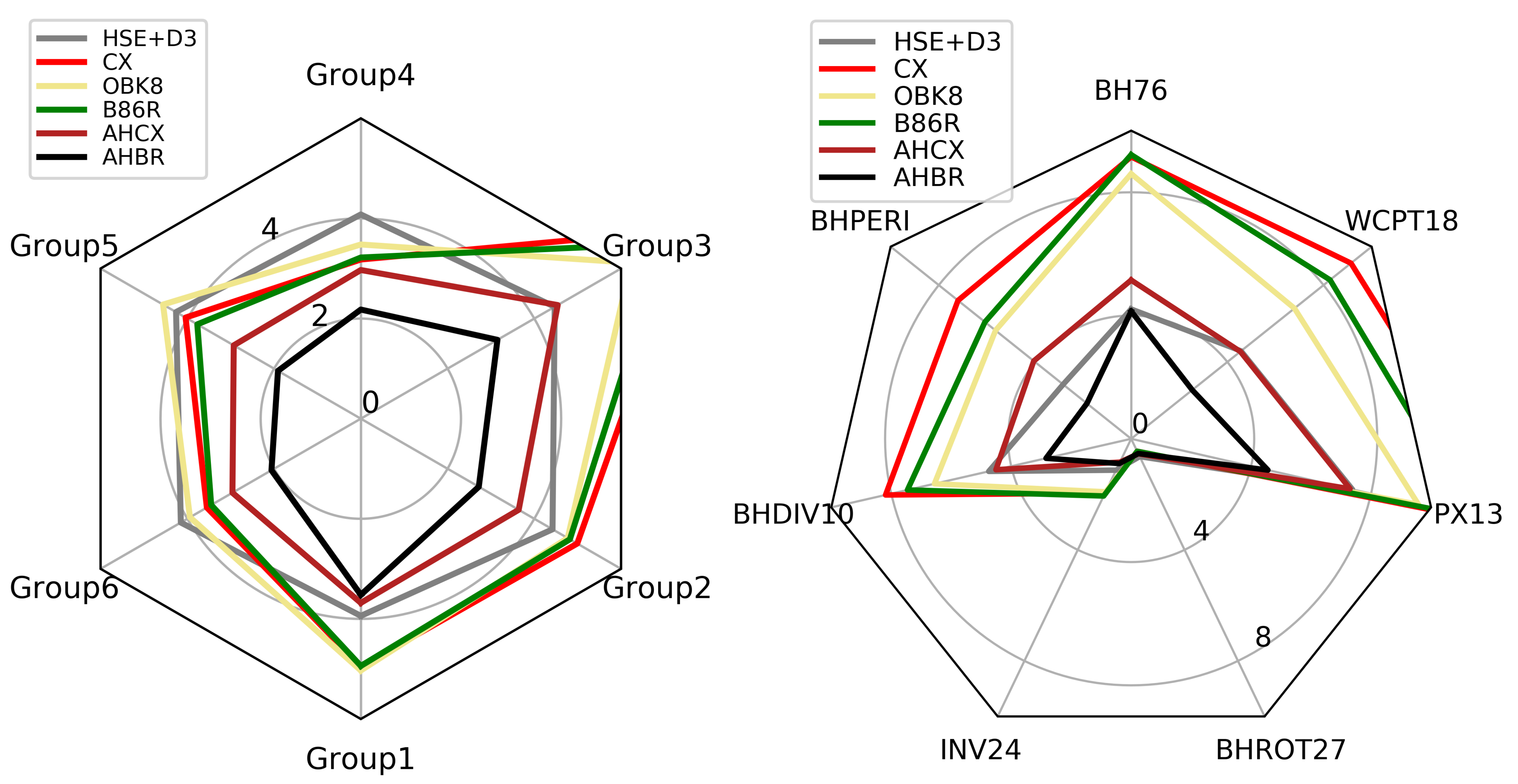}
\caption{Performance comparison of vdW-DFs as averaged over the 
6 groups of molecular benchmarks in the GMTKN55 suite \cite{gmtkn55}
(left panel) and as tracked for individual  transition-state benchmark sets of group 3
(right panel): barrier-height benchmarks for small 
and cyclic molecules (with labels BH76 \& BHPERI), for diverse, inversion 
and rotation processes (with labels BHDIV10, INV24, \& BHROT27, respectively), as well 
as for proton exchange and transfer problems (labels PX13 \& WCPT18).
The GMTKN55 suite also probes performance
on small-system reaction energies (group 1), large-system
reaction energies and isomerizations (group 2) as well as 
on total, inter-, and intra-molecular noncovalent interactions 
(groups 6, 4, and 5, respectively).
\ph{We report mean-absolute 
deviations (MADs,) in kcal/mol, compared 
with coupled-cluster reference energies, at reference geometries \cite{gmtkn55}.}
For comparison, we also include an assessment of dispersion-corrected 
HSE+D3 \cite{HSE03,HSE06,grimme3,gmtkn55} 
\ph{and the recently defined range-separated hybrid
(RSH) vdW-DF-ahcx (abbreviated AHCX) \cite{DefineAHCX}.}
Regular functionals, \ph{exemplified by the three 
vdW-DFs with best overall molecule performance,}
CX 
\ph{\cite{behy14}, B86R \cite{hamada14}, and
vdW-DF-optB88 
(or optB88-vdW,  
abbreviated OBK8)} \cite{optx},
are often challenged by density-driven DFT errors \cite{DDerrorQaA} in such problems.}
\label{fig:TSwDiel}
\end{figure*}

\ph{Figure \ref{fig:TSwDiel}, below, summarizes our assessment of performance over broad molecular properties, illustrating that AHBR is both highly accurate and has an excellent transferability. We find that AHBR can navigate generic density-driven functional errors \cite{BurkeSIE,patra2019rethinking,DDerrorQaA}
that, for example, often affect 
molecular transition states.}

\ph{We propose that the new AHBR be used to test the status of the vdW-DF method, as we also partly illustrate, because it is 
free of a performance bias. The performance of RSH HSE+D3 \cite{HSE03,HSE06,grimme3}
is independent of the benchmark type in the very broad GMTKN55 suite \cite{gmtkn55} on broad molecular properties. The unscreened hybrid B3LYP+D3 \cite{BeckeIII,LYP,grimme3} is an improvement over HSE+D3 on molecular transition states and NOC interactions but not across the board \cite{gmtkn55}, 
SI Tables S.II-III. HSE+D3 and B3LYP+D3 are widely used in materials science and chemistry, respectively, and their transferability sets the bar for the generic 
vdW-DF method testing. The new RSH AHBR (black curve
in Fig.\ \ref{fig:TSwDiel}) outperforms HSE+D3 (gray) across all types of molecular properties and matches (clearly improves) the B3LYP+D3 performance \cite{gmtkn55} for the important group 3 of molecular transition-state benchmarks (for the rest of the GMTKN55 benchmarks). As summarized in Fig.\ S 1 and documented in Tables S II-III of the SI material, these observations holds for either of the weighted-mean-absolute-deviation measures that are 
suggested and used in Ref.\ \onlinecite{gmtkn55}. Unlike AHCX (dark red), AHBR is more successful than HSE+D3 and B3LYP+D3 on, for example, the BH76 benchmark set on molecular barrier heights, problems that are sensitive to density-driven DFT errors  \cite{BurkeSIE,DDerrorQaA}. The AHBR provides
systematic accuracy gains over present-standard
hybrid choices.}

\ph{The specific contributions of the paper can be summarized as follows. 
We first complete a robust planewave-based assessment across the 
full-GMTKN55 benchmark suite \cite{gmtkn55}, documenting that the unscreened
hybrid extension \cite{DFcx02017} of B86R, abbreviated DF2-BR0, provides 
the best performance on molecular properties. This is true for the GMTKN55 suite and among all of the vdW-DFs, including the RSH vdW-DFs, Fig.\ S 2 of the SI material.
Our broad documentation is consistent with 
a very recent independent observation for proton transition barriers \cite{BerlandTrState22}. We  
proceed to define the AHBR, the RSH generalization of DF2-BR0, using an
AH characterization for exchange
effects in B86R.} 
\ph{We validate that AHBR is an exceptional performer across GMTKN55, Fig.\ \ref{fig:TSwDiel}, and retains a strong performance for bulk and some layered
materials. 
Finally, we illustrate the usefulness of AHBR
for DNA assembly and molecular adsorption problems, finding good agreement on
quantum-chemistry reference calculations, the correct site preference 
for the CO/Pt(111) problem, and a good performance for 
characterization of CO$_2$ uptake in two metal organics framwors (MOFs) \cite{CO2MOF,CO2MOF74survey,CO2dobpdc2,MOFdobpdc}. }

The rest of the paper is organized as follows. The theory section II presents 
an overview of the vdW-DF method, analysis of the B86R exchange hole, and
contains the formulation of the new AHBR. 
Section III contains results and discussions, including
illustrations of AHBR accuracy. 
Section IV contains our conclusion and outlook. The paper has two
appendices giving computational details, including
the electrostatic-environment approach used to complete planewave benchmarking
across the GMTKN55 suite.

\section{Theory}

Central in MBPT and in the electron-gas foundation of DFT \cite{lape75,gulu76,lape77,lape80,lameprl1981,lavo87,lavo90}
is the screened density response $\delta n(\omega)$ to some 
external-potential change $\delta \Phi_{\rm ext}^{\omega}$ oscillating
at frequency $\omega$. In MBPT we can, at least in principle, compute 
the nonlocal response function $\chi_{\lambda}(\mathbf{r}, \mathbf{r'};\omega)
\equiv \delta n(\mathbf{r})/\delta \Phi_{\rm ext}^{\omega}(\mathbf{r'})$, often expressed
as a function of a complex frequency $\omega=iu$. 
We can also do that at a range of an assumed
reduced strength $0 < \lambda < 1$ of the electron-electron interaction 
$\lambda \hat{V}$.  Assuming access to this knowledge, the adiabatic connection formula 
(ACF) permits an exact determination
\begin{equation}
E_{\rm xc} = - \int_0^1 \, d\lambda \, \int_0^\infty \, 
	\frac{du}{2\pi} \, \hbox{Tr} \{ \chi_\lambda(iu) V \} 
-E_{\rm self}\, , 
\label{eq:ACF}
\end{equation}
of the XC energy functional $E_{\rm xc}$. Here $V=|\mathbf{r}-\mathbf{r}|^{-1}$ 
denotes the matrix element of the electron-electron interaction $\hat{V}$, $u$
is an imaginary frequency argument in the response description,
while the last term is the electron self energy $E_{\rm self}  =  \hbox{Tr} \{ \hat{n} V\}/2.$ 
The expressions for $E_{\rm xc}$ and $E_{\rm self}$ involve
Coulomb-weighted traces, that is, integrations in spatial coordinates of 
$|\mathbf{r}-\mathbf{r}'|^{-1}$ times $\chi_{\lambda}(\mathbf{r'}, \mathbf{r};\omega)$ 
and times the electron density ${n}(\vec{r'})$, respectively.  Also we have 
(at every coupling-constant value $\lambda$) added an auxiliary
potential that keeps the electron density $n(\mathbf{r})$
unchanged across the implied adiabatic turn on of the
electron-electron interaction $\hat{V}_\lambda = \lambda \hat{V}$.
The actual XC potential used in the Kohn-Sham (KS) scheme for
efficient DFT calculations is simply the $\lambda \to 0$
limit of this auxiliary potential. It is given by a
functional derivative of the XC energy,
as discussed many places elsewhere.

In MBPT, we compute the response functions $\chi_\lambda(\omega)$
as a ground-state expectation value of correlations between density 
fluctuations 
\cite{Note1}.
%
As such, $\chi_\lambda(\omega)$ is directly reflecting the 
Lindhard-type screening that exists in the electron gas at 
assumed coupling constant 
\ph{$\lambda$, as discussed, for example,
in Ref.\ \onlinecite{JPCMreview}}. The screening
is given by the dielectric function
$\kappa_\lambda(\omega)=(1+\lambda V\chi_\lambda)^{-1}$. For practical DFT, we seek XC functional approximations that contain the most pertinent physics contents of the widely complex, many-body interacting processes that define $\chi_\lambda(iu)$. The massive DFT 
usage allows us to get successively more adapt at this as long as we stay systematic and can interpret performance differences, for example, within MBPT.

\begin{figure}
\includegraphics[width=0.95\columnwidth]{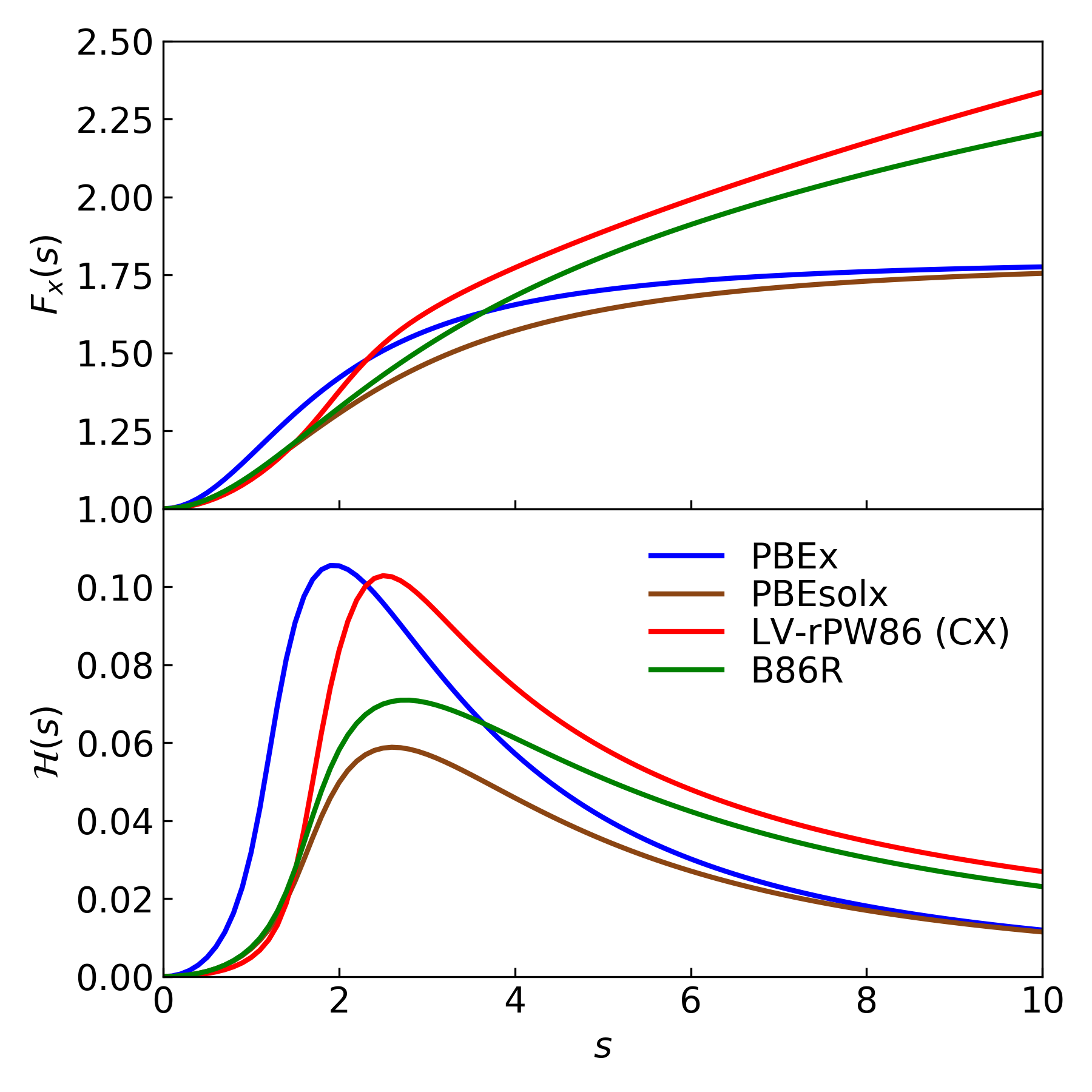}
	\caption{Comparison of exchange enhancement factor $F_x(s)$
	(top panel) and of the Gaussian-exponent prefactor $\mathcal{H}(s)$ 
	that determines the main damping in the AH modeling for the density-density
	correlation defining the underlying exchange holes \cite{HJS08,DefineAHCX}.
}
\label{fig:FxAHplot}
\end{figure}

In the electron-gas 
tradition for XC functional designs \cite{ma,mabr,rasolt,gulu76,lape77,adawda,lape80,lavo87,lavo90,Dion,thonhauser,Berland_2015:van_waals,JPCMreview,DefineAHCX} we focus the discussion on the XC hole $n_{\rm xc}(\mathbf{r};\mathbf{r'})$ that results by 
a complex-frequency integration over $\chi_\lambda(iu)$. The XC hole $n_{\rm xc}(\mathbf{r};\mathbf{r'})$ express the tendency for a given
electron at position $\mathbf{r}$ to suppress the electron occupation at neighboring positions $\mathbf{r'}$.
Importantly, the introduction of this hole permits an ACF-based interpretation of
the XC energy functional \cite{gulu76,lape77,hybesc14},
\begin{equation}
    E_{\rm xc} = \frac{1}{2}\int_\mathbf{r}\int_\mathbf{r'}
    \frac{n(\mathbf{r}) \, n_{\rm xc}(\mathbf{r};\mathbf{r'})}
    {|\mathbf{r}-\mathbf{r'}|} \, ,
    \label{eq:ExcRel}
\end{equation}
that has analogies in electrostatics.
However, the XC hole reflects the impact of zero-point dynamics, i.e., virtual collective (plasmon) excitations in the electron distribution \cite{lu67,lape80,ra,JPCMreview}.

\ph{The exchange-hole component $n_{\rm x}(\mathbf{r};\mathbf{r'})$ of this total XC hole
describes the impact of Pauli exclusion. The Fock-exchange approximation $n^{\rm Fo}_{\rm x}(\mathbf{r};\mathbf{r'})$ to $n_{\rm x}$
results by considering one-particle density 
matrices formed from the KS-wavefunction
solutions, for example, as summarized in the
discussion provided in Ref.\ \onlinecite{DefineAHCX}.}

\subsection{The vdW-DF framework}

To begin a summary of vdW-DF, we note that since LDA
and GGAs are completely set from a modeling of an 
underlying XC hole 
\cite{gulu76,lape77,lape80,pewa86,FullHoles,hybesc14,DefineAHCX},
we are also ready to capture vdW forces as defined
from an electrodynamics coupling of electron-density fluctuations \cite{lo30,jerry65,ma,ra,lavo87,rydberg03p126402,hybesc14,JPCMreview}.
Any XC functional can be seen as the net binding energy of the electrons
and associated XC holes, Eq.\ (\ref{eq:ExcRel}). However,
it is also clear that the electron and the associated (GGA-type) XC hole 
form an antenna of charged parts that have a mutual zero-point energy dynamics.
The electron-XC-hole pairs will interact even across regions that have but a sparse or no electron density \cite{langreth05p599,langrethjpcm2009}. 
In fact, this electron-gas electrodynamics coupling \cite{ra,lavo87,anlalu96,dobdint96,berlandthesis,hybesc14,JPCMreview} is a systematic generalization of the original London picture of  dispersion forces among noble-gas atoms
\cite{lo30,lo37}.

The vdW-DF method achieves a systematic 
extension of MBPT-based GGAs by recasting the 
exact XC functional as an electrodynamics problem 
\cite{ryluladi00,rydbergthesis,dionthesis,hybesc14,JPCMreview}, 
while \ph{counting (via a frequency contour integration)} the coupling-induced shifts in energies
for collective excitations 
\cite{jerry65,lape80,hybesc14}. 
Thus, for the vdW-DF XC energy description, we rely on 
a formally exact recast of the ACF result, \cite{hybesc14,Berland_2015:van_waals,JPCMreview},
\begin{equation}
E_{\rm xc} = \int_0^\infty \, \frac{du}{2\pi} \, 
	\hbox{Tr} \{\ln(\kappa_{\rm ACF}(iu))\} - E_{\rm self}\, . 
\label{eq:newACF}
\end{equation}
In Eq.\ (\ref{eq:newACF}), we have 
introduced an effective, spatially nonlocal, 
dielectric functional function $\kappa_{\rm ACF}(iu)$. 
The formal XC evaluation is given as a weighted 
$\lambda$ average of $\kappa_\lambda(\omega)$ and hence of  
$\chi_{\lambda}$ \cite{hybesc14,Berland_2015:van_waals,JPCMreview} 
and there is full equivalence of Eqs.\ (\ref{eq:ACF}) and (\ref{eq:newACF}), 
given consistent approximations; The recast is also equivalent to the
XC-hole \cite{gulu76,lape77} formulation, Eq.\ (\ref{eq:ExcRel}). The 
electrodynamical recast, Eq.\ (\ref{eq:newACF}), simplifies 
the porting of the ideas of the cumulant expansion \cite{la70,Hedin80}
to the vdW-DF development work. Specifically, we cna link the
Ashcroft-Langreth-Lundqvist picture of vdW-binding 
contributions \cite{adawda,ma,lavo87,ra,anlalu96}, and the MBPT 
input to the constraint-based GGAs \cite{lape80,lavo87,pebuer96,PBEsol,hybesc14}.
In turn this allows the vdW-DF method to provide an 
effective (MBPT-guided) approximation to the $\lambda$-averaged 
response description, Eq.\ (\ref{eq:newACF}), as discussed in 
Refs.\ \onlinecite{Dion,thonhauser,hybesc14,Berland_2015:van_waals,Thonhauser_2015:spin_signature,JPCMreview}.

\begin{figure*}
	\includegraphics[width=0.90\textwidth]{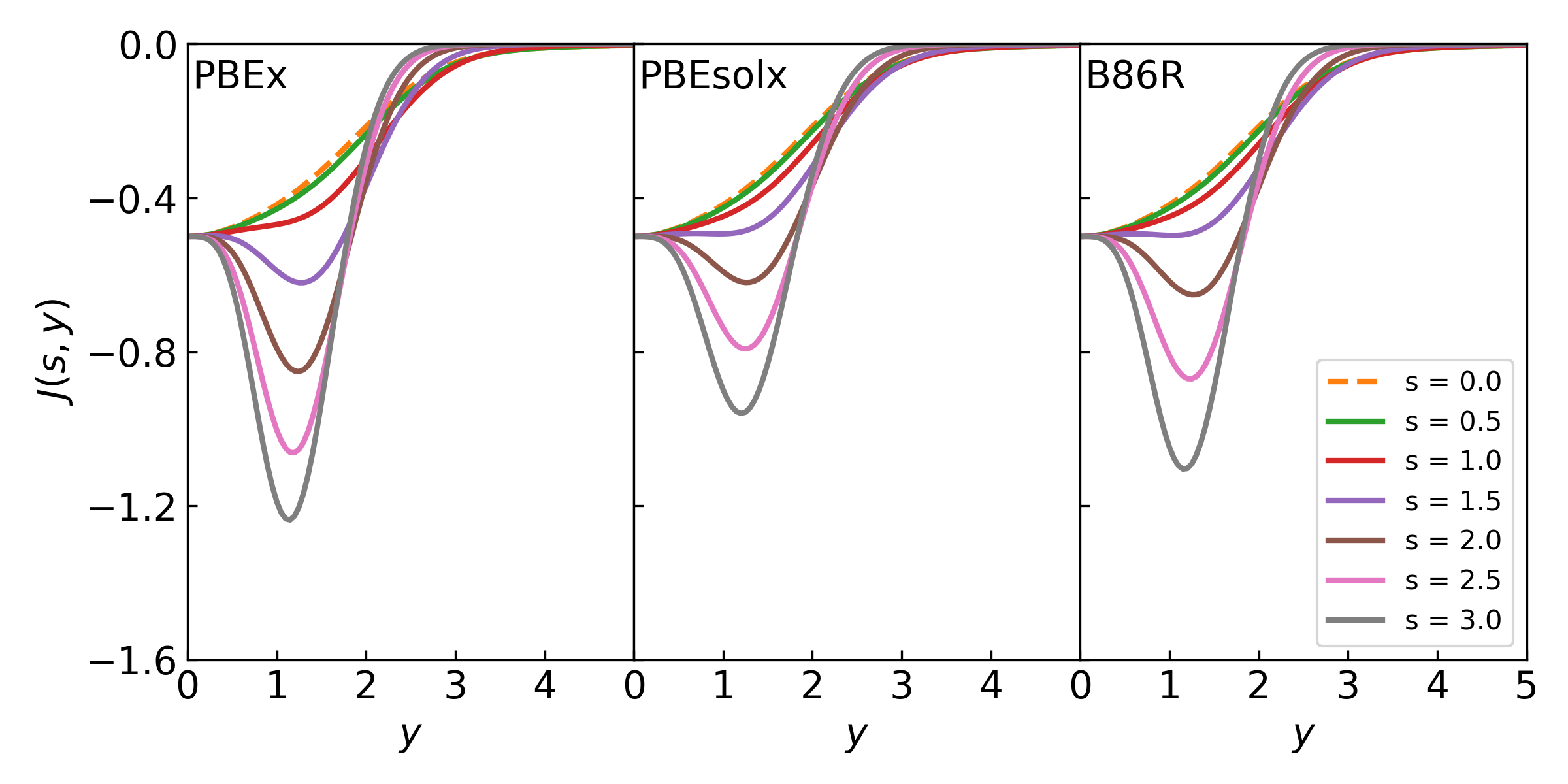}
	\caption{Radial variation in the scaled exchange hole 
	$J(s,y=k_{\rm F}(\mathbf{r})|\mathbf{r}-\mathbf{r'}|)=n_{\rm x}(\mathbf{r},\mathbf{r'})/n(\mathbf{r})$
	for PBE, PBEsol, and B86R, all as described in an
	analytical-hole (AH) model parameterization \cite{HJS08,DefineAHCX}. The shapes
	of these holes define the exchange components of PBE \cite{pebuer96},
	of PBEsol \cite{PBEsol}, and of B86R,
	respectively.  Using the latter exchange-hole model, we define the 
	here-released RSH AHBR, following the design steps that
	we recently documented in crafting AHCX \cite{DefineAHCX}.
	\label{fig:NLrshDesignAH}
	}
\end{figure*}

In the vdW-DF method, we furthermore use a so-called internal functional 
$E_{\rm xc}^{\rm in}$ to first set a lowest-order (GGA-level)
approximation for the screening. The screening is described by a 
truly nonlocal dielectric function $\epsilon(\omega)$ (but we suppress spatial coordinates in this discussion). This dielectric function has collective excitations, plasmons, that sets the
start of the response modeling, given by an effective electron-gas susceptibility $\alpha(\omega)=(\epsilon(\omega)-1)/4\pi$. \cite{lape80,hybesc14,Berland_2015:van_waals}. First, we approximate 
the plasmon propagator as $S_{\rm xc}(\omega)=\ln(\epsilon(\omega))$ 
and rely on an explicit two-pole approximation that reflects plausible assumptions and all plasmon-related conservation laws
\cite{Dion,JPCMreview}. Next, the formal relation
\begin{equation}
	E_{\rm xc}^{\rm in} = \int_0^{\infty} \, \frac{du}{2\pi} \, 
	\hbox{Tr} \{\ln(\epsilon(\omega))\} - E_{\rm self}\, ,
	\label{eq:newEin}
\end{equation}
sets the detail of the plasmon dispersion as the implied contour integration naturally sums the plasmon-pole contributions as $\ln$ 
singularities \cite{jerry65,lape80,hybesc14,JPCMreview}. In summary,
we have a formal link between the GGA-level internal functional
and structure of the starting 
approximation $S_{\rm xc}(\omega)$
for an emerging description of the
actual plasmon dynamics \cite{Dion,JPCMreview}.

We also explicitly enforce a longitudinal projection 
of the response in the dielectrics approximation function 
\begin{equation}
	\kappa_{\rm ACF}(\omega)=-\nabla \cdot \epsilon(\omega) \nabla V/4\pi \, .
	\label{eq:kappaACFapp}
\end{equation}
We note that having this projection inside the
recast ACF, Eq.\ (\ref{eq:newEin}), produces terms that capture the
vdW attraction as described in the presence of electron-gas screening \cite{JPCMreview}.
Moreover, the use of $\epsilon(\omega)=\exp(S_{\rm xc}(\omega))$ implies a cumulant-or-cluster-expansion logic \cite{la70,Hedin80,GunMedSch94,HolAry97,Sokolov13} 
in the response description \cite{JPCMreview}; This allows 
the vdW-DF method to also pick up high-order susceptibility 
and screening effects \cite{hybesc14,JPCMreview},
i.e., to balance the vdW attraction by other nonlocal-correlation
effects \cite{JPCMreview}.

Finally, given the choice of GGA-level plasmon modeling, Eq.\ (\ref{eq:newEin}), we repeat the
contour integration evaluation with Eqs.\ (\ref{eq:newACF})
and (\ref{eq:kappaACFapp}) to secure efficient kernel-based 
evaluation of nonlocal-correlation energies \cite{Dion,dionerratum}. 
This approach means that the vdW-DFs versions
have no discernible cost increase over GGA in planewave codes
\cite{roso09,le12p424210,libxcvdW,jewahy20}. Similarly, the new 
class of range-separated hybrid (RSH) vdW-DFs \cite{DefineAHCX} 
has the same costs as the HSE \cite{HSE03,HSE06,HJS08},
i.e., a RSH that is based on the PBE \cite{pebuer96} GGA. 
However, use of the vdW-DFs sometimes requires a better 
convergence of the electron-density variation 
\cite{optx,vdwsolids,jewahy20}.

The resulting non-empirical vdW-DF description stands out, for 
example, in the class of vdW-inclusive functionals by treating 
all interaction contributions on the same electron-gas footing. We avoid auxiliary inputs beyond ground-state DFT and we avoid
all need for semi-empirical adjustments, for example, to ameliorate
double counting of nonlocal correlations.  The vdW-DF method 
is set up to describe vdW interactions as they emerge 
in concert and competition with covalent and ionic binding \cite{bearcoleluscthhy14} and with orbital-interaction 
shifts produced by wider nonlocal-correlation effects \cite{la70,Hedin80,JPCMreview}. The latter effects are, 
for example, documented to counteract contributions to vdW 
interaction contributions from the high-density regions
near the nuclei as well as from the saddle point of the electron-density variation that exists between fragments \cite{JiScHy18a,JPCMreview}.  
Noting that formal MBPT sets the nature of the exact XC energy, we 
seek to use the MBPT as a guide \cite{Dion,thonhauser}. This is done,  
for example,
by trying to recycle \cite{ma,lavo87,lavo90,ra,Dion,dionerratum,hybesc14,lee10p081101,JPCMreview}
the accuracy gains that were made on the exchange description in the
MBPT-based GGAs \cite{lape80,schwinger,pewa86,lavo87,lavo90,pebuer96,pebuwa96,PBEsol}.

\subsection{vdW-DF versions}

Prior vdW-DF versions \cite{rydberg03p126402,Dion,lee10p081101,Thonhauser_2015:spin_signature,DFcx02017,DFcx02017,DefineAHCX} involve a controlled introduction of systematic design changes.  For example, vdW-DF1 \cite{Dion,dionerratum,thonhauser} and vdW-DF2 \cite{lee10p081101} have the same overall structure but differ in whether we prioritize MBPT or scaling insight on exchange \cite{schwinger,lavo90,thonhauser,lee10p081101} to model 
the collective-excitation response that forms the starting point of the $E_{\rm c}^{\rm nl}$ evaluation. They also differ in how we enforce \cite{rydberg03p126402,pewa86,mulela09,Berland_2015:van_waals}
a method criterion that the actual exchange component of a vdW-DF
should not itself lead to spurious weak binding \cite{Harris85,mulela09}, since the vdW attraction is 
a correlation effect \cite{Note2}.

For a computationally efficient evaluation
\cite{Dion,roso09}, the standard
general-geometry formulation expands the recast ACF, Eq.\ (\ref{eq:newACF}),
to second order in the (nonlocal) plasmon propagator 
$S_{\rm xc}(\omega) = \ln(\epsilon(\omega))$ \cite{Dion,thonhauser,JPCMreview}.
Formally, the expansion is written
\begin{eqnarray}
	E_{\rm xc}^{\rm DFs} & \approx & E_{\rm xc}^{\rm in} + E_{\rm c}^{\rm nl} 
	\label{eq:newExcExpand} \\
	E_{\rm c}^{\rm nl}& = &
	\int_0^{\infty} \, \frac{du}{4\pi} \, \hbox{Tr} \{ S_{\rm xc}^2(iu) - 
	(\nabla S_{\rm xc} \cdot \nabla V/4\pi)^2 \}\, , 
	\label{eq:EcnlExpand}
\end{eqnarray}
where $E_{\rm xc}^{\rm in}$ \ph{and Eq.\ (\ref{eq:newEin}) set the details of $S_{\rm xc}(iu)$ \cite{Dion,Berland_2015:van_waals}.}

In all present vdW-DFs, the internal $E_{\rm xc}^{\rm in}$ 
functional is chosen semi-local (GGA-like), comprising LDA correlation and a simple 
choice of physics-motivated gradient-corrected exchange \cite{lavo87,Dion,
schwinger,lee10p081101}. This choice avoids double-counting of nonlocal 
correlations \cite{Dion,Berland_2015:van_waals,JPCMreview}. The gradient-exchange 
choices used in $E_{\rm xc}^{\rm in}$ are defined by formal-MBPT input. That input
is used through the Langreth-Perdew and Langreth-Vosko (LV) analysis of 
a screened-exchange nature \cite{lape79ssc,lavo90,Dion,thonhauser} 
\ph{(that is natural for bulk and metals)}
in the first 
general-geometry release, vdW-DF1. It is set as the Schwinger exact-exchange 
scaling analysis \cite{schwinger,lee10p081101} 
in the second, vdW-DF2. \ph{The Schwinger MBPT result is instead relevant for capturing exchange effects in molecules \cite{lee10p081101};  This fact is, for example, revealed by a demonstration that it leads to a
non-empirical derivation of Becke-88 exchange \cite{Be88}, when interpret 
in the GGA framework \cite{ElliottBurke2009}.} 

The ACF foundation, Eqs.\ (\ref{eq:newACF}) through
(\ref{eq:kappaACFapp}), does motivate the use of Eq.\ (\ref{eq:newExcExpand})
to pick the exchange \cite{JPCMreview}. That is, we should ideally use 
the internal-exchange formulation, and hence the $S_{\rm xc}(iu)$ form, 
to also define the actual exchange. However, the inner functional is
deliberately kept simple, while the overall exchange design must also
reflect other considerations. Taken together, these observations mean
that it is presently not possible to directly implement this idea. 
The general-geometry vdW-DFs therefore have a looser connection between the XC terms,
\begin{equation}
	E_{\rm xc}^{\rm DFs} \equiv E_{\rm xc}^{0} + E_{\rm c}^{\rm nl}  \, ,
	\label{eq:newExcDF}
\end{equation}
where $E_{\rm xc}^{0}=E_{\rm xc}^{\rm in}+\delta E_{\rm x}^0$, 
permitting a cross-over term $\delta E_{\rm x}^0$, even if it is not compatible with the ACF \cite{JPCMreview}.

The top panel of Fig.\ \ref{fig:FxAHplot} compares the so-called
exchange-enhancement factor $F_{\rm x}$ that defines the nature of the
gradient-corrected exchange in all GGA descriptions as well as in present
vdW-DFs. We show the variation with the the scaled density 
gradient $s= |\nabla n|/(2n k_{\rm F})$ (where $k_{\rm F} = 
(3\pi^2 n)^{1/3}$ is the local Fermi-wavevector)
for the four XC functional cases that are of interest here. 
The exchange energy in any semilocal (GGA-type)
approximation, that the present vdW-DFs also use, must take the form \cite{burke}
\begin{equation}
	E_{\rm x}^{\rm GGA} = \int_\mathbf{r} n(\mathbf{r}) 
	\epsilon_{\rm x}^{\rm LDA}(n(\mathbf{r})) F_{\rm x}(s(\mathbf{r})) \, .
	\label{eq:FxDefRegular}
\end{equation}
Here $\epsilon_{\rm x}^{\rm LDA}= -3k_{\rm F}/4\pi$ denotes the 
LDA exchange result, i.e., the exchange 
energy-per-particle value that characterizes a homogeneous system (at 
density $n=n(\mathbf{r})$).  The variations in this gradient corrected 
exchange is thus set alone by the enhancement form of factor $F_{\rm x}(s)$. 
For $E_{\rm xc}^{\rm in}$ the enhancement
is set as a quadratic expansion, $F_{\rm x}^{\rm in}(s)=1+\mu s^2$, Refs.\ \onlinecite{Dion,thonhauser,lee10p081101}.

Exchange descriptions of the popular constraint-based GGAs arise 
when one imposes an exchange-hole-conservation 
criterion in the modeling of density gradient effects. 
This was first done in the (revised) PW86 
\cite{pewa86,mulela09}. It was repeated in the 
popular PBE \cite{pebuer96} and PBEsol \cite{PBEsol}
designs while then also paying attention to preventing 
the exchange-hole depth from dramatically
exceeding the local-electron density. The PBE and PBEsol also adhere to a local implementation of the so-called Lieb-Oxford bound \cite{LiebOxford81,PeRuSuBu14} on the high-$s$
$F_{\rm x}(s)$ variation, but the actual 
(globally-implemented) bound does not, 
in practice impact this discussion of 
picking a robust GGA-type (or 
hybrid-type) exchange for the vdW-DFs \cite{rydberg03p126402,mulela09,DefineAHCX}. 

For the vdW-DFs we must craft an asymptotic  $F_{\rm x}(s)$ behavior that produces an adequate but not excessive
repulsion by gradient-corrections to exchange for weakly interacting 
molecules \cite{Harris85,rydberg03p126402,mulela09,behy13,MOFdobpdc}.
This is to ideally eliminate (without overcompensating) spurious 
weak-system binding by the errors in the LDA exchange description \cite{Harris85,rydberg03p126402,mulela09}.

The actual exchange descriptions in  vdW-DF1 and vdW-DF2 are set as in the revPBE \cite{ZhYa98} variant 
of PBE and as the refitted PW86 form \cite{mulela09}, respectively. In both cases the selections were made following analysis of the weak binding 
of noble-gas and small-molecule dimers, Refs.\ \onlinecite{rydberg03p126402,mulela09}.
Both of these exchange choices are more repulsive than the PBE exchange, i.e,
they have a gradient-correction to exchange that gives a stronger push 
to separating fragments. That extra repulsion 
is needed \cite{mulela09,MOFdobpdc} since, in the vdW-DFs, we are also upgrading to a truly nonlocal correlation description $E_{\rm c}^{\rm nl}$.
That new term includes vdW forces \cite{JPCMreview} and gives a stronger attraction mechanism than what exists in PBE.

The consistent-exchange CX  \cite{behy14,hybesc14,JPCMreview} 
-- and hence with the CX0P and AHCX hybrid extensions \cite{JiScHy18b} 
-- aligns the two ways that exchange insight underpins the vdW-DF 
details, in the inner-functional $E_{\rm xc} ^0$ and in Eq.\ (\ref{eq:newExcDF}), 
as far as possible.  The idea is to look at the impact of XC-balance on the
binding-energy descriptions
instead of on the total-energy
descriptions \cite{behy14}. The move to reconcile the inner
and actual exchange has the benefit that we use the Lindhard-screening \ph{logic, implied in the
expansion Eqs.\ (\ref{eq:newExcExpand}) and (\ref{eq:EcnlExpand}),} as well as current conservation \cite{hybesc14,JPCMreview}, to effectively balance XC terms.  

Our general design strategy is to maximize the role of MBPT inputs, 
like Lindhard screening, because it is a promising path to securing
high accuracy broadly in one general-purpose XC design \cite{WarLut,lape80,ShaSch}. In formal MBPT, we summarize the 
net impact on the electron-electron interactions in 
terms of a so-called, self-energy term $\Sigma_{\rm xc}(\mathbf{r},
\mathbf{r}',\omega)$. It determines how a single-electron 
excitations propagate in the fully interacting system \cite{WarLut,lape80,ShaSch,JPCMreview}. It plays a similar
role as the DFT exchange-correlation potential 
$v_{\rm xc}(\mathbf{r})=\delta E_{\rm xc}/\delta n(\mathbf{r})$
that defines the independent-particle dynamics of in the DFT
representation of the same system (except that it is both 
frequency dependent and truly nonlocal) \cite{helujpc1971,ShaSch}. 
A key argument for the vdW-DF design strategy is that 
the formal MBPT description of the total energy is tolerant \ph{\cite{WarLut}}, 
i.e., one can get good results even when 
\ph{interaction effects are merely approximated by perturbation theory for the self-energies $\Sigma_{\rm xc}(\mathbf{r},
\mathbf{r}',\omega)$.} \ph{We get a sound
electron-response description as long as we keep those (so-called skeleton) diagrams that capture the essential physics and dominant features of the electron-gas response
\cite{helujpc1971,ma,lavo87,ra,lavo90,lape80,ShaSch,thonhauser,JPCMreview}.} 
Also, in principle, that robustness extents to the choice of
$E_{\rm xc}$, thanks to the Sham-Schl{\"u}ter relation between
$v_{\rm xc}(\mathbf{r})$ and $\Sigma_{\rm xc}(\mathbf{r},
\mathbf{r}',\omega)$ \ph{\cite{ShaSch}.}

It is important to observe, however, that we have not in CX 
(nor in AHCX) enforced a complete exchange alignment for all types of problems. The Lindhard screening and current-conservation are essential parts of the ACF result for the exact XC functional \ph{\cite{JPCMreview}}. The CX and 
associated hybrid designs allow us to use this idea, but only for descriptions of system processes where the important density changes are set by density regions with
moderate values $s \lesssim 2.5$ \cite{behy14}. This criterion, i.e., that relevant binding or process-energy contributions to the XC energy differences should converge
relatively fast with $s$ \cite{behy14,JPCMreview,JeBeTh21,MOFdobpdc}, holds for
typical bulk and surface problems \cite{behy13,Ageo20,JPCMreview}, where the use of the CX/CX0P/AHCX tool chain is suggested \cite{DefineAHCX,Hard2Soft}. It is a welcome bonus for CX and AHCX that the CX criterion (and CX accuracy) often seems to 
hold also for many molecular problems \cite{behy14,RanPRB16,BrownAltvPRB16,JPCMreview,DefineAHCX,MOFdobpdc}. 
However, there are also cases where we can document a large binding 
impact of density changes and where the interaction problem is not completely
set by low-to-moderate $s$ values. This happens for CO$_2$ uptake a diamine-functionalized MOF \cite{MOFdobpdc}. 

The fact that there is a condition on the CX/AHCX implementation of the Lindhard logic also suggests a potential susceptibility to density-driven errors \cite{DDerrorQaA,patra2019rethinking}. Such errors undercut 
overall arguments for a generic XC robustness of CX and AHCX, 
at least for the systems in systems where the sensitivity 
is identified \cite{DDerrorQaA,SiSoVu22}. The translation of MBPT 
robustness \cite{WarLut} into XC-design robustness 
is vulnerable because the Sham-Schl{\"u}ter equation \cite{ShaSch}
also includes factors, i.e., independent-particle Green functions 
$G_0$, that are set by KS energy levels and by the spatial variation in the KS orbitals that arise in the XC approximation. In order for
an XC approximation to inherit the $\Sigma_{\rm xc}(\mathbf{r}, \mathbf{r}',\omega)$ robustness, it should delivers a 
near-exact density variation. Also, density-driven errors emerge, for example, by self-interaction error (SIE) effects in negatively-charged ions \cite{BurkeSIE} and they can, as such, reflect large density changes \cite{MOFdobpdc}. The CX and AHCX rely on \ph{the ACF \cite{JPCMreview},} but once the CX usage is pushed beyond the small-to-moderate-$s$ criterion \cite{behy14,MOFdobpdc}, the consistency benefit is gone. The inherent Lindhard-screening logic and current-conservation mechanism is then no longer able to enforce an automatic XC-hole conversation
on the LV-exchange description \cite{JPCMreview}. 
While the high-$s$ form of CX exchange, i.e., the rPW86, also 
reflects a separately
implemented (older) XC-hole conservation criterion 
\cite{pewa86,mulela09}, the 
strong MBPT connection is lost. In this type of large-$s$ problems, 
we expect that the use of modern constraint-based PBE and PBEsol exchange 
would be the safer approach. In summary, we cannot expect that CX (and hence AHCX) will always remain a robust choice.

\subsection{The logic of the B86R variant}

Improvement in accuracy generally followed from coupling
the vdW-DF1 and vdW-DF2 correlation to other (less repulsive) exchange choices,
for example, with the suggestions for OBK8, C09, OB86, B86R, and vdW-BEEF 
variants  \cite{optx,cooper10p161104,vdwsolids,hamada14,vdwBEEF}. The same is true for the CX release (and formal spin and hybrid extensions 
\cite{Thonhauser_2015:spin_signature,DFcx02017,JiScHy18b,JPCMreview,DefineAHCX,Hard2Soft}) that uses a Lindhard-screening logic to balance the XC components
in typical binding cases \cite{behy14}. The balance question
is also central \ph{for the} 
DF3-opt1 and DF3-opt2 designs  \cite{ChBeTh20}. Some of these vdW-DFs emphasize MBPT input on the gradient correction to exchange \cite{lape80,lavo90,thonhauser,cooper10p161104,hamada14,behy14}.

The introduction of variants has 
advantages for illustrating usefulness
but complicates the search for systematic
further progress. The variants (as well as CX) enhanced the range of applications that can easily be addressed with the vdW-DF method (beyond the reach of vdW-DF1 and vdW-DF2), as summarized in
a number of reviews \cite{langrethjpcm2009,rev8,CO2MOF74survey,Berland_2015:van_waals,JPCMreview}
as well as perspectives \cite{beckeperspective,BurkePerspective,Interface_perspective}. However, flexible variants might effectively be compensating for possible $E_{\rm c}^{\rm nl}$ limitations since they fit the choice of $\delta E_{\rm x}^0$ to a target or 
expected representative application \cite{vv10,sabatini12p424209,optx,vdwBEEF}. 
Having too much flexibility can diffuse the underlying drive for seeking 
increasingly more versatile XC functionals: We could inadvertently be 
hiding an actual method limitation. 

Nevertheless, for our overall XC development goals we need to supplement CX and AHCX by a new 
RSH vdW-DF that is more robust towards density-driven errors. Unfortunately, simply creating a RSH vdW-DF right off of vdW-DF2 (from analysis included in
Ref.\ \onlinecite{DefineAHCX})
in a design termed DF2-AH,
does not meet the need. This is made clear in the SI material with observations summarized in the discussion below.

Fortunately, the B86R variant of vdW-DF2 does offers a realistic
path to craft a new general-purpose 
nonempirical RSH vdW-DF, the AHBR. Importantly, as explained below,
the AHBR and B86R also offers a valuable contrast to AHCX and CX when it comes to prioritizing among possible MBPT inputs. That is, the combination
of AHBR and AHCX gives us
an option for a controlled `functional-derivative' or `functional-contrast' analysis:
We can interpret and learn from observations of performance variations
in terms of well-understood design differences.  A similar idea of making a functional-difference analysis 
was also explored for adsorption studies in Refs.\ \onlinecite{behy13,berlandthesis}. Nicely enough, we discover that the resulting AHBR design  
also has a better-than-AHCX resilience towards the density-driven errors in molecular barrier-height problems as well as in
some large-system isomerization problems, Fig.\ \ref{fig:TSwDiel} and Ref.\ \onlinecite{DDerrorQaA}.

In practice, our AHBR development work starts from inspecting the B86R exchange and by providing an AH modeling of
the B86R exchange hole, adapting Ref.\ \onlinecite{HJS08}. This work is an extension of the analysis presented for CX and for vdW-DF2 (and revisited for PBE and PBEsol) in Ref.\ \onlinecite{DefineAHCX}. Like vdW-DF2 and CX, the B86R respects the 
lesson \cite{mulela09} that the asymptotic form of the 
exchange-enhancement factor should rise as $s^{2/5}$ asymptotically
to appropriately counteract errors in the exchange contributions
to (weak) binding \cite{Harris85}. The B86R accomplishes that 
by relying on a revised Becke86b exchange \cite{becke1986p7184,hamada14}. It is fully characterized by the exchange enhancement
\begin{equation}
	F_{\rm x}^{\rm B86R}(s) = 1 + \frac{\mu_{\rm GEA}s^2}{(1+\mu_{\rm GEA}s^2/\kappa)^{4/5}}\, ,
	\label{eq:fxsB86R}
\end{equation}
where $\mu_{\rm GEA}=10/81$ is the small-$s$ expansion coefficient. This low-$s$ form
is aligned with the correct gradient-expansion result from a diagrammatic MBPT analysis \cite{lape79ssc,lape80,PBEsol}, when
interaction lines are interpreted
as bare Coulomb interactions \cite{ankl85,klle88,lavo90,PBEsol}. 
\ph{We note that the use of 
$\kappa=0.7114 < 1$ in Eq.\ (\ref{eq:fxsB86R}) implies a smaller prefactor on the asymptote, 
$F_{\rm X}(s)\sim s^{2/5}$, than what applies 
for CX and OB86 \cite{vdwsolids,behy14}.}

\ph{Since B86R relies on vdW-DF2 correlation, it has only a weaker consistency in that both exchange and correlation energies are set by MBPT inputs that are valid for molecular-type problems, above.} The exchange enhancement of the internal functional is set as $F_{\rm x}(s) = 1 + 0.2097 s^2$, while the expansion of Eq.\ (\ref{eq:fxsB86R}) is given by $\mu_{\rm GEA} = 10/81$. \ph{This 
means that B86R does not have full alignment of the inner-functional and the actual exchange-energy terms, something that CX maintains up to $s=2-3$ by systematically relying on diagrammatic MBPT (assuming screened interaction lines) \cite{lavo90,thonhauser,JPCMreview}.} 
However, we do find that the use of $\mu_{\rm GEA}=10/81$ \ph{and $\kappa < 1$} in Eq.\ (\ref{eq:fxsB86R}) brings the B86R \cite{hamada14} \ph{exchange enhancement values, $F_{\rm X}(2<s<10)$}, closer to PBEsol exchange \cite{PBEsol}, without giving up the asymptotic $F_{\rm X}(s)\sim s^{2/5}$ behavior that is also necessary \cite{mulela09}. We observe that PBEsol reflects a more modern approach to set exchange by enforcing XC hole conservation \cite{pebuer96,pebuwa96,PBEsol} than the rPW86 \cite{pewa86,mulela09} \ph{(that is, large-$s$)} part of CX.

In summary, switching between CX-AHCX and B86R-AHBR means using different 
assumptions when setting the exchange impact on both the plasmon 
modeling \cite{Dion,thonhauser,lee10p081101} and on the actual XC balance. However, the switching
is still done while staying
within the same framework of MBPT analysis \cite{lape80,schwinger,ankl85,lavo87,lavo90,pebuer96,PBEsol,thonhauser,hybesc14,hamada14,JPCMreview}. There are arguments for and against CX/AHCX and B86R/AHBR 
(just as there are for PBE and PBEsol in the GGA framework).
We shall here employ broad testing to assert which priority brings the greater benefits within the present
range of vdW-DF design ideas.

Since AHBR is intended as a key
part of our performance-contrast
strategy, we must also secure 
and validate a general-purpose capability in this new nonempirical RSH vdW-DF. Here we benefit from
past investments: The logic of the unscreened-hybrid ``vdW-DF+0'' class \cite{DFcx02017,JiScHy18b}
(that generalizes PBE0 \cite{BuErPe97,PBE0}) leads, for example, to the formulation of the B86R-based DF2-BR0. We included a code option for this in the \textsc{Quantum Espresso} (QE) code suite \cite{QE,Giannozzi17,PaoloElStruct1}, while releasing the CX-based hybrid CX0 and CX0P, and we now report a full GMTKN55 assessment for DF2-BR0; See for example Figs. S.1-2 and Tables S.II-X of the SI material. We discover that there are very few outliers in the DF2-BR0 benchmark results 
\cite{Note3}
%
and that the performance gain of DF2-BR0 is particularly strong for the important transition-state problems. 
The fact that DF2-BR0 delivers well-balanced progress across general types of molecular problems is an additional strong motivation to here complete and launch the screened hybrid AHBR. 

\subsection{Analytical-hole design of AHBR}

The bottom panel of Fig.\ \ref{fig:FxAHplot} compares 
the key exponent form $\mathcal{H}(s)$ that defines the
long-separation shape of the exchange hole that is assumed to be
of a modified Gaussian type \cite{pebuwa96,EP98,HJS08,DefineAHCX}. \ph{In 
short, taking inspiration from the
exchange-hole form,
$n_{\rm x}^{\rm LDA} (\mathbf{r};\mathbf{r'})$, that
is known from LDA \cite{gulu76,adawda,pewa92,EP98},
one expects an exponential suppression,
\begin{equation}
    n_{\rm x}(\mathbf{r};\mathbf{r'})\propto
    \exp\left[-\mathcal{H}(s(\mathbf{r})) (s y)^2\right] \, ,
    \label{eq:calHuse}
\end{equation}
where $y=k_{\rm F}(\mathbf{r})|\mathbf{r}-\mathbf{r'}|$ and where
the Gaussian suppression $\mathcal{H}(s)$ depends
on the local value of the scaled density gradient $s(\mathbf{r})$ \cite{pebuwa96}.
The ideas of the
analytical exchange-hole modeling, 
as well as the logic and details
of Eq.\ (\ref{eq:calHuse}), are
presented and discussed in Refs.\ \onlinecite{EP98,HSE03,HJS08,DefineAHCX}.}

\ph{The details of this Gaussian-suppression 
factor $\mathcal{H}(s)$ must be asserted to 
complete this AH model of a given XC functional; Technical details
for vdW-DFs are discussed in Ref.\ \onlinecite{DefineAHCX}.  The $\mathcal{H}(s)$ 
variation is 
given by a rational function} \cite{HJS08} with parameters fitted 
subject to constraints so as to accurately \ph{describe, for example,} the B86R 
exchange behavior without introducing any spurious variation 
(that cannot be ascribed any physical meaning) \cite{EP98,HSE03,HJS08,DefineAHCX}.
The procedure \ph{for setting the parameters of $\mathcal{H}$,} used previously to discuss and understand the PBE, PBEsol, CX and AHCX exchange,
is here \ph{repeated for B86R.} Table XI of the SI material reports \ph{
$\mathcal{H}(s)$ parametrizations that
underpin the now extended
range of RSHs, including AHBR (that is based on understanding B86R
exchange).}

Figure \ref{fig:NLrshDesignAH} contrasts the dependence of 
the exchange hole on the local electron-density environment 
for PBE, PBEsol, B86R exchange.
The panels show spatial variations of the exchange holes 
that result in the AH exchange modeling at a set of 
increasing values for the scaled density gradient $s$. \ph{The exchange hole
$n_{\rm x}$ (of a given XC functional) is 
represented by its so-called
dimensionless form $J(s(\mathbf{r}); \mathbf{r},\mathbf{r'})$, defined by
\begin{equation}
n_{\rm x}(\mathbf{r},\mathbf{r'}) = 
n(\mathbf{r}) \times J(s(\mathbf{r}); \mathbf{r'}-\mathbf{r})\, .
\label{eq:XholeJdef}
\end{equation}
In the GGA-exchange model framework that we work with \cite{pebuwa96,HSE03,HJS08,DefineAHCX}, the
density suppression (by exchange effect) induced
at position $\mathbf{r'}$ by an electron at $\mathbf{r}$, can be completely expressed
in terms of a locally scaled distance
$y=k_{\rm F}(\mathbf{r})|\mathbf{r'}-\mathbf{r}|$
defined by the Fermi wavevector $k_{\rm F}$,
\cite{DefineAHCX}. As indicated in Eq.\ (\ref{eq:XholeJdef}), the shape of 
$J$ depends on the local value of the scaled density gradient $s(\mathbf{r})$. 
However, the entire $J(s,y)$ variation is set by finding  parameterizations of the 
Gaussian-damping functions $\mathcal{H}(s)$, functions, discussed above and 
plotted in the lower panels of Fig.\ \ref{fig:FxAHplot}. Thus, by tracking
the shape of $J(s,y)$, panels of Fig.\ \ref{fig:NLrshDesignAH}, we summarize the
full detail of the exchange modeling \cite{DefineAHCX}. The
actual exchange-hole modeling (for 
PBE, PBEsol, and B86R, respectively) at any
given position $\mathbf{r}$ is revealed
by simply inserting the relevant local values
for $s(\mathbf{r})$ and $k_{\rm F}(\mathbf{r})$.}

\ph{The right panel of Fig.\ \ref{fig:NLrshDesignAH}
represents our new AH modeling for exchange effects in B86R;
It is fully summarized in the $J^{\rm B86R}$ variation that is, in turn, sufficient to both recoup the B86R-exchange term $E_{\rm x}^{\rm B86R}(s)$ and  set AHBR, below, adapting
Refs.\ \onlinecite{HJS08,DefineAHCX}.} We find that, initially (at small $s$ values,) the B86R exchange hole follows the PBEsol-exchange nature but does gradually 
roll over to a more PBE-exchange type behavior; It also 
eventually approaches a CX-like behavior at 
large $s$ values where it respects the lessons of 
the analysis in Ref.\ \onlinecite{mulela09}. Interestingly,
the B86R exchange does, for $s \lesssim 3$, perform better than 
(CX and) PBE exchange in terms of 
avoiding the formation of deep exchange holes: The modelling of the 
B86R exchange hole means that the suppression remains smaller than 
the local value of the electron density \cite{pebuer96,pebuwa96,PBEsol} (at $s \lesssim 3$).

We compare this AH analysis for the B86R exchange hole variation
also with that for CX and AHCX \cite{DefineAHCX}, using the 
lower panel of Fig.\ \ref{fig:FxAHplot} and Fig.\ 1 of
Ref.\ \onlinecite{DefineAHCX}. First, it is clear that setting
the exchange enhancement by $\mu_{\rm GEA}$ brings the B86R 
closer to PBEsol exchange than the CX exchange design.  Since 
the B86R exchange is still constrained by the input from
Ref.\ \onlinecite{mulela09}, the large-$s$ behavior rolls
over towards that of CX.
Moreover, being an intermediate of the PBEsol and of the
CX exchange-hole modeling (Fig.\ 1 of Ref.\ \onlinecite{DefineAHCX},)
the B86R has an mid-$s$-range behavior (around $s\approx 3-4$) 
that is close to the PBE hole form. This is again a more trusted 
behavior than that for rPW86 (that
enters in CX).

From the AH analysis of B86R, Figs.\ \ref{fig:FxAHplot}
and \ref{fig:NLrshDesignAH}, \ph{we complete the RSH nonlocal-correlation functional AHBR, following the same steps as previously described for AHCX \cite{DefineAHCX}.
The key observation is that our knowledge of $J^{\rm B86R}$ variation allow us project out the short-range (SR) exchange-energy component 
$E_{\rm x}^{\rm B86R,SR}[n; \gamma]$ from the exchange term $E_{\rm x}^{\rm B86R}$ of B86R. The projection $E_{\rm x}^{\rm B86R,SR}[n; \gamma]$ is again a density functional. As indicated, however, it also depends on the inverse length scale $\gamma$ that we assume in the RSH design \cite{HSE03,HSE06,DefineAHCX} for the screening in the Fock-exchange term \cite{HJS08,OTRSHalga}
\begin{equation}
    E_{\rm FX}^{\rm SR}(\gamma) = 
    \frac{1}{2}
    \int_{\mathbf{r}}
    \int_{\mathbf{r'}}
    \, 
    \frac{n(\mathbf{r}) n_{\rm x}^{\rm Fo}(\mathbf{r'};
    \mathbf{r})}{|\mathbf{r}-\mathbf{r'}|}\, \hbox{erfc}(\gamma 
    |\mathbf{r'}-\mathbf{r}|) \, ,
\end{equation}
where `erfc' denotes the 
error-function complement.
The overall RSH vdW-DF form is \cite{HSE03,HSE06,DefineAHCX}
\begin{equation}
    E_{\rm xc}^{\rm AHBR}[n]=
    E_{\rm xc}^{\rm B86R}[n]+
    \alpha (E_{\rm FX}^{\rm SR}(\gamma)
    - E_{\rm x}^{\rm B86R,SR}[n;\gamma]) \, ,
    \label{eq:AHBRsetDetails}
\end{equation}
where $\alpha$ denotes the 
extent that we mix in the screened Fock exchange energy $E_{\rm FX}^{\rm SR}(\gamma)$.} 

\ph{A robust and computationally efficient determination of $E_{\rm x}^{\rm SR,B86R}[n; \gamma]$ is a key benefit of working with the AH modeling \cite{HJS08,DefineAHCX}. To complete the RSH vdW-DF construction, Eq.\ (\ref{eq:AHBRsetDetails}), we need $E_{\rm x}^{\rm SR,B86R}[n; \gamma]$. It is given in analogy to Eq.\ (\ref{eq:FxDefRegular}) but set by a modified exchange-enhancement factor 
$F_{\rm x}^{\rm SR}(k_{\rm F}(\mathbf{r}),s(\mathbf{r}))$. Thanks to the AH modeling of the B86R exchange energy, above, we can complete an analytical evaluation of the
formal expression
\begin{equation}
    F_{\rm x}^{\rm B86R,SR}(k_{\rm F},s) = 
    - \frac{8}{9} \int_0^{\infty} yJ^{\rm B86R}(s,y)
    \hbox{\rm erfc}(\gamma y/k_{\rm F}) dy \, .
    \label{eq:FormalFxSR}
\end{equation}
In fact, we get all exchange details of the new AHBR from the corresponding AHCX details, Ref.\ \onlinecite{DefineAHCX}, by simply switching from the CX- to the B86R-specific parametrization of the AH modeling, SI Table S.XI.
}

\ph{Both RSH vdW-DFs, the new AHBR and the AHCX,} can be 
used when screening of the Fock-exchange
is essential, for example, for descriptions 
of adsorption at metal and high-dielectric-constant surfaces \cite{Interface_perspective,catalysis1,DefineAHCX}.

\begin{figure}
	\includegraphics[width=0.95\columnwidth]{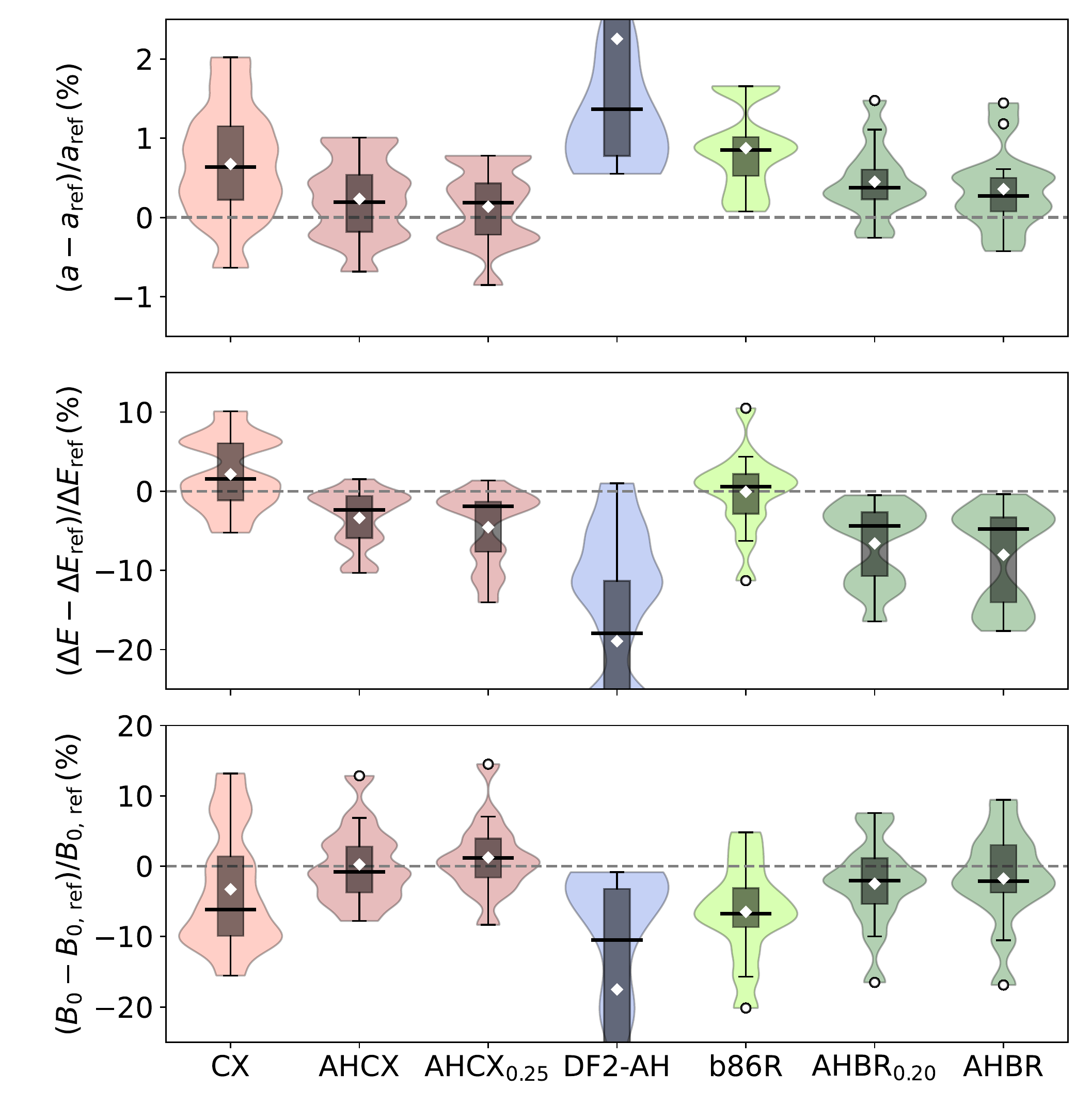}
	\caption{Bulk-system performance as asserted in percentage deviations
	for the CX-AHCX and B86R-AHBR descriptions of lattice constants
	$a$, bulk cohesive energies $E_{\rm coh}$, and bulk moduli $B_0$. We also show the impact of 
	setting the Fock-exchange mixing; The default AHCX and AHBR mixing value is set at 0.20 and 0.25,
	respectively, while a subscript identifies an adjustment. \ph{For completeness, we furthermore include
	a performance assessment for DF2-AH, the RSH form of vdW-DF2 that is implicitly defined by
	analysis in Ref.\ \onlinecite{DefineAHCX}.}
	We compare our results (listed in the SI), computed in the Born-Oppenheimer approximation, 
	with experimental values (back-corrected for vibrational effects) \cite{Tran19}. 
	The violin plots summarize result-statistics data for 13 solids (5 transition metals, 1 simple metal, 
	4 semiconductors, and 3 ionic insulators), with the CX and AHCX results repeated from
	Ref.\ \onlinecite{DefineAHCX}. The set of horizontal bars (of diamonds) reflects the mean 
	(median) deviation in the distributions while the boxes identify the so-called interquartile 
	range, see text. Outliers (identified in the text) are shown by open circles.
	\label{fig:BulkViolin}
	}
\end{figure}

The AHBR is deliberately kept free of fitted parameters.
The extent and nature of the screened-Fock-exchange inclusion could be adjusted but should then be given by physic inputs. Implementation of a formal thermodynamics criterion, in effect that the functional exhibits a piecewise-linearity with the addition of a fractional electron \cite{PePaLe82,GKSstart,KraKro13}, can set the value of 
the inverse screening length $\gamma$ \cite{OTRSHalga,OTRSHgap,OTRSHadsorb17,WiOhHa21}.
Similarly, the extent of Fock-exchange mixing $\alpha$ 
can be set by a coupling-constant analysis \cite{Burke97,JiScHy18a,JiScHy18b} or by demanding 
that the resulting dielectric constant is consistent 
with that implied in the Coulomb range separation \cite{SkGoGa14,SkGoGa16,MiChRe18,BiRePa19,BiWiCh19}.
In this AHBR launching work, we will only 
illustrate and contrast generic RSH vdW-DF usage (that is, 
at fixed, HSE-standard, values for $\alpha$ and $\gamma$).

\section{Results and discussion}

The DFT-result data is obtained using our in-house version of the QE code suite. This code also has generic subroutines for calling RSH vdW-DFs \cite{DefineAHCX} (already released to QE-7.0) and
it benefits from the adaptively compressed exchange
evaluation of Fock exchange \cite{LinACE,PaoloElStruct1,DefineAHCX}.
Appendices A and B provide an overview of the computational details that we use in general demonstrations and for the GMTKN55 assessment work, partly summarized in
Fig.\ \ref{fig:TSwDiel} already.

\begin{table*}
        \caption{\label{tab:TMlat} Comparison of vdW-DF performance on Pt- and noble-metal structure and
	elastic response: Lattice constants $a$ (in {\AA}), cohesive energies $E_{\rm coh}$ (in eV), and 
	bulk moduli $B_0$ (in GPa). The SI material gives full listings for our computed $a$,
	$E_{\rm coh}$, and $B_0$ results
	for these transition metals and 9 other materials. Experimental values, back-corrected for vibrational effects \ph{(as indicated by an asterisk),} are taken from Ref.\ \onlinecite{Tran19}.
        }
\begin{ruledtabular}
\begin{tabular}{r|ccc|ccc|c}
        & CX & AHCX & AHCX$_{0.25}$ & B86R & AHBR$_{0.20}$ & AHBR & Exper.$^*$ \\
 \hline
Cu $a$ & $3.576$  & $3.587$    & $3.592$       & $3.602$             & $3.613$           & $3.617$   & $3.599$ \\
\ph{$E_{\rm coh}$} & $3.781$  & $3.348$    & $3.264$       & $3.582$      & $3.160$           & $3.064$   & $3.513$ \\
$B_0$ & $163$  & $148$    & $146$       & $151$             & $141$  & $136$   & $144$       \\
\hline
Ag $a$ & $4.065$  & $4.078$    & $4.082$       & $4.104$             & $4.115$           & $4.118$   & $4.070$ \\
\ph{$E_{\rm coh}$} & $2.955$  & $2.774$    & $2.737$       & $2.779$      & $2.592$           & $2.549$   & $2.964$ \\
$B_0$ & $115$  & $105$    & $104$       & $102$             & $95$            & $95$    & $106$       \\
\hline
Au $a$ & $4.101$  & $4.098$    & $4.097$       & $4.134$             & $4.127$           & $4.126$   & $4.067$ \\
\ph{$E_{\rm coh}$} & $3.634$  & $3.440$    & $3.398$       & $3.402$      & $3.205$           & $3.158$   & $3.835$ \\
$B_0$ & $171$  & $168$    & $167$       & $153$             & $152$           & $151$      & $182$    \\
\hline
Pt $a$ & $3.929$  & $3.910$    & $3.906$       & $3.952$             & $3.929$           & $3.925$   & $3.917$ \\
\ph{$E_{\rm coh}$} & $6.226$  & $5.524$    & $5.259$       & $5.999$      & $5.131$           & $4.930$   & $5.866$ \\
$B_0$ & $284$  & $298$    & $298$       & $264$             & $278$           & $279$   & $286$       \\
\end{tabular}
\end{ruledtabular}
\end{table*}

The CX and hence AHCX emphasis on screened LV exchange \cite{lape79ssc,lavo87,lavo90} means that they are 
naturally set up for accuracy in metal systems \cite{Gharaee2017,JPCMreview,Hedin80,GunMedSch94}
and, we expect, broadly for bulk and many surface problems, including 
adsorption \cite{DefineAHCX,Hard2Soft}. The new RSH vdW-DF has an
advantage in being a general-purpose choice for molecular properties,
Fig.\ \ref{fig:TSwDiel}. Here we assert and discuss whether the 
new AHBR will remain an option also for bulk and adsorption, and 
whether it also works in a biochemistry and green-technology context. 
Our full-GMTKN55 assessment, see SI material, is part the AHBR documentation and we extract a number of observations also from that mapping.

We note that the move to hybrids, including the RSH vdW-DFs \cite{DefineAHCX}, can help in counteracting excessive charge transfer and hence some density 
driven errors \cite{BurkeSIE,DDerrorQaA}. This is true in the raw, fixed-parameters form presented above and because the AHCX/AHBR
come with an option for $\gamma$ tuning so as to also impose the thermodynamics (fractional-electron) constraint on the designs \cite{PePaLe82,WiOhHa21}.
We expect that such tuned-AHBR usage
will help further on controlling 
(density-driven) errors. However, we have not used that
potential for gaining additional XC-functional consistency in this first assessment.

\subsection{Bulk-structure performance}

Figure \ref{fig:BulkViolin} documents a robust bulk-system performance of the new RSH AHBR.  
More broadly, Fig.\ \ref{fig:BulkViolin} contrasts the performance for bulk of a new, second tool chain (comprising AHBR-B86R) with that of the first 
(AHCX/CX) \cite{Hard2Soft} (and of DF2-AH). 
We do not report data for the unscreened 
hybrid components (DF2-BR0/CX0/CX0P) as we also consider metals \cite{DefineAHCX}.

\begin{table*}
        \caption{\label{tab:Layered} \ph{Comparison of vdW-DF performance on layered materials: Layer-binding $E_{\rm bind}$ (in meV/atom) and optimal layer separation $d_{\rm opt}$ (in {\AA}) for graphite (sp$^2$-hybridized carbon) as well as graphene and $\alpha$-Graphyne bilayers (with sp-sp$^2$-hybridized carbon), hexagonal boron nitride (hBN) and phosphorus; See DMC references \cite{Spanu09p196401,GaKiPa2014,MoDrFa2015,ShKiLe2017,HsChCh2014,ShBaZh2015}
        for summaries of in-plane atomic configurations. Stacking labels AA and AB (Ab) identify geometries with carbon layers in full alignment and displaced one-third (one-ninth) of the sum of in-plane lattice constants, respectively. 
        Stacking label AA' (for hBN) identifies in-plane alignment of 
        boron atoms in one layer with nitrogen atoms in the other layer.}
        }
\begin{ruledtabular}
\begin{tabular}{lcc|ccc|cccc|cc}
        & Stacking & Benchmark & vdW-DF & CX & AHCX & vdW-DF2 & B86R & AHBR$_{0.20}$ & AHBR & RPA & DMC \\
 \hline
Graphite & AB-crystal 
& $E_{\rm bind}$ & $55$ & $67$ & $72$ 
            & $54$ & $62$ 
            & $63$       
            & $64$ 
            & $48^a$/$62^b$ & $60\pm5^c$ \\  
& & $d_{\rm opt}$ & $3.57$ & $3.26$ & $3.27$ 
            & $3.51$ & $3.30$ 
            & $3.31$       
            & $3.31$ 
            & $3.34^a/3.34^b$ & $3.43\pm0.04^d$ \\ 
Graphite & AA-crystal 
& $E_{\rm bind}$ & $50$ &$54$ & $52$ 
            & $47$ & $49$ 
            & $44$       
            & $43$ 
            & - & $36\pm1^e$ \\
& & $d_{\rm opt}$ & $3.73$ & $3.55$ & $3.55$ 
            & $3.67$ & $3.54$ 
            & $3.56$   
            & $3.57$ 
            & - & $3.63^e$ \\  
\hline
Graphene & AB-bilayer 
& $E_{\rm bind}$ & $25$ & $30$ & $33$ 
            & $25$ & $28$ 
            & - & $29$ 
            & $46^f$ & $18\pm1^g$ \\
&  & $d_{\rm opt}$ & $3.61$ & $3.29$ & $3.29$ 
            & $3.53$ & $3.33$ 
            & - & $3.33$ 
            & $3.39^f$ & $3.384^g$ \\
Graphene & AA-bilayer 
& $E_{\rm bind}$ & $22$ & $24$ & $26$ 
            & $21$ & $22$ 
            & - & $23$ 
            & - & $12\pm 1^g$ \\ 
& & $d_{\rm opt}$ & $3.76$ & $3.58$ & $3.57$ 
            & $3.69$ & $3.57$ 
            & - & $3.58$ 
            & - & $3.495^g$ \\  
\hline
$\alpha$-Graphyne & AB-bilayer 
& $E_{\rm bind}$ & $20$ & $20$ & $23$ 
            & $18$ & $17$ 
            & - & $18$ 
            & - & $23^h$ \\
& & $d_{\rm opt}$ & $3.47$ & $3.30$ & $3.26$ 
            & $3.36$ & $3.26$ 
            & - & $3.25$ 
            & - & $3.24^h$ \\
$\alpha$-Graphyne & Ab-bilayer 
& $E_{\rm bind}$ & $19$ & $19$ & $21$ 
            & $18$ & $16$ 
            & - & $17$ 
            & - & $22^h$ \\ 
& & $d_{\rm opt}$ & $3.65$ & $3.49$ & $3.43$ 
            & $3.52$ & $3.42$ 
            & - & $3.41$ 
            & - & $3.43^h$ \\ 
\hline
hBN & AA'-bilayer & $E_{\rm bind}$ & $24$ & $29$ & $32$ 
            & $24$ & $26$ 
            & - & $28$ 
            & $19^f$ & $20 (18)^i$ \\ 
 & & $d_{\rm opt}$ & $3.58$ & $3.26$ & $3.24$ 
            & $3.51$ & $3.31$ 
            & - & $3.28$ 
            & $3.34^f$ & $3.25/3.50^i$ \\  
\hline
Phosphorus  & AB-crystal & $E_{\rm bind}$ 
& $79$ & $127$ & $124$ 
            & $83$ & $117$ 
            & - & $112$ 
            & - & $81\pm 6^j$ \\ 
 & & $d_{\rm opt}$ 
 & $5.69$ & $5.19$ &             $5.26$ 
            & $5.67$ & $5.27$ 
            & - & $5.33$ 
            & - & $5.2^j$ \\ 
\hline
\multicolumn{12}{l}{$^a$ Ref.\ \onlinecite{rpa_graphite}.}\\
\multicolumn{12}{l}{$^b$ Ref.\ \onlinecite{OlTh2013}.}\\
\multicolumn{12}{l}{$^c$ Ref.\ \onlinecite{Spanu09p196401}; We report the raw DMC $E_{\rm bind}$ value (omitting an estimate for vibrational effects) for a relevant comparison.}\\
\multicolumn{12}{l}{$^d$ Ref.\ \onlinecite{Spanu09p196401}; The
authors warn that (in-plane-size) convergence at large layer separations is not sufficient to accurately fit 
$d_{\rm opt}$.}\\
\multicolumn{12}{l}{$^e$ Ref.\ \onlinecite{GaKiPa2014}.}\\
\multicolumn{12}{l}{$^f$ Ref.\ \onlinecite{ZhHaDa2015}.}\\
\multicolumn{12}{l}{$^g$ Ref.\ \onlinecite{MoDrFa2015}.}\\
\multicolumn{12}{l}{$^h$ Ref.\ \onlinecite{ShKiLe2017}.}\\
\multicolumn{12}{l}{$^i$ Ref.\ \onlinecite{HsChCh2014}; $E_{\rm bind}$ value without correction
for infinite-layer extension
is presented in paranthesis. 
No fit for $d_{\rm opt}$ given.}\\
\multicolumn{12}{l}{$^j$ Ref.\ \onlinecite{ShBaZh2015}; The
$d_{\rm opt}$ is extracted
from Fig.\ 1 of that reference.}
\end{tabular}
\end{ruledtabular}
\end{table*}

The violin plots summarize deviations in percentage
of computed results for lattice constants $a$, cohesive energies 
$E_{\rm coh}$, and bulk moduli $B_0$ from back-corrected 
experimental values for five transition metals (Cu, Ag, Au, Pt, Rh), one simple metal (Al), four semiconductors (Si, C, SiC, GaAs)
and three ionic insulators (LiF, MgO, NaCl) as 
in Ref.\ \onlinecite{DefineAHCX}. 
Tables S XIII, S XIV, and
S XV of the SI present a more complete quantitative 
presentation, contrasting values computed in the two
tool chains with \ph{experimental values (that are back corrected for vibrational effects)}
and with those we obtain for the RSH DF2-AH \cite{DefineAHCX}.
The subscript on one AHBR-data (and on one AHCX-data) label
identifies the extent of Fock-exchange mixing; A corresponding
specification is suppressed for `AHBR=AHBR$_{0.25}$'
(`AHCX=AHCX$_{0.20}$') since this mixing reflects a recommended
default, as explained in the following subsection.

To contrast functional performance on bulk, we compare the position
of the mean (median) deviation, shown by a central bar (diamond)
and the so-called interquartile range, shown as a bar. This bar reflects the
difference of positions for the first and third quartile of the
performance distribution (for each functional). We also consider the
presence or absence of outliers (open circles), by which we mean a
performance that lies beyond markers (wiskers) that identify 
1.5 times the interquartile range. Lattice-constant outliers
are Au (Au and Ag) for AHBR$_{0.20}$ (AHBR) while cohesive-energy
outliers are Au and Rh for B86R. There are more outliers for
the bulk modulus: Rh for AHCX/AHCX$_{0.25}$, GaAs for B86R, and
Au for AHBR and AHBR$_{0.20}$. 

Figure \ref{fig:BulkViolin} shows that there is a systematic AHBR 
improvement for lattice constants and bulk moduli compared to B86R, at both 
0.20 and 0.25 Fock-exchange mixings. However, the B86R is more accurate 
than AHBR for predictions of the bulk cohesive energies.

Figure \ref{fig:BulkViolin} shows that the
RSH AHCX is overall a better performer
for bulk than AHBR$_{0.20/0.25}$. This is consistent with 
findings that CX has a small bulk-performance edge over B86R \cite{Gharaee2017,Tran19,DefineAHCX}.  This observation holds, for example,
for the lattice constant. However, AHBR is accurate on structure and hence useful also for substrate 
descriptions in heterogeneous 
system. For example, the lattice-constant accuracy 
on noble-metal and Pt metals remains within 0.5\% deviations relative to
(back-corrected) experimental values, Table \ref{tab:TMlat}.

As an interesting aside, we document that RSH DF2-AH
(implicitly defined in Ref.\ \onlinecite{DefineAHCX})
is not well suited for bulk-system use. This is
not surprizing since there are real issues with using vdW-DF2 for bulk 
systems \cite{vdwsolids}. Tables S XIII,
S XIV, and S XV of the SI document
that there are large deviations between DF2-AH (and vdW-DF2)
predictions and back-corrected experimental values for all of the
investigated bulk properties. In the violin-plot 
Fig.\ \ref{fig:BulkViolin} we do not even depict the
full extent of the interquatile range which for DF2-AH
is set by first/third-quartile relative deviations 
(0.78\% and 3.5\% for $a$, 11\% and 29\% for $E_{\rm coh}$
and -3\% and -34\% for $B_0$). Unlike AHBR, the DF2-AH is
simply not a reliable option for bulk and hence for 
substrates in adsorption studies.

We ascribe the absence of vdW-DF2 and DF2-AH robustness 
(for descriptions across general molecular properties and bulk)
to the fact that their designs depart from some of the 
ideas that emerged in the electron-gas tradition \cite{mabr,lu67,helujpc1971,lape80,pewa86,lavo87,pebuwa96,PBEsol,thonhauser,Berland_2015:van_waals,JPCMreview}. Use of diagrammatic-MBPT input, as
summarized in Refs.\ \onlinecite{lape79ssc,lape80,lavo90},
are prioritized in the design of PBEsol (unscreened) exchange \cite{PBEsol} 
and in CX, with its emphasis on LV screened exchange \cite{lavo90,thonhauser,behy14}.
However, the design of (r)PW86 \cite{pewa86,mulela09}, that defines
vdW-DF2 and DF2-AH exchange, did not prioritize this MBPT input to the
same extent. 

Finally, Fig.\ \ref{fig:BulkViolin} shows that the AHBR and AHCX
bulk-structure characterizations are relatively insensitive to 
the choice of the Fock-exchange mixing. The AHCX design
is set with a default 0.20 Fock exchange mixing \cite{DefineAHCX}
but the overall bulk performance does improve slightly 
by going from AHCX=AHCX$_{0.20}$ to AHCX$_{0.25}$.

\subsection{\ph{Binding in layered matter}} 

\ph{We assess the RSH vdW-DF performance on layered systems using diffusion Monte Carlo (DMC) results \cite{Spanu09p196401,GaKiPa2014,ShKiLe2017,HsChCh2014,ShBaZh2015} as reference data. We also discuss random-phase approximation
(RPA) literature results \cite{rpa_graphite,OlTh2013,ZhHaDa2015} and recent measurements of the cleavage energy in graphite \cite{GrMeas1,GrMeas2}.
The graphite DMC study \cite{Spanu09p196401} came
with both a raw DMC result and
with an estimate for the expected correction (4 meV/atom) for vibrational effects in graphite. Our computed results (as well as the RPA literature results) are obtained in the Born-Oppenheimer approximation and should be compared with the raw DMC values listed in Table
\ref{tab:Layered}.}

\ph{For our RSH and nonhybrid RSH characterizations, we used a 2-step approach, seeking to
compare all functionals 
at a fixed in-plane structure
that is close to experiments. First, we determine the in-plane lattice constants using a set of CX calculations (as CX is strong on structure \cite{JPCMreview,Hard2Soft}). 
This was generally done by variable-cell calculations. For the phosphorus crystal, 
the in-plane $a_1$ lattice
constant is soft \cite{ShBaZh2015} (but the other in-plane lattice constant $a_2$ is still set by covalent bonds); There we kept $a_1$ fixed at the experimental (bulk) value (4.374 {\AA}) while we used CX calculations to determine $a_2$. Further computational details are described in Appendix A.}

\ph{Second, at fixed in-plane lattice vectors, we compute the total unit-cell energy $E(d)$
as a function of layer separation $d$. For 
the RSH and regular vdW-DFs, we thus determine 
the optimal separation value $d_{\rm opt}$ and
the asymptotic system value $E_{\rm asymp}$ (taken
as the system energy at $d=20$ {\AA}). We extract 
the layer binding energy 
\begin{equation}
    E_{\rm bind} = 
    \left[E(d_{\rm opt}) - E_{\rm asymp}\right]/N \,,
    \label{eq:EbindDMCdef}
\end{equation}
where $N$ denotes the number of atoms in the unit cell.} 

\ph{We note that the typical DMC binding-energy 
definition, Eq.\ (\ref{eq:EbindDMCdef}), 
differs from the binding-energy definition (meV-per-layer-atom) that is used for the 
discussion of early bilayer
vdW-DF studies \cite{rydberg03p126402,rydberg03p606,chakarova-kack06p146107}. This has led to confusion in vdW-DF presentations, for example, in Ref.\ \onlinecite{MoDrFa2015}. We include new vdW-DF1 and vdW-DF2 characterization, now using the definition Eq.\ (\ref{eq:EbindDMCdef}) for comparisons.
}

\ph{Table \ref{tab:Layered} reports our 
comparison of CX/AHCX and B86R/AHBR 
performance relative to the DMC reference data. In graphite, the CX and AHCX functionals 
give $E_{\rm bind}$ results 
that are larger than the DMC reference value, even allowing for error
bars reported in Ref.\ \onlinecite{Spanu09p196401}.
However, the B86R and AHBR results for $E_{\rm bind}$,
at 62-64 meV/atom, are in excellent agreement with the relevant 
(that is, the raw) DMC reference value,
$60\pm5$ meV/atom for the regular (AB-stacked) 
graphite crystal \cite{Spanu09p196401}.}

\ph{We observe that the B86R and AHBR results for the optimal layer separation,
at $d_{\rm opt}=3.30-3.31$ {\AA}, are smaller that the reported DMC value, $3.43\pm 0.04$ {\AA} and more in line
with the experimental characterization \cite{ZhSp1989}, at $d_{\rm opt}= 3.35$ {\AA}. 
The graphite-AB DMC study has a lower convergence (with regards to in-plane extension) results for large layer
separation and the authors warn that it impacts the $d_{\rm opt}$ fit \cite{Spanu09p196401}.} 

\ph{At the same time, there are 
good reasons to trust the 
graphite-AB DMC result of Ref.\ \onlinecite{Spanu09p196401} for $E_{\rm bind}$.  The trust comes from recent high-precision measurements of the graphite cleavage energy, giving
binding-energy results of 54 and 55
meV/atom \cite{GrMeas1,GrMeas2}. Including
the estimate of a 4 meV/atom vibrational
correction, there is full alignment with 
the graphite-AB DMC $E_{\rm bind}=60\pm 5$
meV/atom result \cite{Spanu09p196401}. In turn, given the trust in the DMC result, we conclude that AHBR is highly accurate for graphite-AB binding, Table \ref{tab:Layered}.}

\ph{In fact, for the
regular graphite crystal,
CX, B86R, and AHBR provide an 
$E_{\rm bind}$ description 
that is closer to DMC than 
one of the RPA results 
(at $E_{\rm bind}^{\rm RPA}=48$
mev/atom) \cite{rpa_graphite}.
A second RPA study \cite{OlTh2013}
(at $E_{\rm bind}^{\rm RPA}=62$
mev/atom) is closer to the DMC 
result and fully agrees with our 
B86R and AHBR descriptions. However, the authors of that second-RPA study warn that there can be an impact of 
RPA convergence \cite{OlTh2013}.
}

\ph{Considering next the meta-stable graphite
configuration with an AA-stacking \cite{GaKiPa2014},
we find that CX/AHCX and B86R all make larger
errors. However, we also find that AHBR, through its inclusion of a screened Fock-exchange component, is
able to correct much of the B86R overestimations.
The AHBR result lands at about a 7-8 meV/atom deviation from the DMC reference data. 
The AA configuration may be seen as the barrier of an in-plane slip process of graphite (generalizing a picture that applies for polymers \cite{Olsson17,OlHySc18}). We 
therefore interpret the reasonable robustness of the AHBR characterization for graphite-AA as an another example of 
success at characterizing transition 
states.}

\ph{We also assess the performance for bilayer graphene, in AB and AA stackings, 
and for the closely related hexagonal boron nitride (hBN), 
in AA' stacking, against DMC studies \cite{MoDrFa2015,HsChCh2014}.
These studies are again obtained for systems with
a dense in-plane electron distribution, however,
there is significantly less binding contribution in the unit-cells. Part of the 
reason for that is evident by inspecting Fig.\ 8 of Ref.\ \onlinecite{JPCMreview}, 
noting that the dominant contribution in layered materials arises from the moderately-low electron variation that exists between the layers:
In a bilayer form there is only half as many such regions as in the corresponding bulk.
 The
layer binding is further reduced  because bilayer systems lack the coupling to layers that are further away, for example,
as discussed in Refs.\ \cite{bjorkmannlayered1,bjorkmannlayered2}.}

\ph{For the graphene bilayer system, in both AB and AA stackings, we find
that AHBR differs more (by about 11
meV/atom) from the DMC
results \cite{MoDrFa2015}.
The B86R and AHBR do perform better than CX and AHCX and are still significantly
closer than the one RPA description that
we have found \cite{ZhHaDa2015}.} 

\ph{The AHBR description also differs
from the DMC result for a hBN bilayer 
system;  Here it is instead a RPA result \cite{ZhHaDa2015} that is close to the DMC value \cite{HsChCh2014}. Our AHBR result differs 8 meV/atom from the DMC reference, in line with the status for graphite in AA stacking.}  

\ph{We note in passing that vdW-DF and vdW-DF2 are often closer to the DMC 
values for binding energies
than B86R and AHBR (or CX and AHCX). However,
Table \ref{tab:Layered}  makes it clear
that vdW-DF and vdW-DF2 systematically
overestimate the layer-binding separation.}

\ph{We also consider the AHCX and AHBR performance at $\alpha$-graphyne bilayers, cases in which some of the carbon atoms in each plane are in the sp-hybridized form \cite{ShKiLe2017}. Here, interestingly we find that AHCX  slightly
outperforms the AHBR description on both structure
and binding energies. However, the performance of AHBR, as a new general-purpose RSH vdW-DF, is still good, landing 
within 5 meV/atom of the DMC
reference description. We also note that 
both AHCX and AHBR correctly predict the 
1 meV/atom preference that separates the two
competing motifs (AB and Ab stacking \cite{ShKiLe2017}) for these weakly 
bonded systems.}

\ph{Finally, we assert the AHBR and AHCX performance for phosphorus bulk,
in the stable AB stacking. The DMC reference data \cite{ShBaZh2015} has
$E_{\rm bind}$ and we can extract 
an estimated value also for $d_{\rm opt}$. We find, again, that vdW-DF2 and vdW-DF are closer than AHBR and AHCX for the layer binding energy but not for structure. The AHBR (AHCX) description is 2\% or an estimated 0.13 {\AA} (1\% or 0.06 {\AA}) too large on $d_{\rm opt}$ while it overestimates $E_{\rm bind}$ by 31 meV/atom (43 meV/atom)
relative to the DMC result, $81\pm6$ meV/atom \cite{ShBaZh2015}.}

\subsection{Robust molecular benchmarking and setting the Fock-exchange mixing in AHBR} 

Our use of the planewave-code QE gives us the prerequisites for 
delivering a high-quality (in principle, complete-basis-set) assessment.
We can secure a robust  characterization as long as we also include all relevant electrons in the pseudopotentials (PPs) and compensate or control spurious electrostatic and vdW-type intercell interactions in our periodic-cell calculations

However, there are dramatic SIE effects or density-driven errors, and hence challenges with plane-wave benchmarking, in the study of negatively charged atoms 
and radicals \cite{BurkeSIE,DefineAHCX}. The last electron will not necessarily remain bonded, unless we work with small unit cells that artificially raise the vacuum floor in QE \cite{DefineAHCX}. Because we seek to approach the complete-basis set limit, we cannot provide a 
meaningful direct assessment of performance for the G21EA and WATER27 benchmark
sets 
\cite{Note4}
%
of the GMTKN55 suite. Nevertheless, in appendix B, we document that use of a dielectric-environment extension \cite{NicolaENVchem} permits us to circumvent the SIE challenges and reliably complete general functional assessments. 

Figure \ref{fig:TSwDiel} summarizes performance statistics for the top-performing regular vdW-DFs, including the CX and B86R, and both of the corresponding RSH vdW-DFs, the AHCX and 
the new AHBR. Details of this GMTKN55-based assessment are given in the SI material Tables S.II through S.X;
It covers almost all of vdW-DFs that are coded in the QE version 6.7
\cite{Note5},
the related revised VV10 \cite{vv10,Sabatini2013p041108}, as well as the dispersion-corrected \cite{grimme3} 
revPBE+D3 \cite{ZhYa98} and HSE+D3 \cite{HSE03,HSE06}. For comparison, we also list literature 
MAD results \cite{gmtkn55} for revPBE+D3, 
HSE+D3, SCAN+D3 \cite{SCAN,grimme3}, and B3LYP+D3,
as obtained in orbital-based DFT.

The right panel focuses on the seven sets in the important GMTKN55 group 3 of molecular-barrier benchmarks, reporting mean absolute deviations (MADs) relative to
coupled-cluster quantum chemistry calculations \cite{gmtkn55} (in kcal/mol). The left panel presents the broader 
GMTKN55 performance overview (characterized by 
weighted MAD values, i.e., the WTMAD1 measure introduced in 
Ref.\ \onlinecite{gmtkn55}) and thus also covers the GMTKN55 group 1 and group 2 assessments for small- and large-molecules reactions and transformations, as well as groups 6, 4, and 5 of benchmarks covering total, inter- and intra-molecular NOC interactions.

\begin{table*}
        \caption{\label{tab:DDerr} Comparison of PBE, CX, vdW-DF2, HSE+D3, AHCX and AHBR performance on benchmarks sets that 
        have a pronounced density sensitivity, as asserted by a Hartree-Fock (HF) sensitivity
        measure $\tilde{S}_{\rm avg}> 2$ kcal/mol \cite{DDerrorQaA}.
        We also list, where available, literature values for the performance
        of density-corrected PBE, termed DC(HF)-PBE (using HF
        densities to improve the PBE description when the process or reaction
        is found sensitive \cite{DDerrorQaA,SiSoVu22}). For this density-corrected PBE \cite{SaMa21,SiSoVu22}, we list literature results for the performance with a Grimme-D4 dispersion correction \cite{grimme4}. The table section sorts the benchmark comparison
        acording to their inclusion in GMTKN55 Group 3, 1, and 2, respectively. The density sensitivity of G21EA \cite{BurkeSIE} is discussed in Appendix B. Benchmark results are represented in MAD values (in kcal/mol) asserted relative to the coupled-cluster results that
        define the GMTKN55 reference data.}
\begin{ruledtabular}
\begin{tabular}{lcccccccc}
        & $\tilde{S}_{\rm avg}$ & PBE & CX & vdW-DF2 & DC(HF)-PBE/+D4 & HSE+D3 & AHCX & AHBR \\
 \hline
BH76 & $8.0^a$  & $8.46$ & $9.15$ & $6.90$ & $4.4^a/4.7^a$  &  $4.21$ & $5.15$ & $4.14$ \\
PX13 & $4.3$ & $12.12$ & $12.80$ & $1.14$ & - & $7.38$ & $7.30$ & $4.57$ \\ 
\hline
G2RC & $11.3^a$ & $5.85$  & $6.77$ & $9.43$ & $4.3^a/4.1^a$ & $6.48$ & $4.71$  & $3.30$ \\
G21EA & - & $3.07$  & $2.80$ & $9.66$ &  -  & $3.40$ & $2.17$    & $2.32$ \\
\hline
RSE43 & $3.7^a$ & $2.54$  & $2.21$    & $1.13$  & $2.0^a/1.9^a$ & $1.25$ & $1.01$ 
& $0.74$ \\
C60ISO & $5.1$ & $10.06$  & $12.01$    & $10.43$  & - & $2.51$ & $3.99$ & $2.72$ \\
\hline
$^a$ Ref.\ \onlinecite{DDerrorQaA}.
\end{tabular}
\end{ruledtabular}
\end{table*}

Figure \ref{fig:TSwDiel} shows that the new RSH AHBR
design is robust, i.e., shows resilience towards 
density-driven errors \cite{DDerrorQaA}. This class of DFT
problems affects barrier-height problems that in turn
define the GMTKN55 benchmark group 3. 
Accuracy for transition states, 
and hence for predicting reaction rates, is 
considered a challenge even when the
issue is considered in isolation \cite{BK04,DDerrorQaA}. 
It is exciting that AHBR provides a balanced 
progress, i.e., it works just as well (maybe even better)
for transition-state problems as for molecular-reaction energies. 

For the benchmarking summarized in Fig.\ \ref{fig:TSwDiel}, and generally throughout
the paper, we have deliberately kept the RSH parameters fixed at the 
default HSE-specification for the screening-length parameter \cite{HSE06},
0.106 inverse bohr, assuming also a fixed 0.25 (0.20) fraction for the mixing 
of short-range Fock exchange for AHBR (AHCX). These defaults are used throughout 
the paper, although we also sometimes illustrate the impact of switching between 
the two (standard, 0.20 and 0.25) choices for the extent of Fock-exchange mixing,
as marked by subscripts (for example, AHBR$_{0.20}$).

\begin{table*}
        \caption{\label{tab:DDcrossTest} Comparison of revPBE+D3, CX, vdW-DF2, HSE+D3, AHCX and AHBR performance on benchmarks sets that are density-sensitive,
        here as asserted from comparing self-consistent PBE+D4 and HF-PBE+D4
        performance measures.
        For key benchmark examples, the second column asks `Is the MAD value for HF-PBE+D4 (namely, the key element of DC(HF)-PBE+D4) better than for regular (selfconsistent) PBE+D4?', using  a GMTKN55 survey of density-corrected 
        PBE \cite{SaMa21}.  The answers are practical identifications
        of sensitivity. The third column uses that data \cite{SaMa21} to answer the question `Does PBE0+D4 outperform HF-PBE+D4 for this benchmark set?', i.e., it assesses if an unscreened hybrid can itself be expected to correct for the density sensitivity. The first and bottom sections of the table contrast benchmark performances (listed as MAD values in kcal/mol) for GMTKN55 groups 1-3 and in the inter- \& intra-molecular NOC interaction 
        groups 4 \& 5, respectively.
        }
\begin{ruledtabular}
\begin{tabular}{lccccccccc}
        & Sensitive?$^a$ & PBE0+D4?$^a$ & revPBE+D3 & CX & vdW-DF2  & B86R & HSE+D3 & AHCX & AHBR \\
 \hline
BH76 & yes & unclear & $7.38$ & $9.15$ & $6.90$ & $9.22$  & $4.21$ & $5.15$ & $4.14$ \\
PX13 & yes & yes & $9.29$ & $12.80$ & $1.14$ & $11.36$ & $7.38$ & $7.30$ & $4.57$ \\ 
BHPERI & no & yes & $5.74$ & $7.20$ & $3.08$ & $6.08$ & $2.83$ & $4.05$ & $1.84$ \\
W4-11 & yes & yes  & $5.88$ & $8.55$ & $18.69$ & $6.97$ & $6.77$ & $4.99$  & $9.50$ \\
DC13 & yes & unclear  & $9.38$ & $7.88$ & $24.21$ & $7.26$ & $8.24$ & $8.35$  & $6.18$ \\
SIE4x4 & yes & yes  & $21.67$ & $23.80$ & $21.73$ & $ 23.52$ & $13.58$ & $17.00$  & $15.09$ \\
ISOL24 & yes & yes  & $4.82$ & $2.65$ &  $12.69$  & $3.68$ & $2.42$ & $2.25$ & $1.82$ \\
\hline
PNICO23 & yes & no  & $0.84$ & $0.66$ & $0.39$ & $0.56$ & $0.86$ & $0.44$ & $0.25$ \\
HAL59 & yes & no  & $0.82$ & $0.94$ & $0.69$  & $1.00$ & $0.64$ & $0.63$ & $0.58$ \\
WATER27 & yes & unclear  & $2.63$ & $2.88$ & $1.75$ & $5.10$ & $5.73$ & $2.71$ & $2.52$ \\
Amino20x4 & yes & yes & $0.35$ & $0.25$  & $0.39$ & $0.22$ & $0.29$ & $0.22$ & $0.19$ \\
IDISP & - & - & $3.25$ & $2.27$  & $7.89$ & $2.69$ & $2.96$ & $1.61$ & $1.50$ \\
\hline
$^a$ Ref.\ \onlinecite{SaMa21}.
\end{tabular}
\end{ruledtabular}
\end{table*}

In fact, we have set this default recommendation for the AHBR Fock-exchange mixing,
AHBR=AHBR$_{0.25}$, by directly relying on the broad 
GMTKN55 molecular benchmarking for this second generation
RSH vdW-DF.  For individual problems and benchmarks we could proceed to make a system-specific analysis to establish a plausible
choice of the Fock-exchange mixing and screening, for example,
as pursued in Refs.\ \onlinecite{OTRSHalga,OTRSHgap,SkGoGa14,SkGoGa16,OTRSHadsorb17,MiChRe18,BiRePa19,BiWiCh19,SeKrGe21,WiOhHa21}. 
Here we observe that use of AHBR$_{0.25}$ is systematically more 
accurate than AHBR$_{0.20}$ on molecular properties; We pick 
$\alpha=0.25$ to give an impression of the accuracy that we
can hope to get from AHBR usage.

We also note in passing that the suggested default AHCX 0.20 mixing 
came from a coupling-constant analysis of the contribution 
of CX correlation to the atomization energies \cite{gmtkn55,JiScHy18b}.
The logic of that specification needs not hold for CX0P and AHCX 
when it comes to large systems (or bulk), let alone for AHBR.
Looking at the full survey, in the SI material, 
we find that
moving the AHCX to 0.25 Fock-exchange mixing gives 
a small performance gain both overall and for all but
the NOC-interaction benchmark groups 4 and 5. 
The impact is in any case limited.

\subsection{Navigating density-driven errors}

It is natural to discuss the progress of AHBR as a molecule performer 
in terms of the resilience towards density-driven 
errors \cite{BK04,BurkeSIE,SurfChallenge,patra2019rethinking,DDerrorQaA,SaMa21,SiSoVu22}.
The promise of AHBR success on these key challenges
is implied in Fig.\ \ref{fig:TSwDiel} and here we 
provide details.

Any given XC approximation will cause an incorrect XC-energy, and thus DFT-total-energy evaluation, even if we had access to the exact density. 
However, there are additional challenges because the DFT calculations (based on the specific XC functional approximation) can sometimes lead to an electron 
density solution that is far from the exact density. This extra sensitivity causes performance outliers with 
a dramatically reduced accuracy of the system-specific 
DFT study \cite{DDerrorQaA}.  Important examples are the generic-DFT
failure to correctly  confine the last  electrons in some negatively 
charged ions \cite{BurkeSIE,DefineAHCX}, charge trapping in oxide 
defects \cite{LiLiEr2018} and color centers \cite{HaLiKr2021}, molecular-reaction barrier heights
\cite{DDerrorQaA}, the CO-adsorption site-preference 
challenge on Pt(111) \cite{patra2019rethinking}, 
and, we expect, adsorption-induced dissociation in the presence of large charge transfer \cite{SurfChallenge}.

The BH76 molecular barrier-height benchmark has already been 
discussed as a key challenge for securing robustness, i.e.,
a driver for us to complete the design of the AHBR. It is
also clear that the density-driven errors directly impact
negatively charged ions and radicals \cite{BurkeSIE,DefineAHCX},
and hence the performance on, for example, the G21EA and 
the WATER27 sets in groups 2 and 4, see also Appendix B. 
Ref.\ \onlinecite{DDerrorQaA} additionally highlights 
the G2RC and RSE43 benchmark sets as being prone to 
density-driven errors.

For molecules it is possible to use calculations of the Hartree-Fock (HF) electron-density solution to spot when we 
can expect density-driven errors \cite{DDerrorQaA,SiSoVu22}. For such cases, one would, in general, expect that moving to a hybrid is motivated,  but there are also molecular cases where use of the unscreened 
hybrid PBE0 \cite{BuErPe97,PBE0} cannot by itself 
correct the issue \cite{SaMa21}. 

The HF-based tests for density sensitivity \cite{DDerrorQaA}
\ph{are carried out} using non-selfconsistent DFT-energy calculations.
Our \textsc{ppACF} code contribution 
\cite{JiScHy18a,JiScHy18b,JPCMreview,MOFdobpdc} makes this option 
available in the QE code suite, given some manual adjustments
of the xml file that the regular DFT solver provides. 
\ph{For a given problem, we first pursue 
selfconsistent HF calculations (for all
reactants and products) to 
obtain the density variations, denoted 
$n^{\rm HF}(\mathbf{r})$, in that 
approximation. We also compute 
density variations, denoted $n^{\rm LDA}(\mathbf{r})$, using self-consistent 
LDA \cite{pewa92}. Next, we obtain
so-called post-PBE energies  \cite{SiSoVu22}
(for all reactants and products), denoted
$E^{\rm HF-PBE}$ and $E^{\rm LDA-PBE}$, 
by evaluating the total DFT-PBE energies
on the set of fixed $n^{\rm HF}(\mathbf{r})$ 
and $n^{\rm LDA}(\mathbf{r})$ density 
variations, respectively. Finally, we
compute process energy differences, 
denoted $\Delta E_{\rm proc}^{\rm HF-PBE}$
and $\Delta E_{\rm proc}^{\rm LDA-PBE}$
for the set of $E^{\rm HF-PBE}$ and $E^{\rm LDA-PBE}$ energies, and evaluate the
density-error sensitivity measure \cite{SiSoVu22}
\begin{equation}
    \tilde{S}_{\rm proc} = | \Delta E_{\rm proc}^{\rm HF-PBE} - \Delta E_{\rm proc}^{\rm LDA-PBE}|
    \, .
    \label{eq:SprocDef}
\end{equation}
This measure reveals whether there are fundamental  differences between the HF and KS orbitals 
and hence whether we can expect density errors
to significantly affect any given (GGA or vdW-DF)
DFT characterization \cite{SiSoVu22}.}
For a benchmark set one can also define an average
value $\tilde{S}_{\rm avg}$ over case-specific $\tilde{S}_{\rm proc}$ values  \cite{SiSoVu22}, essentially adapting the ideas of benchmark MAD assessments. If $\tilde{S}_{\rm avg}$ ($\tilde{S}_{\rm proc}$) is asserted as larger than 2 kcal/mol, 
then the benchmark set (process) should be considered as
density sensitive \cite{DDerrorQaA,SiSoVu22}.

Using this assessment procedure, we presently add documentation that the PX13 (barrier-height) and the
C60ISO (large-system isomerization) sets are also prone to density-driven errors, i.e., given by characteristic
sensitive measures $\tilde{S}_{\rm avg}> 2$ kcal/mol. The present $\tilde{S}_{\rm avg}$-based mapping supplements the literature identification of pronounced
sensitivities for BH76, G2RC, and RSE43 \cite{DDerrorQaA}.
In the literature cases the measures are comparable even to 
the PBE MAD values themselves; For C60ISO (and to some extent
also for PX13), we find that the measure represents a significant fraction of the PBE MAD value.

Table \ref{tab:DDerr} contrasts the PBE, CX, vdW-DF2, HSE+D3, AHCX, and AHBR performances, asserted as MAD values, on those density-sensitive benchmarks \cite{DDerrorQaA}. The table 
shows that the AHBR here performs at the same level or better 
as HSE-D3, and systematically better than AHCX.
This latter finding is expected because we have given arguments (in Section II) that the condition that exists for CX \ph{(to fully leverage its Lindhard screening foundation, discussed in Section II)} might make the CX more susceptible to density-driven errors. The first observation is important for encouraging broad vdW-DF method testing: The AHBR has a good general-purpose 
capability for recognized challenges.

\begin{figure}
	\includegraphics[width=0.95\columnwidth]{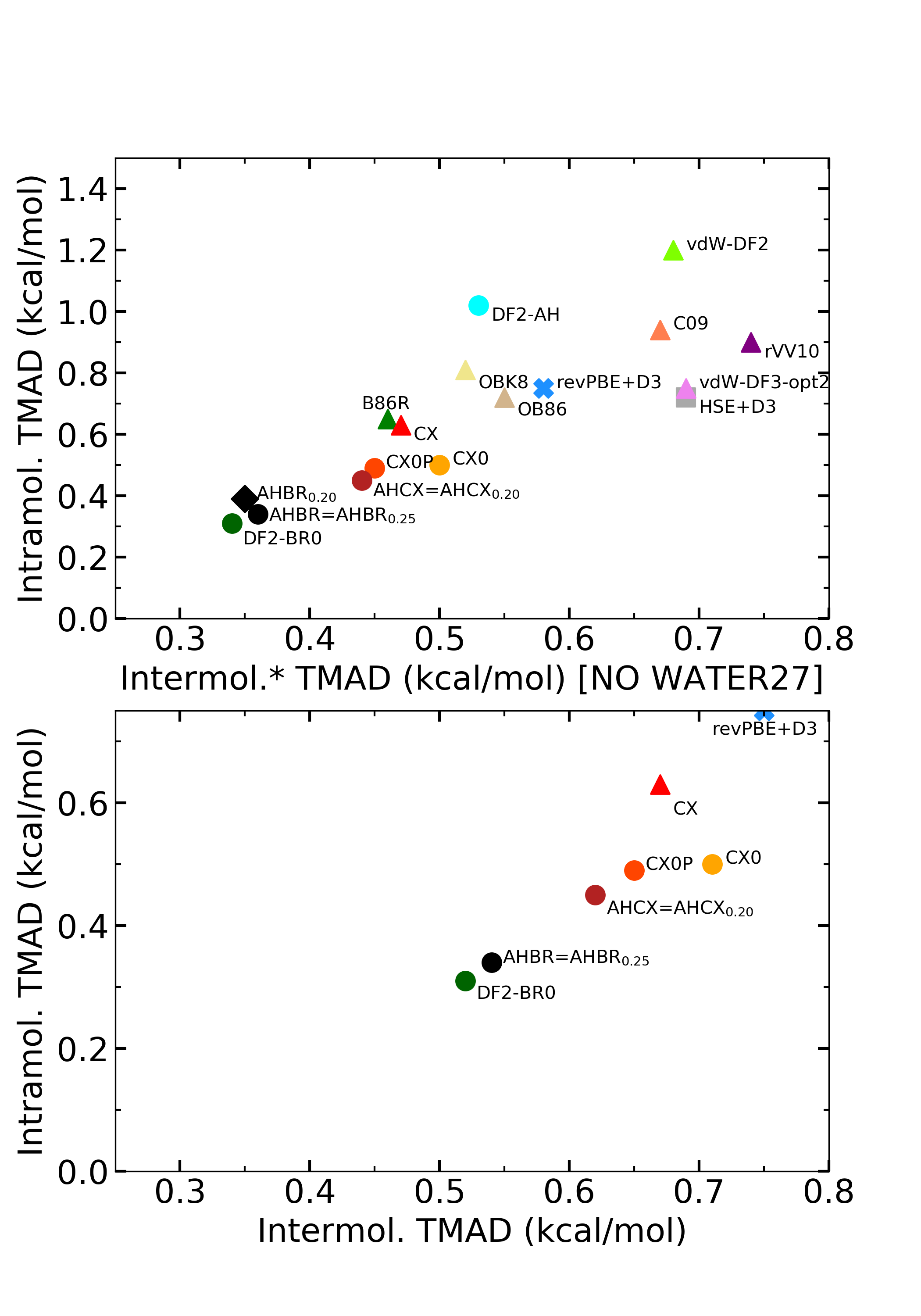}
	\caption{Performance comparisons of vdW-DFs (as well as 
	dispersion-corrected \cite{grimme3} revPBE and HSE) for
	NOC interactions.  We report and correlate 
	total MADs \cite{Jana20,DefineAHCX} (TMADs,) 
	formed as a simple average over benchmark MAD values, 
	of intermolecular and intramolecular NOC interactions; 
	Table S IV of the SI material
	shows a quantitative listing. The survey of RSH vdW-DF
	performance is tracked at two different choices of
	the Fock-exchange mixings (subscripts).  The top panel relies on raw 
	planewave-DFT results, as indicated by the asterisk; We 
	must then omit the WATER27 benchmark set \cite{BurkeSIE,DefineAHCX}.
	The bottom panel presents the corresponding survey
	as it results when we use an electrostatic-handling
	procedure, Appendix B, to also assert the performance for 
	the proton-transfer processes in the WATER27 set.
	}
	\label{fig:NOCwDiel}
\end{figure}

In fact, Table \ref{tab:DDerr} suggests that the AHBR may also compete favorably with PBE-based
density-corrected DFT \cite{DDerrorQaA,SiSoVu22} on density sensitive problems.
That approach is denoted DC(HF)-PBE, or DC(HF)-PBE+D4 
when supplemented with the Grimme-D4 dispersion correction
\cite{grimme4}; The D4 inclusion is not proven relevant
in these GMTKN55 benchmark cases (BH76, G2RC, and RSE43) 
\cite{DDerrorQaA}. 
The DF(HF)-PBE approach takes off from the above-discussed HF-PBE  description, however, in DC(HF)-PBE,
the HF-PBE result only replaces a self-consistent PBE result
when the process-specific measure $\tilde{S}_{\rm proc}$ is larger than 2 kcal/mol.
We find that AHBR performs better than DC(HF)-PBE in all three
cases and it is, for example, more robust for both small-molecule reaction
energies (RC21) and for transition states (as probed in the
barrier-height benchmark BH76).

Table \ref{tab:DDcrossTest} provides further documentation
of the AHBR ability to navigate molecular DFT challenges. 
The table compares our assessments 
for the RSH vdW-DFs with regular functionals (revPBE+D3, CX, vdW-DF2, and 
B86R) and with HSE+D3. The selection to focus our 
additional discussion on these benchmark sets is based on two criteria: 1) the benchmark-specific MAD values are large enough to  allow a reliably interpretation in terms of XC functional trends, and 2) the usage-oriented mapping of the density-sensitivity problems, Ref.\ \onlinecite{SaMa21}, 
suggests that we here face interesting DFT challenges.
That is, we primarily focus on the benchmark sets where a separate assessment \cite{SaMa21} found affirmative answers (listed in the second column) to the question of whether HF-PBE+D4 improves PBE+D4 (and we can again expect an impact of density-driven errors). However, we also direct attention to performance surprises. For all cases we simultaneously report the literature answer (third column) to whether the unscreened hybrid PBE0+D4 in selfconsistent calculations
makes for a further improvement over HF-PBE+D4, i.e.,
whether use of a hybrid can be expected to resolve 
the underlying DFT issues. 

We want the AHBR to succeed well in all such cases as we want it to be general purpose. Note that we supplement the selection in the table with the BHPERI barrier-height set and with the IDISP set. In the case of the BHPERI set, the HF-DFT+D4 was not found to help but the PBE0+D4 was, so something more than density-driven errors could be at play. We have not
found any literature evidence for density sensitivity
in the case of the IDISP set. However, we still include it
in this discussion, for  it is our experience that IDISP and WATER27 are the two NOC-interaction benchmark sets that primarily challenge the vdW-DFs, see SI material.

We find that the RSH vdW-DFs (like HSE+D3) almost always 
improve the description over comparable regular forms, compare AHCX and AHBR MAD values with those for CX and B86R, respectively. For HSE+D3, the 
observation is only infered: HSE+D3 should ideally be compared to 
PBE+D3, which we have not asserted, but Ref.\ \cite{gmtkn55} does 
identify revPBE+D3 as being the best overall GGA+D3 performer. This trend of RSH
strength is generally expected. However, atom descriptions are known to generally challenge hybrids and we do find that the W4-11 set on atomization energies is the exception to this trend. 

\ph{For the W4-11 set, we also find that AHCX performs better than AHBR; This is not surprising
since the two RSH vdW-DFs are born with different
default values of the Fock-exchange mixing.}
We note that picking instead the
same mixing $\alpha=0.2$ (a choice that
we have previously justified for atomization-energy
descriptions using the CX0P and hence AHCX \cite{JiScHy18a}), significantly helps the AHBR$_{0.20}$ performance on W4-11, see SI material.

Table \ref{tab:DDcrossTest} supports Fig.\ \ref{fig:TSwDiel}
in showing that AHBR performs better than the HSE+D3 overall 
also for this set of externally-identified DFT challenges. 
The W4-11 is again an exception, but there are also 
many cases where the AHBR is the top performer for 
the sets that are flagged as density sensitive. The
AHBR is also better on the BHPERI set even though the 
PBE-based gauge found that simply going to the HF-PBE+D4
correction was not a help in that class of problems. It is clear 
that the hybrid benefits for BHPERI is not confined to a selection of dispersion-corrected PBE0. We furthermore 
observe that Ref.\ \onlinecite{SaMa21} found dispersion-corrected PBE0 insufficient by itself to recoup or improve on a dispersion-corrected HF-PBE description for BH76, DC13, PNICO23,  HAL59, and WATER27 sets. It is
encouraging that, in contrast, the AHBR does provide
accuracy gains over both B86R and HSE+D3 in these
special cases.
 
Taken together, tables \ref{tab:DDerr} and \ref{tab:DDcrossTest} also show that vdW-DF2 is able to navigate  density-driven errors 
extremely well in some type of problems, but is overall 
characterized by having highly uneven performance. The vdW-DF2 advantages seems to primarily manifest themselves in group 3 barrier-height problems, such as the BH76, PX13,
and to a lesser extent the BHPERI sets. It is also 
good for the WATER27 set as well as, in fact, on most NOC-interaction problems \cite{DefineAHCX}. 
However, it is a weak performer on the 
IDISP and ISOL24 challenges and the SI material 
shows that the vdW-DF2 also has substantial limitations
for many group 1 and group 2 sets on small- and large-molecule properties.

Some of us have recently documented \cite{MOFdobpdc} that a good vdW-DF1 and vdW-DF2 
performance can sometimes arise because it 
has an XC balance that is better set up to handle the case of more diffusive interactions \cite{MOFdobpdc}.  The CX is considerably more accurate than vdW-DF1 and DF2 for the simpler case of CO$_2$ adsorption in Mg-MOF-74, but the
reverse is true for the more complex cases of CO$_2$ adsorption in diamine-appended Zn$_2$(dobpdc) \cite{MOFdobpdc}. 
\ph{We find it plausible that} the vdW-DF1 
and vdW-DF2 successes (compared to other vdW-DFs) on the more complex m-2-m$-$Zn$_2$(dobpdc) system \ph{in part} result by minimizing density-driven \ph{errors, given the fact that they navigate
such errors in, for example, the BH76 and PX13 sets, above. It is also possible that the vdW-DF1 and vdW-DF2 accuracy on m-2-m$-$Zn$_2$(dobpdc) energies is simply fortuitous as these functionals worsen the description of structure, for example, in m-2-m$-$Zn$_2$(dobpdc) \cite{MOFdobpdc}. However, the relevance 
of using the rPW86 exchange in vdW-DF2 (that, like AHBR, uses a Schwinger-scaling argument to set the nonlocal correlation \cite{lee10p081101}) was asserted by documenting that rPW86 mimics a Fock-exchange description for intermolecular interactions \cite{mulela09}. Since vdW-DF2 does excel at many types of transition-state and  NOC-interaction problems, SI Tables S.V-VII,
we find it wise to respect the lessons 
of vdW-DF2 progress. This is
especially so now that AHBR offers us a chance to
combine its nonlocal-correlation design with 
a screened Fock-exchange form. We therefore include, below, an additional AHBR test, asserting its performance for  m-2-m$-$Zn$_2$(dobpdc).}

\subsection{Further lessons from 
small and large systems} 

\begin{figure*}
	\centering
	\includegraphics[width=0.9\textwidth]{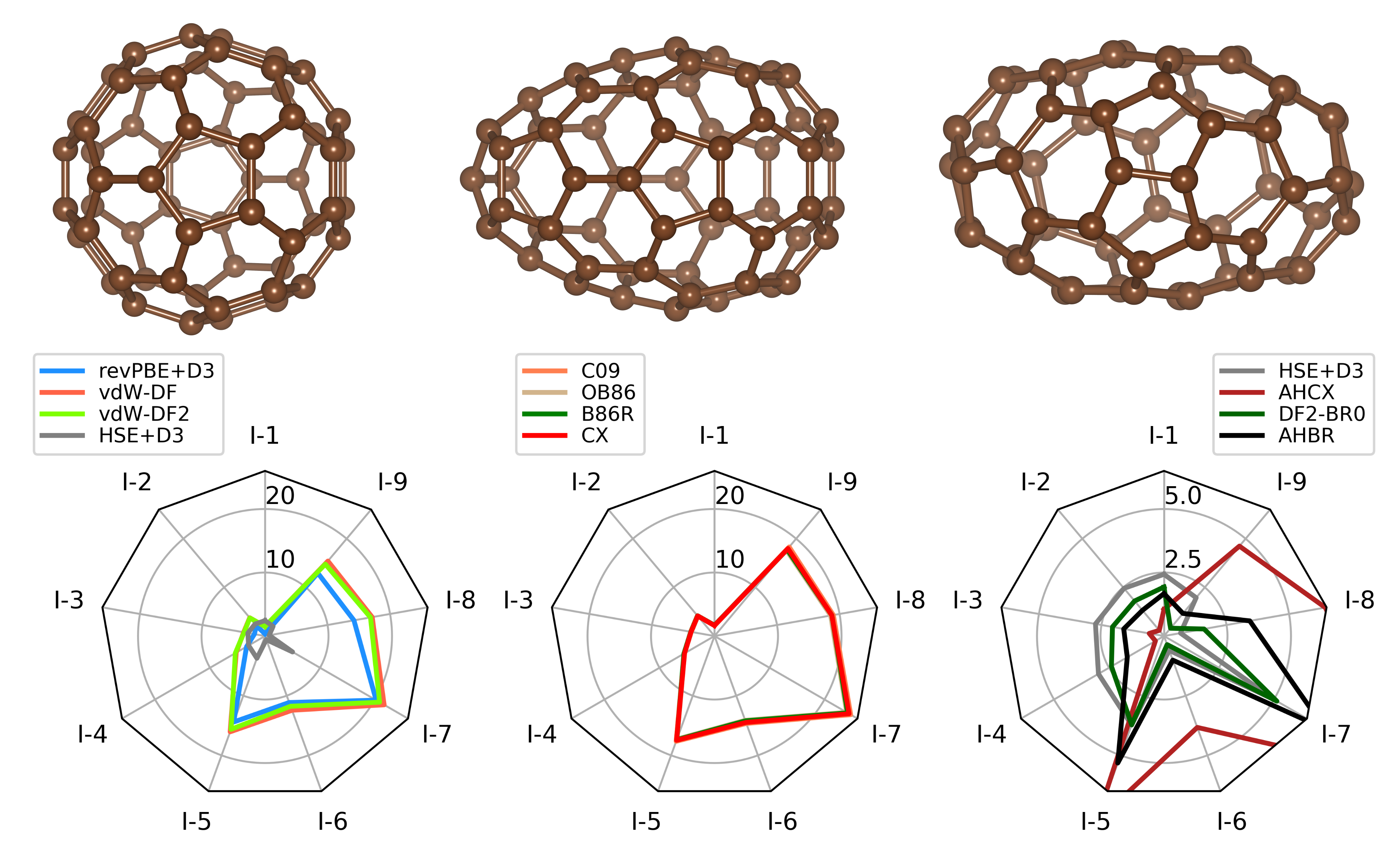}	
	\caption{Performance of original vdW-DF releases (left panel), of recent 
	regular vdW-DFs (middle panel) and of hybrid vdW-DFs (right panel) on a 
	set of C$_{60}$ isomerization problems.  The top left (right) panel 
	shows the regular (highly distorted) configuration. The middle panel
	depicts one of the intermediate configurations in the transformation,
	as tracked in the C60ISO benchmarks set \cite{gmtkn55}. The
	bottom left (middle) panel shows MAD values (in kcal/mol)
	characterizing the performance of the original vdW-DF 
	releases (recent regular vdW-DFs) on describing energy differences
	between such configurations. The bottom right panel shows
	that the hybrid vdW-DFs are needed to substantially improve
	the description.
	\label{fig:C60ISOcompReleases}
	}
\end{figure*}

Figure \ref{fig:NOCwDiel} summarizes the 
inter- and intra-molecular NOC-interaction
parts of our functional-performance comparison.
The figure concentrates on the benchmark groups that give different indications for a convenient AHCX and AHBR choice of default Fock-exchange mixing.
For the vdW-DFs and for rVV10 \cite{vv10,Sabatini2013p041108}, as well as
dispersion-corrected revPBE+D3 and HSE+D3, the group-averaged
performance is represented in a scatter plot that
relies on taking a raw, so-called TMAD, performance indicator \cite{Jana20} on NOC-interaction systems
(from the GMTKN55 suite). This measure is defined by taking a simple average over the MAD values that we obtain for the individual benchmarks, as described in the SI material for Ref.\ \onlinecite{DefineAHCX}.
We use the intermolecular (intramolecular) NOC TMAD value for
group 4 (group 5) to set the abscissa (ordinate); Table 
S IV of the SI materials contains a listing of the TMAD
values that are reported in the panels of Fig.\ \ref{fig:NOCwDiel}
(as well as for those evaluated for the other GMTKN55 benchmark 
groups).

Figure \ref{fig:NOCwDiel} also highlights 
the key impact of excluding (top panel) or including 
(bottom panel) the impact of the WATER27 benchmark.
This benchmark is often excluded in functional comparisons 
on vdW problems \cite{Jana20,DefineAHCX}, because it 
contains the negatively charged small radical OH$^-$
and it is therefore not accessible in a simple
benchmarking \cite{BurkeSIE,DefineAHCX,Hard2Soft}.
Like for the systems in the G21EA benchmark set, 
this radical has pronounced self-interaction 
errors \cite{BurkeSIE} as well as convergence challenges
that prompted us to pursue the more general planewave
benchmarking procedure defined in Appendix B.
Insight on water in general and accuracy in WATER27 benchmarking are essential on science grounds and when pursuing systematic XC development. 

We show that the correct inclusion of the WATER27 benchmark set,
Appendix B and Fig.\ \ref{fig:NOCwDiel}, has dramatic impact on what 
we consider XC functional promise for NOC interactions. The assessment 
map is clearly affected as we switch from the top to the bottom
panel of Fig.\ \ref{fig:NOCwDiel}. In fact, it is alone the 
consistent-exchange class (CX, CX0P/CX0 and
AHCX), the unscreened DF2-BR0, and the 
new AHBR that remain good options for this
challenge, at least as asserted by our planewave
benchmarking of the GMTKN55 suite.  

\begin{figure*}
	\includegraphics[width=0.9\textwidth]{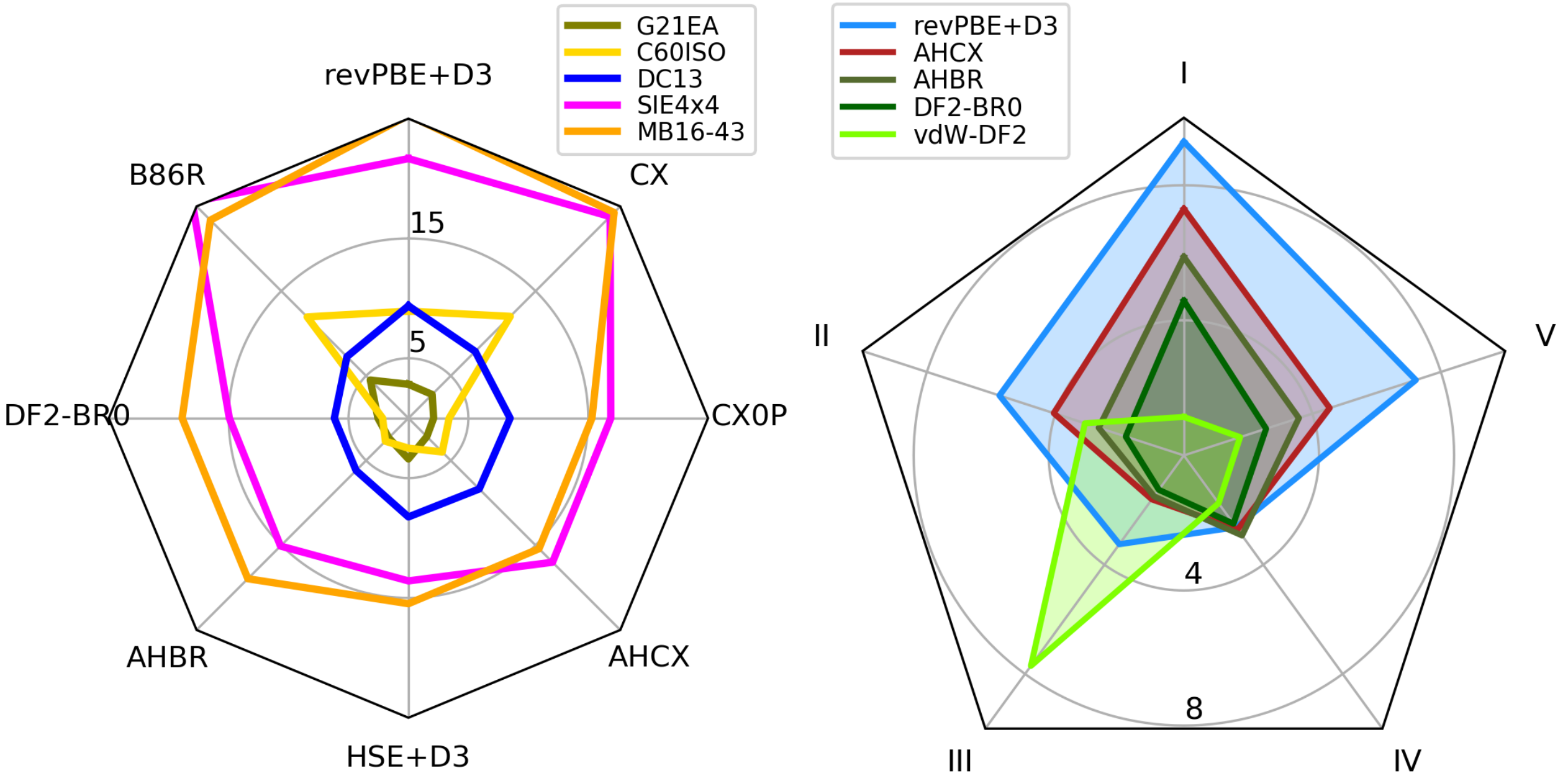}
	\caption{Drivers for XC-development in the GMTKN55 suite of 
	benchmarks \cite{gmtkn55}: Performance of vdW-DF-based hybrids 
	on traditional molecular challenges (identified
	in left panel) and on what we consider key transition-state 
	challenges (right panel) in benchmark set PX13 (I), 
	BHPERI (II), IDISP (III), WATER27 (IV), and  WCPT18 (V). Self-interaction
	errors play an essential role not only in the SIE4x4 set but
	also in G21EA (WATER27) due to the presence of negative ions and
	radicals \cite{BurkeSIE}, requiring an electrostatic-environment
	approach for reliable assessments (appendix B). The traditional
	challenges include the so-called `difficult-for-DFT' (DC13) and
	`mindless benchmarking' (MB16-43) sets, as well as 
	Fullerene isomerization problems (collected in benchmark set C60ISO).  
	\label{fig:Diamond}
	}
\end{figure*}

In a recent paper we suggested that there is value in using
the tool chain of closely related CX/CX0P/AHCX designs for 
a systematic exploration of bulk materials \cite{Hard2Soft}. 
We here propose use of AHBR for explorations on molecules and 
for their adsorption.  It is known that density-corrected 
DFT  should not always be used \cite{DDerrorQaA}. For  
traditional hybrids (like HSE) there is an expectation
that such hybrid-PBE will worsen the thermophysical
description of solids, especially for magnetic elements,  
and for transition-metal adsorption \cite{BurkePerspective,JaYu12,JaLuKo14,GaAbCa16,ShKaMa19}.
The idea of exploring is that we want 
to know what limitations holds for the RSH vdW-DFs and for
AHBR in particular. More broadly, we want to learn to assert, a priori, when one should use both nonlocal-correlation and nonlocal-exchange in combination and when the regular vdW-DF (CX or B86R) suffices. Interestingly, we are finding, here and in Refs.\ \cite{DefineAHCX,Hard2Soft}, that AHCX and AHBR improve 
descriptions of bulk structure and, at least, of
some adsorption problems (over the underlying regular versions, CX and B86R). For molecular properties, it is instead securing a balanced and robust progress that is the challenge for hybrid vdW-DFs.

In this light, it is interesting to note that the
impact of including the WATER27 challenge on the first tool chain (comprising CX-CX0/CX0P-AHCX) is, overall, smaller than that on the B86R design. The B86R itself is, in fact, moved out off the figure range for intermolecular-NOC assessment values. However, the corresponding B86R-based hybrids, DF2-BR0 and the new AHBR, remain exceptionally well suited to meet the full set of NOC challenges, at least as presently asserted. Also, looking at the quantitative measures,  Table S IV
of the SI material, we find that AHBR$_{0.20}$ appears to  perform slightly better than AHBR$_{0.25}$ on intermolecular 
NOC interactions in the approximate 
assessment (top panel),  but the actual status is different (bottom panel).

Overall, we see the robustness of CX/CX0P/AHCX and DF2-BR0/AHBR as an indication of value and usefulness as we seek to map for and understand outstanding DFT challenges \cite{Hard2Soft,gmtkn55}. For NOC-interaction, small-molecular and barrier cases, 
the AHBR is the more robust, error-resilient design, Figs.\ \ref{fig:TSwDiel} and \ref{fig:NOCwDiel}.

Figure \ref{fig:C60ISOcompReleases} considers the C60 isomerization problems,
using the refeence data of the
C60ISO benchmark set for an analysis of 
a large-system transformation (group 2)
case. The C60ISO is a benchmark where all regular vdW-DFs fails and where hybrids is needed, see SI material. It is also a case where we document that there are pronounced density-driven errors at play, compare the C60ISO values that we
compute for $\tilde{S}_{\rm avg}$ and for PBE in Table \ref{tab:DDerr}.

The top panels of Fig.\ \ref{fig:C60ISOcompReleases} illustrate 
the nature of the C60ISO benchmark set in the large-system isomerization group 2 of the GMTKN55 suite. The benchmark 
set considers the energy differences among 10 meta-stable forms 
of C$_{60}$ of energy $E_n$. The top-left panel shows the
stable Fullerene form (`$n=1$') and the top
right panel shows the oblate form (`$n=10$').
Between them are also states of increasing deformations
(denoted `2' to '9') and the GMTKN55 suite
provides reference data for 9 isomerization 
energy-difference problems, denoted `I$n$' (for $n=1$ through $n=9$'). 
These C60ISO problems are specified by 
reference values for the set of differences $E_{n+1}-E_{1}$
($n=1$ through 9).  The 
top middle panel shows the atomic geometry configuration
for the frustrated structure `$n=8$', i.e., a form
that determines isomerization problem `I7' and that 
is often hard to accurately describe.

The set of bottom panels provides a 
radar-plot comparison that reveals both vdW-DF 
limitations and vdW-DF promise.
Specifically, the bottom left
(middle) panel shows that vdW-DF1 and 
vdW-DF2 both fail (that recent releases
and variants including CX and B86R offer
\ph{no} improvements). As the benchmark is prone 
to density-driven errors, Table \ref{tab:DDerr}, it is no surprise that 
HSE+D3 provides an significantly improved 
description over regular vdW-DFs. As shown 
it the bottom-left panel (note the change in scale), the AHCX helps, but AHBR is needed to 
improve the description to the HSE-D3 level.

We interpret the nonhybrid-vdW-DF performance 
issues as arising because we must here describe stretched 
and frustrated binding. The fact that there are density-driven
errors is not surprising, for this isomerization problem
can also be seen as another transition-state case.
In fact, the set of meta-stable 
C$_{60}$ configurations can be seen
as configurations that define some effective
deformation paths taking us from configuration
`1' to '10'. 

We can expect that the DF2-BR0 and AHBR are \ph{accurate} 
also for the C60ISO set (as it is documented in 
the lower-right panel). However, given the status 
for the (barrier-height) benchmark group 3 and 
given that this is a transition-state problem, it 
is perhaps surprising that the vdW-DF1 and vdW-DF2 are 
not acceptable performers here. It appears that there 
are more than one type of challenges in describing 
transition states. We can certainly still learn from 
vdW-DF2 and vdW-DF1, but is also clear that we need to 
cast a wider net to identify good development ideas.
This can, as suggested here, be done  
by use of the more robust AHBR.

\subsection{Need for vdW-DF hybrids: Molecular examples}

Transition-state problems stand out
because they involve a comparison of energy terms 
that must simultaneously reflect several different 
types of binding, for example, relaxed and stretched 
or diffusive.  Some of us have recently characterized functional 
performance in the description of CO$_2$-MOF adsorption 
that happens in concert with a site-specific-reaction
and resulting CO$_2$-insertion in a diamine-appended MOF \cite{MOFdobpdc}.
There, the vdW-DF1 and vdW-DF2 also had an accuracy edge but 
it did not apply for CO$_2$ adsorption in the simpler Mg-MOF-74
system \cite{MOFdobpdc}. The transition-state problems can
be seen as key drivers for XC development \cite{BK04,DDerrorQaA}. 

Figure \ref{fig:TSwDiel} and Figs.\ S.1-2 of the SI material provide a broad illustration that the vdW-DF tool bag must include the set of new
hybrid vdW-DFs. i.e., DF2-BR0, AHCX and AHBR. The set of recent regular vdW-DFs (including CX and B86R) 
do not remain uniformly accurate when tracked across 
the 7 individual barrier-height benchmarks
of GMTKN55, Fig.\ \ref{fig:TSwDiel} (right panel) 
and SI material.  Inspecting SI Tables S V through S X 
makes it clear that problems of a transition-state 
nature (such as we deem those of the C60ISO and MB16-43
sets), in general, challenge the non-hybrid vdW-DFs.
Nevertheless, we find that the DF2-BR0 performance stands out in 
terms of both accuracy and cross-benchmark resilience even 
at transition-state problems. As noted in the introduction,
in turn, the DF2-BR0 resilience and transferability for
molecular problems motivate the AHBR completion.

Meanwhile, the underlying benchmarking data, Tables S II
through S X of the SI material, shows that
the vdW-DF2-ah hybrid \cite{DefineAHCX} is not a reliable 
option for the study of broad molecular properties
(just like it failed for bulk). Like the
vdW-DF2, it has an acceptable performance for 
NOC-interaction and barrier-height problems. However,
vdW-DF2 and vdW-DF2-ah have real shortcomings when the focus 
is instead moved to general properties of small- and 
large-molecule systems.

More broadly, the GMTKN55 survey suggests that we
cannot hope for one gradient-corrected exchange form (as used
in the present type of nonhybrid vdW-DF designs) to 
always succeed. We see that the assessments on GMTKN55 
performance show a spread in performance -- for all nonhybrid 
vdW-DFs -- among different classes of problems, see SI material. 
The same is true, but to a significantly smaller extent, of the 
CX and B86R-based hybrids. There are many problems, and it is 
not trivial to secure a good XC balance in the present type 
of vdW-DF designs, across all of such problems.  Specifically, in the
vdW-DFs, we add an attractive nonlocal-correlation term to 
the spurious LDA-exchange overbinding \cite{Harris85}. \ph{The 
nonlocal correlation contains more 
than pure dispersion 
effects while it also goes beyond GGA
correlation at shorter distances \cite{JPCMreview}.
At the same time, in the nonhybrid vdW-DF} we must rely on nothing but gradient-corrected 
exchange for stabilization
\ph{in the presence of both LDA overbinding and the enhanced attraction of these nonlocal-correlation effects  \cite{rydberg03p126402,mulela09,MOFdobpdc}.}

The fact that the strength of the vdW attraction may enhance in select situations \cite{MOFdobpdc}
suggests an additional mapping role for vdW-DF hybrids. The plan is simply to use and contrast AHCX and AHBR for explorations, for example, in cases with diffusive interactions 
\cite{MOFdobpdc}, when interactions
compete \cite{bearcoleluscthhy14}, and when
phase-transformations compete \cite{PeGrRo20,jewahy20,Hard2Soft}.
In the latter case, the presence of an actual or incipient ferroelectric transformation will itself affect the magnitude of the 
dielectric constant. For example, while the unscreened hybrid CX0P is highly accurate
for the BaZrO$_3$ \cite{PeGrRo20}, we need more to address the 
SrTiO$_3$ \cite{Hard2Soft}.  Access to RSH vdW-DFs means that
we better map the nature of specific challenges, correlate
progress with design choices, and eventually implement new ideas in XC developments. 

\begin{figure}
	\includegraphics[width=0.90\columnwidth]{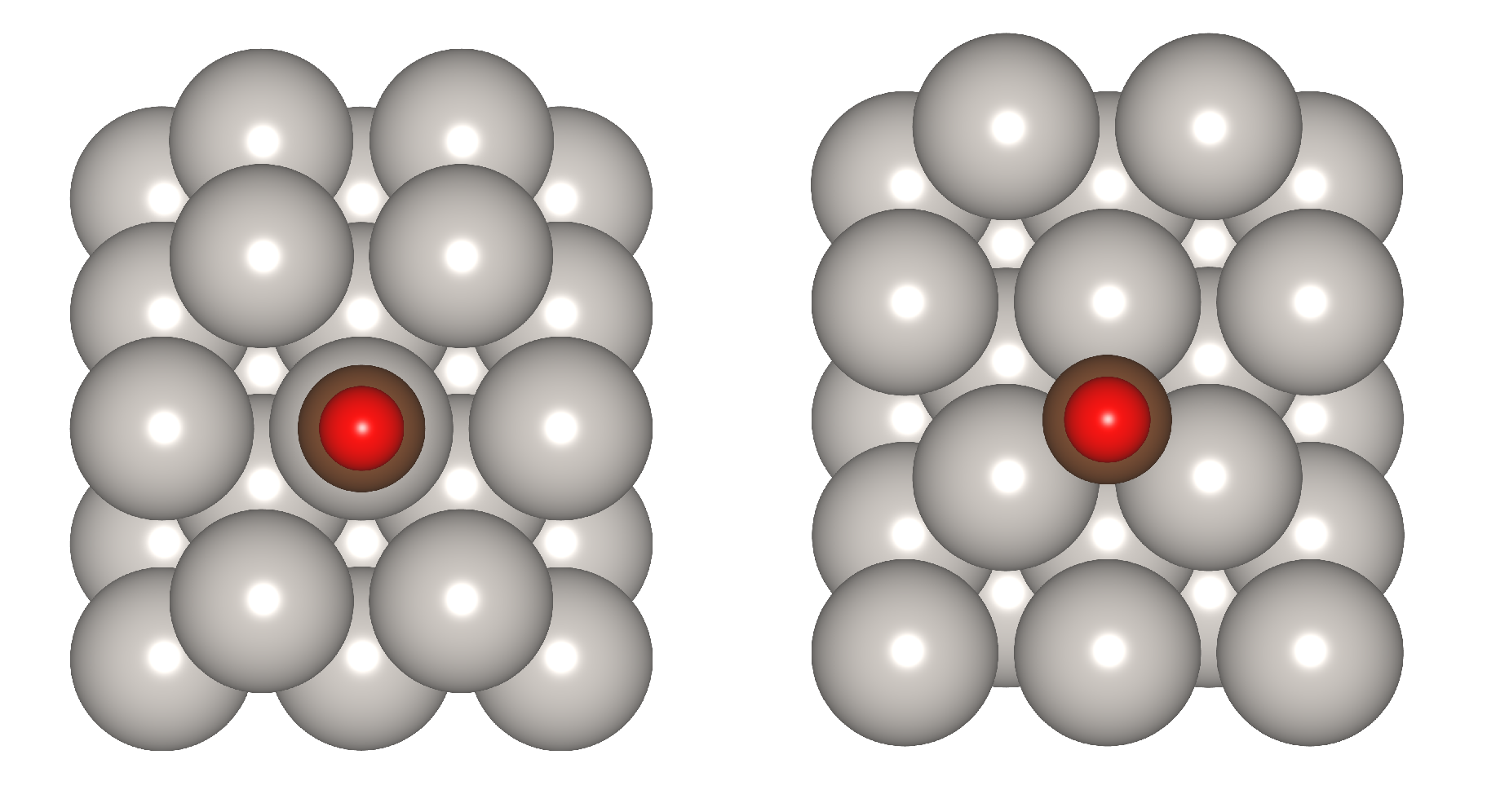}
	\caption{Schematics of the CO-on-Pt(111) problem: Competition between 
	TOP- and FCC-site adsorption. The image is generated using the VESTA program \cite{VESTA3}.
	\label{fig:CO/Pt}
	}
\end{figure}

In Fig. \ref{fig:Diamond} we identify a key set of
molecular-type challenges that we think can be used to drive 
XC development, without going directly to more complex (but technologically relevant) cases. For 
example, an accurate description of some transition-state problems 
(benchmark sets `I', `II', `IV', and `V' in the left panel)
requires a unique XC balance that vdW-DF2 occationally provides,
yet vdW-DF2 often fails spectacularly, as seen in the SI material.  Ref.\ 
\onlinecite{MOFdobpdc} provides an adsorption case where
vdW-DF2 also shines by delivering an exceptionally large repulsion
by gradient-corrected exchange, but such repulsion is not always 
needed \cite{MOFdobpdc}. It is therefore good to seek a simpler way to survey for XC issues, as in Fig.~\ref{fig:Diamond}.  Our proposal 
is based on the experience that we have gained 
by planewave benchmarking across both the entire GMTKN55 
and for CO$_2$ uptake in MOFs, for example, in 
Refs.\ \onlinecite{poloni_understanding_2014,bearcoleluscthhy14,CO2MOF74survey,DefineAHCX,Hard2Soft,MOFdobpdc}.
We mostly echo, but also simplify, the logic that led to the
definition of the full GMTKN55 suite \cite{gmtkn55}. In making 
this identification, we are assuming that we are working with 
a vdW-inclusive functional, like the vdW-DFs, so that the design 
is also capable of dealing with the group 4 and group 5 
types of NOC-interaction problems. 

Well known challenges for any DFT are, of course, still found in the 
SIE4x4 set (on self-interaction errors in of neutral and positively 
charged systems), the MB16-43 (mindless benchmarking) set and the 
DC13 (difficult-for-DFT) set.  To these sets we add the G21EA and WATER27 sets 
because these two sets suffer from pronounced SIE effects, and in the latter
case will also significantly impact the performance of any given candidate
on the group 4 problems. Additionally, we include the IDISP set as it
is almost always the challenge that dominates in setting the performance
of an XC functional on the group 5 of intramolecular NOC interactions.

From the barrier-height class, we find that the performance 
varies prominently when inspecting the BHPERI, PX13, and
WCPT18 sets. At least it is clear that these benchmarks allow the 
hybrid benefits to directly manifest themselves. Finally, we point to the
above-discussed C60ISO set as a supplement to 
the focus
on barrier-height problems: It clearly reflects both a transition-state 
nature but it also points to an additional, yet to be identified aspect. 
This is because the vdW-DF1 and vdW-DF2 barrier-height successes,
at the WCPT18 and PX13 benchmark sets, do not port well to this C60ISO set.

Finally, Fig.\ \ref{fig:Diamond} illustrates the usefulness of simplifying
functional comparisons while focusing on where we can learn more.
The figure compares the performance of the CX/CX0P/AHCX and the B86R/DF2-BR0/AHBR functionals in two types of radar plot.
The left panel makes the comparison based on  G21EA as well as on the sets 
where there are often massive deviations between DFT results and reference data from
quantum-chemistry calculation \cite{gmtkn55}. The right panel makes the comparison
on the selection of barrier problems and of NOC-interaction
problems. An overall impression is that the robustness of the DF2-BR0
and AHBR is confirmed from the testing summary presented in 
Figs.\ \ref{fig:TSwDiel} and S.I. However, we also see that 
the MB16-43 benchmark set identifies an example set of problems where 
we can still learn more from the \ph{Lindhard-screening logic 
(summarized in Section II) and the
screened-exchange gradient expansion
result \cite{lavo87} that underpins CX
and hence AHCX \cite{JPCMreview}.}

\subsection{CO adsorption on Pt}

Figure \ref{fig:CO/Pt} presents a schematics of CO/Pt(111), contrasting the
atomic configurations in two competing adsorption sites, denoted TOP (for being on top of a surface atoms) 
and FCC (for being in a position that corresponds to an extension of the face-centered cubic atomic organization 
of the Pt substrate). 
We note that the CX has a lattice constant that is in close 
agreement with experimental characterizations of Pt, 
Table \ref{tab:TMlat}. This observation motivates our assessment strategy: To keep the adsorption
structure fixed at the CX description. We shall 
concentrate on directly comparing and discussing the 
new-XC-functional accuracy in the CO/Pt(111) adsorption
within a fixed-nuclei framework.

\begin{table}
	\caption{\label{tab:COPt} Comparison of TOP- and FCC-site CO/Pt(111) binding energies $E_{\rm bind}$, 
	site-preference energies $\Delta E_{\rm site}=E_{\rm bind}^{\rm TOP}-E_{\rm bind}^{\rm FCC}$, as well as 
	of molecular-gap results $E_{\rm gap}^{\rm CO}$ (all in eV). The first and second block are for CX-AHCX and for B86R-AHBR, respectively. For both tool chains we also illustrate the impact of the
	choice of the Fock-mixing (as identified in the functional-label subscript). All results are provided for the CX 
    provided substrate lattice constant $a_0= 3.929$ {\AA}, molecular structure, and adsorption-induced deformations. Experimental observations of CO adsorption find TOP site 
	adsorption with binding energy $\Delta E_{\rm site}=-1.32$ eV \cite{MYSHLYAVTSEV201551}.
	}
\begin{ruledtabular}
\begin{tabular}{l|ccc|ccc}
	                               & CX & AHCX & AHCX$_{0.25}$ & B86R   & AHBR$_{0.20}$ & AHBR \\
				     \hline
	$E_{\rm bind}^{\rm TOP}$       & -1.830 & -1.949 & -1.974 & -1.683 & -1.801        & -1.824   \\
	$E_{\rm bind}^{\rm FCC}$       & -1.966 & -1.954 & -1.940 & -1.780 & -1.767        & -1.752   \\      
	$\Delta E_{\rm site}$          &  0.133 &  0.005 & -0.034 &  0.097 & -0.034        & -0.072  \\
\hline
	$E_{\rm gap}^{\rm CO}$         & 7.048  & 8.768  & 9.192   & 7.081  &  8.798        & 9.222   \\
\end{tabular}
\end{ruledtabular}
\end{table}

\begin{figure}
	\includegraphics[width=0.95\columnwidth]{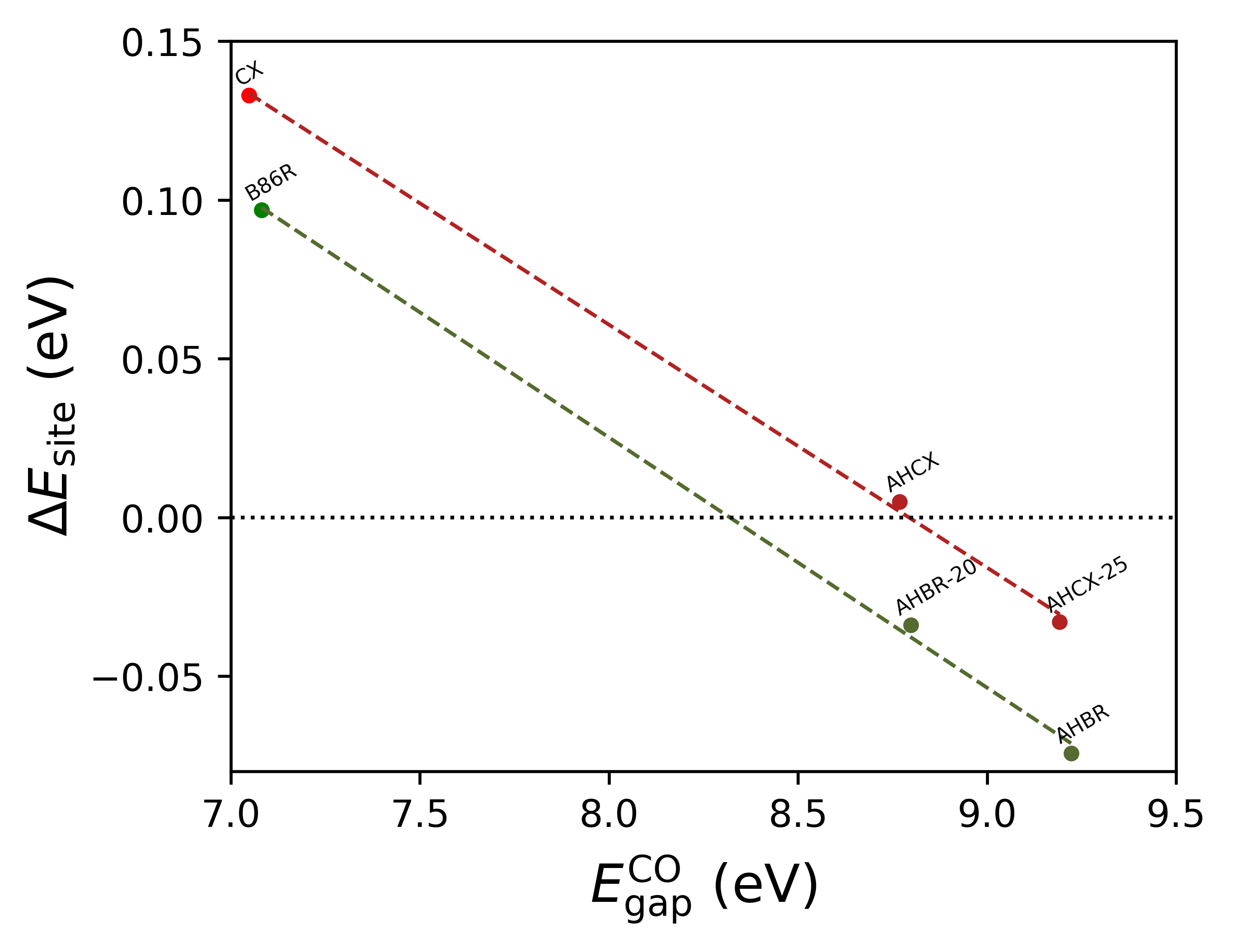}
	\caption{Correlation between results for the CO molecular gap $E_{\rm gap}^{\rm CO}$
	and for the CO/Pt(111) site-preference energy $\Delta E_{\rm site}$ as computed in the CX-AHCX functional chain and in the B86R-AHBR functional chain.
	We keep the adsorption geometry fixed in all studies.
	\label{fig:CO/Pttrend}
	}
\end{figure}

We compute the site variation in the adsorption or binding energy
\begin{equation}
	E_{\rm bind}^{\rm TOP/FCC} = E_{\rm CO/Pt}^{\rm TOP/FCC} 
	- E_{\rm Pt, surf} - E_{\rm CO, mol} \, ,
	\label{eq:COPtbind}
\end{equation}
where $E_{\rm Pt, surf}$ ($E_{\rm mol}^{\rm CO}$) represents 
the energy of the clean surface (isolated molecule) and
where $E_{\rm CO/Pt}^{\rm TOP/FCC}$ denotes the total energy of the 
adsorption configuration, as described with CO at the TOP/FCC site.
We focus on discussing the site-preference energy 
\begin{equation}
\Delta E_{\rm site} = E_{\rm bind}^{\rm TOP}-E_{\rm bind}^{\rm FCC} \, .
	\label{eq:sitepref}
\end{equation}
Experimental studies clearly
indicate that TOP-site adsorption (at binding energy $-1.30$ eV) applies for 
dilute coverage at room temperatures (corresponding to 0.026 eV) \cite{MYSHLYAVTSEV201551}. Results obtained using the random-phase 
approximation concurs with these observations \cite{rpa:ads}. 

Accordingly, DFT calculations with an accurate XC functional should 
find $\Delta E_{\rm site}$ negative and  with a magnitude that exceeds the
0.026 eV value. However, this problem is a long-standing challenge for DFT, in the sense that essentially all non-vdW-DFs (including HSE and PBE0 \cite{BuErPe97,PBE0})
fail, when considered at or close to the actual Pt lattice constant \cite{grinberg2002co,olsen2003co,stroppa2007co,alaei2008c,KresseNJP,lazic10p045401,janthon2017adding,patra2019rethinking,DefineAHCX}. We observe that density-driven errors are expected to complicate the DFT setting of the correct
CO/Pt(111) adsorption-site preference \cite{patra2019rethinking}.

Table \ref{tab:COPt} summarizes our comparison of RSH vdW-DF performance for the classic
CO/Pt(111) problems.  We note that the AHBR lattice constant for Pt is slightly larger
(and further from experiment) than the CX
choice we have used. An AHBR adsorption study at the AHBR lattice constant will have a more narrow Pt $d$ band and therefore yield smaller adsorption energies \cite{HaNo95,Noblest,stroppa2007co,alaei2008c,KresseNJP,janthon2017adding,patra2019rethinking}.

We find that neither CX nor the default AHCX description (using a 0.20 fraction of 
Fock-exchange mixing) offers an improvement in the description of the site-preference challenge. In contrast, the new AHBR works (as do AHBR$_{0.20}$
and AHCX$_{0.25}$), bringing the site-preference description in 
alignment with experimental observations. Still, the AHBR does not offer a complete resolution of the CO/Pt(111) problem because it is over-estimating the
actual adsorption energy $E_{\rm bind}$.

Figure \ref{fig:CO/Pttrend} presents a mapping of correlation between the CO gap and the site-preference energy, as obtained in the CX-AHCX chain
of functionals and in the B86R-AHBR chain.  The adsorption is often discussed in terms of 
the Blyholder model \cite{Blyholder}
and, as such, controlled by the substrate electronic structure (which we must
accurately characterize to correctly describe the 
molecule-to-substrate charge transfer)
and the molecular gap (that we must accurately 
characterize to correctly describe the back donation). 
We keep the adsorption geometry fixed in all calculations, 
and thus contrast the direct 
effects that the functionals have on both the 
molecule gap and substrate electronic 
structure 
\cite{Note6}.

\begin{figure}
\includegraphics[width=0.97\columnwidth]{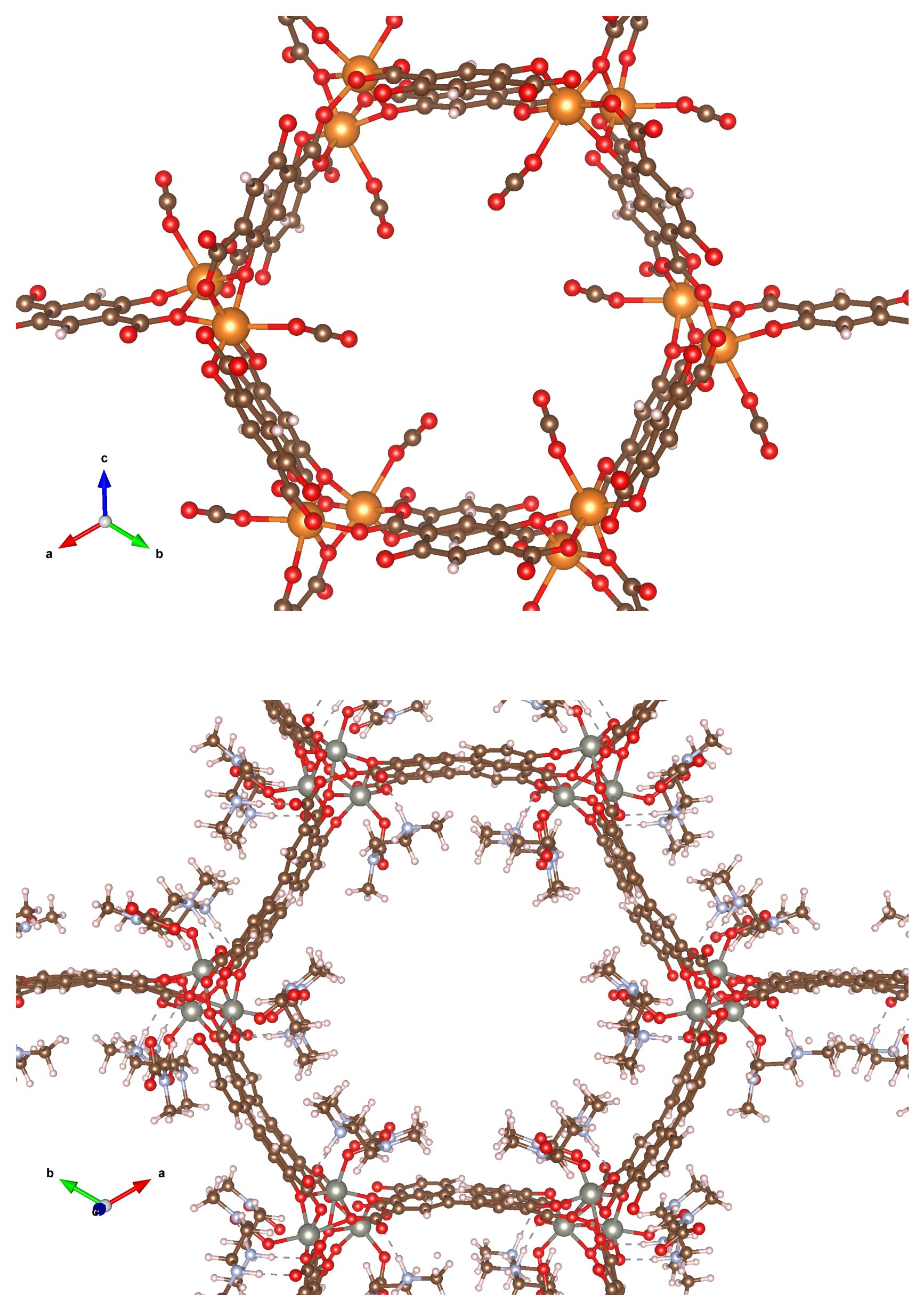}
	\caption{Optimized crystal structures of  CO$_2$-adsorbed Mg-MOF-74 (upper panel) and  CO$_2$-inserted m-2-m$-$Zn$_2$(dobpdc) (lower panel). The gray, orange, red, skyblue, brown, and pink circles represent Zn, Mg, O, N, C, and H atoms, respectively. The images are generated using the VESTA program \cite{VESTA3}. 
}
\label{fig:mof_fig}
\end{figure}

There are several lessons from Fig.\ \ref{fig:CO/Pttrend}. First, there 
is a strong impact on the molecular gap by moving from a regular function 
to the associated RSH. Second, this gap variation 
effectively controls the prediction of site preference within a given 
tool chain. Third, there is also a systematic 
effect of switching between the tool chains and thus
changing the substrate electronic structure. In effect, 
the figure documents that the well-defined AHCX-AHBR 
differences (in terms of nonlocal-correlation- 
and exchange-design details) have a direct impact 
on the Pt(111) electronic description. This impact 
is making the AHBR better at reflecting the true 
site variation in the CO/Pt(111) adsorption energy.

\subsection{CO$_2$ adsorption in Mg-MOF-74 and m-2-m$-$Zn$_2$(dobpdc) MOFs}

\begin{table}
	\caption{\label{tab:MOF} Comparison of performance for AHBR and for the CX-based tool chain on the CO$_2$ adsorption energies in Mg-MOF-74, top section, and in
	m-2-m$-$Zn$_2$(dobpdc),
	bottom section. All calculations are performed at the CX characterizations for the MOF geometries. For Mg-MOF-74, we permit the CO$_2$ to relax as described in the stated 
	functional, and list the results for the characteristic bond length. We presents energy (bond length) results in kJ/mol (in {\AA}) to facilitate an easy comparison with a recent MOF study  covering regular vdW-DFs and revPBE+D3
	\cite{MOFdobpdc}.
	 The subscript `BO' identifies Born-Oppenheimer results, while a superscript
	`room' identifies adsorption free-energy values that include the effects of 
	vibrations as described at room temperature, using Eq.\ (\ref{eq:hads_mof}) to facilitate a direct comparison to experimental values for Mg-MOF-74. The values in
	parenthesis represent an estimate
	for how a back correction for vibrations would impact the measured m-2-m-Zn$_2$(dobpdc)
	$H_{\rm ads}^{\rm room}=-57$ kJ/mol value \cite{CO2MOF74survey}.
	}
\begin{ruledtabular}
\begin{tabular}{l|ccc|c|c}
	                               &    CX$^a$   &   AHCX   & AHCX$_{0.25}$ &  AHBR        & Exper. \\
				       \hline
	Mg$-$O & 2.29 & 2.27 & 2.26 &  2.25 & 2.27$^b$ \\
	$E_{\rm ads, BO}$         &  -53.7$^a$    & -58.5   & -59.7    & -50.3   & -      \\
	$\Delta$ZPE & 2.8 & 2.4 & 2.5 &  2.6 &  \\
	$\Delta$TE & 1.1 & 1.6 & 1.3 &  1.1 & \\
	$H_{\rm ads}^{\rm room}$       &  -49.7  & -54.5 & -55.9  & -46.6  & -43.5$^b$\\
	\hline
	$E_{\rm ads, BO}$         &  -94.6    & -98.7   & -99.7   & -86.8 & -57$^c$ (-63 $-$ -60 )   \\
	\hline
	\multicolumn{6}{l}{$^a$ Ref.\ \onlinecite{MOFdobpdc}}\\
	\multicolumn{6}{l}{$^b$ Ref.\ \onlinecite{CO2MOF74survey}}\\

	\multicolumn{6}{l}{$^c$ Ref.\ \onlinecite{CO2dobpdc2}}
\end{tabular}
\end{ruledtabular}
\end{table}

Figure \ref{fig:mof_fig} contrast the
nature of CO$_2$ adsorption in the simpler Mg$_2$(dobdc) known as Mg-MOF-74, upper panel, and in
the more complex case of the diamine functionalized m-2-m$-$Zn$_2$(dobpdc),
lower panel. The panels show atomic 
coordinates as obtained in fully relaxed CX characterizations that
form the starting points for our performance comparisons and discussion.

We here assert the AHBR (and AHCX) performance by systematically calculating the CO$_2$ binding enthalpy at room temperature $H_{\rm ads}^{\rm room}$ for Mg-MOF-74
and the Born-Oppenheimer (frozen-atom) binding energy
\begin{equation}
E_{\rm ads,BO} = E_{{\rm CO}_2-{\rm MOF}}
-(E_{\rm MOF}+E_{{\rm CO}_2}) \, ,
	\label{eq:eads_mof}
\end{equation}
in the larger (more expensive) case of the diamine-functionalized MOF.
In Eq.\ (\ref{eq:eads_mof}), we evaluate the difference between 
the total energy of CO$_2$-adsorbed MOF ($E_{{\rm CO}_2-{\rm MOF}}$) and the total energy sum of Mg-MOF-74 or m-2-m-Zn$_2$(dobpdc) ($E_{\rm MOF}$) and gas phase CO$_2$ ($E_{{\rm CO}_2}$).
For the gas phase CO$_2$ energy, we optimize the geometry within a 20 {\AA} $\times$ 20 {\AA} $\times$ 20 {\AA} cubic cell. We have verified that our computational set up,
Appendix A, converges the CO$_2$ binding-energy results for Mg-MOF-74 and m-2-m$-$Zn$_2$(dobpdc) adsorptions to 1 kJ/mol. 

To simplify comparison among functionals -- and to limit computational costs -- we provide all calculations with the MOF structures kept fixed at the CX results for the adsorption geometry; This strategy is similar to that we used for discussing the CO/Pt(111) problem, above. 
However, 
in the case of carbon capture in Mg-MOF-74, we permit the CO$_2$ molecules to relax according to the forces that we compute 
in the specific functionals.

We seek to compare with room-temperature measurements of
the CO$_2$ heat of adsorption,
$H_{\rm ads}^{\rm room}$ \cite{CO2MOF74survey,CO2dobpdc1}. The
DFT (and the true) internal-energy binding description are thus affected by vibrational zero-point energy (ZPE) and thermal-energy (TE) corrections:
\begin{equation}
H_{\rm ads}^{\rm room} = E_{\rm ads,BO} + \Delta \hbox{ZPE} + \Delta \hbox{TE} \, .
	\label{eq:hads_mof}
\end{equation}
Accordingly, for the simpler 
Mg-MOF-74 system, we also 
compute and contrast the vibrational frequencies of adsorbed CO$_2$ and free CO$_2$ using a finite difference approach with the Phonopy package \cite{Togo2015}. Specifically,
we displace each atom of CO$_2$ in twelve 
random directions with a constant displacement 
distance (0.03 bohr) to extract corrections $\Delta \hbox{ZPE} + \Delta \hbox{TE}$ that hold at 298 K.

In the larger, more complex diamine-functionalized case, 
we stick with comparing the Born-Oppenheimer results, Eq.\ 
(\ref{eq:eads_mof}), directly
with the measured $H_{\rm ads}^{\rm room} = -57$ kJ/mol value. Based on experience from asserting the performance of
density-explicit vdW-DFs, in Ref.\ \onlinecite{MOFdobpdc}, we expect 
that a plausible back-corrected
experimental value would be 
approximately $-63$ to $-60$ kJ/mol as the computed $\Delta \hbox{ZPE} + \Delta \hbox{TE}$ 
value is about 3--6 kJ/mol (see Ref.~\onlinecite{MOFdobpdc}).

Table \ref{tab:MOF} compares our results,  contrasting the vdW-DF tool chain descriptions
of CO$_2$ uptake in the Mg-MOF-74 
(in the diamine-functionalized MOF) with experiments in the upper (lower) section. In the case of Mg-MOF-74, we find that the 
AHCX result for the Mg$-$O distance
is excellent but that AHBR is also accurate. For adsorption energies
we find that AHCX systematically strengthens the CX binding results in both MOF cases. In the complex 
diamine-functional MOF case, we therefore find that moving to AHCX
does not repair a clear overbinding 
tendency that we have very recently documented for present vdW-DFs \cite{MOFdobpdc}.

In contrast, the results for CO$_2$ adsorption in new RSH vdW-DF shows a
trend of vdW-DF repairing. The AHBR result $H_{\rm ads}^{\rm room}=-46.6$ kJ/mol for the Mg-MOF-74 is in itself excellent, being in close agreement with the experimental value $-43.5$ kJ/mol \cite{CO2MOF74survey}. 
The AHBR outperforms all of the vdW-DFs that we have previously 
tried (see SI material associated with Ref.\ \onlinecite{MOFdobpdc}), for example, lifting the underbinding B86R
$H_{\rm ads}^{\rm room}=-38.9$ kJ/mol value.

Meanwhile, in the case of the diamine-functionalized MOF
\cite{MOFdobpdc}, the AHBR $E_{\rm ads,BO}=-86.8$ kJ/mol result instead
\textit{lowers} the clearly overbinding
B86R $H_{\rm ads}^{\rm room}= -92.1$ kJ/mol
value \cite{MOFdobpdc} towards the \ph{value of the} measurement, at
$-57$ kJ/mol \cite{CO2dobpdc2}. 
According to previous vdW-DF experience
in characterizing vdW-DF vibrational corrections to $E_{\rm ads,BO}$
\cite{MOFdobpdc}, this AHBR characterization leads to the estimate $H_{\rm ads}^{\rm room}\approx -84 $ to $ -81$ kJ/mol. The AHBR therefore outperforms SCAN, SCAN+rVV10 and the recent vdW-DFs (vdW-DF1 and vdW-DF2 are better on energies but have too long binding lengths). The AHBR has an accuracy that matches the semi-empirical rVV10 for this MOF challenge \cite{MOFdobpdc}.
Unlike in the case of Mg-MOF-74, the AHBR does not perform at the revPBE+D3
level for m-2-m$-$Zn$_2$(dobpdc). However, unlike the AHCX, the AHBR is able to move the non-empirical vdW-DFs towards
a binding softening. Robustness, a repairing behavior 
(documented here and for many density-driven challenges), is needed when the vdW-DF method faces significant charge relocations that, in turn, challenge the XC balancing \cite{MOFdobpdc}.

\begin{table}
	\caption{\label{tab:DNA} Comparison of the CX-based and B86R-based tool 
	chain performance for DNA assembly.
	We report MAD values (in kcal/mol) as extracted by averaging the deviations of vdW-DF results relative 
	to coupled-cluster calculations \cite{KrBaSp2019} on the 10 different base-pair
	stacking configurations, using the specific configurations that are provided in 
	Ref.\ \onlinecite{KrBaSp2019}.  We report on ONCV-SG15/160 Ry results, but we also
	compare with literature values as well as an assessment obtained from a separate 
	ultrasoft-PP study (`GBRV').  The reference calculations use
	DLPNO-CCST(T) \cite{RiPiBe2016} to compute the step energy $\Delta E_{\rm WC-step}$ and
	$\Delta E_{\rm B-pair}'$ which is a sum over relevant molecular-pair contributions.
	}
\begin{ruledtabular}
\begin{tabular}{lcc}
XC study & $\Delta E_{\rm WC-step}$ & $\Delta E_{\rm B-pair}'$ \\         
\hline
	B3LYP+D3$^{a}$  & 0.89 & - \\ 
	CX$^{b}$        & 1.73 & 0.44\\ 
	CX$^{\rm GBRV}$ & 1.17 & 1.49\\ 
\hline
	CX              & 1.48 & 1.75\\ 
	AHCX            & 3.06 & 3.30\\ 
    AHCX$_{0.25}$   & 3.47 & 3.69\\ 
\hline
	B86R            & 0.38 & 0.82\\ 
	AHBR$_{0.20}$   & 0.12 & 0.60\\ 
	AHBR            & 0.08 & 0.52\\ 
\hline
$^a$Ref.\ \onlinecite{KrBaSp2019}. \\
$^b$Ref.\ \onlinecite{KiGoRo2020}. \\
\end{tabular}
\end{ruledtabular}
\end{table}

\subsection{Base-pair stacking in a DNA model}

DNA can be seen as a stacking of Watson-Crick (WC) base pairs that are essentially
flat and therefore have a significant (eV-scale) vdW attraction \cite{chakarova-kack06p146107,chakarova-kack06p155402,hooper08p891,cooper08p1304,cooper08p204102,toyoda09p2912,toyoda09p78,berland11p135001,le12p424210} 
from one base pair to the next.  The WC pairs are steps in the resulting
double-helix DNA structure.  There are in total 10 possible combinations for
2 steps, i.e., base-pair combinations that are here denoted ApA, ApT, ApC, ApG, 
CpC, CpG, GpC, TpT, TpC, and TpG; Ref.\ \onlinecite{KrBaSp2019} identifies and illustrates
a set of possible atomic positions for these base-pair combinations. 
We want to compute such base-pair stepping energies \cite{cooper08p1304,cooper08p204102} since the mutual vdW attraction 
might have driven the DNA self-assembly in the first place, as life emerged \cite{Cafferty2013,Smith2018}.

Some of the DNA cohesion comes, of course, from the presence of the sugar-phosphate 
backbones that incorporate and organize the WC bases into two strands, with the 
sequence of WC bases, adenine (A), thymine (T), cytosine (C), or guanine (G),  
setting the genetic code.  The strands are complementary in the sense that the WC-base sequence must be exactly matched, with each of the 
individual bases having only one suitable counter part, i.e., forming steps that must have 
one of the A-T, T-A, C-G, or G-C forms.  The base-pair-combination nomenclature, ApA 
though TpG, reflects the observation that it is sufficient to track the code sequence
on one of the DNA strands.  The DNA strands are mutually bonded, by a combination of 
hydrogen and vdW binding \cite{JiScHy18a}, but the energy of the WC pairings
(A-T, T-A, C-G, G-C, among one base and its counter part directly across)
is not our present focus.  Instead, we seek to understand the extent that the mutual
step-binding energies contribute to the DNA cohesion, using a
DNA model that ignores the backbone but instead relies on hydrogen 
terminations of the bases \cite{cooper08p1304,cooper08p204102,KrBaSp2019}.

To set us up for future, more general DNA explorations, we contrast the
performance of the first and second vdW-DF tool chains relative
to the reference descriptions provided in Ref.\ \onlinecite{KrBaSp2019}. That
is, we compare with so-called domain-based pair natural orbital couple-cluster 
(`DLPNO-CCSD(T)') calculations \cite{RiPiBe2016} at fixed reference base-pair 
combination structures \cite{KrBaSp2019}. The DLPNO-CCST(T) method is also 
used for setting reference energies of the GMTKN55 suite \cite{gmtkn55}. We 
compute the base-pair stepping energies as total-energy differences between 
the full system and the two WC base pairs, for example, in the case 
of the ApC combination
\begin{equation}
	\Delta E_{\rm WC-step}^{\rm ApC} = E_{\rm ApC} - E_{\rm A-T} - E_{\rm C-G} \, .
	\label{eq:WCstep}
\end{equation}
The reference work \cite{KrBaSp2019} also provides data for the sum of pair interactions among the four bases, $\Delta E_{\rm B-pair}'$, excluding the two WC pairings (as indicated by the prime).
This pair summation is illustrated in the abstract figures of Refs.\ \onlinecite{KrBaSp2019, KiGoRo2020}.

Tables S XVI and S XVII of the SI material compares
the performance of CX/AHCX/AHCX$_{0.25}$ and of 
B86R/AHBR$_{0.20}$/AHBR for each of the base-pair combinations,
reporting (in kcal/mol) the $\Delta E_{\rm WC-step}$ 
and $\Delta E_{\rm B-pair}'$, respectively. 
We also report mean deviation (MD)
and MAD values relative to the DLPNO-CCSD(T) calculations \cite{KrBaSp2019}.
Table \ref{tab:DNA} summarises the performance comparison (in terms 
of 10-base-pair-combination averages) and makes it clear that the AHBR is
a strong performer for the description of the DNA base-pair assembly.

We find that B86R performs better that CX and that the AHBR functional
design is in fact very accurate also for descriptions of the DNA stepping energies.
This is especially true when it is used in the suggested default mode 
with a 0.25 Fock-exchange mixing. The description is significantly more
accurate than a standard choice of dispersion-corrected hybrid DFT
provides, as also listed in Table \ref{tab:DNA}.  This result is in 
itself encouraging. 

We also observe that the AHBR is somewhat
less accurate when it is instead used to study 
energies from the sum of pair contributions, $\Delta E_{\rm B-pair}'$.
As in Ref.\ \onlinecite{KiGoRo2020}, we 
find that the vdW-DF based descriptions of the stepping energies 
benefit from a cancellation of errors that
affect the descriptions of pairing between individual bases.

Finally, Table \ref{tab:DNA}  reveals an important difference
between the performance trend in the CX-AHCX and 
B86R-AHBR functional chains for this class of large
molecular problems. We find that moving to a hybrid 
form significantly worsens the CX/AHCX descriptions, whereas
we find that such a step slightly improves the B86R/AHBR-type 
description. In both tool chains, we find that including and 
increasing Fock exchange fraction strengthens the interactions
but the changes are small in the AHBR case. Including Fock
exchange makes AHBR very accurate but the most important
lesson is perhaps that the B86R-AHBR functionals have an
inherent stability here: They start and remain 
accurate on the DNA assembly energies.

\section{Conclusion and outlook}

We have developed a new, accurate, non-empirical RSH vdW-DF, termed vdW-DF2-ahbr, and we have documented general-purpose capabilities for molecular problems 
as well as promise for bulk and adsorption properties. Since AHBR is based on a range separation of the Coulomb interaction, it is setup (and coded) to allow use of a physics-based tuning of both the Fock-exchange
mixing $\alpha$ and of the RSH inverse screening length $\gamma$. This means that it is also possible 
to ameliorate residual RSH vdW-DF errors in charge-transfer descriptions \cite{WiOhHa21}. In this first presentation, however, we have deliberate kept these values fixed to allow a simple demonstration of broad AHBR capabilities and usefulness.

Figures \ref{fig:TSwDiel} and S.1-2 of the SI materials highlight two important take-away messages of this paper, namely that: 1) use of RSH vdW-DFs, and of the new AHBR in particular, provides substantial  accuracy improvements over regular, density explicit vdW-DFs for broad molecular properties, and 2) use of the B86R-based hybrids, i.e., the DF2-BR0 and the here-defined AHBR, provides an evenly robust performance, heightening the accuracy over all types of molecular problems. 

\ph{Figures \ref{fig:TSwDiel} and \ref{fig:NOCwDiel} exemplify a key conclusion about the AHBR advantage: It has a robust ability to navigate so-called density-driven DFT errors \cite{DDerrorQaA}.}

\ph{This vdW-DF2-ahbr resilience, 
in combination with its emphasis on a
MBPT foundation, suggests} that it will be accurate also beyond the successes that we have here documented for a set of DFT challenges in molecule, \ph{layered,} bulk, and surface systems. When using formal MBPT to compute the total energy, there is an inherent robustness towards making approximations \cite{WarLut}. That robustness also extends to the exact XC energy functional \cite{ShaSch}. We rely on MBPT guidance in making XC energy approximations (such as CX, AHCX, and AHBR) so that these vdW-DFs can potentially benefit from that inherent robustness. However, as we also discuss, such benefits can only emerge in practice when the actual XC functional approximation delivers accurate orbitals, for example, as tested on the quality of its density description \cite{DDerrorQaA}.  
That AHBR matches or exceeds B3LYP/HSE+D3 for transition-state problems means that it generally navigates density driven errors and therefore outperforms the B3LYP/HSE+D3 broadly
\ph{ -- for example, as summarized in Figs.\ S 1-2 of the SI material}. Like the HSE+D3 it has both a strong resilience to density errors and the MBPT foundation to benefit from the formal-MBPT robustness towards XC-functional approximations \cite{WarLut,ShaSch}.

We argue that this general-purpose character suggests that AHBR should be used to map strengths and weaknesses of the vdW-DF method on molecules, just like the AHCX can serve us in that role for bulk systems \cite{Hard2Soft}.

An overall outcome of this vdW-DF2-ahbr work is also a 
roadmap with a DFT-usage feedback strategy for making 
further functional improvements in the vdW-DF framework. 
Since both 
the AHCX and AHBR are found to be fairly robust in all 
tested problems, we can contrast performance differences 
over a broad range of systems. Furthermore, since the pair 
of RSH vdW-DFs have systematic design differences, we can correlate the performance 
differences in terms of the nature of the underlying 
physics input to the XC designs. \ph{The 
AHBR and AHCX are particularly useful  
because they are complementary, representing one of two internally-consistent classes of MBPT input on how exchange effects impacts all XC components. Specifically, as explained in Section II.A and II.B, the AHBR and AHCX rely on valid 
but different interpretations of formal MBPT; They correspond to systematic reliance of molecular or a weakly-perturbed-bulk perspective on 
screening, respectively).}
Taken together, 
the observations allow us to interpret the DFT-usage feedback 
and draw development conclusions concerning which 
types of MBPT input to prioritize. This DFT-feedback 
strategy is in many ways just a continuation of the 
electron-gas tradition that, as we see it, has 
pushed MBPT-based DFT from LDA over constraint-based 
GGAs and to the vdW-DF method.

\section*{\label{sec:ack} Acknowledgement}

We thank Carl M.\ Frostenson 
for useful discussions.
Work supported by the Swedish Research Foundation, 
through Grants 2018-03964  and 2020-04997, the 
Swedish Foundation of Strategic research, through 
Grant IMF17-0324, the Sweden’s Innovation Agency 
Vinnova, through project No. 2020-05179, the Chalmers 
Area-of-Advance Production as well as Excellence Initiative Nano,
the U.S.\ Department of Energy, Office of Science, Office of
Basic Energy Sciences under Award DE-SC0019992, the KIST Institutional Program, 
Project No. 2E31801, and the program of Future Hydrogen Original Technology Development, 
Project No.\ 2021M3I3A1083946, through the National Research Foundation of Korea (NRF), funded 
by the Korean government, Ministry of Science and ICT (MSIT), as well as  the Institute
for Information and Communications Technology Promotion, through Project No.\ 2021-0-02076,
funded by the Korean Government (MSIT). The authors also acknowledge computer allocations 
from the Swedish National Infrastructure for Computing (SNIC) 
under Contracts SNIC2019-2-19, SNIC 2020-3-13, SNIC2021-3-18, from
the Chalmers Centre for Computing, Science and Engineering (C3SE),
as well as from the KISTI Supercomputing Center, Project No. KSC-2020-CRE-0189
and the National Energy Research Scientific Computing Center (NERSC), a U.S.\ Department 
of Energy Office of Science User Facility under Contract No.\ DE-AC02-05CH11231.

\appendix

\section{Computational details}

All calculations are carried out using the QE code suite \cite{QE,Giannozzi17,PaoloElStruct1},
using an in-house coding for the AHBR design.
The AHBR code will be released to QE once 
the paper have been accepted for publication.

For hybrids, we used the adaptive-compressed
exchange (ACE) implementation \cite{LinACE,PaoloElStruct1} 
to speed up the Fock-exchange evaluation (except in a few cases
-- ionized Li, Na, and K atoms --  where it seemed to
prevent an easy convergence).  We use
the spin vdW-DF formulation \cite{Thonhauser_2015:spin_signature} (for regular and hybrid vdW-DFs
\cite{Hard2Soft}), when relevant.

We systematically use the ONCV-SG15 \cite{ONCV,sg15} set of 
PPs at a 160 Ry wavefunction-energy cutoff 
for all of the here-reported \ph{molecular,
layered-material,} and bulk benchmarking as well as 
for demonstrator work on base-pair stepping energies in DNA assembly 
and for a description of CO/Pt(111) adsorption. 
To document the PP sensitivity, we also provide a characterization
of the DNA base-pair stepping energies  using the ultrasoft GBRV PP set \cite{GBRV}
at 50 Ry wavefunction energy cutoff and 400 Ry density cutoff.
For a RSH vdW-DF demonstration on 
green-technology problems, 
we calculate the  CO$_2$ binding enthalpy in Mg$_2$(dobdc)
\cite{CO2MOF,CO2MOF74survey}, known as 
Mg-MOF-74, and in the diamine-appended or -functionalized
m-2-m$-$-Zn$_2$(dobpdc) \cite{CO2dobpdc1,CO2dobpdc2,MOFdobpdc} using the ONCV-SG15 PPs with a wavefunction energy cutoff of 220 Ry.
 
\begin{table}
	\caption{Size-convergence of functional-performance assessments: Pilot-study comparison of characteristic
	WTMAD1 values \cite{gmtkn55,DefineAHCX} (in kcal/mol) as obtained for `GMTKN53', a 53-benchmark part \cite{DefineAHCX} of the GMTKN55 suite \cite{gmtkn55}.
	We contrast the planewave performance characterizations that result when using a standard 10 {\AA} and 
	a larger 15 {\AA} choice of vacuum padding (see text). The convergence tests are done for the \textsc{AbInit} PP set \cite{abinit05}, at 80 Ry wavefunction 
	cutoff.	We focus on the `easily accessible GMTKN53' subsuite \cite{DefineAHCX} that excludes (as indicated by 
	asterisks `*') the G21EA benchmark set from the GMTKN55 \cite{gmtkn55} group 1 
	and the WATER27 set from the GMTKN55 group 4. This is done because these sets require a separate 
	dielectric handling \cite{NicolaENVchem,NicolaENVprb}, see Appendix A. 
	\label{tab:size-convergeGMTKN53}
	}
\begin{ruledtabular}
	\begin{tabular}{lcc}
		Tests & CX({10 {\rm {\AA}}}) & CX({15 {\rm {\AA}}}) \\
		\hline
		Group 1*        & 5.19 & 5.20 \\
		Group 2         & 4.93 & 4.92 \\
		Group 3         & 7.22 & 7.22 \\
		Group 4*        & 3.36 & 3.35 \\
		Group 5         & 3.78 & 3.79 \\
		Group 6 (4* and 5) & 3.55 & 3.55 \\
		GMTKN53 (GMTKN55*)   & 4.79 & 4.79 \\
	\end{tabular}
\end{ruledtabular}
\end{table}

For the molecular benchmarking we use a $\Gamma$-point-only 
wavevector sampling and we employ the same cubic unit-cell 
size for all systems within a given benchmark set.  The size 
is determined automatically in a python setup of QE
input files. This is done by first finding the largest 
Cartesian-coordinate extension (among all set-specific 
problems) and then systematically adding an extra 10 {\AA} vacuum in all directions; This vacuum padding ensures that there is at least
10 {\AA} between the largest $x/y/z$ position
of one image to the smallest $x/y/z$ position
in the next in all of the GMTKN55 benchmarks. 
Having a cubic cell,
we can use the Makov-Payne correction \cite{Makov} to help control spurious electrostatics coupling among the periodic
images in our planewave setup.

Appendix B describes the nature of our strategy to secure convergence at the 0.01 kcal/mol level for benchmarking across
the full GMTKN55 suite. The argument has three steps. First we show 
 -- in a pilot survey relying on the
more electron-sparse
\textsc{AbInit} PPs \cite{abinit05} at 80 Ry -- 
that the impact of
false vdW attraction on the per-benchmark MAD values 
never exceeds 0.01 kcal/mol, see Table S XII
of the SI material; Here we use an in house code to
extract the asymptotic vdW-DF interactions (among unit cells)
that, as discussed in Refs.\ \onlinecite{Dion,kleis08p205422,NanovdWScale},
can be formulated in terms of per-unit-cell effective molecular 
$C_6^{\rm mol}$ coefficients and a molecular $E_{\rm vdW}$ attraction-energy
estimate \cite{NanovdWScale}.

Next, Table \ref{tab:size-convergeGMTKN53} summarizes
our survey of the impact caused by spurious intercell
electrostatic couplings, that we find is also limited 
at 0.01 kcal/mol. This documentation was again done in 
a pilot-study approach (using the \textsc{AbInit} PPs \cite{abinit05}) 
by direct QE calculations, using the automatic setups but changing the 
assumed vacuum padding from 10 to 15 {\AA} (as indicated by subscripts). 
The comparison is provided for a subset (the easily-accessible `GMTKN53' \cite{DefineAHCX})
of the GMTK55 suite, that is, excluding (in the GMTKN55 groups 
identified by asterisks `*')  the G21EA and WATER27 benchmark sets
and suitable adapting the WTMAD1 measures \cite{DefineAHCX}.

Finally, the corner-stone of our here-presented
molecular-benchmarking strategy and the main message of 
Appendix B is the following. While negative
ions and small radicals -- in the G21EA and WATER27
benchmarks -- present fundamental computational 
challenges \cite{BurkeSIE}, we can still complete
well-converged full-GMTKN55 assessments in planewave DFT.
The challenges exist because small 
negative systems have pronounced SIEs \cite{BurkeSIE} that make it 
impossible (impractical) to complete a direct planewave
assessment on the G21EA (WATER27) set \cite{DefineAHCX,Hard2Soft}.
Specifically, the highest-occupied level of some small charged systems will 
eventually be pushed above the vacuum level of the planewave-DFT 
potential, exactly because we seek size convergence \cite{NeatonLimit}, see 
Refs.\ \onlinecite{BurkeSIE,DefineAHCX} and 
Appendix B.  However, the electrostatics-plus-SIE nature of the 
challenge \cite{BurkeSIE,DefineAHCX} suggests the workaround: We 
pursue planewave-GMTKN55 benchmarking in the presence 
of fictitious dielectric constants, $\varepsilon_\infty>1$, and then
adiabatically remove this environment perturbation. 

For our survey of the bulk-structure performance 
of AHBR we use again the ONCV-SG15 PPs at 160 Ry,
now with an $8\times8\times8$ $k$-point
sampling (keeping all $k$-point differences
in the ACE-based Fock-exchange evaluation \cite{LinACE,PaoloElStruct1} 
for hybrids).  This choice of setup permits  direct
comparisons with previously reported  CX and AHCX results \cite{DefineAHCX}.

\ph{Comparing with DMC results \cite{Spanu09p196401,GaKiPa2014,MoDrFa2015,ShKiLe2017,HsChCh2014} for the graphite
crystal (graphene, $\alpha$-graphyne, and hBN bilayers), we use
an $8\times8\times6$ ($8\times8\times1$) 
$k$-point sampling, keeping half of the 
$k$-point differences in the Fock-exchange
evaluations, generally using CX to first establish a description of the in-plane atomic structure (that is kept fixed while we vary the layer separation and compute the energy variation in different regular and RSH vdW-DFs).} 

\ph{For a study of the layered phosphorus crystal \cite{ShBaZh2015}, we note that
one of the in-plane lattice constant is sensitive to pressure \cite{Worlton1979}
and hence, likely, to the functional approximation.  We therefore keep
the in-plane lattice constants fixed
at the experimentally observed values \cite{Worlton1979}, $a=3.3133$ {\AA} and $c=4.374$ {\AA}. Additionally,
we first determine the intra-P-layer structure using CX at that given
unit-cell description. This gives an in-layer atom configurations that agrees to within 1\% of the experimental distance and angles reported in Ref.\ \onlinecite{Worlton1979}.
The CX-specified atomic configuration for the individual layer is kept fixed as we subsequently compute the crystal total-energy variation at varying choices of the cell height $2d$ (where $d$ is set by the layer-separation definition that is 
used in the DMC study \cite{ShBaZh2015}).}

For the CO/Pt(111) adsorption demonstrator of AHBR usefulness,
we use a 6-layer surface slab (with a 2-by-2 in-surface
repetition) together with a $6\times 6 \times1$ 
$k$-point sampling (here keeping a $3 \times 3\times 1$
grid of $k$-point differences in hybrids studies), as before with use of the ONCV-SG15 setup.
The B86R (and AHBR) is found to have slightly larger lattice constants
than the CX and AHCX descriptions. That may favor AHBR over AHCX in terms of accurate predictions of the adsorption site
preference, although the lattice constant-differences are small. 
We eliminate this indirect effect by making the tool-chain 
comparison at the frozen geometry that results when CX tracks 
the adsorption-induced relaxations. We note that 
CX is accurate on the Pt lattice constant and on
the elastic-energy description \cite{DefineAHCX}. Our
frozen-geometry approach can therefore be considered 
a good model of the actual CO adsorption problem.

For the carbon-capture-usage illustration, we characterize
CO$_2$-in-Mg-MOF-74 adsorption and CO$_2$ insertion in diamine-functionalized m-2-m$-$Zn$_2$(dobpdc), focusing on
adsorption energies and using a $\Gamma$-point sampling of the Brillouin-zone for Mg-MOF-74 and a $1\times1\times3$ $k$-point grid sampling for m-2-m$-$Zn$_2$(dobpdc). For the Mg-MOF-74 case we
also determine CO$_2$ vibrations
and can thus compare directly to
room-temperature observations of the
heat of CO$_2$ adsorption.
Here we first use CX to compute the unit-cell structure and identify the adsorption geometry. For the Mg-MOF-74 case we 
proceed to determine functional-specific relaxations for the CO$_2$ molecules, using the access to forces; 
For the significantly  larger m-2-m$-$Zn$_2$(dobpdc) system, we 
rely systematically on a CX characterization
of the adsorption. We characterize the CO$_2$ vibrations in the Mg-MOF-74 in a finite-distortion setup using \textsc{phonopy} \cite{Togo2015}. 

Finally, we provide illustrations of using AHBR (and AHBR/B86R 
tool chain) in both a biomolecular and a carbon-capture setting.  
For the biomolecule illustration, we compare 
with coupled-cluster reference calculations for a model 
that replaces the DNA backbone with hydrogen terminations
on the individual bases \cite{cooper08p1304,cooper08p204102,KrBaSp2019}.
Here, we perform $\Gamma$-point calculations for the set of 10 different 
two-base-pair combinations (at representative geometries identified in 
Ref.\ \onlinecite{KrBaSp2019}) using a 30 {\AA}-cubed unit cell and
electrostatic decoupling. In effect, we thus assert the AHBR on energies 
that characterize the DNA assembly.

\section{Planewave molecular benchmarking}

This appendix summarizes our strategy for completing 
a planewave molecular benchmarking across the full GMTKN55 
suite \cite{gmtkn55}. In particular, we explore the nature 
and impact of convergence-related factors that make 
benchmarking challenging, but which, as we show, can 
also systematically be handled and circumvented. As 
such, the appendix motivates and validates the 
molecular benchmarking that we also report in this paper.

\subsection{Impact of spurious vdW attraction}

Our focus on nonlocal-correlation functionals presents, in principle, 
a problem for planewave benchmarking: The vdW attraction is long-ranged and
causes at least some spurious coupling between the periodically repeated images. 
However, we can document that the net impact by false, intercell vdW
coupling on each of the GMTKN55 benchmarks is vanishing, well within our
overall benchmarking target of 0.01 kcal/mol \cite{NeatonLimit}. To that
end, we consider the asymptotic limit of the vdW-DF method \cite{Dion,kleis08p205422,NanovdWScale}.

We proceed as follows. First, for the atomic configurations of molecules or cluster
(in each of the roughly 2450 different atomic configurations that define 
the GMTKN55 suite \cite{gmtkn55}), we evaluate an effective 
molecular/cluster $C^{\rm mol}_6$-interaction coefficient. 
These coefficients characterize the spurious vdW couplings in 
the asymptotic vdW-DF limit that is defined 
and used in Refs.\ \onlinecite{kleis08p205422,NanovdWScale}. Here we use
an in-house extension of the QE code suite, starting on a set of CX 
calculations for the self-consistent electron density variation.
This initial step was done in a pilot study, using 
the more electron-sparse \textsc{AbInit} PPs \cite{abinit05} at 80 Ry 
cutoff, and typically in our standard benchmarking setup, i.e., with 
a minimum of 10 {\AA} vacuum padding in the choice of unit cells. 
However, we used smaller unit-cell sizes in the G21EA set and for the 
negatively charged OH$^-$ radical in the WATER27 set \cite{Hard2Soft},
for reasons that are further discussed below.

We also compute (for every unit-cell problem in GMTKN55) a
corresponding interaction-energy contribution to a given unit cell `i' \cite{NanovdWScale}, 
\begin{equation}
	E_{\rm vdW,i}  = - \sum_{j\neq i} \frac{C_6^{\rm mol}}{|\mathbf{R}_j-\mathbf{R}_i|^6} \, .
	\label{eq:vdWDFasym}
\end{equation}
Here, $\mathbf{R}_j$ denotes a Bravais vector of the periodic computational setup 
in our QE calculations. We observe that this measure of asymptotic vdW attraction,
Eq.\ (\ref{eq:vdWDFasym}), is different from a full vdW-DF method study \cite{kleis08p205422},
on fundamental reasons that are discussed in Refs.\ \onlinecite{hybesc14,NanovdWScale}.
We would, for example, not get exactly the same values if we tracked the
spurious coupling in a super-cell study. However, 
we are setting up of molecular benchmarking with a large vacuum padding, 
see Appendix A. The use of Eq.\ (\ref{eq:vdWDFasym}) is therefore an acceptable
approximation that can gauge the expected magnitude of spurious vdW coupling in our 
planewave calculations.

Table S XII illustrates how we use Eq.\ (\ref{eq:vdWDFasym}) to 
validate that the spurious-vdW impact on our benchmarking is truly neglectable. 
From Eq.\ (\ref{eq:vdWDFasym}) we determine the extent that the spurious 
intercell-vdW coupling
is causing an offset on the periodic-cell vdW-DF results, for each of the 1500  molecular-process energies in the GMTKN55 suite \cite{gmtkn55} (roughly 2450 single-point calculations per functional). 
Next, we define a set of 55 per-benchmark MAD-offset values by tracking
the offset relative to the reference energies. 
Table S XII reports these
benchmark-specific error estimates  in descending order (truncated to five sets).
The largest impact seems to appear for the G21EA set, but that impact-estimate is also an exaggeration. This is
because we dumped the pilot study to 
a reduced unit-cell size 
(8-10 {\AA} total) whenever we faced convergence issues
for the negatively charged ions and small radicals in the 
G21EA set \cite{BurkeSIE,DefineAHCX,Hard2Soft}.
Table S XII makes it clear 
that the false-vdW impact on our planewave molecular 
benchmarking is significantly less that 0.01 kcal/mol.

\subsection{Impact of spurious electrostatic coupling}

Table \ref{tab:size-convergeGMTKN53} provides documentation that we use
a sufficiently large planewave-benchmarking setup.
The table summarizes the following testing that
our use of a standard 10 {\AA} vacuum padding
(in our automatic scripting of QE input files)
is sufficient to ensure an electrostatic decoupling 
among the images in the periodic-cell setup, bringing
the assessment error down to a desired 0.01 kcal/mol
limit \cite{NeatonLimit} for averaged measures.

This validation of our benchmarking strategy was done
brute-force, that is, by repeating a standard CX testing 
with one where instead we use a 15 {\AA} vacuum padding.
Again we use the \textsc{AbInit}-PP setup
(at 80 Ry) but otherwise following the regular
problem specifications for all of the individual problems.
However, the focus is here limited to the
easily-accessible `GMTKN53' subset of the GMTKN55, 
excluding the G21EA and WATER27 sets, as 
was also done in Ref.\ \onlinecite{DefineAHCX}. 

We find, Table \ref{tab:size-convergeGMTKN53}, that
the net impact of going to a truly large vacuum padding
in the setup, is bounded by 0.01 kcal/mol. Here the 
assessment is done in terms of the weighted WTMAD1 measures
as resolved on the GMTKN55 bench groups. As indicated by
asterisks `*', the reported WTMAD1 measures are slightly
adjusted (as described in Ref.\ \onlinecite{DefineAHCX})
due to the `GMTKN53' focus. We deem use of our standard (10 {\AA}) vacuum padding
setup, Appendix A, validated also in terms of securing
sufficient electrostatic decoupling \cite{NeatonLimit}.

\begin{figure}
	\includegraphics[width=0.98\columnwidth]{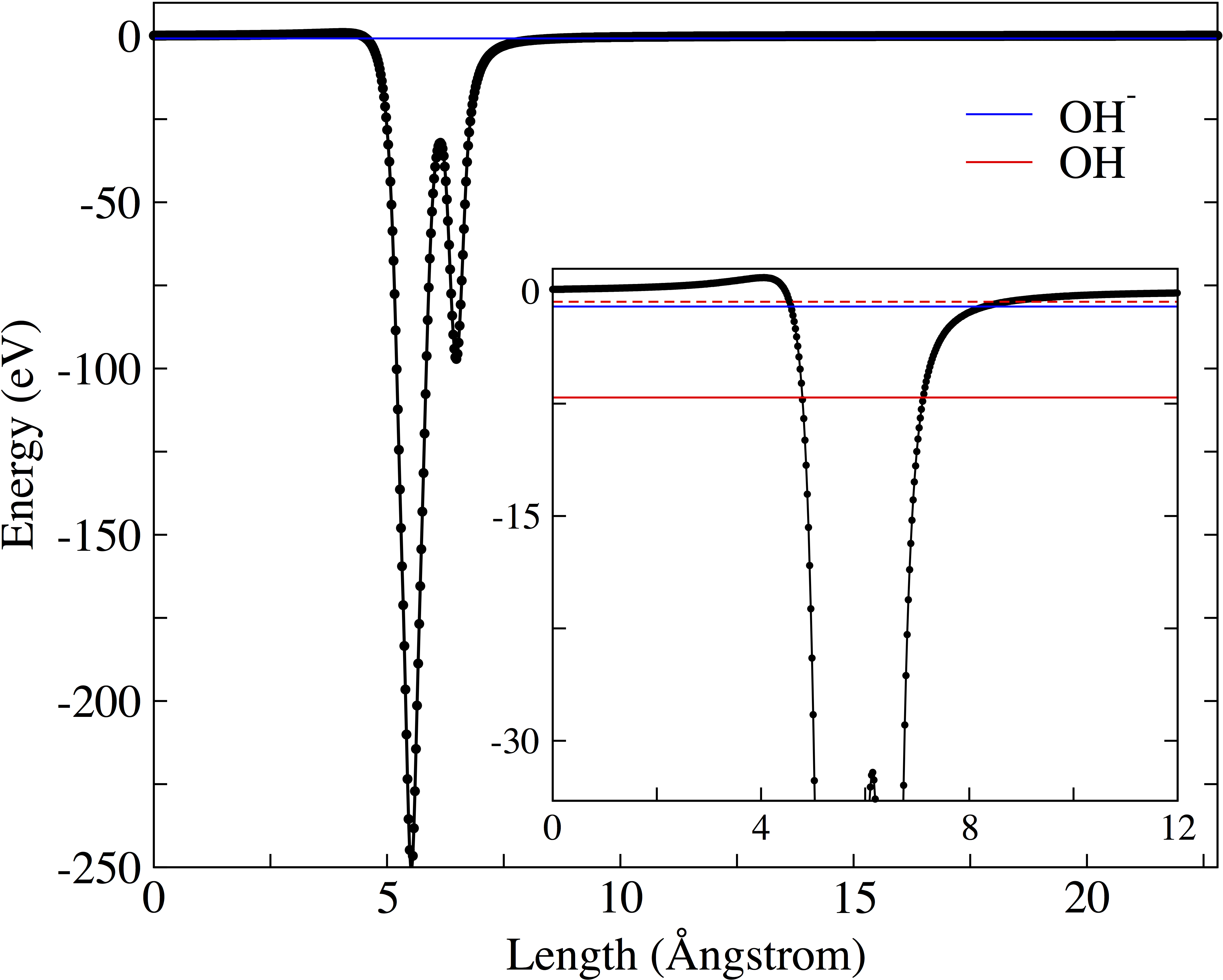} 
	\caption{\label{fig:OHmIllustrate} Convergence challenge for the
	OH$^-$ radical producing potential density-driven DFT errors.
	The instability is here driven by SIE effects \cite{BurkeSIE}
	and the potential error is evident by noting a closeness of the 
	highest-occupied molecular orbital (HOMO) energy position and 
	the vacuum level, i.e., the zero-energy value. Convergence of 
	negatively charged ions and radicals is a challenge not only in 
	this CX calculation but for all XC functionals (even hybrids \cite{DefineAHCX}) 
	since we seek a proper size convergence and therefore push our planewave 
	calculations close to the complete-basis set limit \cite{BurkeSIE}.  The figure shows
	the self-consistently computed electrostatic potential along a ray containing the 
	nuclei (potential dips) and OH$^-$ bonding region.  The \textsc{QE} 
	approximation for the vacuum level is found at the plateau value, here located just above
	the OH$^-$ HOMO level (solid blue horizontal bar). The insert provides a closer 
	look, showing the potential barriers surrounding the molecular regions
	and tracking differences between the neutral OH lowest-unoccupied molecular
	level (broken red bar) and the OH$^-$ HOMO level.
	}
\end{figure}

\subsection{Instabilities driven by self-interaction errors}

A planewave benchmarking setup like ours can technically only reach a true 
complete-basis limit \cite{NeatonLimit} when we 
document size convergence, 
at least in principle, to the infinite-size limit. Effectively, the QE 
planewave code sets the average potential to zero \cite{ChiDFT22}. This 
means that, for any finite unit-cell size, the potential value in  
regions far from the nuclei (in isolated molecule problems) lies 
slightly above the true vacuum floor.
A SIE impact, that  arises in small negative ions \cite{BurkeSIE},
will, in part, be masked by this potential offset in the 
planewave code; For sufficiently small unit-cell sizes one can even
craft a fictitious electron trapping in general XC functionals, even
if non-hybrids may not actually be able to trap it at all in a fully converged 
description \cite{BurkeSIE,DefineAHCX,Hard2Soft}.  Unfortunately, a direct 
discussion of convergence with unit-cell size for all of the GMTKN55 is 
of limited meaning.

Figure \ref{fig:OHmIllustrate} reports a successful (but absurdly
tedious and difficult) electronic-structure convergence that we 
provide for CX (in a plain QE) for the OH$^-$ 
system, as described in our setup for characterizing it in the WATER27 set.
Here the use of our automatic (benchmark-specific) setting of the unit-cell 
size leads to use of a 22.8 {\AA} cell size. This is a highly challenging choice of the unit cell (in terms of convergence)
for a negatively charged system, as represented in a planewave code.  The figure shows the variation 
of the electrostatic potential along the O-H axis, together with the energy position 
of the highest occupied molecular (HOMO) level of the negatively charged radical.  The insert focuses on 
the near-molecule regions and makes it clear that the HOMO level of OH$^-$ is similar to but cannot be approximated 
by the lowest-unoccupied molecular level of the corresponding (spin-polarized but neutral) OH system. 
The vacuum level of the planewave calculation sits here at 0.1 eV above
the vacuum floor for description of a truly isolated molecule; This value is estimated by 
the value that the potential attains at a point furthest from the atoms.  The electronic-structure convergence
indicates a trapping of the HOMO level at $-0.7$ eV, well below the vacuum level. Thus, we can ascribe some, 
but not a complete, trust in this CX-based characterization of the SIE impact \cite{BurkeSIE} on 
the OH$^-$ system.

Still, the convergence of the OH$^-$ HOMO level was a cumbersome process involving a 
convergence ruse: Starting at a low convergence criteria and a significant temperature-like smearing 
\ph{we submitted a sequence of DFT calculations that gradually increased the criteria on density consistency 
and lowered the smearing}. We find that we must protect against electron defections even after the KS convergence 
algorithm began to trust the existence of an actual HOMO level, and the process takes weeks
of human time.  The brute-force ruse approach, giving Fig.\ \ref{fig:OHmIllustrate},
is not practical for a systematic exploration of convergence with unit-cell size for the
OH$^-$ radical.  Worse, an attempt to port such a direct, brute-force strategy to the G21EA set will 
fail on fundamental grounds \cite{BurkeSIE,DefineAHCX}.

\begin{figure}
	\includegraphics[width=0.96\columnwidth]{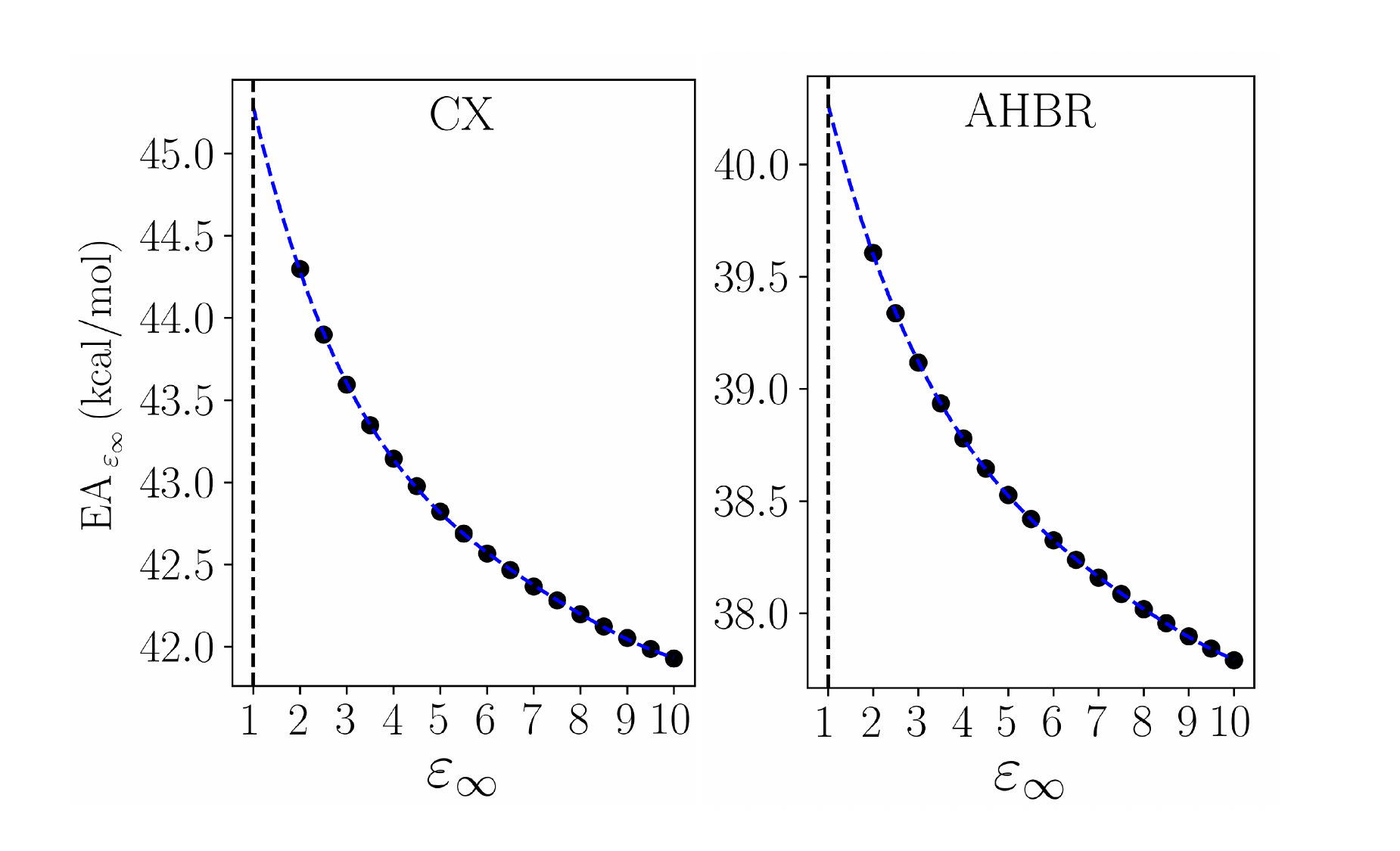}
	\caption{\label{fig:WATERwDiel} 
	Robust, environment-extended \textsc{QE} determination 
	of the OH$^-$ electron-affinity energies `EA' in CX (left panel) and hybrid AHBR (right panel). 
	The choice of a fictitious vacuum-dielectric constant $\varepsilon_\infty>1$ counteracts \cite{NicolaENVchem} 
	the SIE-driven instability \cite{BurkeSIE}. The resulting (environment-adjusted)
	OH$^-$ electron-affinity energies ${\rm `EA'}_{\varepsilon_\infty}$ (shown as dots) vary with $\varepsilon_\infty$ but have a rapid 
	convergence with the choice of the cell dimensions. 
	}
\end{figure}

We note that these SIE-impact problems are important for DFT usage, yet
a general planewave-DFT characterization appears impossible \cite{BurkeSIE}.
We need to track and understand the problems as they relate to charge-transfer processes and 
they point to potentially deeper SIE problems of XC functionals. 
However, to quantify SIE impact on the G21EA and WATER27 benchmark sets, 
we do need to actually trap the HOMO level ---  and that is not directly 
possible in a complete-basis approach for some negative ions \cite{BurkeSIE} (and small radicals).

There is a good solution available from analysis \cite{NicolaENVprb,NicolaENVchem}, given the electrostatic nature of the SIE-impact problems in
G21EA and WATER27 benchmarking
\cite{BurkeSIE}. Trapping the HOMO level in a negatively charged 
small system is difficult, see
Fig.\ \ref{fig:OHmIllustrate}, because the electrostatic potential, on the one hand, 
must overshoot to positive values and, on the 
other hand, will be set by an (unphysical) 
exponential decay in the cross-over 
to that asymptotic-repulsion region \cite{BurkeSIE}. As a consequence, the trapping 
region becomes excessively narrow, pushing the would-be HOMO level above the (true) 
vacuum level, in some cases \cite{BurkeSIE}. Meanwhile, planewave DFT sets the average 
potential to zero, see for example Ref.\ \onlinecite{ChiDFT22}, giving an asymptotic 
potential variation with a floor that will approach the true vacuum (and thus cause QE problems) as we push for size convergence.
Frustratingly, our planewave benchmarking \textit{appears} to be foiled by its 
very strength, namely that we can fairly easily approach the complete-basis-set 
limit. 

Fortunately, the electrostatic nature of the problem also gives us a way to leverage the 
planewave advantages for size-converged G21EA, WATER27 benchmarking, and hence for 
complete GMTKN55 assessments.  The idea \cite{NicolaENVprb,NicolaENVchem} is 
to 1) introduce a control knob that systematically affects the vacuum position 
in QE, 2) obtain well-defined G21EA/WATER27 assessments as a function of 
the control-knob value, and 3) extrapolate these estimates as we turn off the controlled stabilization. We simply do perturbation theory in reverse.

In practice, we rely on an electrostatic-environment extension \cite{NicolaENVchem} of
the QE code suite. Use of a fictitious dielectric-constant $\varepsilon_\infty > 1$ softens the repulsion 
in the electrostatic potential for the HOMO-level trapping. For sufficiently large
$\varepsilon_\infty$ values we reduce the spurious quantum-confinements effects \cite{BurkeSIE} and can thus obtain a SIE-robust G21EA/WATER27 assessment that works at general unit-cell sizes. 

Figure \ref{fig:WATERwDiel} shows the procedure used to assess the OH electron affinity
(as described for the WATER27 setup) for both CX (left panel) and AHBR (right panel). Noting that 
convergence of the neutral OH system is always robust, it is a simple process to converge these 
characterizations to actual WATER27 assessments. For the G21EA set we used this procedure on all 
of the individual electron-affinity problems.  The set of larger dots in Fig.\ \ref{fig:WATERwDiel} 
show actual calculations obtained (at CX and AHBR) at varying $\varepsilon_\infty$ values together with 
fitted approximations (dashed curves). We use those to extrapolate the electron-affinity
descriptions to the $\varepsilon_\infty\to 1$ results
that we actually need for benchmarking.  Comparing those limit values to reference data \cite{gmtkn55} 
gives us a well-defined assessment of SIEs, for WATER27 and G21EA, for all types of XC functionals. 


\begin{thebibliography}{235}%
\makeatletter
\providecommand \@ifxundefined [1]{%
 \@ifx{#1\undefined}
}%
\providecommand \@ifnum [1]{%
 \ifnum #1\expandafter \@firstoftwo
 \else \expandafter \@secondoftwo
 \fi
}%
\providecommand \@ifx [1]{%
 \ifx #1\expandafter \@firstoftwo
 \else \expandafter \@secondoftwo
 \fi
}%
\providecommand \natexlab [1]{#1}%
\providecommand \enquote  [1]{``#1''}%
\providecommand \bibnamefont  [1]{#1}%
\providecommand \bibfnamefont [1]{#1}%
\providecommand \citenamefont [1]{#1}%
\providecommand \href@noop [0]{\@secondoftwo}%
\providecommand \href [0]{\begingroup \@sanitize@url \@href}%
\providecommand \@href[1]{\@@startlink{#1}\@@href}%
\providecommand \@@href[1]{\endgroup#1\@@endlink}%
\providecommand \@sanitize@url [0]{\catcode `\\12\catcode `\$12\catcode
  `\&12\catcode `\#12\catcode `\^12\catcode `\_12\catcode `\%12\relax}%
\providecommand \@@startlink[1]{}%
\providecommand \@@endlink[0]{}%
\providecommand \url  [0]{\begingroup\@sanitize@url \@url }%
\providecommand \@url [1]{\endgroup\@href {#1}{\urlprefix }}%
\providecommand \urlprefix  [0]{URL }%
\providecommand \Eprint [0]{\href }%
\providecommand \doibase [0]{https://doi.org/}%
\providecommand \selectlanguage [0]{\@gobble}%
\providecommand \bibinfo  [0]{\@secondoftwo}%
\providecommand \bibfield  [0]{\@secondoftwo}%
\providecommand \translation [1]{[#1]}%
\providecommand \BibitemOpen [0]{}%
\providecommand \bibitemStop [0]{}%
\providecommand \bibitemNoStop [0]{.\EOS\space}%
\providecommand \EOS [0]{\spacefactor3000\relax}%
\providecommand \BibitemShut  [1]{\csname bibitem#1\endcsname}%
\let\auto@bib@innerbib\@empty
\bibitem [{\citenamefont {Andersson}\ \emph {et~al.}(1996)\citenamefont
  {Andersson}, \citenamefont {Langreth},\ and\ \citenamefont
  {Lundqvist}}]{anlalu96}%
  \BibitemOpen
  \bibfield  {author} {\bibinfo {author} {\bibfnamefont {Y.}~\bibnamefont
  {Andersson}}, \bibinfo {author} {\bibfnamefont {D.~C.}\ \bibnamefont
  {Langreth}},\ and\ \bibinfo {author} {\bibfnamefont {B.~I.}\ \bibnamefont
  {Lundqvist}},\ }\bibfield  {title} {\bibinfo {title} {van der {W}aals
  interactions in density-functional theory},\ }\href@noop {} {\bibfield
  {journal} {\bibinfo  {journal} {Phys. Rev. Lett.}\ }\textbf {\bibinfo
  {volume} {76}},\ \bibinfo {pages} {102} (\bibinfo {year} {1996})}\BibitemShut
  {NoStop}%
\bibitem [{\citenamefont {Dobson}\ and\ \citenamefont
  {Dinte}(1996)}]{dobdint96}%
  \BibitemOpen
  \bibfield  {author} {\bibinfo {author} {\bibfnamefont {J.~F.}\ \bibnamefont
  {Dobson}}\ and\ \bibinfo {author} {\bibfnamefont {B.~P.}\ \bibnamefont
  {Dinte}},\ }\bibfield  {title} {\bibinfo {title} {Constraint satisfaction in
  local and gradient susceptibility approximations: Application to a van der
  {W}aals density functional},\ }\href@noop {} {\bibfield  {journal} {\bibinfo
  {journal} {Phys. Rev. Lett.}\ }\textbf {\bibinfo {volume} {76}},\ \bibinfo
  {pages} {1780} (\bibinfo {year} {1996})}\BibitemShut {NoStop}%
\bibitem [{\citenamefont {Rydberg}\ \emph {et~al.}(2000)\citenamefont
  {Rydberg}, \citenamefont {Lundqvist}, \citenamefont {Langreth},\ and\
  \citenamefont {Dion}}]{ryluladi00}%
  \BibitemOpen
  \bibfield  {author} {\bibinfo {author} {\bibfnamefont {H.}~\bibnamefont
  {Rydberg}}, \bibinfo {author} {\bibfnamefont {B.~I.}\ \bibnamefont
  {Lundqvist}}, \bibinfo {author} {\bibfnamefont {D.~C.}\ \bibnamefont
  {Langreth}},\ and\ \bibinfo {author} {\bibfnamefont {M.}~\bibnamefont
  {Dion}},\ }\bibfield  {title} {\bibinfo {title} {Tractable nonlocal
  correlation density functionals for flat surfaces and slabs},\ }\href@noop {}
  {\bibfield  {journal} {\bibinfo  {journal} {Phys. Rev. B}\ }\textbf {\bibinfo
  {volume} {62}},\ \bibinfo {pages} {6997} (\bibinfo {year}
  {2000})}\BibitemShut {NoStop}%
\bibitem [{\citenamefont {Rydberg}\ \emph
  {et~al.}(2003{\natexlab{a}})\citenamefont {Rydberg}, \citenamefont {Dion},
  \citenamefont {Jacobson}, \citenamefont {Schr{\"o}der}, \citenamefont
  {Hyldgaard}, \citenamefont {Simak}, \citenamefont {Langreth},\ and\
  \citenamefont {Lundqvist}}]{rydberg03p126402}%
  \BibitemOpen
  \bibfield  {author} {\bibinfo {author} {\bibfnamefont {H.}~\bibnamefont
  {Rydberg}}, \bibinfo {author} {\bibfnamefont {M.}~\bibnamefont {Dion}},
  \bibinfo {author} {\bibfnamefont {N.}~\bibnamefont {Jacobson}}, \bibinfo
  {author} {\bibfnamefont {E.}~\bibnamefont {Schr{\"o}der}}, \bibinfo {author}
  {\bibfnamefont {P.}~\bibnamefont {Hyldgaard}}, \bibinfo {author}
  {\bibfnamefont {S.~I.}\ \bibnamefont {Simak}}, \bibinfo {author}
  {\bibfnamefont {D.~C.}\ \bibnamefont {Langreth}},\ and\ \bibinfo {author}
  {\bibfnamefont {B.~I.}\ \bibnamefont {Lundqvist}},\ }\bibfield  {title}
  {\bibinfo {title} {van der {W}aals density functional for layered
  structures},\ }\href@noop {} {\bibfield  {journal} {\bibinfo  {journal}
  {Phys. Rev. Lett.}\ }\textbf {\bibinfo {volume} {91}},\ \bibinfo {pages}
  {126402} (\bibinfo {year} {2003}{\natexlab{a}})}\BibitemShut {NoStop}%
\bibitem [{\citenamefont {Dion}\ \emph {et~al.}(2004)\citenamefont {Dion},
  \citenamefont {Rydberg}, \citenamefont {Schr{\"o}der}, \citenamefont
  {Langreth},\ and\ \citenamefont {Lundqvist}}]{Dion}%
  \BibitemOpen
  \bibfield  {author} {\bibinfo {author} {\bibfnamefont {M.}~\bibnamefont
  {Dion}}, \bibinfo {author} {\bibfnamefont {H.}~\bibnamefont {Rydberg}},
  \bibinfo {author} {\bibfnamefont {E.}~\bibnamefont {Schr{\"o}der}}, \bibinfo
  {author} {\bibfnamefont {D.~C.}\ \bibnamefont {Langreth}},\ and\ \bibinfo
  {author} {\bibfnamefont {B.~I.}\ \bibnamefont {Lundqvist}},\ }\bibfield
  {title} {\bibinfo {title} {van der {W}aals density functional for general
  geometries},\ }\href@noop {} {\bibfield  {journal} {\bibinfo  {journal}
  {Phys. Rev. Lett.}\ }\textbf {\bibinfo {volume} {92}},\ \bibinfo {pages}
  {246401} (\bibinfo {year} {2004})}\BibitemShut {NoStop}%
\bibitem [{\citenamefont {Dion}\ \emph {et~al.}(2005)\citenamefont {Dion},
  \citenamefont {Rydberg}, \citenamefont {Schr{\"o}der}, \citenamefont
  {Langreth},\ and\ \citenamefont {Lundqvist}}]{dionerratum}%
  \BibitemOpen
  \bibfield  {author} {\bibinfo {author} {\bibfnamefont {M.}~\bibnamefont
  {Dion}}, \bibinfo {author} {\bibfnamefont {H.}~\bibnamefont {Rydberg}},
  \bibinfo {author} {\bibfnamefont {E.}~\bibnamefont {Schr{\"o}der}}, \bibinfo
  {author} {\bibfnamefont {D.~C.}\ \bibnamefont {Langreth}},\ and\ \bibinfo
  {author} {\bibfnamefont {B.~I.}\ \bibnamefont {Lundqvist}},\ }\bibfield
  {title} {\bibinfo {title} {Erratum: van der {W}aals density functional for
  general geometries [{P}hys. {R}ev. {L}ett. {\bf 92}, 246401 (2004)]},\
  }\href@noop {} {\bibfield  {journal} {\bibinfo  {journal} {Phys. Rev. Lett.}\
  }\textbf {\bibinfo {volume} {95}},\ \bibinfo {pages} {109902(E)} (\bibinfo
  {year} {2005})}\BibitemShut {NoStop}%
\bibitem [{\citenamefont {Thonhauser}\ \emph {et~al.}(2007)\citenamefont
  {Thonhauser}, \citenamefont {Cooper}, \citenamefont {Li}, \citenamefont
  {Puzder}, \citenamefont {Hyldgaard},\ and\ \citenamefont
  {Langreth}}]{thonhauser}%
  \BibitemOpen
  \bibfield  {author} {\bibinfo {author} {\bibfnamefont {T.}~\bibnamefont
  {Thonhauser}}, \bibinfo {author} {\bibfnamefont {V.~R.}\ \bibnamefont
  {Cooper}}, \bibinfo {author} {\bibfnamefont {S.}~\bibnamefont {Li}}, \bibinfo
  {author} {\bibfnamefont {A.}~\bibnamefont {Puzder}}, \bibinfo {author}
  {\bibfnamefont {P.}~\bibnamefont {Hyldgaard}},\ and\ \bibinfo {author}
  {\bibfnamefont {D.~C.}\ \bibnamefont {Langreth}},\ }\bibfield  {title}
  {\bibinfo {title} {{v}an der {W}aals density functional: {S}elf-consistent
  potential and the nature of the van der {W}aals bond},\ }\href@noop {}
  {\bibfield  {journal} {\bibinfo  {journal} {Phys. Rev. B.}\ }\textbf
  {\bibinfo {volume} {76}},\ \bibinfo {pages} {125112} (\bibinfo {year}
  {2007})}\BibitemShut {NoStop}%
\bibitem [{\citenamefont {Lee}\ \emph {et~al.}(2010{\natexlab{a}})\citenamefont
  {Lee}, \citenamefont {Murray}, \citenamefont {Kong}, \citenamefont
  {Lundqvist},\ and\ \citenamefont {Langreth}}]{lee10p081101}%
  \BibitemOpen
  \bibfield  {author} {\bibinfo {author} {\bibfnamefont {K.}~\bibnamefont
  {Lee}}, \bibinfo {author} {\bibfnamefont {{\`E}.~D.}\ \bibnamefont {Murray}},
  \bibinfo {author} {\bibfnamefont {L.}~\bibnamefont {Kong}}, \bibinfo {author}
  {\bibfnamefont {B.~I.}\ \bibnamefont {Lundqvist}},\ and\ \bibinfo {author}
  {\bibfnamefont {D.~C.}\ \bibnamefont {Langreth}},\ }\bibfield  {title}
  {\bibinfo {title} {Higher-accuracy van der {W}aals density functional},\
  }\href@noop {} {\bibfield  {journal} {\bibinfo  {journal} {Phys. Rev. B}\
  }\textbf {\bibinfo {volume} {82}},\ \bibinfo {pages} {081101(R)} (\bibinfo
  {year} {2010}{\natexlab{a}})}\BibitemShut {NoStop}%
\bibitem [{\citenamefont {Hyldgaard}\ \emph {et~al.}(2014)\citenamefont
  {Hyldgaard}, \citenamefont {Berland},\ and\ \citenamefont
  {Schr{\"o}der}}]{hybesc14}%
  \BibitemOpen
  \bibfield  {author} {\bibinfo {author} {\bibfnamefont {P.}~\bibnamefont
  {Hyldgaard}}, \bibinfo {author} {\bibfnamefont {K.}~\bibnamefont {Berland}},\
  and\ \bibinfo {author} {\bibfnamefont {E.}~\bibnamefont {Schr{\"o}der}},\
  }\bibfield  {title} {\bibinfo {title} {Interpretation of van der {W}aals
  density functionals},\ }\href@noop {} {\bibfield  {journal} {\bibinfo
  {journal} {Phys. Rev. B}\ }\textbf {\bibinfo {volume} {90}},\ \bibinfo
  {pages} {075148} (\bibinfo {year} {2014})}\BibitemShut {NoStop}%
\bibitem [{\citenamefont {Berland}\ and\ \citenamefont
  {Hyldgaard}(2014)}]{behy14}%
  \BibitemOpen
  \bibfield  {author} {\bibinfo {author} {\bibfnamefont {K.}~\bibnamefont
  {Berland}}\ and\ \bibinfo {author} {\bibfnamefont {P.}~\bibnamefont
  {Hyldgaard}},\ }\bibfield  {title} {\bibinfo {title} {Exchange functional
  that tests the robustness of the plasmon description of the van der {W}aals
  density functional},\ }\href@noop {} {\bibfield  {journal} {\bibinfo
  {journal} {Phys. Rev. B}\ }\textbf {\bibinfo {volume} {89}},\ \bibinfo
  {pages} {035412} (\bibinfo {year} {2014})}\BibitemShut {NoStop}%
\bibitem [{\citenamefont {Berland}\ \emph {et~al.}(2014)\citenamefont
  {Berland}, \citenamefont {Arter}, \citenamefont {Cooper}, \citenamefont
  {Lee}, \citenamefont {Lundqvist}, \citenamefont {Schr{\"o}der}, \citenamefont
  {Thonhauser},\ and\ \citenamefont {Hyldgaard}}]{bearcoleluscthhy14}%
  \BibitemOpen
  \bibfield  {author} {\bibinfo {author} {\bibfnamefont {K.}~\bibnamefont
  {Berland}}, \bibinfo {author} {\bibfnamefont {C.~A.}\ \bibnamefont {Arter}},
  \bibinfo {author} {\bibfnamefont {V.~R.}\ \bibnamefont {Cooper}}, \bibinfo
  {author} {\bibfnamefont {K.}~\bibnamefont {Lee}}, \bibinfo {author}
  {\bibfnamefont {B.~I.}\ \bibnamefont {Lundqvist}}, \bibinfo {author}
  {\bibfnamefont {E.}~\bibnamefont {Schr{\"o}der}}, \bibinfo {author}
  {\bibfnamefont {T.}~\bibnamefont {Thonhauser}},\ and\ \bibinfo {author}
  {\bibfnamefont {P.}~\bibnamefont {Hyldgaard}},\ }\bibfield  {title} {\bibinfo
  {title} {{v}an der {W}aals density functionals built upon the electron-gas
  tradition: Facing the challenge of competing interactions},\ }\href@noop {}
  {\bibfield  {journal} {\bibinfo  {journal} {J. Chem. Phys.}\ }\textbf
  {\bibinfo {volume} {140}},\ \bibinfo {pages} {18A539} (\bibinfo {year}
  {2014})}\BibitemShut {NoStop}%
\bibitem [{\citenamefont {Berland}\ \emph {et~al.}(2015)\citenamefont
  {Berland}, \citenamefont {Cooper}, \citenamefont {Lee}, \citenamefont
  {Schr\"{o}der}, \citenamefont {Thonhauser}, \citenamefont {Hyldgaard},\ and\
  \citenamefont {Lundqvist}}]{Berland_2015:van_waals}%
  \BibitemOpen
  \bibfield  {author} {\bibinfo {author} {\bibfnamefont {K.}~\bibnamefont
  {Berland}}, \bibinfo {author} {\bibfnamefont {V.~R.}\ \bibnamefont {Cooper}},
  \bibinfo {author} {\bibfnamefont {K.}~\bibnamefont {Lee}}, \bibinfo {author}
  {\bibfnamefont {E.}~\bibnamefont {Schr\"{o}der}}, \bibinfo {author}
  {\bibfnamefont {T.}~\bibnamefont {Thonhauser}}, \bibinfo {author}
  {\bibfnamefont {P.}~\bibnamefont {Hyldgaard}},\ and\ \bibinfo {author}
  {\bibfnamefont {B.~I.}\ \bibnamefont {Lundqvist}},\ }\bibfield  {title}
  {\bibinfo {title} {{v}an der {W}aals forces in density functional theory: {A}
  review of the {vdW-DF} method},\ }\href
  {https://doi.org/10.1088/0034-4885/78/6/066501} {\bibfield  {journal}
  {\bibinfo  {journal} {Rep. Prog. Phys.}\ }\textbf {\bibinfo {volume} {78}},\
  \bibinfo {pages} {066501} (\bibinfo {year} {2015})}\BibitemShut {NoStop}%
\bibitem [{\citenamefont {Thonhauser}\ \emph {et~al.}(2015)\citenamefont
  {Thonhauser}, \citenamefont {Zuluaga}, \citenamefont {Arter}, \citenamefont
  {Berland}, \citenamefont {Schr{\"{o}}der},\ and\ \citenamefont
  {Hyldgaard}}]{Thonhauser_2015:spin_signature}%
  \BibitemOpen
  \bibfield  {author} {\bibinfo {author} {\bibfnamefont {T.}~\bibnamefont
  {Thonhauser}}, \bibinfo {author} {\bibfnamefont {S.}~\bibnamefont {Zuluaga}},
  \bibinfo {author} {\bibfnamefont {C.~A.}\ \bibnamefont {Arter}}, \bibinfo
  {author} {\bibfnamefont {K.}~\bibnamefont {Berland}}, \bibinfo {author}
  {\bibfnamefont {E.}~\bibnamefont {Schr{\"{o}}der}},\ and\ \bibinfo {author}
  {\bibfnamefont {P.}~\bibnamefont {Hyldgaard}},\ }\bibfield  {title} {\bibinfo
  {title} {Spin signature of nonlocal correlation binding in metal-organic
  frameworks},\ }\href {https://doi.org/10.1103/PhysRevLett.115.136402}
  {\bibfield  {journal} {\bibinfo  {journal} {Phys. Rev. Lett.}\ }\textbf
  {\bibinfo {volume} {115}},\ \bibinfo {pages} {136402} (\bibinfo {year}
  {2015})}\BibitemShut {NoStop}%
\bibitem [{\citenamefont {Berland}\ \emph {et~al.}(2017)\citenamefont
  {Berland}, \citenamefont {Jiao}, \citenamefont {Lee}, \citenamefont {Rangel},
  \citenamefont {Neaton},\ and\ \citenamefont {Hyldgaard}}]{DFcx02017}%
  \BibitemOpen
  \bibfield  {author} {\bibinfo {author} {\bibfnamefont {K.}~\bibnamefont
  {Berland}}, \bibinfo {author} {\bibfnamefont {Y.}~\bibnamefont {Jiao}},
  \bibinfo {author} {\bibfnamefont {J.-H.}\ \bibnamefont {Lee}}, \bibinfo
  {author} {\bibfnamefont {T.}~\bibnamefont {Rangel}}, \bibinfo {author}
  {\bibfnamefont {J.~B.}\ \bibnamefont {Neaton}},\ and\ \bibinfo {author}
  {\bibfnamefont {P.}~\bibnamefont {Hyldgaard}},\ }\bibfield  {title} {\bibinfo
  {title} {Assessment of two hybrid van der {Waals} density functionals for
  covalent and noncovalent binding of molecules},\ }\href@noop {} {\bibfield
  {journal} {\bibinfo  {journal} {J. Chem. Phys.}\ }\textbf {\bibinfo {volume}
  {146}},\ \bibinfo {pages} {234106} (\bibinfo {year} {2017})}\BibitemShut
  {NoStop}%
\bibitem [{\citenamefont {Jiao}\ \emph
  {et~al.}(2018{\natexlab{a}})\citenamefont {Jiao}, \citenamefont
  {Schr{\"o}der},\ and\ \citenamefont {Hyldgaard}}]{JiScHy18b}%
  \BibitemOpen
  \bibfield  {author} {\bibinfo {author} {\bibfnamefont {Y.}~\bibnamefont
  {Jiao}}, \bibinfo {author} {\bibfnamefont {E.}~\bibnamefont {Schr{\"o}der}},\
  and\ \bibinfo {author} {\bibfnamefont {P.}~\bibnamefont {Hyldgaard}},\
  }\bibfield  {title} {\bibinfo {title} {{Extent of Fock-exchange mixing for a
  hybrid van der Waals density functional?}},\ }\href@noop {} {\bibfield
  {journal} {\bibinfo  {journal} {J. Chem. Phys.}\ }\textbf {\bibinfo {volume}
  {148}},\ \bibinfo {pages} {194115} (\bibinfo {year}
  {2018}{\natexlab{a}})}\BibitemShut {NoStop}%
\bibitem [{\citenamefont {Chakraborty}\ \emph {et~al.}(2020)\citenamefont
  {Chakraborty}, \citenamefont {Berland},\ and\ \citenamefont
  {Thonhauser}}]{ChBeTh20}%
  \BibitemOpen
  \bibfield  {author} {\bibinfo {author} {\bibfnamefont {D.}~\bibnamefont
  {Chakraborty}}, \bibinfo {author} {\bibfnamefont {K.}~\bibnamefont
  {Berland}},\ and\ \bibinfo {author} {\bibfnamefont {T.}~\bibnamefont
  {Thonhauser}},\ }\bibfield  {title} {\bibinfo {title} {{Next-Generation
  Nonlocal van der Waals Density Functional}},\ }\href@noop {} {\bibfield
  {journal} {\bibinfo  {journal} {J. Chem. Theory Comput.}\ }\textbf {\bibinfo
  {volume} {16}},\ \bibinfo {pages} {5893} (\bibinfo {year}
  {2020})}\BibitemShut {NoStop}%
\bibitem [{\citenamefont {Shukla}\ \emph {et~al.}(2022)\citenamefont {Shukla},
  \citenamefont {Jiao}, \citenamefont {Frostenson},\ and\ \citenamefont
  {Hyldgaard}}]{DefineAHCX}%
  \BibitemOpen
  \bibfield  {author} {\bibinfo {author} {\bibfnamefont {V.}~\bibnamefont
  {Shukla}}, \bibinfo {author} {\bibfnamefont {Y.}~\bibnamefont {Jiao}},
  \bibinfo {author} {\bibfnamefont {C.~M.}\ \bibnamefont {Frostenson}},\ and\
  \bibinfo {author} {\bibfnamefont {P.}~\bibnamefont {Hyldgaard}},\ }\bibfield
  {title} {\bibinfo {title} {{vdW-DF-ahcx: a range-separated van der Waals
  density functional hybrid}},\ }\href@noop {} {\bibfield  {journal} {\bibinfo
  {journal} {J. Phys.: Condens. Matter}\ }\textbf {\bibinfo {volume} {34}},\
  \bibinfo {pages} {025902} (\bibinfo {year} {2022})}\BibitemShut {NoStop}%
\bibitem [{\citenamefont {Frostenson}\ \emph {et~al.}(2022)\citenamefont
  {Frostenson}, \citenamefont {Granhed}, \citenamefont {Shukla}, \citenamefont
  {Olsson}, \citenamefont {Schr{\"o}der},\ and\ \citenamefont
  {Hyldgaard}}]{Hard2Soft}%
  \BibitemOpen
  \bibfield  {author} {\bibinfo {author} {\bibfnamefont {C.~M.}\ \bibnamefont
  {Frostenson}}, \bibinfo {author} {\bibfnamefont {E.~J.}\ \bibnamefont
  {Granhed}}, \bibinfo {author} {\bibfnamefont {V.}~\bibnamefont {Shukla}},
  \bibinfo {author} {\bibfnamefont {P.~A.~T.}\ \bibnamefont {Olsson}}, \bibinfo
  {author} {\bibfnamefont {E.}~\bibnamefont {Schr{\"o}der}},\ and\ \bibinfo
  {author} {\bibfnamefont {P.}~\bibnamefont {Hyldgaard}},\ }\bibfield  {title}
  {\bibinfo {title} {{Hard and soft materials: Putting consistent van der Waals
  density functionals to work}},\ }\href@noop {} {\bibfield  {journal}
  {\bibinfo  {journal} {Electron. Struct.}\ }\textbf {\bibinfo {volume} {4}},\
  \bibinfo {pages} {014001} (\bibinfo {year} {2022})}\BibitemShut {NoStop}%
\bibitem [{\citenamefont {Lundqvist}\ \emph {et~al.}(2001)\citenamefont
  {Lundqvist}, \citenamefont {Bogicevic}, \citenamefont {Carling},
  \citenamefont {Dudiy}, \citenamefont {Gao}, \citenamefont {Hartford},
  \citenamefont {Hyldgaard}, \citenamefont {Jacobson}, \citenamefont
  {Langreth}, \citenamefont {Lorente}, \citenamefont {Ovesson}, \citenamefont
  {Razaznejad}, \citenamefont {Ruberto}, \citenamefont {Rydberg}, \citenamefont
  {Schr{\"o}der}, \citenamefont {Simak}, \citenamefont {Wahnstr{\"o}m},\ and\
  \citenamefont {Yourdshahyan}}]{2001surfscience}%
  \BibitemOpen
  \bibfield  {author} {\bibinfo {author} {\bibfnamefont {B.~I.}\ \bibnamefont
  {Lundqvist}}, \bibinfo {author} {\bibfnamefont {A.}~\bibnamefont
  {Bogicevic}}, \bibinfo {author} {\bibfnamefont {K.}~\bibnamefont {Carling}},
  \bibinfo {author} {\bibfnamefont {S.~V.}\ \bibnamefont {Dudiy}}, \bibinfo
  {author} {\bibfnamefont {S.}~\bibnamefont {Gao}}, \bibinfo {author}
  {\bibfnamefont {J.}~\bibnamefont {Hartford}}, \bibinfo {author}
  {\bibfnamefont {P.}~\bibnamefont {Hyldgaard}}, \bibinfo {author}
  {\bibfnamefont {N.}~\bibnamefont {Jacobson}}, \bibinfo {author}
  {\bibfnamefont {D.~C.}\ \bibnamefont {Langreth}}, \bibinfo {author}
  {\bibfnamefont {N.}~\bibnamefont {Lorente}}, \bibinfo {author} {\bibfnamefont
  {S.}~\bibnamefont {Ovesson}}, \bibinfo {author} {\bibfnamefont
  {B.}~\bibnamefont {Razaznejad}}, \bibinfo {author} {\bibfnamefont
  {C.}~\bibnamefont {Ruberto}}, \bibinfo {author} {\bibfnamefont
  {H.}~\bibnamefont {Rydberg}}, \bibinfo {author} {\bibfnamefont
  {E.}~\bibnamefont {Schr{\"o}der}}, \bibinfo {author} {\bibfnamefont {S.~I.}\
  \bibnamefont {Simak}}, \bibinfo {author} {\bibfnamefont {G.}~\bibnamefont
  {Wahnstr{\"o}m}},\ and\ \bibinfo {author} {\bibfnamefont {Y.}~\bibnamefont
  {Yourdshahyan}},\ }\bibfield  {title} {\bibinfo {title} {Density-functional
  bridge between surfaces and interfaces},\ }\href@noop {} {\bibfield
  {journal} {\bibinfo  {journal} {Surf. Sci.}\ }\textbf {\bibinfo {volume}
  {493}},\ \bibinfo {pages} {253} (\bibinfo {year} {2001})}\BibitemShut
  {NoStop}%
\bibitem [{\citenamefont {Rydberg}\ \emph
  {et~al.}(2003{\natexlab{b}})\citenamefont {Rydberg}, \citenamefont
  {Jacobson}, \citenamefont {Hyldgaard}, \citenamefont {Simak}, \citenamefont
  {Lundqvist},\ and\ \citenamefont {Langreth}}]{rydberg03p606}%
  \BibitemOpen
  \bibfield  {author} {\bibinfo {author} {\bibfnamefont {H.}~\bibnamefont
  {Rydberg}}, \bibinfo {author} {\bibfnamefont {N.}~\bibnamefont {Jacobson}},
  \bibinfo {author} {\bibfnamefont {P.}~\bibnamefont {Hyldgaard}}, \bibinfo
  {author} {\bibfnamefont {S.~I.}\ \bibnamefont {Simak}}, \bibinfo {author}
  {\bibfnamefont {B.~I.}\ \bibnamefont {Lundqvist}},\ and\ \bibinfo {author}
  {\bibfnamefont {D.~C.}\ \bibnamefont {Langreth}},\ }\bibfield  {title}
  {\bibinfo {title} {Hard numbers on soft matter},\ }\href@noop {} {\bibfield
  {journal} {\bibinfo  {journal} {Surf. Sci.}\ }\textbf {\bibinfo {volume}
  {532--535}},\ \bibinfo {pages} {606} (\bibinfo {year}
  {2003}{\natexlab{b}})}\BibitemShut {NoStop}%
\bibitem [{\citenamefont {Kleis}\ \emph {et~al.}(2005)\citenamefont {Kleis},
  \citenamefont {Hyldgaard},\ and\ \citenamefont {Schr\"oder}}]{kleis05p192}%
  \BibitemOpen
  \bibfield  {author} {\bibinfo {author} {\bibfnamefont {J.}~\bibnamefont
  {Kleis}}, \bibinfo {author} {\bibfnamefont {P.}~\bibnamefont {Hyldgaard}},\
  and\ \bibinfo {author} {\bibfnamefont {E.}~\bibnamefont {Schr\"oder}},\
  }\bibfield  {title} {\bibinfo {title} {van der {W}aals interaction of
  parallel polymers and nanotubes},\ }\href@noop {} {\bibfield  {journal}
  {\bibinfo  {journal} {Comp. Mater. Sci.}\ }\textbf {\bibinfo {volume} {33}},\
  \bibinfo {pages} {192} (\bibinfo {year} {2005})}\BibitemShut {NoStop}%
\bibitem [{\citenamefont {Kleis}\ and\ \citenamefont
  {Schr\"{o}der}(2005)}]{kleis05p164902}%
  \BibitemOpen
  \bibfield  {author} {\bibinfo {author} {\bibfnamefont {J.}~\bibnamefont
  {Kleis}}\ and\ \bibinfo {author} {\bibfnamefont {E.}~\bibnamefont
  {Schr\"{o}der}},\ }\bibfield  {title} {\bibinfo {title} {van der {W}aals
  interaction of simple, parallel polymers},\ }\href@noop {} {\bibfield
  {journal} {\bibinfo  {journal} {J. Chem. Phys.}\ }\textbf {\bibinfo {volume}
  {122}},\ \bibinfo {pages} {164902} (\bibinfo {year} {2005})}\BibitemShut
  {NoStop}%
\bibitem [{\citenamefont {Kleis}\ \emph {et~al.}(2007)\citenamefont {Kleis},
  \citenamefont {Lundqvist}, \citenamefont {Langreth},\ and\ \citenamefont
  {Schr\"{o}der}}]{kleis07p100201}%
  \BibitemOpen
  \bibfield  {author} {\bibinfo {author} {\bibfnamefont {J.}~\bibnamefont
  {Kleis}}, \bibinfo {author} {\bibfnamefont {B.~I.}\ \bibnamefont
  {Lundqvist}}, \bibinfo {author} {\bibfnamefont {D.~C.}\ \bibnamefont
  {Langreth}},\ and\ \bibinfo {author} {\bibfnamefont {E.}~\bibnamefont
  {Schr\"{o}der}},\ }\bibfield  {title} {\bibinfo {title} {Towards a working
  density-functional theory for polymers: {F}irst-principles determination of
  the polyethylene crystal structure},\ }\href@noop {} {\bibfield  {journal}
  {\bibinfo  {journal} {Phys. Rev. B.}\ }\textbf {\bibinfo {volume} {76}},\
  \bibinfo {pages} {100201} (\bibinfo {year} {2007})}\BibitemShut {NoStop}%
\bibitem [{\citenamefont {Kleis}\ \emph {et~al.}(2008)\citenamefont {Kleis},
  \citenamefont {Schr\"{o}der},\ and\ \citenamefont
  {Hyldgaard}}]{kleis08p205422}%
  \BibitemOpen
  \bibfield  {author} {\bibinfo {author} {\bibfnamefont {J.}~\bibnamefont
  {Kleis}}, \bibinfo {author} {\bibfnamefont {E.}~\bibnamefont
  {Schr\"{o}der}},\ and\ \bibinfo {author} {\bibfnamefont {P.}~\bibnamefont
  {Hyldgaard}},\ }\bibfield  {title} {\bibinfo {title} {Nature and strength of
  bonding in a crystal of semiconducting nanotubes: van der {W}aals density
  functional calculations and analytical results},\ }\href@noop {} {\bibfield
  {journal} {\bibinfo  {journal} {Phys. Rev. B.}\ }\textbf {\bibinfo {volume}
  {77}},\ \bibinfo {pages} {205422} (\bibinfo {year} {2008})}\BibitemShut
  {NoStop}%
\bibitem [{\citenamefont {Hooper}\ \emph {et~al.}(2008)\citenamefont {Hooper},
  \citenamefont {Cooper}, \citenamefont {Thonhauser}, \citenamefont {Romero},
  \citenamefont {Zerilli},\ and\ \citenamefont {Langreth}}]{hooper08p891}%
  \BibitemOpen
  \bibfield  {author} {\bibinfo {author} {\bibfnamefont {J.}~\bibnamefont
  {Hooper}}, \bibinfo {author} {\bibfnamefont {V.~R.}\ \bibnamefont {Cooper}},
  \bibinfo {author} {\bibfnamefont {T.}~\bibnamefont {Thonhauser}}, \bibinfo
  {author} {\bibfnamefont {N.~A.}\ \bibnamefont {Romero}}, \bibinfo {author}
  {\bibfnamefont {F.}~\bibnamefont {Zerilli}},\ and\ \bibinfo {author}
  {\bibfnamefont {D.~C.}\ \bibnamefont {Langreth}},\ }\bibfield  {title}
  {\bibinfo {title} {Predicting {C}--{H}/$\pi$ interactions with nonlocal
  density functional theory},\ }\href@noop {} {\bibfield  {journal} {\bibinfo
  {journal} {ChemPhysChem}\ }\textbf {\bibinfo {volume} {9}},\ \bibinfo {pages}
  {891} (\bibinfo {year} {2008})}\BibitemShut {NoStop}%
\bibitem [{\citenamefont {Cooper}\ \emph
  {et~al.}(2008{\natexlab{a}})\citenamefont {Cooper}, \citenamefont
  {Thonhauser}, \citenamefont {Puzder}, \citenamefont {Schr\"{o}der},
  \citenamefont {Lundqvist},\ and\ \citenamefont {Langreth}}]{cooper08p1304}%
  \BibitemOpen
  \bibfield  {author} {\bibinfo {author} {\bibfnamefont {V.~R.}\ \bibnamefont
  {Cooper}}, \bibinfo {author} {\bibfnamefont {T.}~\bibnamefont {Thonhauser}},
  \bibinfo {author} {\bibfnamefont {A.}~\bibnamefont {Puzder}}, \bibinfo
  {author} {\bibfnamefont {E.}~\bibnamefont {Schr\"{o}der}}, \bibinfo {author}
  {\bibfnamefont {B.~I.}\ \bibnamefont {Lundqvist}},\ and\ \bibinfo {author}
  {\bibfnamefont {D.~C.}\ \bibnamefont {Langreth}},\ }\bibfield  {title}
  {\bibinfo {title} {Stacking interactions and the twist of {DNA}},\
  }\href@noop {} {\bibfield  {journal} {\bibinfo  {journal} {J. Am. Chem.
  Soc.}\ }\textbf {\bibinfo {volume} {130}},\ \bibinfo {pages} {1304} (\bibinfo
  {year} {2008}{\natexlab{a}})}\BibitemShut {NoStop}%
\bibitem [{\citenamefont {Cooper}\ \emph
  {et~al.}(2008{\natexlab{b}})\citenamefont {Cooper}, \citenamefont
  {Thonhauser},\ and\ \citenamefont {Langreth}}]{cooper08p204102}%
  \BibitemOpen
  \bibfield  {author} {\bibinfo {author} {\bibfnamefont {V.~R.}\ \bibnamefont
  {Cooper}}, \bibinfo {author} {\bibfnamefont {T.}~\bibnamefont {Thonhauser}},\
  and\ \bibinfo {author} {\bibfnamefont {D.~C.}\ \bibnamefont {Langreth}},\
  }\bibfield  {title} {\bibinfo {title} {An application of the van der {W}aals
  density functional: Hydrogen bonding and stacking interactions between
  nucleobases},\ }\href@noop {} {\bibfield  {journal} {\bibinfo  {journal} {J.
  Chem. Phys.}\ }\textbf {\bibinfo {volume} {128}},\ \bibinfo {pages} {204102}
  (\bibinfo {year} {2008}{\natexlab{b}})}\BibitemShut {NoStop}%
\bibitem [{\citenamefont {Chakarova-K\"{a}ck}\ \emph
  {et~al.}(2006)\citenamefont {Chakarova-K\"{a}ck}, \citenamefont
  {Schr\"{o}der}, \citenamefont {Lundqvist},\ and\ \citenamefont
  {Langreth}}]{chakarova-kack06p146107}%
  \BibitemOpen
  \bibfield  {author} {\bibinfo {author} {\bibfnamefont {S.~D.}\ \bibnamefont
  {Chakarova-K\"{a}ck}}, \bibinfo {author} {\bibfnamefont {E.}~\bibnamefont
  {Schr\"{o}der}}, \bibinfo {author} {\bibfnamefont {B.~I.}\ \bibnamefont
  {Lundqvist}},\ and\ \bibinfo {author} {\bibfnamefont {D.~C.}\ \bibnamefont
  {Langreth}},\ }\bibfield  {title} {\bibinfo {title} {{Application of van der
  {W}aals Density Functional to an Extended System: Adsorption of Benzene and
  Naphthalene on Graphite}},\ }\href@noop {} {\bibfield  {journal} {\bibinfo
  {journal} {Phys. Rev. Lett.}\ }\textbf {\bibinfo {volume} {96}},\ \bibinfo
  {pages} {146107} (\bibinfo {year} {2006})}\BibitemShut {NoStop}%
\bibitem [{\citenamefont {Chakarova-K{\"a}ck}\ \emph
  {et~al.}(2006)\citenamefont {Chakarova-K{\"a}ck}, \citenamefont {Borck},
  \citenamefont {Schr\"{o}der},\ and\ \citenamefont
  {Lundqvist}}]{chakarova-kack06p155402}%
  \BibitemOpen
  \bibfield  {author} {\bibinfo {author} {\bibfnamefont {S.~D.}\ \bibnamefont
  {Chakarova-K{\"a}ck}}, \bibinfo {author} {\bibfnamefont {{\O}.}~\bibnamefont
  {Borck}}, \bibinfo {author} {\bibfnamefont {E.}~\bibnamefont
  {Schr\"{o}der}},\ and\ \bibinfo {author} {\bibfnamefont {B.~I.}\ \bibnamefont
  {Lundqvist}},\ }\bibfield  {title} {\bibinfo {title} {Adsorption of phenol on
  graphite (0001) and $\alpha$-{Al}$_2${O}$_3$ (0001): Nature of van der
  {W}aals bonds from first-principles calculations},\ }\href@noop {} {\bibfield
   {journal} {\bibinfo  {journal} {Phys. Rev. B.}\ }\textbf {\bibinfo {volume}
  {74}},\ \bibinfo {pages} {155402} (\bibinfo {year} {2006})}\BibitemShut
  {NoStop}%
\bibitem [{\citenamefont {Chakarova-K{\"a}ck}\ \emph
  {et~al.}(2010)\citenamefont {Chakarova-K{\"a}ck}, \citenamefont {Vojvodic},
  \citenamefont {Kleis}, \citenamefont {Hyldgaard},\ and\ \citenamefont
  {Schr{\"o}der}}]{chakarova-kack10p013017}%
  \BibitemOpen
  \bibfield  {author} {\bibinfo {author} {\bibfnamefont {S.~D.}\ \bibnamefont
  {Chakarova-K{\"a}ck}}, \bibinfo {author} {\bibfnamefont {A.}~\bibnamefont
  {Vojvodic}}, \bibinfo {author} {\bibfnamefont {J.}~\bibnamefont {Kleis}},
  \bibinfo {author} {\bibfnamefont {P.}~\bibnamefont {Hyldgaard}},\ and\
  \bibinfo {author} {\bibfnamefont {E.}~\bibnamefont {Schr{\"o}der}},\
  }\bibfield  {title} {\bibinfo {title} {Binding of polycyclic aromatic
  hydrocarbons and graphene dimers in density functional theory},\ }\href@noop
  {} {\bibfield  {journal} {\bibinfo  {journal} {New J. Phys.}\ }\textbf
  {\bibinfo {volume} {12}},\ \bibinfo {pages} {013017} (\bibinfo {year}
  {2010})}\BibitemShut {NoStop}%
\bibitem [{\citenamefont {Grimme}(2004)}]{grimme1}%
  \BibitemOpen
  \bibfield  {author} {\bibinfo {author} {\bibfnamefont {S.}~\bibnamefont
  {Grimme}},\ }\bibfield  {title} {\bibinfo {title} {Accurate description of
  van der {W}aals complexes by density functional theory including empirical
  corrections},\ }\href@noop {} {\bibfield  {journal} {\bibinfo  {journal} {J.
  Comput. Chem.}\ }\textbf {\bibinfo {volume} {25}},\ \bibinfo {pages} {1463}
  (\bibinfo {year} {2004})}\BibitemShut {NoStop}%
\bibitem [{\citenamefont {Becke}\ and\ \citenamefont
  {Johnson}(2005)}]{becke05p154101}%
  \BibitemOpen
  \bibfield  {author} {\bibinfo {author} {\bibfnamefont {A.~D.}\ \bibnamefont
  {Becke}}\ and\ \bibinfo {author} {\bibfnamefont {E.~R.}\ \bibnamefont
  {Johnson}},\ }\bibfield  {title} {\bibinfo {title} {A density-functional
  model of the dispersion interaction},\ }\href@noop {} {\bibfield  {journal}
  {\bibinfo  {journal} {J. Chem. Phys.}\ }\textbf {\bibinfo {volume} {123}},\
  \bibinfo {pages} {154101} (\bibinfo {year} {2005})}\BibitemShut {NoStop}%
\bibitem [{\citenamefont {Becke}\ and\ \citenamefont
  {Johnson}(2007)}]{becke07p154108}%
  \BibitemOpen
  \bibfield  {author} {\bibinfo {author} {\bibfnamefont {A.~D.}\ \bibnamefont
  {Becke}}\ and\ \bibinfo {author} {\bibfnamefont {E.~R.}\ \bibnamefont
  {Johnson}},\ }\bibfield  {title} {\bibinfo {title} {Exchange-hole dipole
  moment and the dispersion interaction revisited},\ }\href@noop {} {\bibfield
  {journal} {\bibinfo  {journal} {J. Chem. Phys.}\ }\textbf {\bibinfo {volume}
  {127}},\ \bibinfo {pages} {154108} (\bibinfo {year} {2007})}\BibitemShut
  {NoStop}%
\bibitem [{\citenamefont {Grimme}(2006)}]{grimme2}%
  \BibitemOpen
  \bibfield  {author} {\bibinfo {author} {\bibfnamefont {S.}~\bibnamefont
  {Grimme}},\ }\bibfield  {title} {\bibinfo {title} {Semiempirical hybrid
  density functional with perturbative second-order correlation},\ }\href@noop
  {} {\bibfield  {journal} {\bibinfo  {journal} {J. Chem. Phys.}\ }\textbf
  {\bibinfo {volume} {124}},\ \bibinfo {pages} {034108} (\bibinfo {year}
  {2006})}\BibitemShut {NoStop}%
\bibitem [{\citenamefont {Silvestrelli}(2008)}]{silvestrelli08p53002}%
  \BibitemOpen
  \bibfield  {author} {\bibinfo {author} {\bibfnamefont {P.~L.}\ \bibnamefont
  {Silvestrelli}},\ }\bibfield  {title} {\bibinfo {title} {{v}an der {W}aals
  interactions in {DFT} made easy by {W}annier functions},\ }\href@noop {}
  {\bibfield  {journal} {\bibinfo  {journal} {Phys. Rev. Lett.}\ }\textbf
  {\bibinfo {volume} {100}},\ \bibinfo {pages} {053002} (\bibinfo {year}
  {2008})}\BibitemShut {NoStop}%
\bibitem [{\citenamefont {Tkatchenko}\ and\ \citenamefont
  {Scheffler}(2009)}]{ts09}%
  \BibitemOpen
  \bibfield  {author} {\bibinfo {author} {\bibfnamefont {A.}~\bibnamefont
  {Tkatchenko}}\ and\ \bibinfo {author} {\bibfnamefont {M.}~\bibnamefont
  {Scheffler}},\ }\bibfield  {title} {\bibinfo {title} {Accurate molecular van
  der {W}aals interactions from ground-state electron density and free-atom
  reference data},\ }\href@noop {} {\bibfield  {journal} {\bibinfo  {journal}
  {Phys. Rev. Lett.}\ }\textbf {\bibinfo {volume} {102}},\ \bibinfo {pages}
  {073005} (\bibinfo {year} {2009})}\BibitemShut {NoStop}%
\bibitem [{\citenamefont {Ruiz}\ \emph {et~al.}(2012)\citenamefont {Ruiz},
  \citenamefont {Liu}, \citenamefont {Zojer}, \citenamefont {Scheffler},\ and\
  \citenamefont {Tkatchenko}}]{ts2}%
  \BibitemOpen
  \bibfield  {author} {\bibinfo {author} {\bibfnamefont {V.~G.}\ \bibnamefont
  {Ruiz}}, \bibinfo {author} {\bibfnamefont {W.}~\bibnamefont {Liu}}, \bibinfo
  {author} {\bibfnamefont {E.}~\bibnamefont {Zojer}}, \bibinfo {author}
  {\bibfnamefont {M.}~\bibnamefont {Scheffler}},\ and\ \bibinfo {author}
  {\bibfnamefont {A.}~\bibnamefont {Tkatchenko}},\ }\bibfield  {title}
  {\bibinfo {title} {Density-functional theory with screened van der {W}aals
  interactions for the modeling of hybrid inorganic-organic systems},\
  }\href@noop {} {\bibfield  {journal} {\bibinfo  {journal} {Phys. Rev. Lett.}\
  }\textbf {\bibinfo {volume} {108}},\ \bibinfo {pages} {146103} (\bibinfo
  {year} {2012})}\BibitemShut {NoStop}%
\bibitem [{\citenamefont {Tkatchenko}\ \emph {et~al.}(2012)\citenamefont
  {Tkatchenko}, \citenamefont {DiStasio}, \citenamefont {Car},\ and\
  \citenamefont {Scheffler}}]{ts-mbd}%
  \BibitemOpen
  \bibfield  {author} {\bibinfo {author} {\bibfnamefont {A.}~\bibnamefont
  {Tkatchenko}}, \bibinfo {author} {\bibfnamefont {R.~A.}\ \bibnamefont
  {DiStasio}}, \bibinfo {author} {\bibfnamefont {R.}~\bibnamefont {Car}},\ and\
  \bibinfo {author} {\bibfnamefont {M.}~\bibnamefont {Scheffler}},\ }\bibfield
  {title} {\bibinfo {title} {Accurate and efficient method for many-body van
  der {W}aals interactions},\ }\href@noop {} {\bibfield  {journal} {\bibinfo
  {journal} {Phys. Rev. Lett.}\ }\textbf {\bibinfo {volume} {108}},\ \bibinfo
  {pages} {236402} (\bibinfo {year} {2012})}\BibitemShut {NoStop}%
\bibitem [{\citenamefont {Ambrosetti}\ and\ \citenamefont
  {Silvestrelli}(2012)}]{ambrosetti12p73101}%
  \BibitemOpen
  \bibfield  {author} {\bibinfo {author} {\bibfnamefont {A.}~\bibnamefont
  {Ambrosetti}}\ and\ \bibinfo {author} {\bibfnamefont {P.~L.}\ \bibnamefont
  {Silvestrelli}},\ }\bibfield  {title} {\bibinfo {title} {van der {W}aals
  interactions in density functional theory using {W}annier functions: Improved
  {C}$_6$ and {C}$_3$ coefficients by a different approach},\ }\href@noop {}
  {\bibfield  {journal} {\bibinfo  {journal} {Phys. Rev. B}\ }\textbf {\bibinfo
  {volume} {85}},\ \bibinfo {pages} {073101} (\bibinfo {year}
  {2012})}\BibitemShut {NoStop}%
\bibitem [{\citenamefont {Caldeweyher}\ \emph {et~al.}(2019)\citenamefont
  {Caldeweyher}, \citenamefont {Ehlert}, \citenamefont {Hansen}, \citenamefont
  {Neugebauer}, \citenamefont {Spicher}, \citenamefont {Bannwarth},\ and\
  \citenamefont {Grimme}}]{grimme4}%
  \BibitemOpen
  \bibfield  {author} {\bibinfo {author} {\bibfnamefont {E.}~\bibnamefont
  {Caldeweyher}}, \bibinfo {author} {\bibfnamefont {S.}~\bibnamefont {Ehlert}},
  \bibinfo {author} {\bibfnamefont {A.}~\bibnamefont {Hansen}}, \bibinfo
  {author} {\bibfnamefont {H.}~\bibnamefont {Neugebauer}}, \bibinfo {author}
  {\bibfnamefont {S.}~\bibnamefont {Spicher}}, \bibinfo {author} {\bibfnamefont
  {C.}~\bibnamefont {Bannwarth}},\ and\ \bibinfo {author} {\bibfnamefont
  {S.~A.}\ \bibnamefont {Grimme}},\ }\bibfield  {title} {\bibinfo {title} {{A
  Generally Applicable Atomic-Charge Dependent London Dispersion Correction}},\
  }\href@noop {} {\bibfield  {journal} {\bibinfo  {journal} {Journal of
  Chemical Physics}\ }\textbf {\bibinfo {volume} {150}},\ \bibinfo {pages}
  {154122} (\bibinfo {year} {2019})}\BibitemShut {NoStop}%
\bibitem [{\citenamefont {Kim}\ \emph {et~al.}(2020{\natexlab{a}})\citenamefont
  {Kim}, \citenamefont {Kim}, \citenamefont {Gould}, \citenamefont {Lee},
  \citenamefont {Lebeque},\ and\ \citenamefont {Kim}}]{KiKiGo20}%
  \BibitemOpen
  \bibfield  {author} {\bibinfo {author} {\bibfnamefont {M.}~\bibnamefont
  {Kim}}, \bibinfo {author} {\bibfnamefont {W.~J.}\ \bibnamefont {Kim}},
  \bibinfo {author} {\bibfnamefont {T.}~\bibnamefont {Gould}}, \bibinfo
  {author} {\bibfnamefont {E.~K.}\ \bibnamefont {Lee}}, \bibinfo {author}
  {\bibfnamefont {S.}~\bibnamefont {Lebeque}},\ and\ \bibinfo {author}
  {\bibfnamefont {H.}~\bibnamefont {Kim}},\ }\bibfield  {title} {\bibinfo
  {title} {{uMBD: A Materials-Ready Dispersion Correction That Uniformly Treats
  Metallic, Ionic, and van der Waals Bonding}},\ }\href@noop {} {\bibfield
  {journal} {\bibinfo  {journal} {J. Am. Chem. Soc.}\ }\textbf {\bibinfo
  {volume} {142}},\ \bibinfo {pages} {2346} (\bibinfo {year}
  {2020}{\natexlab{a}})}\BibitemShut {NoStop}%
\bibitem [{\citenamefont {Jiao}\ \emph
  {et~al.}(2018{\natexlab{b}})\citenamefont {Jiao}, \citenamefont
  {Schr{\"o}der},\ and\ \citenamefont {Hyldgaard}}]{JiScHy18a}%
  \BibitemOpen
  \bibfield  {author} {\bibinfo {author} {\bibfnamefont {Y.}~\bibnamefont
  {Jiao}}, \bibinfo {author} {\bibfnamefont {E.}~\bibnamefont {Schr{\"o}der}},\
  and\ \bibinfo {author} {\bibfnamefont {P.}~\bibnamefont {Hyldgaard}},\
  }\bibfield  {title} {\bibinfo {title} {{Signatures of van der Waals binding:
  a coupling-constant scaling analysis}},\ }\href@noop {} {\bibfield  {journal}
  {\bibinfo  {journal} {Phys. Rev. B}\ }\textbf {\bibinfo {volume} {97}},\
  \bibinfo {pages} {085115} (\bibinfo {year} {2018}{\natexlab{b}})}\BibitemShut
  {NoStop}%
\bibitem [{\citenamefont {Hyldgaard}\ \emph {et~al.}(2020)\citenamefont
  {Hyldgaard}, \citenamefont {Jiao},\ and\ \citenamefont
  {Shukla}}]{JPCMreview}%
  \BibitemOpen
  \bibfield  {author} {\bibinfo {author} {\bibfnamefont {P.}~\bibnamefont
  {Hyldgaard}}, \bibinfo {author} {\bibfnamefont {Y.}~\bibnamefont {Jiao}},\
  and\ \bibinfo {author} {\bibfnamefont {V.}~\bibnamefont {Shukla}},\
  }\bibfield  {title} {\bibinfo {title} {{Screening nature of the van der Waals
  density functional method: A review and analysis of the many-body physics
  foundation}},\ }\href@noop {} {\bibfield  {journal} {\bibinfo  {journal} {J.
  Phys.: Condens. Matter}\ }\textbf {\bibinfo {volume} {32}},\ \bibinfo {pages}
  {393001} (\bibinfo {year} {2020})}\BibitemShut {NoStop}%
\bibitem [{\citenamefont {Lee}\ \emph {et~al.}(2022)\citenamefont {Lee},
  \citenamefont {Hyldgaard},\ and\ \citenamefont {Neaton}}]{MOFdobpdc}%
  \BibitemOpen
  \bibfield  {author} {\bibinfo {author} {\bibfnamefont {J.-H.}\ \bibnamefont
  {Lee}}, \bibinfo {author} {\bibfnamefont {P.}~\bibnamefont {Hyldgaard}},\
  and\ \bibinfo {author} {\bibfnamefont {J.~B.}\ \bibnamefont {Neaton}},\
  }\bibfield  {title} {\bibinfo {title} {{An Assessment of Density Functionals
  for Predicting CO$_2$ Adsorption in Diamine-Functionalized Metal-Organic
  Frameworks}},\ }\href@noop {} {\bibfield  {journal} {\bibinfo  {journal} {J.
  Chem. Phys.}\ }\textbf {\bibinfo {volume} {156}},\ \bibinfo {pages} {154113}
  (\bibinfo {year} {2022})}\BibitemShut {NoStop}%
\bibitem [{\citenamefont {Rapcewicz}\ and\ \citenamefont
  {Ashcroft}(1991)}]{ra}%
  \BibitemOpen
  \bibfield  {author} {\bibinfo {author} {\bibfnamefont {K.}~\bibnamefont
  {Rapcewicz}}\ and\ \bibinfo {author} {\bibfnamefont {N.~W.}\ \bibnamefont
  {Ashcroft}},\ }\bibfield  {title} {\bibinfo {title} {Fluctuation attraction
  in condensed matter: {A} nonlocal functional approach},\ }\href@noop {}
  {\bibfield  {journal} {\bibinfo  {journal} {Phys. Rev. B}\ }\textbf {\bibinfo
  {volume} {44}},\ \bibinfo {pages} {4032} (\bibinfo {year}
  {1991})}\BibitemShut {NoStop}%
\bibitem [{\citenamefont {Langreth}\ and\ \citenamefont
  {Vosko}(1987)}]{lavo87}%
  \BibitemOpen
  \bibfield  {author} {\bibinfo {author} {\bibfnamefont {D.~C.}\ \bibnamefont
  {Langreth}}\ and\ \bibinfo {author} {\bibfnamefont {S.~H.}\ \bibnamefont
  {Vosko}},\ }\bibfield  {title} {\bibinfo {title} {Exact electron-gas response
  functions at high density},\ }\href@noop {} {\bibfield  {journal} {\bibinfo
  {journal} {Phys. Rev. Lett.}\ }\textbf {\bibinfo {volume} {59}},\ \bibinfo
  {pages} {497} (\bibinfo {year} {1987})}\BibitemShut {NoStop}%
\bibitem [{\citenamefont {Hedin}\ and\ \citenamefont
  {Lundqvist}(1971)}]{helujpc1971}%
  \BibitemOpen
  \bibfield  {author} {\bibinfo {author} {\bibfnamefont {L.}~\bibnamefont
  {Hedin}}\ and\ \bibinfo {author} {\bibfnamefont {B.~I.}\ \bibnamefont
  {Lundqvist}},\ }\bibfield  {title} {\bibinfo {title} {Explicit local
  exchange-correlation potentials},\ }\href@noop {} {\bibfield  {journal}
  {\bibinfo  {journal} {J. Phys. C}\ }\textbf {\bibinfo {volume} {4}},\
  \bibinfo {pages} {2064} (\bibinfo {year} {1971})}\BibitemShut {NoStop}%
\bibitem [{\citenamefont {Gunnarsson}\ and\ \citenamefont
  {Lundqvist}(1976)}]{gulu76}%
  \BibitemOpen
  \bibfield  {author} {\bibinfo {author} {\bibfnamefont {O.}~\bibnamefont
  {Gunnarsson}}\ and\ \bibinfo {author} {\bibfnamefont {B.~I.}\ \bibnamefont
  {Lundqvist}},\ }\bibfield  {title} {\bibinfo {title} {Exchange and
  correlation in atoms, molecules, and solids by the spin-density-functional
  formalism},\ }\href@noop {} {\bibfield  {journal} {\bibinfo  {journal} {Phys.
  Rev. B}\ }\textbf {\bibinfo {volume} {13}},\ \bibinfo {pages} {4274}
  (\bibinfo {year} {1976})}\BibitemShut {NoStop}%
\bibitem [{\citenamefont {Perdew}\ and\ \citenamefont {Wang}(1992)}]{pewa92}%
  \BibitemOpen
  \bibfield  {author} {\bibinfo {author} {\bibfnamefont {J.~P.}\ \bibnamefont
  {Perdew}}\ and\ \bibinfo {author} {\bibfnamefont {Y.}~\bibnamefont {Wang}},\
  }\bibfield  {title} {\bibinfo {title} {Pair-distribution function and its
  coupling-constant average for the spin-polarized electron gas},\ }\href@noop
  {} {\bibfield  {journal} {\bibinfo  {journal} {Phys. Rev. B}\ }\textbf
  {\bibinfo {volume} {46}},\ \bibinfo {pages} {12947} (\bibinfo {year}
  {1992})}\BibitemShut {NoStop}%
\bibitem [{\citenamefont {Perdew}\ \emph
  {et~al.}(1996{\natexlab{a}})\citenamefont {Perdew}, \citenamefont {Burke},\
  and\ \citenamefont {Ernzerhof}}]{pebuer96}%
  \BibitemOpen
  \bibfield  {author} {\bibinfo {author} {\bibfnamefont {J.~P.}\ \bibnamefont
  {Perdew}}, \bibinfo {author} {\bibfnamefont {K.}~\bibnamefont {Burke}},\ and\
  \bibinfo {author} {\bibfnamefont {M.}~\bibnamefont {Ernzerhof}},\ }\bibfield
  {title} {\bibinfo {title} {Generalized gradient approximation made simple},\
  }\href@noop {} {\bibfield  {journal} {\bibinfo  {journal} {Phys. Rev. Lett.}\
  }\textbf {\bibinfo {volume} {77}},\ \bibinfo {pages} {3865} (\bibinfo {year}
  {1996}{\natexlab{a}})}\BibitemShut {NoStop}%
\bibitem [{\citenamefont {Perdew}\ \emph {et~al.}(2008)\citenamefont {Perdew},
  \citenamefont {Ruzsinszky}, \citenamefont {Csonka}, \citenamefont {Vydrov},
  \citenamefont {Scuseria}, \citenamefont {Constantin}, \citenamefont {Zhou},\
  and\ \citenamefont {Burke}}]{PBEsol}%
  \BibitemOpen
  \bibfield  {author} {\bibinfo {author} {\bibfnamefont {J.~P.}\ \bibnamefont
  {Perdew}}, \bibinfo {author} {\bibfnamefont {A.}~\bibnamefont {Ruzsinszky}},
  \bibinfo {author} {\bibfnamefont {G.~I.}\ \bibnamefont {Csonka}}, \bibinfo
  {author} {\bibfnamefont {O.~A.}\ \bibnamefont {Vydrov}}, \bibinfo {author}
  {\bibfnamefont {G.~E.}\ \bibnamefont {Scuseria}}, \bibinfo {author}
  {\bibfnamefont {L.~A.}\ \bibnamefont {Constantin}}, \bibinfo {author}
  {\bibfnamefont {X.}~\bibnamefont {Zhou}},\ and\ \bibinfo {author}
  {\bibfnamefont {K.}~\bibnamefont {Burke}},\ }\bibfield  {title} {\bibinfo
  {title} {Restoring the density-gradient expansion for exchange in solids and
  surfaces},\ }\href@noop {} {\bibfield  {journal} {\bibinfo  {journal} {Phys.
  Rev. Lett.}\ }\textbf {\bibinfo {volume} {100}},\ \bibinfo {pages} {136406}
  (\bibinfo {year} {2008})}\BibitemShut {NoStop}%
\bibitem [{\citenamefont {Sun}\ \emph {et~al.}(2015)\citenamefont {Sun},
  \citenamefont {Ruzsinszky},\ and\ \citenamefont {Perdew}}]{SCAN}%
  \BibitemOpen
  \bibfield  {author} {\bibinfo {author} {\bibfnamefont {J.}~\bibnamefont
  {Sun}}, \bibinfo {author} {\bibfnamefont {A.}~\bibnamefont {Ruzsinszky}},\
  and\ \bibinfo {author} {\bibfnamefont {J.~P.}\ \bibnamefont {Perdew}},\
  }\bibfield  {title} {\bibinfo {title} {Strongly constrained and appropriately
  normed semilocal density functional},\ }\href@noop {} {\bibfield  {journal}
  {\bibinfo  {journal} {Phys. Rev. Lett.}\ }\textbf {\bibinfo {volume} {115}},\
  \bibinfo {pages} {036402} (\bibinfo {year} {2015})}\BibitemShut {NoStop}%
\bibitem [{\citenamefont {Peng}\ \emph {et~al.}(2016)\citenamefont {Peng},
  \citenamefont {Yang}, \citenamefont {Perdew},\ and\ \citenamefont
  {Sun}}]{SCANvdW}%
  \BibitemOpen
  \bibfield  {author} {\bibinfo {author} {\bibfnamefont {H.}~\bibnamefont
  {Peng}}, \bibinfo {author} {\bibfnamefont {Z.-H.}\ \bibnamefont {Yang}},
  \bibinfo {author} {\bibfnamefont {J.~P.}\ \bibnamefont {Perdew}},\ and\
  \bibinfo {author} {\bibfnamefont {J.}~\bibnamefont {Sun}},\ }\bibfield
  {title} {\bibinfo {title} {{Versatile van der Waals Density Functional Based
  on a Meta-Generalized Gradient Approximation}},\ }\href@noop {} {\bibfield
  {journal} {\bibinfo  {journal} {Phys. Rev. X}\ }\textbf {\bibinfo {volume}
  {6}},\ \bibinfo {pages} {041005} (\bibinfo {year} {2016})}\BibitemShut
  {NoStop}%
\bibitem [{\citenamefont {Langreth}\ and\ \citenamefont
  {Perdew}(1979)}]{lape79ssc}%
  \BibitemOpen
  \bibfield  {author} {\bibinfo {author} {\bibfnamefont {D.~C.}\ \bibnamefont
  {Langreth}}\ and\ \bibinfo {author} {\bibfnamefont {J.~P.}\ \bibnamefont
  {Perdew}},\ }\bibfield  {title} {\bibinfo {title} {The gradient approximation
  to the exchange-correlation energy functional: a generalization that works},\
  }\href@noop {} {\bibfield  {journal} {\bibinfo  {journal} {Sol. State.
  Comm.}\ }\textbf {\bibinfo {volume} {31}},\ \bibinfo {pages} {567} (\bibinfo
  {year} {1979})}\BibitemShut {NoStop}%
\bibitem [{\citenamefont {Tran}\ \emph {et~al.}(2019)\citenamefont {Tran},
  \citenamefont {Kalantari}, \citenamefont {Traor{\'e}}, \citenamefont
  {Rocquefelte},\ and\ \citenamefont {Blaha}}]{Tran19}%
  \BibitemOpen
  \bibfield  {author} {\bibinfo {author} {\bibfnamefont {F.}~\bibnamefont
  {Tran}}, \bibinfo {author} {\bibfnamefont {L.}~\bibnamefont {Kalantari}},
  \bibinfo {author} {\bibfnamefont {B.}~\bibnamefont {Traor{\'e}}}, \bibinfo
  {author} {\bibfnamefont {X.}~\bibnamefont {Rocquefelte}},\ and\ \bibinfo
  {author} {\bibfnamefont {P.}~\bibnamefont {Blaha}},\ }\bibfield  {title}
  {\bibinfo {title} {{Nonlocal van der Waals functionals for solids: Choosing
  an appropriate one}},\ }\href@noop {} {\bibfield  {journal} {\bibinfo
  {journal} {Phys. Rev. M}\ }\textbf {\bibinfo {volume} {3}},\ \bibinfo {pages}
  {063602} (\bibinfo {year} {2019})}\BibitemShut {NoStop}%
\bibitem [{\citenamefont {Hamada}(2014)}]{hamada14}%
  \BibitemOpen
  \bibfield  {author} {\bibinfo {author} {\bibfnamefont {I.}~\bibnamefont
  {Hamada}},\ }\bibfield  {title} {\bibinfo {title} {van der {W}aals density
  functional made accurate},\ }\href@noop {} {\bibfield  {journal} {\bibinfo
  {journal} {Phys. Rev. B}\ }\textbf {\bibinfo {volume} {89}},\ \bibinfo
  {pages} {121103(R)} (\bibinfo {year} {2014})}\BibitemShut {NoStop}%
\bibitem [{\citenamefont {Klime\v{s}}\ \emph {et~al.}(2010)\citenamefont
  {Klime\v{s}}, \citenamefont {Bowler},\ and\ \citenamefont
  {Michaelides}}]{optx}%
  \BibitemOpen
  \bibfield  {author} {\bibinfo {author} {\bibfnamefont {J.}~\bibnamefont
  {Klime\v{s}}}, \bibinfo {author} {\bibfnamefont {D.~R.}\ \bibnamefont
  {Bowler}},\ and\ \bibinfo {author} {\bibfnamefont {A.}~\bibnamefont
  {Michaelides}},\ }\bibfield  {title} {\bibinfo {title} {Chemical accuracy for
  the van der {W}aals density functional},\ }\href@noop {} {\bibfield
  {journal} {\bibinfo  {journal} {J. Phys.: Condens. Matter}\ }\textbf
  {\bibinfo {volume} {22}},\ \bibinfo {pages} {022201} (\bibinfo {year}
  {2010})}\BibitemShut {NoStop}%
\bibitem [{\citenamefont {Vydrov}\ and\ \citenamefont {{Van
  Voorhis}}(2010)}]{vv10}%
  \BibitemOpen
  \bibfield  {author} {\bibinfo {author} {\bibfnamefont {O.~A.}\ \bibnamefont
  {Vydrov}}\ and\ \bibinfo {author} {\bibfnamefont {T.}~\bibnamefont {{Van
  Voorhis}}},\ }\bibfield  {title} {\bibinfo {title} {Nonlocal van der {W}aals
  density functional: {T}he simpler the better},\ }\href@noop {} {\bibfield
  {journal} {\bibinfo  {journal} {J. Chem. Phys.}\ }\textbf {\bibinfo {volume}
  {133}},\ \bibinfo {pages} {244103} (\bibinfo {year} {2010})}\BibitemShut
  {NoStop}%
\bibitem [{\citenamefont {Cooper}(2010)}]{cooper10p161104}%
  \BibitemOpen
  \bibfield  {author} {\bibinfo {author} {\bibfnamefont {V.~R.}\ \bibnamefont
  {Cooper}},\ }\bibfield  {title} {\bibinfo {title} {van der {W}aals density
  functional: An appropriate exchange functional},\ }\href@noop {} {\bibfield
  {journal} {\bibinfo  {journal} {Phys. Rev. B}\ }\textbf {\bibinfo {volume}
  {81}},\ \bibinfo {pages} {161104(R)} (\bibinfo {year} {2010})}\BibitemShut
  {NoStop}%
\bibitem [{\citenamefont {Klime\v{s}}\ \emph {et~al.}(2011)\citenamefont
  {Klime\v{s}}, \citenamefont {Bowler},\ and\ \citenamefont
  {Michaelides}}]{vdwsolids}%
  \BibitemOpen
  \bibfield  {author} {\bibinfo {author} {\bibfnamefont {J.}~\bibnamefont
  {Klime\v{s}}}, \bibinfo {author} {\bibfnamefont {D.~R.}\ \bibnamefont
  {Bowler}},\ and\ \bibinfo {author} {\bibfnamefont {A.}~\bibnamefont
  {Michaelides}},\ }\bibfield  {title} {\bibinfo {title} {{v}an der {W}aals
  density functionals applied to solids},\ }\href@noop {} {\bibfield  {journal}
  {\bibinfo  {journal} {Phys. Rev. B}\ }\textbf {\bibinfo {volume} {83}},\
  \bibinfo {pages} {195131} (\bibinfo {year} {2011})}\BibitemShut {NoStop}%
\bibitem [{\citenamefont {Sabatini}\ \emph {et~al.}(2013)\citenamefont
  {Sabatini}, \citenamefont {Gorni},\ and\ \citenamefont
  {de~Gironcoli}}]{Sabatini2013p041108}%
  \BibitemOpen
  \bibfield  {author} {\bibinfo {author} {\bibfnamefont {R.}~\bibnamefont
  {Sabatini}}, \bibinfo {author} {\bibfnamefont {T.}~\bibnamefont {Gorni}},\
  and\ \bibinfo {author} {\bibfnamefont {S.}~\bibnamefont {de~Gironcoli}},\
  }\bibfield  {title} {\bibinfo {title} {Nonlocal van der {W}aals density
  functional made simple and efficient},\ }\href@noop {} {\bibfield  {journal}
  {\bibinfo  {journal} {Phys. Rev. B}\ }\textbf {\bibinfo {volume} {87}},\
  \bibinfo {pages} {041108(R)} (\bibinfo {year} {2013})}\BibitemShut {NoStop}%
\bibitem [{\citenamefont {Harris}(1985)}]{Harris85}%
  \BibitemOpen
  \bibfield  {author} {\bibinfo {author} {\bibfnamefont {J.}~\bibnamefont
  {Harris}},\ }\bibfield  {title} {\bibinfo {title} {Simplified method for
  calculating the energy of weakly interacting fragments},\ }\href@noop {}
  {\bibfield  {journal} {\bibinfo  {journal} {Phys. Rev. B}\ }\textbf {\bibinfo
  {volume} {31}},\ \bibinfo {pages} {1770} (\bibinfo {year}
  {1985})}\BibitemShut {NoStop}%
\bibitem [{\citenamefont {Murray}\ \emph {et~al.}(2009)\citenamefont {Murray},
  \citenamefont {Lee},\ and\ \citenamefont {Langreth}}]{mulela09}%
  \BibitemOpen
  \bibfield  {author} {\bibinfo {author} {\bibfnamefont {{\'E}.~D.}\
  \bibnamefont {Murray}}, \bibinfo {author} {\bibfnamefont {K.}~\bibnamefont
  {Lee}},\ and\ \bibinfo {author} {\bibfnamefont {D.~C.}\ \bibnamefont
  {Langreth}},\ }\bibfield  {title} {\bibinfo {title} {Investigation of
  exchange energy density functional accuracy for interacting molecules},\
  }\href@noop {} {\bibfield  {journal} {\bibinfo  {journal} {J. Chem. Theory
  Comput.}\ }\textbf {\bibinfo {volume} {5}},\ \bibinfo {pages} {2754}
  (\bibinfo {year} {2009})}\BibitemShut {NoStop}%
\bibitem [{\citenamefont {Becke}(2014)}]{beckeperspective}%
  \BibitemOpen
  \bibfield  {author} {\bibinfo {author} {\bibfnamefont {A.~D.}\ \bibnamefont
  {Becke}},\ }\bibfield  {title} {\bibinfo {title} {Perspective: Fifty years of
  density-functional theory in chemical physics},\ }\href@noop {} {\bibfield
  {journal} {\bibinfo  {journal} {J. Chem. Phys.}\ }\textbf {\bibinfo {volume}
  {140}},\ \bibinfo {pages} {18A301} (\bibinfo {year} {2014})}\BibitemShut
  {NoStop}%
\bibitem [{\citenamefont {Burke}(2012)}]{BurkePerspective}%
  \BibitemOpen
  \bibfield  {author} {\bibinfo {author} {\bibfnamefont {K.}~\bibnamefont
  {Burke}},\ }\bibfield  {title} {\bibinfo {title} {Perspective on density
  functional theory},\ }\href@noop {} {\bibfield  {journal} {\bibinfo
  {journal} {J. Chem. Phys.}\ }\textbf {\bibinfo {volume} {136}},\ \bibinfo
  {pages} {150901} (\bibinfo {year} {2012})}\BibitemShut {NoStop}%
\bibitem [{\citenamefont {Pople}\ \emph {et~al.}(1989)\citenamefont {Pople},
  \citenamefont {Head‐Gordon}, \citenamefont {Fox}, \citenamefont
  {Raghavachari},\ and\ \citenamefont {Curtiss}}]{G1set}%
  \BibitemOpen
  \bibfield  {author} {\bibinfo {author} {\bibfnamefont {J.~A.}\ \bibnamefont
  {Pople}}, \bibinfo {author} {\bibfnamefont {M.}~\bibnamefont
  {Head‐Gordon}}, \bibinfo {author} {\bibfnamefont {D.~J.}\ \bibnamefont
  {Fox}}, \bibinfo {author} {\bibfnamefont {K.}~\bibnamefont {Raghavachari}},\
  and\ \bibinfo {author} {\bibfnamefont {L.~A.}\ \bibnamefont {Curtiss}},\
  }\bibfield  {title} {\bibinfo {title} {Gaussian‐1 theory: {A} general
  procedure for prediction of molecular energies},\ }\href@noop {} {\bibfield
  {journal} {\bibinfo  {journal} {J. Chem. Phys.}\ }\textbf {\bibinfo {volume}
  {90}},\ \bibinfo {pages} {5622} (\bibinfo {year} {1989})}\BibitemShut
  {NoStop}%
\bibitem [{\citenamefont {Goerigk}\ and\ \citenamefont {Grimme}(2010)}]{G2RC}%
  \BibitemOpen
  \bibfield  {author} {\bibinfo {author} {\bibfnamefont {L.}~\bibnamefont
  {Goerigk}}\ and\ \bibinfo {author} {\bibfnamefont {S.}~\bibnamefont
  {Grimme}},\ }\bibfield  {title} {\bibinfo {title} {{A General Database for
  Main Group Thermochemistry, Kinetics, and Noncovalent Interactions -
  Assessment of common and Reparameterized (meta-)GGA Density Functionals}},\
  }\href {https://doi.org/10.1021/ct900489g} {\bibfield  {journal} {\bibinfo
  {journal} {J. Chem. Theory Comput.}\ }\textbf {\bibinfo {volume} {6}},\
  \bibinfo {pages} {107} (\bibinfo {year} {2010})}\BibitemShut {NoStop}%
\bibitem [{\citenamefont {Goerigk}\ \emph {et~al.}(2017)\citenamefont
  {Goerigk}, \citenamefont {Hansen}, \citenamefont {Bayer}, \citenamefont
  {Ehrlich}, \citenamefont {Najibi},\ and\ \citenamefont {Grimme}}]{gmtkn55}%
  \BibitemOpen
  \bibfield  {author} {\bibinfo {author} {\bibfnamefont {L.}~\bibnamefont
  {Goerigk}}, \bibinfo {author} {\bibfnamefont {A.}~\bibnamefont {Hansen}},
  \bibinfo {author} {\bibfnamefont {C.}~\bibnamefont {Bayer}}, \bibinfo
  {author} {\bibfnamefont {S.}~\bibnamefont {Ehrlich}}, \bibinfo {author}
  {\bibfnamefont {A.}~\bibnamefont {Najibi}},\ and\ \bibinfo {author}
  {\bibfnamefont {S.}~\bibnamefont {Grimme}},\ }\bibfield  {title} {\bibinfo
  {title} {{A look at the density functional theory zoo with the advanced
  GMTKN55 database for general main group thermochemistry, kinetics and
  noncovalent interactions}},\ }\href@noop {} {\bibfield  {journal} {\bibinfo
  {journal} {Phys. Chem. Chem. Phys.}\ }\textbf {\bibinfo {volume} {19}},\
  \bibinfo {pages} {32184} (\bibinfo {year} {2017})}\BibitemShut {NoStop}%
\bibitem [{\citenamefont {Rosi}\ \emph {et~al.}(2003)\citenamefont {Rosi},
  \citenamefont {Eckert}, \citenamefont {Eddaoudi}, \citenamefont {Vodak},
  \citenamefont {Kim}, \citenamefont {O’Keeffe},\ and\ \citenamefont
  {Yaghi}}]{HydMOF}%
  \BibitemOpen
  \bibfield  {author} {\bibinfo {author} {\bibfnamefont {N.}~\bibnamefont
  {Rosi}}, \bibinfo {author} {\bibfnamefont {J.}~\bibnamefont {Eckert}},
  \bibinfo {author} {\bibfnamefont {M.}~\bibnamefont {Eddaoudi}}, \bibinfo
  {author} {\bibfnamefont {D.}~\bibnamefont {Vodak}}, \bibinfo {author}
  {\bibfnamefont {J.}~\bibnamefont {Kim}}, \bibinfo {author} {\bibfnamefont
  {M.}~\bibnamefont {O’Keeffe}},\ and\ \bibinfo {author} {\bibfnamefont
  {O.}~\bibnamefont {Yaghi}},\ }\bibfield  {title} {\bibinfo {title} {{Hydrogen
  Storage in Microporous Metal-Organic Frameworks}},\ }\href@noop {} {\bibfield
   {journal} {\bibinfo  {journal} {Science}\ }\textbf {\bibinfo {volume}
  {300}},\ \bibinfo {pages} {1127} (\bibinfo {year} {2003})}\BibitemShut
  {NoStop}%
\bibitem [{\citenamefont {Langreth}\ \emph {et~al.}(2009)\citenamefont
  {Langreth}, \citenamefont {Lundqvist}, \citenamefont {Chakarova-K{\"a}ck},
  \citenamefont {Cooper}, \citenamefont {Dion}, \citenamefont {Hyldgaard},
  \citenamefont {Kelkkanen}, \citenamefont {Kleis}, \citenamefont {Kong},
  \citenamefont {Li}, \citenamefont {Moses}, \citenamefont {Murray},
  \citenamefont {Puzder}, \citenamefont {Rydberg}, \citenamefont
  {Schr{\"o}der},\ and\ \citenamefont {Thonhauser}}]{langrethjpcm2009}%
  \BibitemOpen
  \bibfield  {author} {\bibinfo {author} {\bibfnamefont {D.~C.}\ \bibnamefont
  {Langreth}}, \bibinfo {author} {\bibfnamefont {B.~I.}\ \bibnamefont
  {Lundqvist}}, \bibinfo {author} {\bibfnamefont {S.~D.}\ \bibnamefont
  {Chakarova-K{\"a}ck}}, \bibinfo {author} {\bibfnamefont {V.~R.}\ \bibnamefont
  {Cooper}}, \bibinfo {author} {\bibfnamefont {M.}~\bibnamefont {Dion}},
  \bibinfo {author} {\bibfnamefont {P.}~\bibnamefont {Hyldgaard}}, \bibinfo
  {author} {\bibfnamefont {A.}~\bibnamefont {Kelkkanen}}, \bibinfo {author}
  {\bibfnamefont {J.}~\bibnamefont {Kleis}}, \bibinfo {author} {\bibfnamefont
  {L.}~\bibnamefont {Kong}}, \bibinfo {author} {\bibfnamefont {S.}~\bibnamefont
  {Li}}, \bibinfo {author} {\bibfnamefont {P.~G.}\ \bibnamefont {Moses}},
  \bibinfo {author} {\bibfnamefont {E.}~\bibnamefont {Murray}}, \bibinfo
  {author} {\bibfnamefont {A.}~\bibnamefont {Puzder}}, \bibinfo {author}
  {\bibfnamefont {H.}~\bibnamefont {Rydberg}}, \bibinfo {author} {\bibfnamefont
  {E.}~\bibnamefont {Schr{\"o}der}},\ and\ \bibinfo {author} {\bibfnamefont
  {T.}~\bibnamefont {Thonhauser}},\ }\bibfield  {title} {\bibinfo {title} {A
  density functional for sparse matter},\ }\href@noop {} {\bibfield  {journal}
  {\bibinfo  {journal} {J. Phys.: Condens. Matter}\ }\textbf {\bibinfo {volume}
  {21}},\ \bibinfo {pages} {084203} (\bibinfo {year} {2009})}\BibitemShut
  {NoStop}%
\bibitem [{\citenamefont {Kong}\ \emph {et~al.}(2009)\citenamefont {Kong},
  \citenamefont {Cooper}, \citenamefont {Nijem}, \citenamefont {Li},
  \citenamefont {Li}, \citenamefont {Chabal},\ and\ \citenamefont
  {Langreth}}]{kong09p081407}%
  \BibitemOpen
  \bibfield  {author} {\bibinfo {author} {\bibfnamefont {L.}~\bibnamefont
  {Kong}}, \bibinfo {author} {\bibfnamefont {V.~R.}\ \bibnamefont {Cooper}},
  \bibinfo {author} {\bibfnamefont {N.}~\bibnamefont {Nijem}}, \bibinfo
  {author} {\bibfnamefont {K.}~\bibnamefont {Li}}, \bibinfo {author}
  {\bibfnamefont {J.}~\bibnamefont {Li}}, \bibinfo {author} {\bibfnamefont
  {Y.~J.}\ \bibnamefont {Chabal}},\ and\ \bibinfo {author} {\bibfnamefont
  {D.~C.}\ \bibnamefont {Langreth}},\ }\bibfield  {title} {\bibinfo {title}
  {Theoretical and experimental analysis of {H}$_2$ binding in a prototypical
  metal-organic framework material},\ }\href@noop {} {\bibfield  {journal}
  {\bibinfo  {journal} {Phys. Rev. B}\ }\textbf {\bibinfo {volume} {79}},\
  \bibinfo {pages} {081407(R)} (\bibinfo {year} {2009})}\BibitemShut {NoStop}%
\bibitem [{\citenamefont {Li}\ and\ \citenamefont
  {Thonhauser}(2012)}]{li12p424204}%
  \BibitemOpen
  \bibfield  {author} {\bibinfo {author} {\bibfnamefont {Q.}~\bibnamefont
  {Li}}\ and\ \bibinfo {author} {\bibfnamefont {T.}~\bibnamefont
  {Thonhauser}},\ }\bibfield  {title} {\bibinfo {title} {A theoretical study of
  the hydrogen-storage potential of ({H}$_2$)$_4${CH}$_4$ in metal organic
  framework materials and carbon nanotubes},\ }\href@noop {} {\bibfield
  {journal} {\bibinfo  {journal} {J. Phys.: Condens. Matter}\ }\textbf
  {\bibinfo {volume} {24}},\ \bibinfo {pages} {424204} (\bibinfo {year}
  {2012})}\BibitemShut {NoStop}%
\bibitem [{\citenamefont {Cooper}\ \emph {et~al.}(2012)\citenamefont {Cooper},
  \citenamefont {Ihm},\ and\ \citenamefont {Morris}}]{cooper12p34}%
  \BibitemOpen
  \bibfield  {author} {\bibinfo {author} {\bibfnamefont {V.~R.}\ \bibnamefont
  {Cooper}}, \bibinfo {author} {\bibfnamefont {Y.}~\bibnamefont {Ihm}},\ and\
  \bibinfo {author} {\bibfnamefont {J.~R.}\ \bibnamefont {Morris}},\ }\bibfield
   {title} {\bibinfo {title} {Hydrogen adsorption at the graphene surface: A
  vd{W-DF} perspective},\ }\href@noop {} {\bibfield  {journal} {\bibinfo
  {journal} {Phys. Proc.}\ }\textbf {\bibinfo {volume} {34}},\ \bibinfo {pages}
  {34} (\bibinfo {year} {2012})}\BibitemShut {NoStop}%
\bibitem [{\citenamefont {Millward}\ and\ \citenamefont
  {Yaghi}(2005)}]{CO2MOF}%
  \BibitemOpen
  \bibfield  {author} {\bibinfo {author} {\bibfnamefont {A.~R.}\ \bibnamefont
  {Millward}}\ and\ \bibinfo {author} {\bibfnamefont {O.~M.}\ \bibnamefont
  {Yaghi}},\ }\bibfield  {title} {\bibinfo {title} {{Metal-Organic Frameworks
  with Exceptionally High Capacity for Storage of Carbon Dioxide at Room
  Temperature}},\ }\href@noop {} {\bibfield  {journal} {\bibinfo  {journal} {J.
  Am. Chem. Soc.}\ }\textbf {\bibinfo {volume} {127}},\ \bibinfo {pages}
  {17998} (\bibinfo {year} {2005})}\BibitemShut {NoStop}%
\bibitem [{\citenamefont {McDonald}\ \emph {et~al.}(2012)\citenamefont
  {McDonald}, \citenamefont {Lee}, \citenamefont {Mason}, \citenamefont
  {Wiers}, \citenamefont {Hong},\ and\ \citenamefont {Long}}]{CO2dobpdc1}%
  \BibitemOpen
  \bibfield  {author} {\bibinfo {author} {\bibfnamefont {T.~M.}\ \bibnamefont
  {McDonald}}, \bibinfo {author} {\bibfnamefont {W.}~\bibnamefont {Lee}},
  \bibinfo {author} {\bibfnamefont {J.~A.}\ \bibnamefont {Mason}}, \bibinfo
  {author} {\bibfnamefont {B.~M.}\ \bibnamefont {Wiers}}, \bibinfo {author}
  {\bibfnamefont {C.~S.}\ \bibnamefont {Hong}},\ and\ \bibinfo {author}
  {\bibfnamefont {J.~R.}\ \bibnamefont {Long}},\ }\bibfield  {title} {\bibinfo
  {title} {{Capture of Carbon Dioxide from Air and Flue Gas in the
  Alkylamine-Appended Metal-Organic Framework mmen-Mg2(dobpdc)}},\ }\href@noop
  {} {\bibfield  {journal} {\bibinfo  {journal} {J. Am. Chem. Soc.}\ }\textbf
  {\bibinfo {volume} {134}},\ \bibinfo {pages} {7056} (\bibinfo {year}
  {2012})}\BibitemShut {NoStop}%
\bibitem [{\citenamefont {Bae}\ and\ \citenamefont {Long}(2013)}]{CO2sepMOF}%
  \BibitemOpen
  \bibfield  {author} {\bibinfo {author} {\bibfnamefont {T.-H.}\ \bibnamefont
  {Bae}}\ and\ \bibinfo {author} {\bibfnamefont {J.~R.}\ \bibnamefont {Long}},\
  }\bibfield  {title} {\bibinfo {title} {{CO2/N2 separations with mixed-matrix
  membranes containing Mg2(dobdc) nanocrystals}},\ }\href@noop {} {\bibfield
  {journal} {\bibinfo  {journal} {Energy Environ. Sci.}\ }\textbf {\bibinfo
  {volume} {6}},\ \bibinfo {pages} {3565} (\bibinfo {year} {2013})}\BibitemShut
  {NoStop}%
\bibitem [{\citenamefont {McDonald}\ \emph {et~al.}(2015)\citenamefont
  {McDonald}, \citenamefont {Mason}, \citenamefont {Kong}, \citenamefont
  {Bloch}, \citenamefont {Gygi}, \citenamefont {Dani}, \citenamefont
  {Crocella}, \citenamefont {Giordanino}, \citenamefont {Odoh}, \citenamefont
  {Drisdell}, \citenamefont {Vlaisavljevich}, \citenamefont {Dzubak},
  \citenamefont {Poloni}, \citenamefont {Schnell}, \citenamefont {Planas},
  \citenamefont {Lee}, \citenamefont {Pascal}, \citenamefont {Wan},
  \citenamefont {Prendergast}, \citenamefont {Neaton}, \citenamefont {Smit},
  \citenamefont {Kortright}, \citenamefont {Gagliardi}, \citenamefont
  {Bordiga}, \citenamefont {Reimer},\ and\ \citenamefont {Long}}]{CO2dobpdc2}%
  \BibitemOpen
  \bibfield  {author} {\bibinfo {author} {\bibfnamefont {T.~M.}\ \bibnamefont
  {McDonald}}, \bibinfo {author} {\bibfnamefont {J.~A.}\ \bibnamefont {Mason}},
  \bibinfo {author} {\bibfnamefont {X.}~\bibnamefont {Kong}}, \bibinfo {author}
  {\bibfnamefont {E.~D.}\ \bibnamefont {Bloch}}, \bibinfo {author}
  {\bibfnamefont {D.}~\bibnamefont {Gygi}}, \bibinfo {author} {\bibfnamefont
  {A.}~\bibnamefont {Dani}}, \bibinfo {author} {\bibfnamefont {V.}~\bibnamefont
  {Crocella}}, \bibinfo {author} {\bibfnamefont {F.}~\bibnamefont
  {Giordanino}}, \bibinfo {author} {\bibfnamefont {S.~O.}\ \bibnamefont
  {Odoh}}, \bibinfo {author} {\bibfnamefont {W.~S.}\ \bibnamefont {Drisdell}},
  \bibinfo {author} {\bibfnamefont {B.}~\bibnamefont {Vlaisavljevich}},
  \bibinfo {author} {\bibfnamefont {A.~L.}\ \bibnamefont {Dzubak}}, \bibinfo
  {author} {\bibfnamefont {R.}~\bibnamefont {Poloni}}, \bibinfo {author}
  {\bibfnamefont {S.~K.}\ \bibnamefont {Schnell}}, \bibinfo {author}
  {\bibfnamefont {N.}~\bibnamefont {Planas}}, \bibinfo {author} {\bibfnamefont
  {K.}~\bibnamefont {Lee}}, \bibinfo {author} {\bibfnamefont {T.}~\bibnamefont
  {Pascal}}, \bibinfo {author} {\bibfnamefont {L.~F.}\ \bibnamefont {Wan}},
  \bibinfo {author} {\bibfnamefont {D.}~\bibnamefont {Prendergast}}, \bibinfo
  {author} {\bibfnamefont {J.~B.}\ \bibnamefont {Neaton}}, \bibinfo {author}
  {\bibfnamefont {B.}~\bibnamefont {Smit}}, \bibinfo {author} {\bibfnamefont
  {J.~B.}\ \bibnamefont {Kortright}}, \bibinfo {author} {\bibfnamefont
  {L.}~\bibnamefont {Gagliardi}}, \bibinfo {author} {\bibfnamefont
  {S.}~\bibnamefont {Bordiga}}, \bibinfo {author} {\bibfnamefont {J.~A.}\
  \bibnamefont {Reimer}},\ and\ \bibinfo {author} {\bibfnamefont {J.~R.}\
  \bibnamefont {Long}},\ }\bibfield  {title} {\bibinfo {title} {{Cooperative
  insertion of CO2 in diamine-appended metal-organic frameworks}},\ }\href@noop
  {} {\bibfield  {journal} {\bibinfo  {journal} {Nature}\ }\textbf {\bibinfo
  {volume} {519}},\ \bibinfo {pages} {303} (\bibinfo {year}
  {2015})}\BibitemShut {NoStop}%
\bibitem [{\citenamefont {Zuluaga}\ \emph {et~al.}(2014)\citenamefont
  {Zuluaga}, \citenamefont {Canepa}, \citenamefont {Tan}, \citenamefont
  {Chabal},\ and\ \citenamefont {Thonhauser}}]{zuluaga2014}%
  \BibitemOpen
  \bibfield  {author} {\bibinfo {author} {\bibfnamefont {S.}~\bibnamefont
  {Zuluaga}}, \bibinfo {author} {\bibfnamefont {P.}~\bibnamefont {Canepa}},
  \bibinfo {author} {\bibfnamefont {K.}~\bibnamefont {Tan}}, \bibinfo {author}
  {\bibfnamefont {Y.~J.}\ \bibnamefont {Chabal}},\ and\ \bibinfo {author}
  {\bibfnamefont {T.}~\bibnamefont {Thonhauser}},\ }\bibfield  {title}
  {\bibinfo {title} {Study of van der {W}aals bonding and interactions in metal
  organic framework materials},\ }\href@noop {} {\bibfield  {journal} {\bibinfo
   {journal} {J. Phys.: Condens. Matter}\ }\textbf {\bibinfo {volume} {26}},\
  \bibinfo {pages} {133002} (\bibinfo {year} {2014})}\BibitemShut {NoStop}%
\bibitem [{\citenamefont {Poloni}\ \emph {et~al.}(2014)\citenamefont {Poloni},
  \citenamefont {Lee}, \citenamefont {Berger}, \citenamefont {Smit},\ and\
  \citenamefont {Neaton}}]{poloni_understanding_2014}%
  \BibitemOpen
  \bibfield  {author} {\bibinfo {author} {\bibfnamefont {R.}~\bibnamefont
  {Poloni}}, \bibinfo {author} {\bibfnamefont {K.}~\bibnamefont {Lee}},
  \bibinfo {author} {\bibfnamefont {R.~F.}\ \bibnamefont {Berger}}, \bibinfo
  {author} {\bibfnamefont {B.}~\bibnamefont {Smit}},\ and\ \bibinfo {author}
  {\bibfnamefont {J.~B.}\ \bibnamefont {Neaton}},\ }\bibfield  {title}
  {\bibinfo {title} {Understanding trends in {CO}$_2$ adsorption in metal
  organic frameworks with open-metal sites},\ }\href
  {https://doi.org/10.1021/jz500202x} {\bibfield  {journal} {\bibinfo
  {journal} {J. Phys. Chem. Lett.}\ }\textbf {\bibinfo {volume} {5}},\ \bibinfo
  {pages} {861} (\bibinfo {year} {2014})}\BibitemShut {NoStop}%
\bibitem [{\citenamefont {Queen}\ \emph {et~al.}(2014)\citenamefont {Queen},
  \citenamefont {Hudson}, \citenamefont {Bloch}, \citenamefont {Mason},
  \citenamefont {Gonzalez}, \citenamefont {Lee}, \citenamefont {Gygi},
  \citenamefont {Howe}, \citenamefont {Lee}, \citenamefont {Darwish},
  \citenamefont {James}, \citenamefont {Peterson}, \citenamefont {Teat},
  \citenamefont {Smit}, \citenamefont {Neaton}, \citenamefont {Long},\ and\
  \citenamefont {Brown}}]{CO2MOF74survey}%
  \BibitemOpen
  \bibfield  {author} {\bibinfo {author} {\bibfnamefont {W.~L.}\ \bibnamefont
  {Queen}}, \bibinfo {author} {\bibfnamefont {M.~R.}\ \bibnamefont {Hudson}},
  \bibinfo {author} {\bibfnamefont {E.~D.}\ \bibnamefont {Bloch}}, \bibinfo
  {author} {\bibfnamefont {J.~A.}\ \bibnamefont {Mason}}, \bibinfo {author}
  {\bibfnamefont {M.~I.}\ \bibnamefont {Gonzalez}}, \bibinfo {author}
  {\bibfnamefont {J.~S.}\ \bibnamefont {Lee}}, \bibinfo {author} {\bibfnamefont
  {D.}~\bibnamefont {Gygi}}, \bibinfo {author} {\bibfnamefont {J.~D.}\
  \bibnamefont {Howe}}, \bibinfo {author} {\bibfnamefont {K.}~\bibnamefont
  {Lee}}, \bibinfo {author} {\bibfnamefont {T.~A.}\ \bibnamefont {Darwish}},
  \bibinfo {author} {\bibfnamefont {M.}~\bibnamefont {James}}, \bibinfo
  {author} {\bibfnamefont {V.~K.}\ \bibnamefont {Peterson}}, \bibinfo {author}
  {\bibfnamefont {S.~J.}\ \bibnamefont {Teat}}, \bibinfo {author}
  {\bibfnamefont {B.}~\bibnamefont {Smit}}, \bibinfo {author} {\bibfnamefont
  {J.~B.}\ \bibnamefont {Neaton}}, \bibinfo {author} {\bibfnamefont {J.~R.}\
  \bibnamefont {Long}},\ and\ \bibinfo {author} {\bibfnamefont {C.~M.}\
  \bibnamefont {Brown}},\ }\bibfield  {title} {\bibinfo {title} {{Comprehensive
  study of carbon dioxide adsorption in the metal–organic frameworks
  M2(dobdc) (M = Mg, Mn, Fe, Co, Ni, Cu, Zn)}},\ }\href@noop {} {\bibfield
  {journal} {\bibinfo  {journal} {Chem. Sci.}\ }\textbf {\bibinfo {volume}
  {5}},\ \bibinfo {pages} {4569} (\bibinfo {year} {2014})}\BibitemShut
  {NoStop}%
\bibitem [{\citenamefont {Lee}\ and\ \citenamefont
  {Persson}(2012)}]{lee2012li}%
  \BibitemOpen
  \bibfield  {author} {\bibinfo {author} {\bibfnamefont {E.}~\bibnamefont
  {Lee}}\ and\ \bibinfo {author} {\bibfnamefont {K.~A.}\ \bibnamefont
  {Persson}},\ }\bibfield  {title} {\bibinfo {title} {Li absorption and
  intercalation in single layer graphene and few layer graphene by first
  principles},\ }\href@noop {} {\bibfield  {journal} {\bibinfo  {journal} {Nano
  letters}\ }\textbf {\bibinfo {volume} {12}},\ \bibinfo {pages} {4624}
  (\bibinfo {year} {2012})}\BibitemShut {NoStop}%
\bibitem [{\citenamefont {Persson}\ \emph {et~al.}(2010)\citenamefont
  {Persson}, \citenamefont {Hinuma}, \citenamefont {Meng}, \citenamefont
  {Van~der Ven},\ and\ \citenamefont {Ceder}}]{persson2010thermodynamic}%
  \BibitemOpen
  \bibfield  {author} {\bibinfo {author} {\bibfnamefont {K.}~\bibnamefont
  {Persson}}, \bibinfo {author} {\bibfnamefont {Y.}~\bibnamefont {Hinuma}},
  \bibinfo {author} {\bibfnamefont {Y.~S.}\ \bibnamefont {Meng}}, \bibinfo
  {author} {\bibfnamefont {A.}~\bibnamefont {Van~der Ven}},\ and\ \bibinfo
  {author} {\bibfnamefont {G.}~\bibnamefont {Ceder}},\ }\bibfield  {title}
  {\bibinfo {title} {{Thermodynamic and kinetic properties of the Li-graphite
  system from first-principles calculations}},\ }\href@noop {} {\bibfield
  {journal} {\bibinfo  {journal} {Physical Review B}\ }\textbf {\bibinfo
  {volume} {82}},\ \bibinfo {pages} {125416} (\bibinfo {year}
  {2010})}\BibitemShut {NoStop}%
\bibitem [{\citenamefont {Shukla}\ \emph {et~al.}(2019)\citenamefont {Shukla},
  \citenamefont {Jena}, \citenamefont {Naqvi}, \citenamefont {Luo},\ and\
  \citenamefont {Ahuja}}]{shukla2019modelling}%
  \BibitemOpen
  \bibfield  {author} {\bibinfo {author} {\bibfnamefont {V.}~\bibnamefont
  {Shukla}}, \bibinfo {author} {\bibfnamefont {N.~K.}\ \bibnamefont {Jena}},
  \bibinfo {author} {\bibfnamefont {S.~R.}\ \bibnamefont {Naqvi}}, \bibinfo
  {author} {\bibfnamefont {W.}~\bibnamefont {Luo}},\ and\ \bibinfo {author}
  {\bibfnamefont {R.}~\bibnamefont {Ahuja}},\ }\bibfield  {title} {\bibinfo
  {title} {{Modelling high-performing batteries with MXenes: The case of
  S-functionalized two-dimensional nitride MXene electrode}},\ }\href@noop {}
  {\bibfield  {journal} {\bibinfo  {journal} {Nano Energy}\ }\textbf {\bibinfo
  {volume} {58}},\ \bibinfo {pages} {877} (\bibinfo {year} {2019})}\BibitemShut
  {NoStop}%
\bibitem [{\citenamefont {Shukla}\ \emph {et~al.}(2018)\citenamefont {Shukla},
  \citenamefont {Araujo}, \citenamefont {Jena},\ and\ \citenamefont
  {Ahuja}}]{shukla2018borophene}%
  \BibitemOpen
  \bibfield  {author} {\bibinfo {author} {\bibfnamefont {V.}~\bibnamefont
  {Shukla}}, \bibinfo {author} {\bibfnamefont {R.~B.}\ \bibnamefont {Araujo}},
  \bibinfo {author} {\bibfnamefont {N.~K.}\ \bibnamefont {Jena}},\ and\
  \bibinfo {author} {\bibfnamefont {R.}~\bibnamefont {Ahuja}},\ }\bibfield
  {title} {\bibinfo {title} {{Borophene's tryst with stability: exploring 2D
  hydrogen boride as an electrode for rechargeable batteries}},\ }\href@noop {}
  {\bibfield  {journal} {\bibinfo  {journal} {Physical Chemistry Chemical
  Physics}\ }\textbf {\bibinfo {volume} {20}},\ \bibinfo {pages} {22008}
  (\bibinfo {year} {2018})}\BibitemShut {NoStop}%
\bibitem [{\citenamefont {Le}\ \emph {et~al.}(2012)\citenamefont {Le},
  \citenamefont {Kara}, \citenamefont {Schr{\"o}der}, \citenamefont
  {Hyldgaard},\ and\ \citenamefont {Rahman}}]{le12p424210}%
  \BibitemOpen
  \bibfield  {author} {\bibinfo {author} {\bibfnamefont {D.}~\bibnamefont
  {Le}}, \bibinfo {author} {\bibfnamefont {A.}~\bibnamefont {Kara}}, \bibinfo
  {author} {\bibfnamefont {E.}~\bibnamefont {Schr{\"o}der}}, \bibinfo {author}
  {\bibfnamefont {P.}~\bibnamefont {Hyldgaard}},\ and\ \bibinfo {author}
  {\bibfnamefont {T.~S.}\ \bibnamefont {Rahman}},\ }\bibfield  {title}
  {\bibinfo {title} {Physisorption of nucleobases on graphene: A comparative
  van der {W}aals study},\ }\href@noop {} {\bibfield  {journal} {\bibinfo
  {journal} {J. Phys.: Condens. Matter}\ }\textbf {\bibinfo {volume} {24}},\
  \bibinfo {pages} {424210} (\bibinfo {year} {2012})}\BibitemShut {NoStop}%
\bibitem [{\citenamefont {Umrao}\ \emph {et~al.}(2019)\citenamefont {Umrao},
  \citenamefont {Maurya}, \citenamefont {Shukla}, \citenamefont {Grigoriev},
  \citenamefont {Ahuja}, \citenamefont {Vinayak}, \citenamefont {Srivastava},
  \citenamefont {Saxena}, \citenamefont {Oh},\ and\ \citenamefont
  {Srivastava}}]{umrao2019anticarcinogenic}%
  \BibitemOpen
  \bibfield  {author} {\bibinfo {author} {\bibfnamefont {S.}~\bibnamefont
  {Umrao}}, \bibinfo {author} {\bibfnamefont {A.}~\bibnamefont {Maurya}},
  \bibinfo {author} {\bibfnamefont {V.}~\bibnamefont {Shukla}}, \bibinfo
  {author} {\bibfnamefont {A.}~\bibnamefont {Grigoriev}}, \bibinfo {author}
  {\bibfnamefont {R.}~\bibnamefont {Ahuja}}, \bibinfo {author} {\bibfnamefont
  {M.}~\bibnamefont {Vinayak}}, \bibinfo {author} {\bibfnamefont
  {R.}~\bibnamefont {Srivastava}}, \bibinfo {author} {\bibfnamefont
  {P.}~\bibnamefont {Saxena}}, \bibinfo {author} {\bibfnamefont {I.-K.}\
  \bibnamefont {Oh}},\ and\ \bibinfo {author} {\bibfnamefont {A.}~\bibnamefont
  {Srivastava}},\ }\bibfield  {title} {\bibinfo {title} {{Anticarcinogenic
  Activity of Blue Fluorescent Hexagonal Boron Nitride Quantum Dots: As an
  Effective Enhancer for DNA Cleavage Activity of Anticancer Drug
  doxorubicin}},\ }\href@noop {} {\bibfield  {journal} {\bibinfo  {journal}
  {Mater. Today Bio}\ }\textbf {\bibinfo {volume} {1}},\ \bibinfo {pages}
  {100001} (\bibinfo {year} {2019})}\BibitemShut {NoStop}%
\bibitem [{\citenamefont {Berland}\ and\ \citenamefont
  {Hyldgaard}(2013)}]{behy13}%
  \BibitemOpen
  \bibfield  {author} {\bibinfo {author} {\bibfnamefont {K.}~\bibnamefont
  {Berland}}\ and\ \bibinfo {author} {\bibfnamefont {P.}~\bibnamefont
  {Hyldgaard}},\ }\bibfield  {title} {\bibinfo {title} {Analysis of van der
  {W}aals density functional components: Binding and corrugation of benzene and
  {C}$_{60}$ on boron nitride and graphene},\ }\href@noop {} {\bibfield
  {journal} {\bibinfo  {journal} {Phys. Rev. B}\ }\textbf {\bibinfo {volume}
  {87}},\ \bibinfo {pages} {205421} (\bibinfo {year} {2013})}\BibitemShut
  {NoStop}%
\bibitem [{\citenamefont {Lee}\ \emph {et~al.}(2010{\natexlab{b}})\citenamefont
  {Lee}, \citenamefont {Furche},\ and\ \citenamefont {Burke}}]{BurkeSIE}%
  \BibitemOpen
  \bibfield  {author} {\bibinfo {author} {\bibfnamefont {D.}~\bibnamefont
  {Lee}}, \bibinfo {author} {\bibfnamefont {F.}~\bibnamefont {Furche}},\ and\
  \bibinfo {author} {\bibfnamefont {K.}~\bibnamefont {Burke}},\ }\bibfield
  {title} {\bibinfo {title} {Accuracy of electron affinities of atoms in
  approximatedensity functional theory},\ }\href@noop {} {\bibfield  {journal}
  {\bibinfo  {journal} {J. Phys. Chem. Lett.}\ }\textbf {\bibinfo {volume}
  {1}},\ \bibinfo {pages} {2124} (\bibinfo {year}
  {2010}{\natexlab{b}})}\BibitemShut {NoStop}%
\bibitem [{\citenamefont {Song}\ \emph
  {et~al.}(2022{\natexlab{a}})\citenamefont {Song}, \citenamefont {Vuckovic},
  \citenamefont {Sim},\ and\ \citenamefont {Burke}}]{DDerrorQaA}%
  \BibitemOpen
  \bibfield  {author} {\bibinfo {author} {\bibfnamefont {S.}~\bibnamefont
  {Song}}, \bibinfo {author} {\bibfnamefont {S.}~\bibnamefont {Vuckovic}},
  \bibinfo {author} {\bibfnamefont {E.}~\bibnamefont {Sim}},\ and\ \bibinfo
  {author} {\bibfnamefont {K.}~\bibnamefont {Burke}},\ }\bibfield  {title}
  {\bibinfo {title} {{Density-Corrected DFT Explained: Questions and
  Answers}},\ }\href@noop {} {\bibfield  {journal} {\bibinfo  {journal}
  {Journal of Chemical Theory and Computations}\ }\textbf {\bibinfo {volume}
  {18}},\ \bibinfo {pages} {817} (\bibinfo {year}
  {2022}{\natexlab{a}})}\BibitemShut {NoStop}%
\bibitem [{\citenamefont {Henderson}\ \emph {et~al.}(2008)\citenamefont
  {Henderson}, \citenamefont {Janesko},\ and\ \citenamefont
  {Scusseria}}]{HJS08}%
  \BibitemOpen
  \bibfield  {author} {\bibinfo {author} {\bibfnamefont {T.~M.}\ \bibnamefont
  {Henderson}}, \bibinfo {author} {\bibfnamefont {B.~G.}\ \bibnamefont
  {Janesko}},\ and\ \bibinfo {author} {\bibfnamefont {G.~E.}\ \bibnamefont
  {Scusseria}},\ }\bibfield  {title} {\bibinfo {title} {Generalized gradient
  approximation model exchange holes for range-separated hybrids},\ }\href@noop
  {} {\bibfield  {journal} {\bibinfo  {journal} {J. Chem. Phys}\ }\textbf
  {\bibinfo {volume} {128}},\ \bibinfo {pages} {194105} (\bibinfo {year}
  {2008})}\BibitemShut {NoStop}%
\bibitem [{\citenamefont {Patra}\ \emph {et~al.}(2019)\citenamefont {Patra},
  \citenamefont {Peng}, \citenamefont {Sun},\ and\ \citenamefont
  {Perdew}}]{patra2019rethinking}%
  \BibitemOpen
  \bibfield  {author} {\bibinfo {author} {\bibfnamefont {A.}~\bibnamefont
  {Patra}}, \bibinfo {author} {\bibfnamefont {H.}~\bibnamefont {Peng}},
  \bibinfo {author} {\bibfnamefont {J.}~\bibnamefont {Sun}},\ and\ \bibinfo
  {author} {\bibfnamefont {J.~P.}\ \bibnamefont {Perdew}},\ }\bibfield  {title}
  {\bibinfo {title} {Rethinking {CO} adsorption on transition-metal surfaces:
  Effect of density-driven self-interaction errors},\ }\href@noop {} {\bibfield
   {journal} {\bibinfo  {journal} {Phys. Rev. B}\ }\textbf {\bibinfo {volume}
  {100}},\ \bibinfo {pages} {035442} (\bibinfo {year} {2019})}\BibitemShut
  {NoStop}%
\bibitem [{\citenamefont {Heyd}\ \emph {et~al.}(2003)\citenamefont {Heyd},
  \citenamefont {Scuseria},\ and\ \citenamefont {Ernzerhof}}]{HSE03}%
  \BibitemOpen
  \bibfield  {author} {\bibinfo {author} {\bibfnamefont {J.}~\bibnamefont
  {Heyd}}, \bibinfo {author} {\bibfnamefont {G.~E.}\ \bibnamefont {Scuseria}},\
  and\ \bibinfo {author} {\bibfnamefont {M.}~\bibnamefont {Ernzerhof}},\
  }\bibfield  {title} {\bibinfo {title} {{Hybrid} functionals based on a
  screened {C}oulomb potential},\ }\href
  {https://doi.org/https://doi.org/10.1063/1.1564060} {\bibfield  {journal}
  {\bibinfo  {journal} {J. Chem. Phys.}\ }\textbf {\bibinfo {volume} {118}},\
  \bibinfo {pages} {8207} (\bibinfo {year} {2003})}\BibitemShut {NoStop}%
\bibitem [{\citenamefont {Heyd}\ \emph {et~al.}(2006)\citenamefont {Heyd},
  \citenamefont {Scuseria},\ and\ \citenamefont {Ernzerhof}}]{HSE06}%
  \BibitemOpen
  \bibfield  {author} {\bibinfo {author} {\bibfnamefont {J.}~\bibnamefont
  {Heyd}}, \bibinfo {author} {\bibfnamefont {G.~E.}\ \bibnamefont {Scuseria}},\
  and\ \bibinfo {author} {\bibfnamefont {M.}~\bibnamefont {Ernzerhof}},\
  }\bibfield  {title} {\bibinfo {title} {{Erratum:} "{Hybrid} functionals based
  on a screened {C}oulomb potential" [{J. Chem. Phys. 118, 8207 (2003)}]},\
  }\href {https://doi.org/https://doi.org/10.1063/1.2204597} {\bibfield
  {journal} {\bibinfo  {journal} {J. Chem. Phys.}\ }\textbf {\bibinfo {volume}
  {124}},\ \bibinfo {pages} {219906} (\bibinfo {year} {2006})}\BibitemShut
  {NoStop}%
\bibitem [{\citenamefont {Grimme}\ \emph {et~al.}(2010)\citenamefont {Grimme},
  \citenamefont {Antony}, \citenamefont {Ehrlich},\ and\ \citenamefont
  {Krieg}}]{grimme3}%
  \BibitemOpen
  \bibfield  {author} {\bibinfo {author} {\bibfnamefont {S.}~\bibnamefont
  {Grimme}}, \bibinfo {author} {\bibfnamefont {J.}~\bibnamefont {Antony}},
  \bibinfo {author} {\bibfnamefont {S.}~\bibnamefont {Ehrlich}},\ and\ \bibinfo
  {author} {\bibfnamefont {H.}~\bibnamefont {Krieg}},\ }\bibfield  {title}
  {\bibinfo {title} {A consistent and accurate ab initio parametrization of
  density functional dispersion correction ({DFT-D}) for the 94 elements
  {H}--{P}u},\ }\href@noop {} {\bibfield  {journal} {\bibinfo  {journal} {J.
  Chem. Phys.}\ }\textbf {\bibinfo {volume} {132}},\ \bibinfo {pages} {154104}
  (\bibinfo {year} {2010})}\BibitemShut {NoStop}%
\bibitem [{\citenamefont {Becke}(1993)}]{BeckeIII}%
  \BibitemOpen
  \bibfield  {author} {\bibinfo {author} {\bibfnamefont {A.~D.}\ \bibnamefont
  {Becke}},\ }\bibfield  {title} {\bibinfo {title} {{Density‐functional
  thermochemistry. III. The role of exact exchange}},\ }\href@noop {}
  {\bibfield  {journal} {\bibinfo  {journal} {J. Chem. Phys.}\ }\textbf
  {\bibinfo {volume} {98}},\ \bibinfo {pages} {5648} (\bibinfo {year}
  {1993})}\BibitemShut {NoStop}%
\bibitem [{\citenamefont {Lee}\ \emph {et~al.}(1988)\citenamefont {Lee},
  \citenamefont {Yang},\ and\ \citenamefont {Parr}}]{LYP}%
  \BibitemOpen
  \bibfield  {author} {\bibinfo {author} {\bibfnamefont {C.}~\bibnamefont
  {Lee}}, \bibinfo {author} {\bibfnamefont {W.}~\bibnamefont {Yang}},\ and\
  \bibinfo {author} {\bibfnamefont {R.~G.}\ \bibnamefont {Parr}},\ }\bibfield
  {title} {\bibinfo {title} {{Development of the Coll-Salvetti
  correlation-energy formula into a functional of the electron density}},\
  }\href@noop {} {\bibfield  {journal} {\bibinfo  {journal} {Phys. Rev. B}\
  }\textbf {\bibinfo {volume} {37}},\ \bibinfo {pages} {785} (\bibinfo {year}
  {1988})}\BibitemShut {NoStop}%
\bibitem [{\citenamefont {Seyedraoufi}\ and\ \citenamefont
  {Berland}(2022)}]{BerlandTrState22}%
  \BibitemOpen
  \bibfield  {author} {\bibinfo {author} {\bibfnamefont {S.}~\bibnamefont
  {Seyedraoufi}}\ and\ \bibinfo {author} {\bibfnamefont {K.}~\bibnamefont
  {Berland}},\ }\bibfield  {title} {\bibinfo {title} {{Improved proton-transfer
  barriers with van der Waals density functionals: Role of repulsive non-local
  correlation}},\ }\href@noop {} {\bibfield  {journal} {\bibinfo  {journal} {J.
  Chem. Phys.}\ }\textbf {\bibinfo {volume} {156}},\ \bibinfo {pages} {244106}
  (\bibinfo {year} {2022})}\BibitemShut {NoStop}%
\bibitem [{\citenamefont {Langreth}\ and\ \citenamefont
  {Perdew}(1975)}]{lape75}%
  \BibitemOpen
  \bibfield  {author} {\bibinfo {author} {\bibfnamefont {D.~C.}\ \bibnamefont
  {Langreth}}\ and\ \bibinfo {author} {\bibfnamefont {J.~P.}\ \bibnamefont
  {Perdew}},\ }\bibfield  {title} {\bibinfo {title} {The exchange-correlation
  energy of a metallic surface},\ }\href@noop {} {\bibfield  {journal}
  {\bibinfo  {journal} {Solid State Commun.}\ }\textbf {\bibinfo {volume}
  {17}},\ \bibinfo {pages} {1425} (\bibinfo {year} {1975})}\BibitemShut
  {NoStop}%
\bibitem [{\citenamefont {Langreth}\ and\ \citenamefont
  {Perdew}(1977)}]{lape77}%
  \BibitemOpen
  \bibfield  {author} {\bibinfo {author} {\bibfnamefont {D.~C.}\ \bibnamefont
  {Langreth}}\ and\ \bibinfo {author} {\bibfnamefont {J.~P.}\ \bibnamefont
  {Perdew}},\ }\bibfield  {title} {\bibinfo {title} {Exchange-correlation
  energy of a metallic surface: {W}ave-vector analysis},\ }\href@noop {}
  {\bibfield  {journal} {\bibinfo  {journal} {Phys. Rev. B}\ }\textbf {\bibinfo
  {volume} {15}},\ \bibinfo {pages} {2884} (\bibinfo {year}
  {1977})}\BibitemShut {NoStop}%
\bibitem [{\citenamefont {Langreth}\ and\ \citenamefont
  {Perdew}(1980)}]{lape80}%
  \BibitemOpen
  \bibfield  {author} {\bibinfo {author} {\bibfnamefont {D.~C.}\ \bibnamefont
  {Langreth}}\ and\ \bibinfo {author} {\bibfnamefont {J.~P.}\ \bibnamefont
  {Perdew}},\ }\bibfield  {title} {\bibinfo {title} {Theory of nonuniform
  electronic systems. {I}. {A}nalysis of the gradient approximation and a
  generalization that works},\ }\href@noop {} {\bibfield  {journal} {\bibinfo
  {journal} {Phys. Rev. B}\ }\textbf {\bibinfo {volume} {21}},\ \bibinfo
  {pages} {5469} (\bibinfo {year} {1980})}\BibitemShut {NoStop}%
\bibitem [{\citenamefont {Langreth}\ and\ \citenamefont
  {Mehl}(1981)}]{lameprl1981}%
  \BibitemOpen
  \bibfield  {author} {\bibinfo {author} {\bibfnamefont {D.~C.}\ \bibnamefont
  {Langreth}}\ and\ \bibinfo {author} {\bibfnamefont {M.~J.}\ \bibnamefont
  {Mehl}},\ }\bibfield  {title} {\bibinfo {title} {Easily implementable
  nonlocal exchange-correlation energy functional},\ }\href@noop {} {\bibfield
  {journal} {\bibinfo  {journal} {Phys. Rev. Lett.}\ }\textbf {\bibinfo
  {volume} {47}},\ \bibinfo {pages} {446} (\bibinfo {year} {1981})}\BibitemShut
  {NoStop}%
\bibitem [{\citenamefont {Langreth}\ and\ \citenamefont
  {Vosko}(1990)}]{lavo90}%
  \BibitemOpen
  \bibfield  {author} {\bibinfo {author} {\bibfnamefont {D.~C.}\ \bibnamefont
  {Langreth}}\ and\ \bibinfo {author} {\bibfnamefont {S.~H.}\ \bibnamefont
  {Vosko}},\ }\bibfield  {title} {\bibinfo {title} {Response functions and
  non-local functionals},\ }\href@noop {} {\bibfield  {journal} {\bibinfo
  {journal} {Adv. Quantum. Chem.}\ }\textbf {\bibinfo {volume} {21}},\ \bibinfo
  {pages} {175} (\bibinfo {year} {1990})}\BibitemShut {NoStop}%
\bibitem [{Note1()}]{Note1}%
  \BibitemOpen
  \bibinfo {note} {Here and below, we suppress the double spatial coordinates
  of this nonlocal function for brevity.}\BibitemShut {Stop}%
\bibitem [{\citenamefont {Maggs}\ and\ \citenamefont {Ashcroft}(1987)}]{ma}%
  \BibitemOpen
  \bibfield  {author} {\bibinfo {author} {\bibfnamefont {A.~C.}\ \bibnamefont
  {Maggs}}\ and\ \bibinfo {author} {\bibfnamefont {N.~W.}\ \bibnamefont
  {Ashcroft}},\ }\bibfield  {title} {\bibinfo {title} {Electronic fluctuation
  and cohesion in metals},\ }\href@noop {} {\bibfield  {journal} {\bibinfo
  {journal} {Phys. Rev. Lett.}\ }\textbf {\bibinfo {volume} {59}},\ \bibinfo
  {pages} {113} (\bibinfo {year} {1987})}\BibitemShut {NoStop}%
\bibitem [{\citenamefont {Ma}\ and\ \citenamefont {Brueckner}(1968)}]{mabr}%
  \BibitemOpen
  \bibfield  {author} {\bibinfo {author} {\bibfnamefont {S.}~\bibnamefont
  {Ma}}\ and\ \bibinfo {author} {\bibfnamefont {K.}~\bibnamefont {Brueckner}},\
  }\bibfield  {title} {\bibinfo {title} {Correlation energy of an electron gas
  with a slowly varying high density},\ }\href@noop {} {\bibfield  {journal}
  {\bibinfo  {journal} {Phys. Rev.}\ }\textbf {\bibinfo {volume} {165}},\
  \bibinfo {pages} {18} (\bibinfo {year} {1968})}\BibitemShut {NoStop}%
\bibitem [{\citenamefont {Rasolt}\ and\ \citenamefont
  {Geldart}(1975)}]{rasolt}%
  \BibitemOpen
  \bibfield  {author} {\bibinfo {author} {\bibfnamefont {M.}~\bibnamefont
  {Rasolt}}\ and\ \bibinfo {author} {\bibfnamefont {D.~J.~W.}\ \bibnamefont
  {Geldart}},\ }\bibfield  {title} {\bibinfo {title} {Gradient corrections in
  the exchange and correlation energy of an inhomogeneous electron gas},\
  }\href@noop {} {\bibfield  {journal} {\bibinfo  {journal} {Phys. Rev. Lett.}\
  }\textbf {\bibinfo {volume} {35}},\ \bibinfo {pages} {1234} (\bibinfo {year}
  {1975})}\BibitemShut {NoStop}%
\bibitem [{\citenamefont {Gunnarsson}\ \emph {et~al.}(1979)\citenamefont
  {Gunnarsson}, \citenamefont {Jonson},\ and\ \citenamefont
  {Lundqvist}}]{adawda}%
  \BibitemOpen
  \bibfield  {author} {\bibinfo {author} {\bibfnamefont {O.}~\bibnamefont
  {Gunnarsson}}, \bibinfo {author} {\bibfnamefont {M.}~\bibnamefont {Jonson}},\
  and\ \bibinfo {author} {\bibfnamefont {B.~I.}\ \bibnamefont {Lundqvist}},\
  }\bibfield  {title} {\bibinfo {title} {Descriptions of exchange and
  correlation effects in inhomogeneous electron systems},\ }\href@noop {}
  {\bibfield  {journal} {\bibinfo  {journal} {Phys. Rev. B}\ }\textbf {\bibinfo
  {volume} {20}},\ \bibinfo {pages} {3136} (\bibinfo {year}
  {1979})}\BibitemShut {NoStop}%
\bibitem [{\citenamefont {Lundqvist}(1967)}]{lu67}%
  \BibitemOpen
  \bibfield  {author} {\bibinfo {author} {\bibfnamefont {B.~I.}\ \bibnamefont
  {Lundqvist}},\ }\bibfield  {title} {\bibinfo {title} {Single-particle
  spectrum of degenerate electron gas. {I}. {S}tructure of spectral weight
  function},\ }\href@noop {} {\bibfield  {journal} {\bibinfo  {journal} {Phys.
  Kondens. Mater.}\ }\textbf {\bibinfo {volume} {6}},\ \bibinfo {pages} {193}
  (\bibinfo {year} {1967})}\BibitemShut {NoStop}%
\bibitem [{\citenamefont {Perdew}\ and\ \citenamefont {Wang}(1986)}]{pewa86}%
  \BibitemOpen
  \bibfield  {author} {\bibinfo {author} {\bibfnamefont {J.~P.}\ \bibnamefont
  {Perdew}}\ and\ \bibinfo {author} {\bibfnamefont {Y.}~\bibnamefont {Wang}},\
  }\bibfield  {title} {\bibinfo {title} {Accurate and simple density functional
  for the electronic exchange energy: {G}eneralized gradient approximation},\
  }\href@noop {} {\bibfield  {journal} {\bibinfo  {journal} {Phys. Rev. B}\
  }\textbf {\bibinfo {volume} {33}},\ \bibinfo {pages} {8800} (\bibinfo {year}
  {1986})}\BibitemShut {NoStop}%
\bibitem [{\citenamefont {Pribram-Jones}\ \emph {et~al.}(2015)\citenamefont
  {Pribram-Jones}, \citenamefont {Gross},\ and\ \citenamefont
  {Burke}}]{FullHoles}%
  \BibitemOpen
  \bibfield  {author} {\bibinfo {author} {\bibfnamefont {A.}~\bibnamefont
  {Pribram-Jones}}, \bibinfo {author} {\bibfnamefont {D.~A.}\ \bibnamefont
  {Gross}},\ and\ \bibinfo {author} {\bibfnamefont {K.}~\bibnamefont {Burke}},\
  }\bibfield  {title} {\bibinfo {title} {{DFT: A theory full of holes}},\
  }\href@noop {} {\bibfield  {journal} {\bibinfo  {journal} {Annual Review of
  Physical Chemistry}\ }\textbf {\bibinfo {volume} {66}},\ \bibinfo {pages}
  {283} (\bibinfo {year} {2015})}\BibitemShut {NoStop}%
\bibitem [{\citenamefont {London}(1930)}]{lo30}%
  \BibitemOpen
  \bibfield  {author} {\bibinfo {author} {\bibfnamefont {F.}~\bibnamefont
  {London}},\ }\bibfield  {title} {\bibinfo {title} {{Z}ur {T}heorie und
  {S}ystematik der {M}olekularkr\"afte},\ }\href@noop {} {\bibfield  {journal}
  {\bibinfo  {journal} {Z. Phys.}\ }\textbf {\bibinfo {volume} {63}},\ \bibinfo
  {pages} {245} (\bibinfo {year} {1930})}\BibitemShut {NoStop}%
\bibitem [{\citenamefont {Mahan}(1965)}]{jerry65}%
  \BibitemOpen
  \bibfield  {author} {\bibinfo {author} {\bibfnamefont {G.~D.}\ \bibnamefont
  {Mahan}},\ }\bibfield  {title} {\bibinfo {title} {van der {W}aals forces in
  solids},\ }\href@noop {} {\bibfield  {journal} {\bibinfo  {journal} {J. Chem.
  Phys.}\ }\textbf {\bibinfo {volume} {43}},\ \bibinfo {pages} {1569} (\bibinfo
  {year} {1965})}\BibitemShut {NoStop}%
\bibitem [{\citenamefont {Langreth}\ \emph {et~al.}(2005)\citenamefont
  {Langreth}, \citenamefont {Dion}, \citenamefont {Rydberg}, \citenamefont
  {Schr{\"o}der}, \citenamefont {Hyldgaard},\ and\ \citenamefont
  {Lundqvist}}]{langreth05p599}%
  \BibitemOpen
  \bibfield  {author} {\bibinfo {author} {\bibfnamefont {D.~C.}\ \bibnamefont
  {Langreth}}, \bibinfo {author} {\bibfnamefont {M.}~\bibnamefont {Dion}},
  \bibinfo {author} {\bibfnamefont {H.}~\bibnamefont {Rydberg}}, \bibinfo
  {author} {\bibfnamefont {E.}~\bibnamefont {Schr{\"o}der}}, \bibinfo {author}
  {\bibfnamefont {P.}~\bibnamefont {Hyldgaard}},\ and\ \bibinfo {author}
  {\bibfnamefont {B.~I.}\ \bibnamefont {Lundqvist}},\ }\bibfield  {title}
  {\bibinfo {title} {{v}an der {W}aals density functional theory with
  applications},\ }\href@noop {} {\bibfield  {journal} {\bibinfo  {journal}
  {Int. J. Quan. Chem.}\ }\textbf {\bibinfo {volume} {101}},\ \bibinfo {pages}
  {599} (\bibinfo {year} {2005})}\BibitemShut {NoStop}%
\bibitem [{\citenamefont {Berland}(2012)}]{berlandthesis}%
  \BibitemOpen
  \bibfield  {author} {\bibinfo {author} {\bibfnamefont {K.}~\bibnamefont
  {Berland}},\ }\emph {\bibinfo {title} {Connected by voids: {I}nteractions and
  screening in sparse matter}},\ \href@noop {} {Ph.D. thesis},\ \bibinfo
  {school} {Department of Microtechnology and Nanoscience -- MC2, Chalmers
  University of Technology}, \bibinfo {address} {G\"oteborg, Sweden} (\bibinfo
  {year} {2012})\BibitemShut {NoStop}%
\bibitem [{\citenamefont {London}(1937)}]{lo37}%
  \BibitemOpen
  \bibfield  {author} {\bibinfo {author} {\bibfnamefont {F.}~\bibnamefont
  {London}},\ }\bibfield  {title} {\bibinfo {title} {The general theory of
  molecular forces},\ }\href@noop {} {\bibfield  {journal} {\bibinfo  {journal}
  {Z. Physik. Chemie}\ }\textbf {\bibinfo {volume} {33}},\ \bibinfo {pages} {8}
  (\bibinfo {year} {1937})},\ \bibinfo {note} {{E}nglish translations in H.
  Hettema, \textit{Quantum Chemistry, Classic Scientific Papers}, World
  Scientific, Singapore (2000); F. London, Trans. Faraday Soc. \textbf{33}, 8
  (1937).}\BibitemShut {Stop}%
\bibitem [{\citenamefont {Rydberg}(2001)}]{rydbergthesis}%
  \BibitemOpen
  \bibfield  {author} {\bibinfo {author} {\bibfnamefont {H.}~\bibnamefont
  {Rydberg}},\ }\emph {\bibinfo {title} {Nonlocal correlations in density
  functional theory}},\ \href@noop {} {Ph.D. thesis},\ \bibinfo  {school}
  {Department of Applied Physics, Chalmers University of Technology}, \bibinfo
  {address} {G\"oteborg, Sweden} (\bibinfo {year} {2001}),\ \bibinfo {note}
  {http://bitmath.se/rydberg/Thesis}\BibitemShut {NoStop}%
\bibitem [{\citenamefont {Dion}(2004)}]{dionthesis}%
  \BibitemOpen
  \bibfield  {author} {\bibinfo {author} {\bibfnamefont {M.}~\bibnamefont
  {Dion}},\ }\emph {\bibinfo {title} {{v}an der {W}aals forces in density
  functional theory}},\ \href@noop {} {Ph.D. thesis},\ \bibinfo  {school}
  {Rutgers University}, \bibinfo {address} {Piscataway, NC, USA} (\bibinfo
  {year} {2004})\BibitemShut {NoStop}%
\bibitem [{\citenamefont {Langreth}(1970)}]{la70}%
  \BibitemOpen
  \bibfield  {author} {\bibinfo {author} {\bibfnamefont {D.~C.}\ \bibnamefont
  {Langreth}},\ }\bibfield  {title} {\bibinfo {title} {{Singularities in the
  X-Ray Spectra of Metals}},\ }\href@noop {} {\bibfield  {journal} {\bibinfo
  {journal} {Phys. Rev. B}\ }\textbf {\bibinfo {volume} {1}},\ \bibinfo {pages}
  {471} (\bibinfo {year} {1970})}\BibitemShut {NoStop}%
\bibitem [{\citenamefont {Hedin}(1980)}]{Hedin80}%
  \BibitemOpen
  \bibfield  {author} {\bibinfo {author} {\bibfnamefont {L.}~\bibnamefont
  {Hedin}},\ }\bibfield  {title} {\bibinfo {title} {Effects of recoil on
  shake-up spectra in metals},\ }\href@noop {} {\bibfield  {journal} {\bibinfo
  {journal} {Phys. Scr.}\ }\textbf {\bibinfo {volume} {21}},\ \bibinfo {pages}
  {477} (\bibinfo {year} {1980})}\BibitemShut {NoStop}%
\bibitem [{\citenamefont {Gunnarsson}\ \emph {et~al.}(1994)\citenamefont
  {Gunnarsson}, \citenamefont {Meden},\ and\ \citenamefont
  {Sch{\"o}nhammer}}]{GunMedSch94}%
  \BibitemOpen
  \bibfield  {author} {\bibinfo {author} {\bibfnamefont {O.}~\bibnamefont
  {Gunnarsson}}, \bibinfo {author} {\bibfnamefont {V.}~\bibnamefont {Meden}},\
  and\ \bibinfo {author} {\bibfnamefont {K.}~\bibnamefont {Sch{\"o}nhammer}},\
  }\bibfield  {title} {\bibinfo {title} {{Corrections to Migdal's theorem for
  spectral functions: A cumulant treatment of the time-dependent Green's
  function}},\ }\href {https://doi.org/10.1103/PhysRevB.50.10462} {\bibfield
  {journal} {\bibinfo  {journal} {Phys. Rev. B}\ }\textbf {\bibinfo {volume}
  {50}},\ \bibinfo {pages} {10462} (\bibinfo {year} {1994})}\BibitemShut
  {NoStop}%
\bibitem [{\citenamefont {Holm}\ and\ \citenamefont
  {Aryasetiawan}(1997)}]{HolAry97}%
  \BibitemOpen
  \bibfield  {author} {\bibinfo {author} {\bibfnamefont {B.}~\bibnamefont
  {Holm}}\ and\ \bibinfo {author} {\bibfnamefont {F.}~\bibnamefont
  {Aryasetiawan}},\ }\bibfield  {title} {\bibinfo {title} {Self-consistent
  cumulant expansion for the electron gas},\ }\href
  {https://doi.org/10.1103/PhysRevB.56.12825} {\bibfield  {journal} {\bibinfo
  {journal} {Phys. Rev. B}\ }\textbf {\bibinfo {volume} {56}},\ \bibinfo
  {pages} {12825} (\bibinfo {year} {1997})}\BibitemShut {NoStop}%
\bibitem [{\citenamefont {Sokolov}\ \emph {et~al.}(2013)\citenamefont
  {Sokolov}, \citenamefont {Simmonett},\ and\ \citenamefont {{Schaefer
  III}}}]{Sokolov13}%
  \BibitemOpen
  \bibfield  {author} {\bibinfo {author} {\bibfnamefont {A.~Y.}\ \bibnamefont
  {Sokolov}}, \bibinfo {author} {\bibfnamefont {A.~C.}\ \bibnamefont
  {Simmonett}},\ and\ \bibinfo {author} {\bibfnamefont {H.~F.}\ \bibnamefont
  {{Schaefer III}}},\ }\bibfield  {title} {\bibinfo {title} {Density cumulant
  functional theory: {T}he {DC-12} method, an improved description of the
  one-particle density matrix},\ }\href {https://doi.org/10.1063/1.4773580}
  {\bibfield  {journal} {\bibinfo  {journal} {J. Chem. Phys.}\ }\textbf
  {\bibinfo {volume} {138}},\ \bibinfo {pages} {024107} (\bibinfo {year}
  {2013})}\BibitemShut {NoStop}%
\bibitem [{\citenamefont {Rom\'{a}n-P\'{e}rez}\ and\ \citenamefont
  {Soler}(2009)}]{roso09}%
  \BibitemOpen
  \bibfield  {author} {\bibinfo {author} {\bibfnamefont {G.}~\bibnamefont
  {Rom\'{a}n-P\'{e}rez}}\ and\ \bibinfo {author} {\bibfnamefont {J.~M.}\
  \bibnamefont {Soler}},\ }\bibfield  {title} {\bibinfo {title} {Efficient
  implementation of a van der {W}aals density functional: {A}pplication to
  double-wall carbon nanotubes},\ }\href@noop {} {\bibfield  {journal}
  {\bibinfo  {journal} {Phys. Rev. Lett.}\ }\textbf {\bibinfo {volume} {103}},\
  \bibinfo {pages} {096102} (\bibinfo {year} {2009})}\BibitemShut {NoStop}%
\bibitem [{\citenamefont {{Hjorth Larsen}}\ \emph {et~al.}(2017)\citenamefont
  {{Hjorth Larsen}}, \citenamefont {{Kuisma}}, \citenamefont {{L{\"o}fgren}},
  \citenamefont {{Pouillon}}, \citenamefont {{Erhart}},\ and\ \citenamefont
  {{Hyldgaard}}}]{libxcvdW}%
  \BibitemOpen
  \bibfield  {author} {\bibinfo {author} {\bibfnamefont {A.}~\bibnamefont
  {{Hjorth Larsen}}}, \bibinfo {author} {\bibfnamefont {M.}~\bibnamefont
  {{Kuisma}}}, \bibinfo {author} {\bibfnamefont {J.}~\bibnamefont
  {{L{\"o}fgren}}}, \bibinfo {author} {\bibfnamefont {Y.}~\bibnamefont
  {{Pouillon}}}, \bibinfo {author} {\bibfnamefont {P.}~\bibnamefont
  {{Erhart}}},\ and\ \bibinfo {author} {\bibfnamefont {P.}~\bibnamefont
  {{Hyldgaard}}},\ }\bibfield  {title} {\bibinfo {title} {{libvdwxc: {A}
  library for exchange-correlation functionals in the vd{W}-{DF} family}},\
  }\href@noop {} {\bibfield  {journal} {\bibinfo  {journal} {Modell. Simul.
  Mater. Sci. Eng.}\ }\textbf {\bibinfo {volume} {25}},\ \bibinfo {pages}
  {065004} (\bibinfo {year} {2017})}\BibitemShut {NoStop}%
\bibitem [{\citenamefont {Granhed}\ \emph {et~al.}(2020)\citenamefont
  {Granhed}, \citenamefont {Wahnstr{\"o}m},\ and\ \citenamefont
  {Hyldgaard}}]{jewahy20}%
  \BibitemOpen
  \bibfield  {author} {\bibinfo {author} {\bibfnamefont {E.~J.}\ \bibnamefont
  {Granhed}}, \bibinfo {author} {\bibfnamefont {G.}~\bibnamefont
  {Wahnstr{\"o}m}},\ and\ \bibinfo {author} {\bibfnamefont {P.}~\bibnamefont
  {Hyldgaard}},\ }\bibfield  {title} {\bibinfo {title} {{BaZrO$_3$} stability
  under pressure: the role of non-local exchange and correlation},\ }\href@noop
  {} {\bibfield  {journal} {\bibinfo  {journal} {Phys. Rev. B}\ }\textbf
  {\bibinfo {volume} {101}},\ \bibinfo {pages} {224105} (\bibinfo {year}
  {2020})}\BibitemShut {NoStop}%
\bibitem [{\citenamefont {Schwinger}(1981)}]{schwinger}%
  \BibitemOpen
  \bibfield  {author} {\bibinfo {author} {\bibfnamefont {J.}~\bibnamefont
  {Schwinger}},\ }\bibfield  {title} {\bibinfo {title} {Thomas-{F}ermi model:
  {T}he second correction},\ }\href@noop {} {\bibfield  {journal} {\bibinfo
  {journal} {Phys. Rev. A}\ }\textbf {\bibinfo {volume} {24}},\ \bibinfo
  {pages} {2353} (\bibinfo {year} {1981})}\BibitemShut {NoStop}%
\bibitem [{\citenamefont {Perdew}\ \emph
  {et~al.}(1996{\natexlab{b}})\citenamefont {Perdew}, \citenamefont {Burke},\
  and\ \citenamefont {Wang}}]{pebuwa96}%
  \BibitemOpen
  \bibfield  {author} {\bibinfo {author} {\bibfnamefont {J.~P.}\ \bibnamefont
  {Perdew}}, \bibinfo {author} {\bibfnamefont {K.}~\bibnamefont {Burke}},\ and\
  \bibinfo {author} {\bibfnamefont {Y.}~\bibnamefont {Wang}},\ }\bibfield
  {title} {\bibinfo {title} {Generalized gradient approximation for the
  exchange-correlation hole of a many-electron system},\ }\href@noop {}
  {\bibfield  {journal} {\bibinfo  {journal} {Phys. Rev. B}\ }\textbf {\bibinfo
  {volume} {54}},\ \bibinfo {pages} {16533} (\bibinfo {year}
  {1996}{\natexlab{b}})}\BibitemShut {NoStop}%
\bibitem [{Note2()}]{Note2}%
  \BibitemOpen
  \bibinfo {note} {We see the vdW attraction as arising from electrodynamic
  coupling among electron-XC-hole pairs \cite
  {lo37,ra,lavo87,hybesc14,Berland_2015:van_waals,JPCMreview}. The
  nonlocal-correlation component $E_{\protect \rm c}^{\protect \rm nl}$ must
  therefore vanish if we could suppress the electron charge. Meanwhile, the
  vdW-DF method has no auxhiliary inputs so it follows from a a simple
  coupling-constant scaling argument that the exchange-energy component must
  exactly equal this vanishing-electron-charge limit of the vdW-DF $E_{\protect
  \rm xc}$ \cite {Burke97,JiScHy18a}.}\BibitemShut {Stop}%
\bibitem [{\citenamefont {Becke}(1988)}]{Be88}%
  \BibitemOpen
  \bibfield  {author} {\bibinfo {author} {\bibfnamefont {A.~D.}\ \bibnamefont
  {Becke}},\ }\bibfield  {title} {\bibinfo {title} {Density-functional
  exchange-energy approximation with correct asymptotic behavior},\ }\href@noop
  {} {\bibfield  {journal} {\bibinfo  {journal} {Phys. Rev. A}\ }\textbf
  {\bibinfo {volume} {38}},\ \bibinfo {pages} {3098} (\bibinfo {year}
  {1988})}\BibitemShut {NoStop}%
\bibitem [{\citenamefont {Elliott}\ and\ \citenamefont
  {Burke}(2009)}]{ElliottBurke2009}%
  \BibitemOpen
  \bibfield  {author} {\bibinfo {author} {\bibfnamefont {P.}~\bibnamefont
  {Elliott}}\ and\ \bibinfo {author} {\bibfnamefont {K.}~\bibnamefont
  {Burke}},\ }\bibfield  {title} {\bibinfo {title} {{Non-empirical derivation
  of the parameter in the B88 exchange functional.}},\ }\href@noop {}
  {\bibfield  {journal} {\bibinfo  {journal} {Can. J. Chem}\ }\textbf {\bibinfo
  {volume} {87}},\ \bibinfo {pages} {1485} (\bibinfo {year}
  {2009})}\BibitemShut {NoStop}%
\bibitem [{\citenamefont {Burke}\ and\ \citenamefont {Wagner}(2013)}]{burke}%
  \BibitemOpen
  \bibfield  {author} {\bibinfo {author} {\bibfnamefont {K.}~\bibnamefont
  {Burke}}\ and\ \bibinfo {author} {\bibfnamefont {L.~O.}\ \bibnamefont
  {Wagner}},\ }\bibfield  {title} {\bibinfo {title} {{DFT} in a nutshell},\
  }\href@noop {} {\bibfield  {journal} {\bibinfo  {journal} {Int. J. Quantum
  Chem.}\ }\textbf {\bibinfo {volume} {113}},\ \bibinfo {pages} {96} (\bibinfo
  {year} {2013})}\BibitemShut {NoStop}%
\bibitem [{\citenamefont {Lieb}\ and\ \citenamefont
  {Oxford}(1981)}]{LiebOxford81}%
  \BibitemOpen
  \bibfield  {author} {\bibinfo {author} {\bibfnamefont {E.~H.}\ \bibnamefont
  {Lieb}}\ and\ \bibinfo {author} {\bibfnamefont {S.}~\bibnamefont {Oxford}},\
  }\bibfield  {title} {\bibinfo {title} {Strongly constrained and appropriately
  normed semilocal density functional},\ }\href@noop {} {\bibfield  {journal}
  {\bibinfo  {journal} {Int. J. Quantum Chem.}\ }\textbf {\bibinfo {volume}
  {19}},\ \bibinfo {pages} {427} (\bibinfo {year} {1981})}\BibitemShut
  {NoStop}%
\bibitem [{\citenamefont {Perdew}\ \emph {et~al.}(2014)\citenamefont {Perdew},
  \citenamefont {Ruzsinszky}, \citenamefont {Sun},\ and\ \citenamefont
  {Burke}}]{PeRuSuBu14}%
  \BibitemOpen
  \bibfield  {author} {\bibinfo {author} {\bibfnamefont {J.~P.}\ \bibnamefont
  {Perdew}}, \bibinfo {author} {\bibfnamefont {A.}~\bibnamefont {Ruzsinszky}},
  \bibinfo {author} {\bibfnamefont {J.}~\bibnamefont {Sun}},\ and\ \bibinfo
  {author} {\bibfnamefont {K.}~\bibnamefont {Burke}},\ }\bibfield  {title}
  {\bibinfo {title} {Gedanken densities and exact constraints in density
  functional theory},\ }\href@noop {} {\bibfield  {journal} {\bibinfo
  {journal} {J. Chem. Phys.}\ }\textbf {\bibinfo {volume} {140}},\ \bibinfo
  {pages} {18A533} (\bibinfo {year} {2014})}\BibitemShut {NoStop}%
\bibitem [{\citenamefont {Zhang}\ and\ \citenamefont {Yang}(1998)}]{ZhYa98}%
  \BibitemOpen
  \bibfield  {author} {\bibinfo {author} {\bibfnamefont {Y.}~\bibnamefont
  {Zhang}}\ and\ \bibinfo {author} {\bibfnamefont {W.}~\bibnamefont {Yang}},\
  }\bibfield  {title} {\bibinfo {title} {Comment on generalized gradient
  approximation made simple},\ }\href@noop {} {\bibfield  {journal} {\bibinfo
  {journal} {Phys. Rev. Lett.}\ }\textbf {\bibinfo {volume} {80}},\ \bibinfo
  {pages} {890} (\bibinfo {year} {1998})}\BibitemShut {NoStop}%
\bibitem [{\citenamefont {Luttinger}\ and\ \citenamefont
  {Ward}(1960)}]{WarLut}%
  \BibitemOpen
  \bibfield  {author} {\bibinfo {author} {\bibfnamefont {J.~M.}\ \bibnamefont
  {Luttinger}}\ and\ \bibinfo {author} {\bibfnamefont {J.~C.}\ \bibnamefont
  {Ward}},\ }\bibfield  {title} {\bibinfo {title} {{Ground-State Energy of a
  Many-Fermion System. II}},\ }\href@noop {} {\bibfield  {journal} {\bibinfo
  {journal} {Phys. Rev.}\ }\textbf {\bibinfo {volume} {118}},\ \bibinfo {pages}
  {1417} (\bibinfo {year} {1960})}\BibitemShut {NoStop}%
\bibitem [{\citenamefont {Sham}\ and\ \citenamefont
  {Sch{\"u}ter}(1983)}]{ShaSch}%
  \BibitemOpen
  \bibfield  {author} {\bibinfo {author} {\bibfnamefont {L.~J.}\ \bibnamefont
  {Sham}}\ and\ \bibinfo {author} {\bibfnamefont {M.}~\bibnamefont
  {Sch{\"u}ter}},\ }\bibfield  {title} {\bibinfo {title} {{Density-Functional
  Theory of the Energy Gap}},\ }\href@noop {} {\bibfield  {journal} {\bibinfo
  {journal} {Phys. Rev. Lett.}\ }\textbf {\bibinfo {volume} {51}},\ \bibinfo
  {pages} {1888} (\bibinfo {year} {1983})}\BibitemShut {NoStop}%
\bibitem [{\citenamefont {Jenkins}\ \emph {et~al.}(2021)\citenamefont
  {Jenkins}, \citenamefont {Berland},\ and\ \citenamefont
  {Thonhauser}}]{JeBeTh21}%
  \BibitemOpen
  \bibfield  {author} {\bibinfo {author} {\bibfnamefont {T.}~\bibnamefont
  {Jenkins}}, \bibinfo {author} {\bibfnamefont {K.}~\bibnamefont {Berland}},\
  and\ \bibinfo {author} {\bibfnamefont {T.}~\bibnamefont {Thonhauser}},\
  }\bibfield  {title} {\bibinfo {title} {{Reduced-gradient analysis of vdW
  complexes}},\ }\href@noop {} {\bibfield  {journal} {\bibinfo  {journal}
  {Electron. Struct.}\ }\textbf {\bibinfo {volume} {3}},\ \bibinfo {pages}
  {034009} (\bibinfo {year} {2021})}\BibitemShut {NoStop}%
\bibitem [{\citenamefont {de~Andrade}\ \emph {et~al.}(2020)\citenamefont
  {de~Andrade}, \citenamefont {Kullgren},\ and\ \citenamefont
  {Broqvist}}]{Ageo20}%
  \BibitemOpen
  \bibfield  {author} {\bibinfo {author} {\bibfnamefont {A.~M.}\ \bibnamefont
  {de~Andrade}}, \bibinfo {author} {\bibfnamefont {J.}~\bibnamefont
  {Kullgren}},\ and\ \bibinfo {author} {\bibfnamefont {P.}~\bibnamefont
  {Broqvist}},\ }\bibfield  {title} {\bibinfo {title} {{Quantitative and
  qualitative performance of density functional theory rationalized by reduced
  density gradient distributions}},\ }\href@noop {} {\bibfield  {journal}
  {\bibinfo  {journal} {Phys. Rev. B}\ }\textbf {\bibinfo {volume} {102}},\
  \bibinfo {pages} {075115} (\bibinfo {year} {2020})}\BibitemShut {NoStop}%
\bibitem [{\citenamefont {Rangel}\ \emph {et~al.}(2016)\citenamefont {Rangel},
  \citenamefont {Berland}, \citenamefont {Sharifzadeh}, \citenamefont
  {Brown-Altvater}, \citenamefont {Lee}, \citenamefont {Hyldgaard},
  \citenamefont {Kronik},\ and\ \citenamefont {Neaton}}]{RanPRB16}%
  \BibitemOpen
  \bibfield  {author} {\bibinfo {author} {\bibfnamefont {T.}~\bibnamefont
  {Rangel}}, \bibinfo {author} {\bibfnamefont {K.}~\bibnamefont {Berland}},
  \bibinfo {author} {\bibfnamefont {S.}~\bibnamefont {Sharifzadeh}}, \bibinfo
  {author} {\bibfnamefont {F.}~\bibnamefont {Brown-Altvater}}, \bibinfo
  {author} {\bibfnamefont {K.}~\bibnamefont {Lee}}, \bibinfo {author}
  {\bibfnamefont {P.}~\bibnamefont {Hyldgaard}}, \bibinfo {author}
  {\bibfnamefont {L.}~\bibnamefont {Kronik}},\ and\ \bibinfo {author}
  {\bibfnamefont {J.~B.}\ \bibnamefont {Neaton}},\ }\bibfield  {title}
  {\bibinfo {title} {Structural and excited-state properties of oligoacene
  crystals from first principles},\ }\href@noop {} {\bibfield  {journal}
  {\bibinfo  {journal} {Phys. Rev. B}\ }\textbf {\bibinfo {volume} {93}},\
  \bibinfo {pages} {115206} (\bibinfo {year} {2016})}\BibitemShut {NoStop}%
\bibitem [{\citenamefont {Brown-Altvater}\ \emph {et~al.}(2016)\citenamefont
  {Brown-Altvater}, \citenamefont {Rangel},\ and\ \citenamefont
  {Neaton}}]{BrownAltvPRB16}%
  \BibitemOpen
  \bibfield  {author} {\bibinfo {author} {\bibfnamefont {F.}~\bibnamefont
  {Brown-Altvater}}, \bibinfo {author} {\bibfnamefont {T.}~\bibnamefont
  {Rangel}},\ and\ \bibinfo {author} {\bibfnamefont {J.~B.}\ \bibnamefont
  {Neaton}},\ }\bibfield  {title} {\bibinfo {title} {{\em Ab initio} phonon
  dispersion in crystalline naphthalene using van der {W}aals density
  functionals},\ }\href@noop {} {\bibfield  {journal} {\bibinfo  {journal}
  {Phys. Rev. B}\ }\textbf {\bibinfo {volume} {93}},\ \bibinfo {pages} {195206}
  (\bibinfo {year} {2016})}\BibitemShut {NoStop}%
\bibitem [{\citenamefont {Song}\ \emph
  {et~al.}(2022{\natexlab{b}})\citenamefont {Song}, \citenamefont {Song},
  \citenamefont {Vuckovic},\ and\ \citenamefont {Burke}}]{SiSoVu22}%
  \BibitemOpen
  \bibfield  {author} {\bibinfo {author} {\bibfnamefont {S.}~\bibnamefont
  {Song}}, \bibinfo {author} {\bibfnamefont {A.}~\bibnamefont {Song}}, \bibinfo
  {author} {\bibfnamefont {S.}~\bibnamefont {Vuckovic}},\ and\ \bibinfo
  {author} {\bibfnamefont {K.}~\bibnamefont {Burke}},\ }\bibfield  {title}
  {\bibinfo {title} {{Improving Results by Improving Densities:
  Density-Corrected Density Functional Theory}},\ }\href@noop {} {\bibfield
  {journal} {\bibinfo  {journal} {Journal of the American Chemical Society}\
  }\textbf {\bibinfo {volume} {144}},\ \bibinfo {pages} {6625} (\bibinfo {year}
  {2022}{\natexlab{b}})}\BibitemShut {NoStop}%
\bibitem [{\citenamefont {Wellendorff}\ \emph {et~al.}(2012)\citenamefont
  {Wellendorff}, \citenamefont {Lundgaard}, \citenamefont {M{\o}gelh{\o}j},
  \citenamefont {Petzold}, \citenamefont {Landis}, \citenamefont {N{\o}rskov},
  \citenamefont {Bligaard},\ and\ \citenamefont {Jacobsen}}]{vdwBEEF}%
  \BibitemOpen
  \bibfield  {author} {\bibinfo {author} {\bibfnamefont {J.}~\bibnamefont
  {Wellendorff}}, \bibinfo {author} {\bibfnamefont {K.~T.}\ \bibnamefont
  {Lundgaard}}, \bibinfo {author} {\bibfnamefont {A.}~\bibnamefont
  {M{\o}gelh{\o}j}}, \bibinfo {author} {\bibfnamefont {V.}~\bibnamefont
  {Petzold}}, \bibinfo {author} {\bibfnamefont {D.~D.}\ \bibnamefont {Landis}},
  \bibinfo {author} {\bibfnamefont {J.~K.}\ \bibnamefont {N{\o}rskov}},
  \bibinfo {author} {\bibfnamefont {T.}~\bibnamefont {Bligaard}},\ and\
  \bibinfo {author} {\bibfnamefont {K.~W.}\ \bibnamefont {Jacobsen}},\
  }\bibfield  {title} {\bibinfo {title} {Density functionals for surface
  science: {E}xchange-correlation model development with {B}ayesian error
  estimation},\ }\href@noop {} {\bibfield  {journal} {\bibinfo  {journal}
  {Phys. Rev. B}\ }\textbf {\bibinfo {volume} {85}},\ \bibinfo {pages} {235149}
  (\bibinfo {year} {2012})}\BibitemShut {NoStop}%
\bibitem [{\citenamefont {Klime\v{s}}\ and\ \citenamefont
  {Michaelides}(2012)}]{rev8}%
  \BibitemOpen
  \bibfield  {author} {\bibinfo {author} {\bibfnamefont {J.}~\bibnamefont
  {Klime\v{s}}}\ and\ \bibinfo {author} {\bibfnamefont {A.}~\bibnamefont
  {Michaelides}},\ }\bibfield  {title} {\bibinfo {title} {Perspective:
  {A}dvances and challenges in treating van der {W}aals dispersion forces in
  density functional theory},\ }\href@noop {} {\bibfield  {journal} {\bibinfo
  {journal} {J. Chem. Phys.}\ }\textbf {\bibinfo {volume} {137}},\ \bibinfo
  {pages} {120901} (\bibinfo {year} {2012})}\BibitemShut {NoStop}%
\bibitem [{\citenamefont {Hofmann}\ \emph {et~al.}(2021)\citenamefont
  {Hofmann}, \citenamefont {Zojer}, \citenamefont {H{\"o}rmann}, \citenamefont
  {Jeindi},\ and\ \citenamefont {Maurer}}]{Interface_perspective}%
  \BibitemOpen
  \bibfield  {author} {\bibinfo {author} {\bibfnamefont {O.~T.}\ \bibnamefont
  {Hofmann}}, \bibinfo {author} {\bibfnamefont {E.}~\bibnamefont {Zojer}},
  \bibinfo {author} {\bibfnamefont {L.}~\bibnamefont {H{\"o}rmann}}, \bibinfo
  {author} {\bibfnamefont {A.}~\bibnamefont {Jeindi}},\ and\ \bibinfo {author}
  {\bibfnamefont {R.~J.}\ \bibnamefont {Maurer}},\ }\bibfield  {title}
  {\bibinfo {title} {First-principles calculations of hybrid inorganic-organic
  interfaces: from state-of-the-art to best practice},\ }\href@noop {}
  {\bibfield  {journal} {\bibinfo  {journal} {Phys. Chem. Chem. Phys., Advance
  Article}\ }\textbf {\bibinfo {volume} {23}},\ \bibinfo {pages} {8132}
  (\bibinfo {year} {2021})}\BibitemShut {NoStop}%
\bibitem [{\citenamefont {Sabatini}\ \emph {et~al.}(2012)\citenamefont
  {Sabatini}, \citenamefont {K\"{u}\c{c}\"{u}kbenli}, \citenamefont {Kolb},
  \citenamefont {Thonhauser},\ and\ \citenamefont
  {de~Gironcoli}}]{sabatini12p424209}%
  \BibitemOpen
  \bibfield  {author} {\bibinfo {author} {\bibfnamefont {R.}~\bibnamefont
  {Sabatini}}, \bibinfo {author} {\bibfnamefont {E.}~\bibnamefont
  {K\"{u}\c{c}\"{u}kbenli}}, \bibinfo {author} {\bibfnamefont {B.}~\bibnamefont
  {Kolb}}, \bibinfo {author} {\bibfnamefont {T.}~\bibnamefont {Thonhauser}},\
  and\ \bibinfo {author} {\bibfnamefont {S.}~\bibnamefont {de~Gironcoli}},\
  }\bibfield  {title} {\bibinfo {title} {Structural evolution of amino acid
  crystals under stress from a non-empirical density functional},\ }\href@noop
  {} {\bibfield  {journal} {\bibinfo  {journal} {J. Phys.: Condens. Matter}\
  }\textbf {\bibinfo {volume} {24}},\ \bibinfo {pages} {424209} (\bibinfo
  {year} {2012})}\BibitemShut {NoStop}%
\bibitem [{\citenamefont {Becke}(1986)}]{becke1986p7184}%
  \BibitemOpen
  \bibfield  {author} {\bibinfo {author} {\bibfnamefont {A.~D.}\ \bibnamefont
  {Becke}},\ }\bibfield  {title} {\bibinfo {title} {On the large-gradient
  behavior of the density functional exchange energy},\ }\href@noop {}
  {\bibfield  {journal} {\bibinfo  {journal} {J. Chem. Phys.}\ }\textbf
  {\bibinfo {volume} {85}},\ \bibinfo {pages} {7184} (\bibinfo {year}
  {1986})}\BibitemShut {NoStop}%
\bibitem [{\citenamefont {Antoniewicz}\ and\ \citenamefont
  {Kleinman}(1985)}]{ankl85}%
  \BibitemOpen
  \bibfield  {author} {\bibinfo {author} {\bibfnamefont {P.~R.}\ \bibnamefont
  {Antoniewicz}}\ and\ \bibinfo {author} {\bibfnamefont {L.}~\bibnamefont
  {Kleinman}},\ }\bibfield  {title} {\bibinfo {title} {{Kohn-Sham exchange
  potential exact to first order in $\rho(K)/\rho_0$}},\ }\href@noop {}
  {\bibfield  {journal} {\bibinfo  {journal} {Phys. Rev. B}\ }\textbf {\bibinfo
  {volume} {31}},\ \bibinfo {pages} {6779} (\bibinfo {year}
  {1985})}\BibitemShut {NoStop}%
\bibitem [{\citenamefont {Kleinman}\ and\ \citenamefont {Lee}(1988)}]{klle88}%
  \BibitemOpen
  \bibfield  {author} {\bibinfo {author} {\bibfnamefont {L.}~\bibnamefont
  {Kleinman}}\ and\ \bibinfo {author} {\bibfnamefont {S.}~\bibnamefont {Lee}},\
  }\bibfield  {title} {\bibinfo {title} {Gradient expansion of the
  exchange-energy density functional: {E}ffect of taking limits in the wrong
  order},\ }\href@noop {} {\bibfield  {journal} {\bibinfo  {journal} {Phys.
  Rev. B}\ }\textbf {\bibinfo {volume} {37}},\ \bibinfo {pages} {4634}
  (\bibinfo {year} {1988})}\BibitemShut {NoStop}%
\bibitem [{\citenamefont {Burke}\ \emph
  {et~al.}(1997{\natexlab{a}})\citenamefont {Burke}, \citenamefont
  {Ernzerhof},\ and\ \citenamefont {Perdew}}]{BuErPe97}%
  \BibitemOpen
  \bibfield  {author} {\bibinfo {author} {\bibfnamefont {K.}~\bibnamefont
  {Burke}}, \bibinfo {author} {\bibfnamefont {M.}~\bibnamefont {Ernzerhof}},\
  and\ \bibinfo {author} {\bibfnamefont {J.~P.}\ \bibnamefont {Perdew}},\
  }\bibfield  {title} {\bibinfo {title} {The adiabatic connection method: a
  non-empirical hybrid},\ }\href@noop {} {\bibfield  {journal} {\bibinfo
  {journal} {Chem. Phys. Lett.}\ }\textbf {\bibinfo {volume} {265}},\ \bibinfo
  {pages} {115} (\bibinfo {year} {1997}{\natexlab{a}})}\BibitemShut {NoStop}%
\bibitem [{\citenamefont {Adamo}\ and\ \citenamefont {Barone}(1999)}]{PBE0}%
  \BibitemOpen
  \bibfield  {author} {\bibinfo {author} {\bibfnamefont {C.}~\bibnamefont
  {Adamo}}\ and\ \bibinfo {author} {\bibfnamefont {V.}~\bibnamefont {Barone}},\
  }\bibfield  {title} {\bibinfo {title} {Towards reliable density functional
  methods without adjustable parameters: {T}he {PBE}0 model},\ }\href@noop {}
  {\bibfield  {journal} {\bibinfo  {journal} {J. Chem. Phys.}\ }\textbf
  {\bibinfo {volume} {110}},\ \bibinfo {pages} {6158} (\bibinfo {year}
  {1999})}\BibitemShut {NoStop}%
\bibitem [{\citenamefont {Giannozzi}\ \emph {et~al.}(2009)\citenamefont
  {Giannozzi}, \citenamefont {Baroni}, \citenamefont {Bonini}, \citenamefont
  {Calandra}, \citenamefont {Car}, \citenamefont {Cavazzoni}, \citenamefont
  {Ceresoli}, \citenamefont {Chiarotti}, \citenamefont {Cococcioni},
  \citenamefont {Dabo}, \citenamefont {Corso}, \citenamefont {de~Gironcoli},
  \citenamefont {Fabris}, \citenamefont {Fratesi}, \citenamefont {Gebauer},
  \citenamefont {Gerstmann}, \citenamefont {Gougoussis}, \citenamefont
  {Kokalj}, \citenamefont {Lazzeri}, \citenamefont {Martin-Samos},
  \citenamefont {Marzari}, \citenamefont {Mauri}, \citenamefont {Mazzarello},
  \citenamefont {Paolini}, \citenamefont {Pasquarello}, \citenamefont
  {Paulatto}, \citenamefont {Sbraccia}, \citenamefont {Scandolo}, \citenamefont
  {Sclauzero}, \citenamefont {Seitsonen}, \citenamefont {Smogunov},
  \citenamefont {Umari},\ and\ \citenamefont {Wentzcovitch}}]{QE}%
  \BibitemOpen
  \bibfield  {author} {\bibinfo {author} {\bibfnamefont {P.}~\bibnamefont
  {Giannozzi}}, \bibinfo {author} {\bibfnamefont {S.}~\bibnamefont {Baroni}},
  \bibinfo {author} {\bibfnamefont {N.}~\bibnamefont {Bonini}}, \bibinfo
  {author} {\bibfnamefont {M.}~\bibnamefont {Calandra}}, \bibinfo {author}
  {\bibfnamefont {R.}~\bibnamefont {Car}}, \bibinfo {author} {\bibfnamefont
  {C.}~\bibnamefont {Cavazzoni}}, \bibinfo {author} {\bibfnamefont
  {D.}~\bibnamefont {Ceresoli}}, \bibinfo {author} {\bibfnamefont {G.~L.}\
  \bibnamefont {Chiarotti}}, \bibinfo {author} {\bibfnamefont {M.}~\bibnamefont
  {Cococcioni}}, \bibinfo {author} {\bibfnamefont {I.}~\bibnamefont {Dabo}},
  \bibinfo {author} {\bibfnamefont {A.~D.}\ \bibnamefont {Corso}}, \bibinfo
  {author} {\bibfnamefont {S.}~\bibnamefont {de~Gironcoli}}, \bibinfo {author}
  {\bibfnamefont {S.}~\bibnamefont {Fabris}}, \bibinfo {author} {\bibfnamefont
  {G.}~\bibnamefont {Fratesi}}, \bibinfo {author} {\bibfnamefont
  {R.}~\bibnamefont {Gebauer}}, \bibinfo {author} {\bibfnamefont
  {U.}~\bibnamefont {Gerstmann}}, \bibinfo {author} {\bibfnamefont
  {C.}~\bibnamefont {Gougoussis}}, \bibinfo {author} {\bibfnamefont
  {A.}~\bibnamefont {Kokalj}}, \bibinfo {author} {\bibfnamefont
  {M.}~\bibnamefont {Lazzeri}}, \bibinfo {author} {\bibfnamefont
  {L.}~\bibnamefont {Martin-Samos}}, \bibinfo {author} {\bibfnamefont
  {N.}~\bibnamefont {Marzari}}, \bibinfo {author} {\bibfnamefont
  {F.}~\bibnamefont {Mauri}}, \bibinfo {author} {\bibfnamefont
  {R.}~\bibnamefont {Mazzarello}}, \bibinfo {author} {\bibfnamefont
  {S.}~\bibnamefont {Paolini}}, \bibinfo {author} {\bibfnamefont
  {A.}~\bibnamefont {Pasquarello}}, \bibinfo {author} {\bibfnamefont
  {L.}~\bibnamefont {Paulatto}}, \bibinfo {author} {\bibfnamefont
  {C.}~\bibnamefont {Sbraccia}}, \bibinfo {author} {\bibfnamefont
  {S.}~\bibnamefont {Scandolo}}, \bibinfo {author} {\bibfnamefont
  {G.}~\bibnamefont {Sclauzero}}, \bibinfo {author} {\bibfnamefont {A.~P.}\
  \bibnamefont {Seitsonen}}, \bibinfo {author} {\bibfnamefont {A.}~\bibnamefont
  {Smogunov}}, \bibinfo {author} {\bibfnamefont {P.}~\bibnamefont {Umari}},\
  and\ \bibinfo {author} {\bibfnamefont {R.~M.}\ \bibnamefont {Wentzcovitch}},\
  }\bibfield  {title} {\bibinfo {title} {{QUANTUM ESPRESSO}: a modular and
  open-source software project for quantum simulations of materials},\
  }\href@noop {} {\bibfield  {journal} {\bibinfo  {journal} {J. Phys.: Condens.
  Matter}\ }\textbf {\bibinfo {volume} {21}},\ \bibinfo {pages} {395502}
  (\bibinfo {year} {2009})}\BibitemShut {NoStop}%
\bibitem [{\citenamefont {Giannozzi}\ \emph {et~al.}(2017)\citenamefont
  {Giannozzi}, \citenamefont {Andreussi}, \citenamefont {Brumme}, \citenamefont
  {Bunau}, \citenamefont {Buongiorno~Nardelli}, \citenamefont {Calandra},
  \citenamefont {Car}, \citenamefont {Cavazzoni}, \citenamefont {Ceresoli},
  \citenamefont {Cococcioni}, \citenamefont {Collonna}, \citenamefont
  {Carnimeo}, \citenamefont {Dal~Corso}, \citenamefont {{de Gironcoli}},
  \citenamefont {Delugas}, \citenamefont {{DiStasio Jr}}, \citenamefont
  {Feretti}, \citenamefont {Floris}, \citenamefont {Fratesi}, \citenamefont
  {Fugalio}, \citenamefont {Gebauer}, \citenamefont {Gerstmann}, \citenamefont
  {Giustino}, \citenamefont {Gorni}, \citenamefont {Jia}, \citenamefont
  {Kawamura}, \citenamefont {Ko}, \citenamefont {Kokalj}, \citenamefont
  {K{\"u}c{\"u}kbenli}, \citenamefont {Lazzeri}, \citenamefont {Marseli},
  \citenamefont {Marzari}, \citenamefont {Mauri}, \citenamefont {Nguyen},
  \citenamefont {Nguyen}, \citenamefont {Otero-de-la Roza}, \citenamefont
  {Paulatto}, \citenamefont {Ponc{\'e}}, \citenamefont {Rocca}, \citenamefont
  {Sabatini}, \citenamefont {Santra}, \citenamefont {Schlipf}, \citenamefont
  {Seitsonen}, \citenamefont {Smogunov}, \citenamefont {Timrov}, \citenamefont
  {Thonhauser}, \citenamefont {Umari}, \citenamefont {Vast}, \citenamefont
  {Wu},\ and\ \citenamefont {Baroni}}]{Giannozzi17}%
  \BibitemOpen
  \bibfield  {author} {\bibinfo {author} {\bibfnamefont {P.}~\bibnamefont
  {Giannozzi}}, \bibinfo {author} {\bibfnamefont {O.}~\bibnamefont
  {Andreussi}}, \bibinfo {author} {\bibfnamefont {T.}~\bibnamefont {Brumme}},
  \bibinfo {author} {\bibfnamefont {O.}~\bibnamefont {Bunau}}, \bibinfo
  {author} {\bibfnamefont {M.}~\bibnamefont {Buongiorno~Nardelli}}, \bibinfo
  {author} {\bibfnamefont {M.}~\bibnamefont {Calandra}}, \bibinfo {author}
  {\bibfnamefont {R.}~\bibnamefont {Car}}, \bibinfo {author} {\bibfnamefont
  {C.}~\bibnamefont {Cavazzoni}}, \bibinfo {author} {\bibfnamefont
  {D.}~\bibnamefont {Ceresoli}}, \bibinfo {author} {\bibfnamefont
  {M.}~\bibnamefont {Cococcioni}}, \bibinfo {author} {\bibfnamefont
  {N.}~\bibnamefont {Collonna}}, \bibinfo {author} {\bibfnamefont
  {I.}~\bibnamefont {Carnimeo}}, \bibinfo {author} {\bibfnamefont
  {A.}~\bibnamefont {Dal~Corso}}, \bibinfo {author} {\bibfnamefont
  {S.}~\bibnamefont {{de Gironcoli}}}, \bibinfo {author} {\bibfnamefont
  {P.}~\bibnamefont {Delugas}}, \bibinfo {author} {\bibfnamefont {R.~A.}\
  \bibnamefont {{DiStasio Jr}}}, \bibinfo {author} {\bibfnamefont
  {A.}~\bibnamefont {Feretti}}, \bibinfo {author} {\bibfnamefont
  {A.}~\bibnamefont {Floris}}, \bibinfo {author} {\bibfnamefont
  {G.}~\bibnamefont {Fratesi}}, \bibinfo {author} {\bibfnamefont
  {G.}~\bibnamefont {Fugalio}}, \bibinfo {author} {\bibfnamefont
  {R.}~\bibnamefont {Gebauer}}, \bibinfo {author} {\bibfnamefont
  {U.}~\bibnamefont {Gerstmann}}, \bibinfo {author} {\bibfnamefont
  {F.}~\bibnamefont {Giustino}}, \bibinfo {author} {\bibfnamefont
  {T.}~\bibnamefont {Gorni}}, \bibinfo {author} {\bibfnamefont
  {J.}~\bibnamefont {Jia}}, \bibinfo {author} {\bibfnamefont {M.}~\bibnamefont
  {Kawamura}}, \bibinfo {author} {\bibfnamefont {H.-Y.}\ \bibnamefont {Ko}},
  \bibinfo {author} {\bibfnamefont {A.}~\bibnamefont {Kokalj}}, \bibinfo
  {author} {\bibfnamefont {E.}~\bibnamefont {K{\"u}c{\"u}kbenli}}, \bibinfo
  {author} {\bibfnamefont {M.}~\bibnamefont {Lazzeri}}, \bibinfo {author}
  {\bibfnamefont {M.}~\bibnamefont {Marseli}}, \bibinfo {author} {\bibfnamefont
  {N.}~\bibnamefont {Marzari}}, \bibinfo {author} {\bibfnamefont
  {F.}~\bibnamefont {Mauri}}, \bibinfo {author} {\bibfnamefont {N.~L.}\
  \bibnamefont {Nguyen}}, \bibinfo {author} {\bibfnamefont {H.-V.}\
  \bibnamefont {Nguyen}}, \bibinfo {author} {\bibfnamefont {A.}~\bibnamefont
  {Otero-de-la Roza}}, \bibinfo {author} {\bibfnamefont {L.}~\bibnamefont
  {Paulatto}}, \bibinfo {author} {\bibfnamefont {S.}~\bibnamefont {Ponc{\'e}}},
  \bibinfo {author} {\bibfnamefont {D.}~\bibnamefont {Rocca}}, \bibinfo
  {author} {\bibfnamefont {R.}~\bibnamefont {Sabatini}}, \bibinfo {author}
  {\bibfnamefont {B.}~\bibnamefont {Santra}}, \bibinfo {author} {\bibfnamefont
  {M.}~\bibnamefont {Schlipf}}, \bibinfo {author} {\bibfnamefont
  {A.}~\bibnamefont {Seitsonen}}, \bibinfo {author} {\bibfnamefont
  {A.}~\bibnamefont {Smogunov}}, \bibinfo {author} {\bibfnamefont
  {I.}~\bibnamefont {Timrov}}, \bibinfo {author} {\bibfnamefont
  {T.}~\bibnamefont {Thonhauser}}, \bibinfo {author} {\bibfnamefont
  {P.}~\bibnamefont {Umari}}, \bibinfo {author} {\bibfnamefont
  {N.}~\bibnamefont {Vast}}, \bibinfo {author} {\bibfnamefont {X.}~\bibnamefont
  {Wu}},\ and\ \bibinfo {author} {\bibfnamefont {S.}~\bibnamefont {Baroni}},\
  }\bibfield  {title} {\bibinfo {title} {Advanced capabilities for materials
  modelling with quantum espresso},\ }\href@noop {} {\bibfield  {journal}
  {\bibinfo  {journal} {J.Phys.: Condens. Matter}\ }\textbf {\bibinfo {volume}
  {29}},\ \bibinfo {pages} {465901} (\bibinfo {year} {2017})}\BibitemShut
  {NoStop}%
\bibitem [{\citenamefont {Carnimeo}\ \emph {et~al.}(2019)\citenamefont
  {Carnimeo}, \citenamefont {Baroni},\ and\ \citenamefont
  {Giannozzi}}]{PaoloElStruct1}%
  \BibitemOpen
  \bibfield  {author} {\bibinfo {author} {\bibfnamefont {I.}~\bibnamefont
  {Carnimeo}}, \bibinfo {author} {\bibfnamefont {S.}~\bibnamefont {Baroni}},\
  and\ \bibinfo {author} {\bibfnamefont {P.}~\bibnamefont {Giannozzi}},\
  }\bibfield  {title} {\bibinfo {title} {{Fast hybrid density-functional
  computations using plane-wave beasis sets}},\ }\href@noop {} {\bibfield
  {journal} {\bibinfo  {journal} {Electron. Struct.}\ }\textbf {\bibinfo
  {volume} {1}},\ \bibinfo {pages} {015009} (\bibinfo {year}
  {2019})}\BibitemShut {NoStop}%
\bibitem [{Note3()}]{Note3}%
  \BibitemOpen
  \bibinfo {note} {The DF2-BR0 performance on the 55 specific benchmarks is, in
  almost all cases, as good as or better than that for CX0 or for CX0P (and
  they are already strong overall performers,) as shown in the SI; The W4-11
  benchmark set (and, to a lesser extent, the DIPCS10 and MB16-43 sets,) forms
  an exception to this general observation.}\BibitemShut {Stop}%
\bibitem [{\citenamefont {Ernzerhof}\ and\ \citenamefont
  {Perdew}(1998)}]{EP98}%
  \BibitemOpen
  \bibfield  {author} {\bibinfo {author} {\bibfnamefont {M.}~\bibnamefont
  {Ernzerhof}}\ and\ \bibinfo {author} {\bibfnamefont {J.~P.}\ \bibnamefont
  {Perdew}},\ }\bibfield  {title} {\bibinfo {title} {Generalized gradient
  approximation to the angle- and system-averaged exchange hole},\ }\href
  {https://doi.org/http://dx.doi.org/10.1063/1.476928} {\bibfield  {journal}
  {\bibinfo  {journal} {J. Chem. Phys.}\ }\textbf {\bibinfo {volume} {109}},\
  \bibinfo {pages} {3313} (\bibinfo {year} {1998})}\BibitemShut {NoStop}%
\bibitem [{\citenamefont {Rafaely-Abramson}\ \emph {et~al.}(2012)\citenamefont
  {Rafaely-Abramson}, \citenamefont {Sharifzadeh}, \citenamefont {Govind},
  \citenamefont {Autschbach}, \citenamefont {Neaton}, \citenamefont {Baer},\
  and\ \citenamefont {Kronik}}]{OTRSHalga}%
  \BibitemOpen
  \bibfield  {author} {\bibinfo {author} {\bibfnamefont {S.}~\bibnamefont
  {Rafaely-Abramson}}, \bibinfo {author} {\bibfnamefont {S.}~\bibnamefont
  {Sharifzadeh}}, \bibinfo {author} {\bibfnamefont {N.}~\bibnamefont {Govind}},
  \bibinfo {author} {\bibfnamefont {J.}~\bibnamefont {Autschbach}}, \bibinfo
  {author} {\bibfnamefont {J.~B.}\ \bibnamefont {Neaton}}, \bibinfo {author}
  {\bibfnamefont {R.}~\bibnamefont {Baer}},\ and\ \bibinfo {author}
  {\bibfnamefont {L.}~\bibnamefont {Kronik}},\ }\bibfield  {title} {\bibinfo
  {title} {{Quasiparticle Spectra from a Nonempirical Optimally Tuned
  Range-Separated Hybrid Density Functional}},\ }\href@noop {} {\bibfield
  {journal} {\bibinfo  {journal} {Phys. Rev. Lett.}\ }\textbf {\bibinfo
  {volume} {109}},\ \bibinfo {pages} {226405} (\bibinfo {year}
  {2012})}\BibitemShut {NoStop}%
\bibitem [{\citenamefont {Wijzenbroek}\ \emph {et~al.}(2015)\citenamefont
  {Wijzenbroek}, \citenamefont {Klein}, \citenamefont {Smits}, \citenamefont
  {Somers},\ and\ \citenamefont {Kroes}}]{catalysis1}%
  \BibitemOpen
  \bibfield  {author} {\bibinfo {author} {\bibfnamefont {M.}~\bibnamefont
  {Wijzenbroek}}, \bibinfo {author} {\bibfnamefont {D.~M.}\ \bibnamefont
  {Klein}}, \bibinfo {author} {\bibfnamefont {B.}~\bibnamefont {Smits}},
  \bibinfo {author} {\bibfnamefont {M.~F.}\ \bibnamefont {Somers}},\ and\
  \bibinfo {author} {\bibfnamefont {G.-J.}\ \bibnamefont {Kroes}},\ }\bibfield
  {title} {\bibinfo {title} {Performance of a non-local {van der Waals} density
  functional on the dissociation of {H}2 on metal surfaces},\ }\href@noop {}
  {\bibfield  {journal} {\bibinfo  {journal} {J. Phys. Chem. A}\ }\textbf
  {\bibinfo {volume} {119}},\ \bibinfo {pages} {12146} (\bibinfo {year}
  {2015})},\ \bibinfo {note} {pMID: 26258988}\BibitemShut {NoStop}%
\bibitem [{\citenamefont {Perdew}\ \emph {et~al.}(1982)\citenamefont {Perdew},
  \citenamefont {Parr}, \citenamefont {Levy},\ and\ \citenamefont
  {J.~L.~Balduz}}]{PePaLe82}%
  \BibitemOpen
  \bibfield  {author} {\bibinfo {author} {\bibfnamefont {J.~P.}\ \bibnamefont
  {Perdew}}, \bibinfo {author} {\bibfnamefont {R.~G.}\ \bibnamefont {Parr}},
  \bibinfo {author} {\bibfnamefont {M.}~\bibnamefont {Levy}},\ and\ \bibinfo
  {author} {\bibfnamefont {J.}~\bibnamefont {J.~L.~Balduz}},\ }\bibfield
  {title} {\bibinfo {title} {{Density-Functional Theory for Fractional Particle
  Number: Derivative Discontinutities of the Energy}},\ }\href@noop {}
  {\bibfield  {journal} {\bibinfo  {journal} {Phys. Rev. Lett.}\ }\textbf
  {\bibinfo {volume} {49}},\ \bibinfo {pages} {1691} (\bibinfo {year}
  {1982})}\BibitemShut {NoStop}%
\bibitem [{\citenamefont {Seidl}\ \emph {et~al.}(1996)\citenamefont {Seidl},
  \citenamefont {G{\"o}rling}, \citenamefont {Vogl}, \citenamefont {Majewski},\
  and\ \citenamefont {Levy}}]{GKSstart}%
  \BibitemOpen
  \bibfield  {author} {\bibinfo {author} {\bibfnamefont {A.}~\bibnamefont
  {Seidl}}, \bibinfo {author} {\bibfnamefont {A.}~\bibnamefont {G{\"o}rling}},
  \bibinfo {author} {\bibfnamefont {P.}~\bibnamefont {Vogl}}, \bibinfo {author}
  {\bibfnamefont {J.~A.}\ \bibnamefont {Majewski}},\ and\ \bibinfo {author}
  {\bibfnamefont {M.}~\bibnamefont {Levy}},\ }\bibfield  {title} {\bibinfo
  {title} {{Generalized Kohn-Sham schemes and the band-gap problem}},\
  }\href@noop {} {\bibfield  {journal} {\bibinfo  {journal} {Phys. Rev. B}\
  }\textbf {\bibinfo {volume} {53}},\ \bibinfo {pages} {3764} (\bibinfo {year}
  {1996})}\BibitemShut {NoStop}%
\bibitem [{\citenamefont {Kraisler}\ and\ \citenamefont
  {Kronik}(2013)}]{KraKro13}%
  \BibitemOpen
  \bibfield  {author} {\bibinfo {author} {\bibfnamefont {E.}~\bibnamefont
  {Kraisler}}\ and\ \bibinfo {author} {\bibfnamefont {L.}~\bibnamefont
  {Kronik}},\ }\bibfield  {title} {\bibinfo {title} {{Piecewise Linearity of
  Approximate Density Functionals Revisited: Implications for Frontier Orbital
  Energies}},\ }\href@noop {} {\bibfield  {journal} {\bibinfo  {journal} {Phys.
  Rev. Lett.}\ }\textbf {\bibinfo {volume} {110}},\ \bibinfo {pages} {126403}
  (\bibinfo {year} {2013})}\BibitemShut {NoStop}%
\bibitem [{\citenamefont {Rafaely-Abramson}\ \emph {et~al.}(2013)\citenamefont
  {Rafaely-Abramson}, \citenamefont {Sharifzadeh}, \citenamefont {Jain},
  \citenamefont {Baer}, \citenamefont {Neaton},\ and\ \citenamefont
  {Kronik}}]{OTRSHgap}%
  \BibitemOpen
  \bibfield  {author} {\bibinfo {author} {\bibfnamefont {S.}~\bibnamefont
  {Rafaely-Abramson}}, \bibinfo {author} {\bibfnamefont {S.}~\bibnamefont
  {Sharifzadeh}}, \bibinfo {author} {\bibfnamefont {M.}~\bibnamefont {Jain}},
  \bibinfo {author} {\bibfnamefont {R.}~\bibnamefont {Baer}}, \bibinfo {author}
  {\bibfnamefont {J.~B.}\ \bibnamefont {Neaton}},\ and\ \bibinfo {author}
  {\bibfnamefont {L.}~\bibnamefont {Kronik}},\ }\bibfield  {title} {\bibinfo
  {title} {Gab renormalization of molecular crystals from density functional
  theory},\ }\href@noop {} {\bibfield  {journal} {\bibinfo  {journal} {Phys.
  Rev. B}\ }\textbf {\bibinfo {volume} {88}},\ \bibinfo {pages} {081204(R)}
  (\bibinfo {year} {2013})}\BibitemShut {NoStop}%
\bibitem [{\citenamefont {Liu}\ \emph {et~al.}(2017)\citenamefont {Liu},
  \citenamefont {Egger}, \citenamefont {Rafaely-Abramson}, \citenamefont
  {Kronik},\ and\ \citenamefont {Neaton}}]{OTRSHadsorb17}%
  \BibitemOpen
  \bibfield  {author} {\bibinfo {author} {\bibfnamefont {Z.-F.}\ \bibnamefont
  {Liu}}, \bibinfo {author} {\bibfnamefont {D.~A.}\ \bibnamefont {Egger}},
  \bibinfo {author} {\bibfnamefont {S.}~\bibnamefont {Rafaely-Abramson}},
  \bibinfo {author} {\bibfnamefont {L.}~\bibnamefont {Kronik}},\ and\ \bibinfo
  {author} {\bibfnamefont {J.~B.}\ \bibnamefont {Neaton}},\ }\bibfield  {title}
  {\bibinfo {title} {Energy level alignment at molecule-metal interfaces from
  an optimally tuned range-separated hybrid functional},\ }\href@noop {}
  {\bibfield  {journal} {\bibinfo  {journal} {J. Chem. Phys.}\ }\textbf
  {\bibinfo {volume} {146}},\ \bibinfo {pages} {092326} (\bibinfo {year}
  {2017})}\BibitemShut {NoStop}%
\bibitem [{\citenamefont {Wing}\ \emph {et~al.}(2021)\citenamefont {Wing},
  \citenamefont {Ohad}, \citenamefont {Haber}, \citenamefont {Filip},
  \citenamefont {Gant}, \citenamefont {Neaton},\ and\ \citenamefont
  {Kronik}}]{WiOhHa21}%
  \BibitemOpen
  \bibfield  {author} {\bibinfo {author} {\bibfnamefont {D.}~\bibnamefont
  {Wing}}, \bibinfo {author} {\bibfnamefont {G.}~\bibnamefont {Ohad}}, \bibinfo
  {author} {\bibfnamefont {J.~B.}\ \bibnamefont {Haber}}, \bibinfo {author}
  {\bibfnamefont {M.~R.}\ \bibnamefont {Filip}}, \bibinfo {author}
  {\bibfnamefont {S.~E.}\ \bibnamefont {Gant}}, \bibinfo {author}
  {\bibfnamefont {J.~B.}\ \bibnamefont {Neaton}},\ and\ \bibinfo {author}
  {\bibfnamefont {L.}~\bibnamefont {Kronik}},\ }\bibfield  {title} {\bibinfo
  {title} {{Band gaps of crystalline solids from Wannier-localization–based
  optimal tuning of a screened range-separated hybrid functional}},\
  }\href@noop {} {\bibfield  {journal} {\bibinfo  {journal} {PNAS}\ }\textbf
  {\bibinfo {volume} {118}},\ \bibinfo {pages} {e2104556118} (\bibinfo {year}
  {2021})}\BibitemShut {NoStop}%
\bibitem [{\citenamefont {Burke}\ \emph
  {et~al.}(1997{\natexlab{b}})\citenamefont {Burke}, \citenamefont
  {Ernzerhof},\ and\ \citenamefont {Perdew}}]{Burke97}%
  \BibitemOpen
  \bibfield  {author} {\bibinfo {author} {\bibfnamefont {K.}~\bibnamefont
  {Burke}}, \bibinfo {author} {\bibfnamefont {M.}~\bibnamefont {Ernzerhof}},\
  and\ \bibinfo {author} {\bibfnamefont {J.~P.}\ \bibnamefont {Perdew}},\
  }\bibfield  {title} {\bibinfo {title} {The adiabatic connection method: a
  non-empirical hybrid},\ }\href
  {https://doi.org/http://dx.doi.org/10.1016/S0009-2614(96)01373-5} {\bibfield
  {journal} {\bibinfo  {journal} {Chem. Phys. Lett.}\ }\textbf {\bibinfo
  {volume} {265}},\ \bibinfo {pages} {115 } (\bibinfo {year}
  {1997}{\natexlab{b}})}\BibitemShut {NoStop}%
\bibitem [{\citenamefont {Skone}\ \emph {et~al.}(2014)\citenamefont {Skone},
  \citenamefont {Govoni},\ and\ \citenamefont {Galli}}]{SkGoGa14}%
  \BibitemOpen
  \bibfield  {author} {\bibinfo {author} {\bibfnamefont {J.~H.}\ \bibnamefont
  {Skone}}, \bibinfo {author} {\bibfnamefont {M.}~\bibnamefont {Govoni}},\ and\
  \bibinfo {author} {\bibfnamefont {G.}~\bibnamefont {Galli}},\ }\bibfield
  {title} {\bibinfo {title} {Self-consistent hybrid functional for condensed
  systems},\ }\href@noop {} {\bibfield  {journal} {\bibinfo  {journal} {Phys.
  Rev. B}\ }\textbf {\bibinfo {volume} {89}},\ \bibinfo {pages} {195112}
  (\bibinfo {year} {2014})}\BibitemShut {NoStop}%
\bibitem [{\citenamefont {Skone}\ \emph {et~al.}(2016)\citenamefont {Skone},
  \citenamefont {Govoni},\ and\ \citenamefont {Galli}}]{SkGoGa16}%
  \BibitemOpen
  \bibfield  {author} {\bibinfo {author} {\bibfnamefont {J.~H.}\ \bibnamefont
  {Skone}}, \bibinfo {author} {\bibfnamefont {M.}~\bibnamefont {Govoni}},\ and\
  \bibinfo {author} {\bibfnamefont {G.}~\bibnamefont {Galli}},\ }\bibfield
  {title} {\bibinfo {title} {Nonempirical range-separated hybrid functionals
  for solids and molecules},\ }\href@noop {} {\bibfield  {journal} {\bibinfo
  {journal} {Phys. Rev. B}\ }\textbf {\bibinfo {volume} {93}},\ \bibinfo
  {pages} {235106} (\bibinfo {year} {2016})}\BibitemShut {NoStop}%
\bibitem [{\citenamefont {Miceli}\ \emph {et~al.}(2018)\citenamefont {Miceli},
  \citenamefont {Chen}, \citenamefont {Reshetnyak},\ and\ \citenamefont
  {Pasquarello}}]{MiChRe18}%
  \BibitemOpen
  \bibfield  {author} {\bibinfo {author} {\bibfnamefont {G.}~\bibnamefont
  {Miceli}}, \bibinfo {author} {\bibfnamefont {W.}~\bibnamefont {Chen}},
  \bibinfo {author} {\bibfnamefont {I.}~\bibnamefont {Reshetnyak}},\ and\
  \bibinfo {author} {\bibfnamefont {A.}~\bibnamefont {Pasquarello}},\
  }\bibfield  {title} {\bibinfo {title} {Nonempirical hybrid functionals for
  band gaps and polaronic distortions in solids},\ }\href@noop {} {\bibfield
  {journal} {\bibinfo  {journal} {Phys. Rev. B}\ }\textbf {\bibinfo {volume}
  {97}},\ \bibinfo {pages} {121112(R)} (\bibinfo {year} {2018})}\BibitemShut
  {NoStop}%
\bibitem [{\citenamefont {Bischoff}\ \emph
  {et~al.}(2019{\natexlab{a}})\citenamefont {Bischoff}, \citenamefont
  {Reshetnyak},\ and\ \citenamefont {Pasquarello}}]{BiRePa19}%
  \BibitemOpen
  \bibfield  {author} {\bibinfo {author} {\bibfnamefont {T.}~\bibnamefont
  {Bischoff}}, \bibinfo {author} {\bibfnamefont {I.}~\bibnamefont
  {Reshetnyak}},\ and\ \bibinfo {author} {\bibfnamefont {A.}~\bibnamefont
  {Pasquarello}},\ }\bibfield  {title} {\bibinfo {title} {Adjustable potential
  probes for band-gab predictions of extended systems through nonempirical
  hybrid functionals},\ }\href@noop {} {\bibfield  {journal} {\bibinfo
  {journal} {Phys. Rev. B}\ }\textbf {\bibinfo {volume} {99}},\ \bibinfo
  {pages} {201114(R)} (\bibinfo {year} {2019}{\natexlab{a}})}\BibitemShut
  {NoStop}%
\bibitem [{\citenamefont {Bischoff}\ \emph
  {et~al.}(2019{\natexlab{b}})\citenamefont {Bischoff}, \citenamefont {Wiktor},
  \citenamefont {Chen},\ and\ \citenamefont {Pasquarello}}]{BiWiCh19}%
  \BibitemOpen
  \bibfield  {author} {\bibinfo {author} {\bibfnamefont {T.}~\bibnamefont
  {Bischoff}}, \bibinfo {author} {\bibfnamefont {J.}~\bibnamefont {Wiktor}},
  \bibinfo {author} {\bibfnamefont {W.}~\bibnamefont {Chen}},\ and\ \bibinfo
  {author} {\bibfnamefont {A.}~\bibnamefont {Pasquarello}},\ }\bibfield
  {title} {\bibinfo {title} {Nonemperical hybrid functionals for band gaps of
  inorganic metal-halide perovskites},\ }\href@noop {} {\bibfield  {journal}
  {\bibinfo  {journal} {Phys. Rev. M}\ }\textbf {\bibinfo {volume} {3}},\
  \bibinfo {pages} {123802} (\bibinfo {year} {2019}{\natexlab{b}})}\BibitemShut
  {NoStop}%
\bibitem [{\citenamefont {Lin}(2016)}]{LinACE}%
  \BibitemOpen
  \bibfield  {author} {\bibinfo {author} {\bibfnamefont {L.}~\bibnamefont
  {Lin}},\ }\bibfield  {title} {\bibinfo {title} {Adaptively compressed
  exchange operator},\ }\href@noop {} {\bibfield  {journal} {\bibinfo
  {journal} {J. Chem. Theory Comput.}\ }\textbf {\bibinfo {volume} {12}},\
  \bibinfo {pages} {2242} (\bibinfo {year} {2016})}\BibitemShut {NoStop}%
\bibitem [{\citenamefont {{Gharaee}}\ \emph {et~al.}(2017)\citenamefont
  {{Gharaee}}, \citenamefont {{Erhart}},\ and\ \citenamefont
  {{Hyldgaard}}}]{Gharaee2017}%
  \BibitemOpen
  \bibfield  {author} {\bibinfo {author} {\bibfnamefont {L.}~\bibnamefont
  {{Gharaee}}}, \bibinfo {author} {\bibfnamefont {P.}~\bibnamefont
  {{Erhart}}},\ and\ \bibinfo {author} {\bibfnamefont {P.}~\bibnamefont
  {{Hyldgaard}}},\ }\bibfield  {title} {\bibinfo {title} {{Finite-temperature
  properties of non-magnetic transition metals: {C}omparison of the performance
  of constraint-based semi and nonlocal functionals}},\ }\href@noop {}
  {\bibfield  {journal} {\bibinfo  {journal} {Phys. Rev. B}\ }\textbf {\bibinfo
  {volume} {95}},\ \bibinfo {pages} {085147} (\bibinfo {year}
  {2017})}\BibitemShut {NoStop}%
\bibitem [{\citenamefont {Spanu}\ \emph {et~al.}(2009)\citenamefont {Spanu},
  \citenamefont {Sorella},\ and\ \citenamefont {Galli}}]{Spanu09p196401}%
  \BibitemOpen
  \bibfield  {author} {\bibinfo {author} {\bibfnamefont {L.}~\bibnamefont
  {Spanu}}, \bibinfo {author} {\bibfnamefont {S.}~\bibnamefont {Sorella}},\
  and\ \bibinfo {author} {\bibfnamefont {G.}~\bibnamefont {Galli}},\ }\bibfield
   {title} {\bibinfo {title} {Nature and strength of interlayer binding in
  graphite},\ }\href {https://doi.org/10.1103/PhysRevLett.103.196401}
  {\bibfield  {journal} {\bibinfo  {journal} {Phys. Rev. Lett.}\ }\textbf
  {\bibinfo {volume} {103}},\ \bibinfo {pages} {196401} (\bibinfo {year}
  {2009})}\BibitemShut {NoStop}%
\bibitem [{\citenamefont {Ganesh}\ \emph {et~al.}(2014)\citenamefont {Ganesh},
  \citenamefont {Kim}, \citenamefont {Park}, \citenamefont {Yoon},
  \citenamefont {Reboredo},\ and\ \citenamefont {Kent}}]{GaKiPa2014}%
  \BibitemOpen
  \bibfield  {author} {\bibinfo {author} {\bibfnamefont {P.}~\bibnamefont
  {Ganesh}}, \bibinfo {author} {\bibfnamefont {J.}~\bibnamefont {Kim}},
  \bibinfo {author} {\bibfnamefont {C.}~\bibnamefont {Park}}, \bibinfo {author}
  {\bibfnamefont {M.}~\bibnamefont {Yoon}}, \bibinfo {author} {\bibfnamefont
  {F.~A.}\ \bibnamefont {Reboredo}},\ and\ \bibinfo {author} {\bibfnamefont
  {P.~R.~C.}\ \bibnamefont {Kent}},\ }\bibfield  {title} {\bibinfo {title}
  {{Binding and Diffusion of Lithium in Graphite: Quantum Monte Carlo
  Benchmarks and Validation of van der Waals Density Functional Methods}},\
  }\href@noop {} {\bibfield  {journal} {\bibinfo  {journal} {J. Chem. Theory
  Comput.}\ }\textbf {\bibinfo {volume} {10}},\ \bibinfo {pages} {5318}
  (\bibinfo {year} {2014})}\BibitemShut {NoStop}%
\bibitem [{\citenamefont {Mostaani}\ \emph {et~al.}(2015)\citenamefont
  {Mostaani}, \citenamefont {Drummond},\ and\ \citenamefont
  {Fal'ko}}]{MoDrFa2015}%
  \BibitemOpen
  \bibfield  {author} {\bibinfo {author} {\bibfnamefont {E.}~\bibnamefont
  {Mostaani}}, \bibinfo {author} {\bibfnamefont {N.~D.}\ \bibnamefont
  {Drummond}},\ and\ \bibinfo {author} {\bibfnamefont {V.~I.}\ \bibnamefont
  {Fal'ko}},\ }\bibfield  {title} {\bibinfo {title} {{Quantum Monte Carlo
  Calculation of the Binding Energy of Bilayer Graphene}},\ }\href@noop {}
  {\bibfield  {journal} {\bibinfo  {journal} {Phys. Rev. Lett.}\ }\textbf
  {\bibinfo {volume} {115}},\ \bibinfo {pages} {115501} (\bibinfo {year}
  {2015})}\BibitemShut {NoStop}%
\bibitem [{\citenamefont {Shin}\ \emph {et~al.}(2017)\citenamefont {Shin},
  \citenamefont {Kim}, \citenamefont {Lee}, \citenamefont {Heinonen},
  \citenamefont {Benali},\ and\ \citenamefont {Kwon}}]{ShKiLe2017}%
  \BibitemOpen
  \bibfield  {author} {\bibinfo {author} {\bibfnamefont {H.}~\bibnamefont
  {Shin}}, \bibinfo {author} {\bibfnamefont {J.}~\bibnamefont {Kim}}, \bibinfo
  {author} {\bibfnamefont {H.}~\bibnamefont {Lee}}, \bibinfo {author}
  {\bibfnamefont {O.}~\bibnamefont {Heinonen}}, \bibinfo {author}
  {\bibfnamefont {A.}~\bibnamefont {Benali}},\ and\ \bibinfo {author}
  {\bibfnamefont {Y.}~\bibnamefont {Kwon}},\ }\bibfield  {title} {\bibinfo
  {title} {{Nature of interlayer Binding and Stacking of sp-sp$^2$ Hybridized
  Carbon Layers: A Quantum Monte Carlo Study}},\ }\href@noop {} {\bibfield
  {journal} {\bibinfo  {journal} {J. Chem. Theory Comput.}\ }\textbf {\bibinfo
  {volume} {13}},\ \bibinfo {pages} {5639} (\bibinfo {year}
  {2017})}\BibitemShut {NoStop}%
\bibitem [{\citenamefont {Hsing}\ \emph {et~al.}(2015)\citenamefont {Hsing},
  \citenamefont {Chen}, \citenamefont {Chou}, \citenamefont {Chang},\ and\
  \citenamefont {Wei}}]{HsChCh2014}%
  \BibitemOpen
  \bibfield  {author} {\bibinfo {author} {\bibfnamefont {C.-R.}\ \bibnamefont
  {Hsing}}, \bibinfo {author} {\bibfnamefont {C.}~\bibnamefont {Chen}},
  \bibinfo {author} {\bibfnamefont {J.-P.}\ \bibnamefont {Chou}}, \bibinfo
  {author} {\bibfnamefont {C.-M.}\ \bibnamefont {Chang}},\ and\ \bibinfo
  {author} {\bibfnamefont {C.-M.}\ \bibnamefont {Wei}},\ }\bibfield  {title}
  {\bibinfo {title} {{Van der Waals interaction in a boron nitride bilayer}},\
  }\href@noop {} {\bibfield  {journal} {\bibinfo  {journal} {New J. Phys.}\
  }\textbf {\bibinfo {volume} {16}},\ \bibinfo {pages} {113015} (\bibinfo
  {year} {2015})}\BibitemShut {NoStop}%
\bibitem [{\citenamefont {Shulenburger}\ \emph {et~al.}(2015)\citenamefont
  {Shulenburger}, \citenamefont {Baczewski}, \citenamefont {Zhu}, \citenamefont
  {Guan},\ and\ \citenamefont {Tomanek}}]{ShBaZh2015}%
  \BibitemOpen
  \bibfield  {author} {\bibinfo {author} {\bibfnamefont {L.}~\bibnamefont
  {Shulenburger}}, \bibinfo {author} {\bibfnamefont {A.~D.}\ \bibnamefont
  {Baczewski}}, \bibinfo {author} {\bibfnamefont {Z.}~\bibnamefont {Zhu}},
  \bibinfo {author} {\bibfnamefont {J.}~\bibnamefont {Guan}},\ and\ \bibinfo
  {author} {\bibfnamefont {D.}~\bibnamefont {Tomanek}},\ }\bibfield  {title}
  {\bibinfo {title} {{The Nature of the Interlayer Interaction in Bulk and
  Few-Layer Phosphorus}},\ }\href@noop {} {\bibfield  {journal} {\bibinfo
  {journal} {Nano Lett.}\ }\textbf {\bibinfo {volume} {15}},\ \bibinfo {pages}
  {8170} (\bibinfo {year} {2015})}\BibitemShut {NoStop}%
\bibitem [{\citenamefont {Leb\'{e}gue}\ \emph {et~al.}(2010)\citenamefont
  {Leb\'{e}gue}, \citenamefont {Harl}, \citenamefont {Gould}, \citenamefont
  {\'Angy\'an}, \citenamefont {Kresse},\ and\ \citenamefont
  {Dobson}}]{rpa_graphite}%
  \BibitemOpen
  \bibfield  {author} {\bibinfo {author} {\bibfnamefont {S.}~\bibnamefont
  {Leb\'{e}gue}}, \bibinfo {author} {\bibfnamefont {J.}~\bibnamefont {Harl}},
  \bibinfo {author} {\bibfnamefont {T.}~\bibnamefont {Gould}}, \bibinfo
  {author} {\bibfnamefont {J.~G.}\ \bibnamefont {\'Angy\'an}}, \bibinfo
  {author} {\bibfnamefont {G.}~\bibnamefont {Kresse}},\ and\ \bibinfo {author}
  {\bibfnamefont {J.~F.}\ \bibnamefont {Dobson}},\ }\bibfield  {title}
  {\bibinfo {title} {Cohesive properties and asymptotics of the dispersion
  interaction in graphite by the random phase approximation},\ }\href@noop {}
  {\bibfield  {journal} {\bibinfo  {journal} {Phys. Rev. Lett.}\ }\textbf
  {\bibinfo {volume} {105}},\ \bibinfo {pages} {196401} (\bibinfo {year}
  {2010})}\BibitemShut {NoStop}%
\bibitem [{\citenamefont {Olsen}\ and\ \citenamefont
  {Thygesen}(2013)}]{OlTh2013}%
  \BibitemOpen
  \bibfield  {author} {\bibinfo {author} {\bibfnamefont {T.}~\bibnamefont
  {Olsen}}\ and\ \bibinfo {author} {\bibfnamefont {K.~S.}\ \bibnamefont
  {Thygesen}},\ }\bibfield  {title} {\bibinfo {title} {{Random phase
  approximation applied to solids, molecules, and graphene-metal interfaces:
  From van der Waals to covalent bonding}},\ }\href@noop {} {\bibfield
  {journal} {\bibinfo  {journal} {Phys. Rev. B}\ }\textbf {\bibinfo {volume}
  {87}},\ \bibinfo {pages} {075111} (\bibinfo {year} {2013})}\BibitemShut
  {NoStop}%
\bibitem [{\citenamefont {Zhou}\ \emph {et~al.}(2015)\citenamefont {Zhou},
  \citenamefont {Han}, \citenamefont {Dai}, \citenamefont {Sun},\ and\
  \citenamefont {Srolovitz}}]{ZhHaDa2015}%
  \BibitemOpen
  \bibfield  {author} {\bibinfo {author} {\bibfnamefont {S.}~\bibnamefont
  {Zhou}}, \bibinfo {author} {\bibfnamefont {J.}~\bibnamefont {Han}}, \bibinfo
  {author} {\bibfnamefont {S.}~\bibnamefont {Dai}}, \bibinfo {author}
  {\bibfnamefont {J.}~\bibnamefont {Sun}},\ and\ \bibinfo {author}
  {\bibfnamefont {D.~J.}\ \bibnamefont {Srolovitz}},\ }\bibfield  {title}
  {\bibinfo {title} {{van der Waals bilayer energetics: Generalized
  stacking-fault energy of graphene, boron nitride, and graphene/boran nitride
  bilayers}},\ }\href@noop {} {\bibfield  {journal} {\bibinfo  {journal} {Phys.
  Rev. B}\ }\textbf {\bibinfo {volume} {92}},\ \bibinfo {pages} {155438}
  (\bibinfo {year} {2015})}\BibitemShut {NoStop}%
\bibitem [{\citenamefont {Wang}\ \emph {et~al.}(2015)\citenamefont {Wang},
  \citenamefont {Dai}, \citenamefont {Li}, \citenamefont {Yang}, \citenamefont
  {Srolovitz},\ and\ \citenamefont {Zheng}}]{GrMeas1}%
  \BibitemOpen
  \bibfield  {author} {\bibinfo {author} {\bibfnamefont {W.}~\bibnamefont
  {Wang}}, \bibinfo {author} {\bibfnamefont {S.}~\bibnamefont {Dai}}, \bibinfo
  {author} {\bibfnamefont {X.}~\bibnamefont {Li}}, \bibinfo {author}
  {\bibfnamefont {J.}~\bibnamefont {Yang}}, \bibinfo {author} {\bibfnamefont
  {D.~J.}\ \bibnamefont {Srolovitz}},\ and\ \bibinfo {author} {\bibfnamefont
  {Q.}~\bibnamefont {Zheng}},\ }\bibfield  {title} {\bibinfo {title}
  {{Measurement of the cleavage energy in graphite}},\ }\href@noop {}
  {\bibfield  {journal} {\bibinfo  {journal} {Nat. Commun.}\ }\textbf {\bibinfo
  {volume} {6}},\ \bibinfo {pages} {7853} (\bibinfo {year} {2015})}\BibitemShut
  {NoStop}%
\bibitem [{\citenamefont {Rasche}\ \emph {et~al.}(2022)\citenamefont {Rasche},
  \citenamefont {Brunner}, \citenamefont {Schramm}, \citenamefont {Ghimire},
  \citenamefont {Nitzsche}, \citenamefont {B{\"u}chner}, \citenamefont
  {Giraud}, \citenamefont {Richter}, ,\ and\ \citenamefont
  {Dufouleur}}]{GrMeas2}%
  \BibitemOpen
  \bibfield  {author} {\bibinfo {author} {\bibfnamefont {B.}~\bibnamefont
  {Rasche}}, \bibinfo {author} {\bibfnamefont {J.}~\bibnamefont {Brunner}},
  \bibinfo {author} {\bibfnamefont {T.}~\bibnamefont {Schramm}}, \bibinfo
  {author} {\bibfnamefont {M.~P.}\ \bibnamefont {Ghimire}}, \bibinfo {author}
  {\bibfnamefont {U.}~\bibnamefont {Nitzsche}}, \bibinfo {author}
  {\bibfnamefont {B.}~\bibnamefont {B{\"u}chner}}, \bibinfo {author}
  {\bibfnamefont {R.}~\bibnamefont {Giraud}}, \bibinfo {author} {\bibfnamefont
  {M.}~\bibnamefont {Richter}}, ,\ and\ \bibinfo {author} {\bibfnamefont
  {J.}~\bibnamefont {Dufouleur}},\ }\bibfield  {title} {\bibinfo {title}
  {{Determination of Cleavage Energy and Efficient Nanostructuring of Layered
  Materials by Atomic Force Microscopy}},\ }\href@noop {} {\bibfield  {journal}
  {\bibinfo  {journal} {Nano Lett.}\ }\textbf {\bibinfo {volume} {22}},\
  \bibinfo {pages} {3550} (\bibinfo {year} {2022})}\BibitemShut {NoStop}%
\bibitem [{\citenamefont {Zhao}\ and\ \citenamefont {Spain}(1989)}]{ZhSp1989}%
  \BibitemOpen
  \bibfield  {author} {\bibinfo {author} {\bibfnamefont {Y.~X.}\ \bibnamefont
  {Zhao}}\ and\ \bibinfo {author} {\bibfnamefont {I.~L.}\ \bibnamefont
  {Spain}},\ }\bibfield  {title} {\bibinfo {title} {{X-ray diffraction data for
  graphite to 20 GPa}},\ }\href@noop {} {\bibfield  {journal} {\bibinfo
  {journal} {Phys. Rev. B}\ }\textbf {\bibinfo {volume} {40}},\ \bibinfo
  {pages} {993} (\bibinfo {year} {1989})}\BibitemShut {NoStop}%
\bibitem [{\citenamefont {Olsson}\ \emph {et~al.}(2017)\citenamefont {Olsson},
  \citenamefont {Schr{\"o}der}, \citenamefont {Hyldgaard}, \citenamefont
  {Kroon}, \citenamefont {Andreasson},\ and\ \citenamefont
  {Bergvall}}]{Olsson17}%
  \BibitemOpen
  \bibfield  {author} {\bibinfo {author} {\bibfnamefont {P.~A.~T.}\
  \bibnamefont {Olsson}}, \bibinfo {author} {\bibfnamefont {E.}~\bibnamefont
  {Schr{\"o}der}}, \bibinfo {author} {\bibfnamefont {P.}~\bibnamefont
  {Hyldgaard}}, \bibinfo {author} {\bibfnamefont {M.}~\bibnamefont {Kroon}},
  \bibinfo {author} {\bibfnamefont {E.}~\bibnamefont {Andreasson}},\ and\
  \bibinfo {author} {\bibfnamefont {E.}~\bibnamefont {Bergvall}},\ }\bibfield
  {title} {\bibinfo {title} {Ab initio and classical atomistic modelling of
  structure and defects in crystalline orthorhombic polythylene: Twin
  boundaries, slip interfaces, and nature of barriers},\ }\href@noop {}
  {\bibfield  {journal} {\bibinfo  {journal} {Polymer}\ }\textbf {\bibinfo
  {volume} {121}},\ \bibinfo {pages} {234} (\bibinfo {year}
  {2017})}\BibitemShut {NoStop}%
\bibitem [{\citenamefont {Olsson}\ \emph {et~al.}(2018)\citenamefont {Olsson},
  , \citenamefont {Hyldgaard}, \citenamefont {Schr{\"o}der}, \citenamefont
  {Jutemar}, \citenamefont {Andreasson},\ and\ \citenamefont
  {Kroon}}]{OlHySc18}%
  \BibitemOpen
  \bibfield  {author} {\bibinfo {author} {\bibfnamefont {P.~A.~T.}\
  \bibnamefont {Olsson}}, , \bibinfo {author} {\bibfnamefont {P.}~\bibnamefont
  {Hyldgaard}}, \bibinfo {author} {\bibfnamefont {E.}~\bibnamefont
  {Schr{\"o}der}}, \bibinfo {author} {\bibfnamefont {E.~P.}\ \bibnamefont
  {Jutemar}}, \bibinfo {author} {\bibfnamefont {E.}~\bibnamefont
  {Andreasson}},\ and\ \bibinfo {author} {\bibfnamefont {M.}~\bibnamefont
  {Kroon}},\ }\bibfield  {title} {\bibinfo {title} {Ab initio investigation of
  martensitic transformation in crystalline polyethylene},\ }\href@noop {}
  {\bibfield  {journal} {\bibinfo  {journal} {Phys. Rev. M}\ }\textbf {\bibinfo
  {volume} {2}},\ \bibinfo {pages} {075602} (\bibinfo {year}
  {2018})}\BibitemShut {NoStop}%
\bibitem [{\citenamefont {Bj{\"o}rkman}\ \emph
  {et~al.}(2012{\natexlab{a}})\citenamefont {Bj{\"o}rkman}, \citenamefont
  {Gulans}, \citenamefont {Krasheninnikov},\ and\ \citenamefont
  {Nieminen}}]{bjorkmannlayered1}%
  \BibitemOpen
  \bibfield  {author} {\bibinfo {author} {\bibfnamefont {T.}~\bibnamefont
  {Bj{\"o}rkman}}, \bibinfo {author} {\bibfnamefont {A.}~\bibnamefont
  {Gulans}}, \bibinfo {author} {\bibfnamefont {A.~V.}\ \bibnamefont
  {Krasheninnikov}},\ and\ \bibinfo {author} {\bibfnamefont {R.~M.}\
  \bibnamefont {Nieminen}},\ }\bibfield  {title} {\bibinfo {title} {van der
  {W}aals bonding in layered compounds from advanced density-functional
  first-principles calculations},\ }\href@noop {} {\bibfield  {journal}
  {\bibinfo  {journal} {Phys. Rev. Lett.}\ }\textbf {\bibinfo {volume} {108}},\
  \bibinfo {pages} {235502} (\bibinfo {year} {2012}{\natexlab{a}})}\BibitemShut
  {NoStop}%
\bibitem [{\citenamefont {Bj{\"o}rkman}\ \emph
  {et~al.}(2012{\natexlab{b}})\citenamefont {Bj{\"o}rkman}, \citenamefont
  {Gulans}, \citenamefont {Krasheninnikov},\ and\ \citenamefont
  {Nieminen}}]{bjorkmannlayered2}%
  \BibitemOpen
  \bibfield  {author} {\bibinfo {author} {\bibfnamefont {T.}~\bibnamefont
  {Bj{\"o}rkman}}, \bibinfo {author} {\bibfnamefont {A.}~\bibnamefont
  {Gulans}}, \bibinfo {author} {\bibfnamefont {A.~V.}\ \bibnamefont
  {Krasheninnikov}},\ and\ \bibinfo {author} {\bibfnamefont {R.~M.}\
  \bibnamefont {Nieminen}},\ }\bibfield  {title} {\bibinfo {title} {Are we van
  der {W}aals ready?},\ }\href@noop {} {\bibfield  {journal} {\bibinfo
  {journal} {J. Phys.: Condens. Matter}\ }\textbf {\bibinfo {volume} {24}},\
  \bibinfo {pages} {424218} (\bibinfo {year} {2012}{\natexlab{b}})}\BibitemShut
  {NoStop}%
\bibitem [{Note4()}]{Note4}%
  \BibitemOpen
  \bibinfo {note} {The AHB21 and BH76 sets also contain the OH$^-$ radical,
  (and the IL16 set contains negatively charged ions) but sitting in benchmark
  sets that we find can use a small unit cell, the convergence problems do not
  manifest themselves even in raw QE calculations. We have still, in out
  present GMTKN55 survey used the electrostatic-environment approach to
  reliably assert also the performance for the AHB21 benchmark set; However,
  finding no relevant correction for AHB21 relative to the default brute-force
  (that is, native or environment-free) QE exploration (within the present
  selection for unit-cell sizes), we focus the discussion of
  self-interaction-error impact on G21EA and WATER27.}\BibitemShut {Stop}%
\bibitem [{\citenamefont {Andreussi}\ \emph {et~al.}(2012)\citenamefont
  {Andreussi}, \citenamefont {Dabo},\ and\ \citenamefont
  {Marzari}}]{NicolaENVchem}%
  \BibitemOpen
  \bibfield  {author} {\bibinfo {author} {\bibfnamefont {O.}~\bibnamefont
  {Andreussi}}, \bibinfo {author} {\bibfnamefont {I.}~\bibnamefont {Dabo}},\
  and\ \bibinfo {author} {\bibfnamefont {N.}~\bibnamefont {Marzari}},\
  }\bibfield  {title} {\bibinfo {title} {Revised self-consistent continuum
  solvation in electronic-structure calculations},\ }\href@noop {} {\bibfield
  {journal} {\bibinfo  {journal} {J. Chem. Phys.}\ }\textbf {\bibinfo {volume}
  {136}},\ \bibinfo {pages} {064102} (\bibinfo {year} {2012})}\BibitemShut
  {NoStop}%
\bibitem [{Note5()}]{Note5}%
  \BibitemOpen
  \bibinfo {note} {The vdW-DF-BEEF performance is not asserted, even if it is
  also now implemented in the QE code suite. The idea of that design is to seek
  system-optimized exchange partners for the nonlocal-correlation description
  but that idea does not lend itself easily to broad testing and a fair
  comparison in the fixed-parameter benchmarking strategy that we
  use.}\BibitemShut {Stop}%
\bibitem [{\citenamefont {Santra}\ and\ \citenamefont {Martin}(2021)}]{SaMa21}%
  \BibitemOpen
  \bibfield  {author} {\bibinfo {author} {\bibfnamefont {G.}~\bibnamefont
  {Santra}}\ and\ \bibinfo {author} {\bibfnamefont {J.~M.~L.}\ \bibnamefont
  {Martin}},\ }\bibfield  {title} {\bibinfo {title} {{What Types of Chemical
  Problems Benefit from Density-Corrected DFT? A Probe Using an Extensive and
  Chemically Diverse Test Suite}},\ }\href@noop {} {\bibfield  {journal}
  {\bibinfo  {journal} {Journal of Chemical Theory and Computation}\ }\textbf
  {\bibinfo {volume} {17}},\ \bibinfo {pages} {1368} (\bibinfo {year}
  {2021})}\BibitemShut {NoStop}%
\bibitem [{\citenamefont {Boese}\ and\ \citenamefont {Martin}(2004)}]{BK04}%
  \BibitemOpen
  \bibfield  {author} {\bibinfo {author} {\bibfnamefont {A.~D.}\ \bibnamefont
  {Boese}}\ and\ \bibinfo {author} {\bibfnamefont {J.~M.~L.}\ \bibnamefont
  {Martin}},\ }\bibfield  {title} {\bibinfo {title} {{Development of density
  functionals for thermochemical kinetics}},\ }\href@noop {} {\bibfield
  {journal} {\bibinfo  {journal} {Journal of Chemical Physics}\ }\textbf
  {\bibinfo {volume} {121}},\ \bibinfo {pages} {3405} (\bibinfo {year}
  {2004})}\BibitemShut {NoStop}%
\bibitem [{\citenamefont {Seidl}\ \emph {et~al.}(2021)\citenamefont {Seidl},
  \citenamefont {Kretz}, \citenamefont {Gehrmann},\ and\ \citenamefont
  {Egger}}]{SeKrGe21}%
  \BibitemOpen
  \bibfield  {author} {\bibinfo {author} {\bibfnamefont {S.~A.}\ \bibnamefont
  {Seidl}}, \bibinfo {author} {\bibfnamefont {B.}~\bibnamefont {Kretz}},
  \bibinfo {author} {\bibfnamefont {C.}~\bibnamefont {Gehrmann}},\ and\
  \bibinfo {author} {\bibfnamefont {D.~A.}\ \bibnamefont {Egger}},\ }\bibfield
  {title} {\bibinfo {title} {Assessing the accuracy of screened range-separated
  hybrids for bulk properties of semiconductors},\ }\href@noop {} {\bibfield
  {journal} {\bibinfo  {journal} {Phys. Rev. M}\ }\textbf {\bibinfo {volume}
  {5}},\ \bibinfo {pages} {034602} (\bibinfo {year} {2021})}\BibitemShut
  {NoStop}%
\bibitem [{\citenamefont {Gerrits}\ \emph {et~al.}(2020)\citenamefont
  {Gerrits}, \citenamefont {Egidius W. F.~Smeets}, \citenamefont {Powell},
  \citenamefont {Doblhoff-Dier},\ and\ \citenamefont {Kroes}}]{SurfChallenge}%
  \BibitemOpen
  \bibfield  {author} {\bibinfo {author} {\bibfnamefont {N.}~\bibnamefont
  {Gerrits}}, \bibinfo {author} {\bibfnamefont {S.~V.}\ \bibnamefont {Egidius
  W. F.~Smeets}}, \bibinfo {author} {\bibfnamefont {A.~D.}\ \bibnamefont
  {Powell}}, \bibinfo {author} {\bibfnamefont {K.}~\bibnamefont
  {Doblhoff-Dier}},\ and\ \bibinfo {author} {\bibfnamefont {G.-J.}\
  \bibnamefont {Kroes}},\ }\bibfield  {title} {\bibinfo {title} {{Density
  Functional Theory for Molecule-Metal Surface Reactions: When Does the
  Generalized Gradient Approximation Get It Right and What to Do If It Does
  Not}},\ }\href@noop {} {\bibfield  {journal} {\bibinfo  {journal} {J. Phys.
  Chem. Lett.}\ }\textbf {\bibinfo {volume} {11}},\ \bibinfo {pages} {10552}
  (\bibinfo {year} {2020})}\BibitemShut {NoStop}%
\bibitem [{\citenamefont {Linder{\"a}lv}\ \emph {et~al.}(2018)\citenamefont
  {Linder{\"a}lv}, \citenamefont {Lindman},\ and\ \citenamefont
  {Erhart}}]{LiLiEr2018}%
  \BibitemOpen
  \bibfield  {author} {\bibinfo {author} {\bibfnamefont {C.}~\bibnamefont
  {Linder{\"a}lv}}, \bibinfo {author} {\bibfnamefont {A.}~\bibnamefont
  {Lindman}},\ and\ \bibinfo {author} {\bibfnamefont {P.}~\bibnamefont
  {Erhart}},\ }\bibfield  {title} {\bibinfo {title} {{A Unifying Perspective on
  Oxygen Vacancies in Wide Band Gap Oxides}},\ }\href@noop {} {\bibfield
  {journal} {\bibinfo  {journal} {J. Phys. Chem. Lett.}\ }\textbf {\bibinfo
  {volume} {9}},\ \bibinfo {pages} {222} (\bibinfo {year} {2018})}\BibitemShut
  {NoStop}%
\bibitem [{\citenamefont {Hashemi}\ \emph {et~al.}(2021)\citenamefont
  {Hashemi}, \citenamefont {Linder{\"a}lv}, \citenamefont {Krascheninnikov},
  \citenamefont {Ala-Nissila}, \citenamefont {Erhart},\ and\ \citenamefont
  {Komsa}}]{HaLiKr2021}%
  \BibitemOpen
  \bibfield  {author} {\bibinfo {author} {\bibfnamefont {H.}~\bibnamefont
  {Hashemi}}, \bibinfo {author} {\bibfnamefont {C.}~\bibnamefont
  {Linder{\"a}lv}}, \bibinfo {author} {\bibfnamefont {A.~V.}\ \bibnamefont
  {Krascheninnikov}}, \bibinfo {author} {\bibfnamefont {T.}~\bibnamefont
  {Ala-Nissila}}, \bibinfo {author} {\bibfnamefont {P.}~\bibnamefont
  {Erhart}},\ and\ \bibinfo {author} {\bibfnamefont {H.-P.}\ \bibnamefont
  {Komsa}},\ }\bibfield  {title} {\bibinfo {title} {{Photoluminescence line
  shapes for color centers in silicon-carbide from density functional theory
  calculations}},\ }\href@noop {} {\bibfield  {journal} {\bibinfo  {journal}
  {Phys. Rev. B}\ }\textbf {\bibinfo {volume} {103}},\ \bibinfo {pages}
  {125203} (\bibinfo {year} {2021})}\BibitemShut {NoStop}%
\bibitem [{\citenamefont {Patra}\ \emph {et~al.}(2020)\citenamefont {Patra},
  \citenamefont {Jana}, \citenamefont {Constantin},\ and\ \citenamefont
  {Samal}}]{Jana20}%
  \BibitemOpen
  \bibfield  {author} {\bibinfo {author} {\bibfnamefont {A.}~\bibnamefont
  {Patra}}, \bibinfo {author} {\bibfnamefont {S.}~\bibnamefont {Jana}},
  \bibinfo {author} {\bibfnamefont {L.~A.}\ \bibnamefont {Constantin}},\ and\
  \bibinfo {author} {\bibfnamefont {P.}~\bibnamefont {Samal}},\ }\bibfield
  {title} {\bibinfo {title} {Efficient yet accurate dispersion-corrected
  semilocal exchange–correlation functionals for non-covalent interactions},\
  }\href@noop {} {\bibfield  {journal} {\bibinfo  {journal} {J. Chem. Phys.}\
  }\textbf {\bibinfo {volume} {153}},\ \bibinfo {pages} {084117} (\bibinfo
  {year} {2020})}\BibitemShut {NoStop}%
\bibitem [{\citenamefont {Y.-R-Jang}\ and\ \citenamefont {Yu}(2012)}]{JaYu12}%
  \BibitemOpen
  \bibfield  {author} {\bibinfo {author} {\bibnamefont {Y.-R-Jang}}\ and\
  \bibinfo {author} {\bibfnamefont {B.~D.}\ \bibnamefont {Yu}},\ }\bibfield
  {title} {\bibinfo {title} {{Hybrid Functional Study of the Structural and
  Electronic Properties of Co and Ni}},\ }\href@noop {} {\bibfield  {journal}
  {\bibinfo  {journal} {J. Phys. Soc. Jpn.}\ }\textbf {\bibinfo {volume}
  {81}},\ \bibinfo {pages} {114715} (\bibinfo {year} {2012})}\BibitemShut
  {NoStop}%
\bibitem [{\citenamefont {Janthon}\ \emph {et~al.}(2014)\citenamefont
  {Janthon}, \citenamefont {Luo}, \citenamefont {Kozlov}, \citenamefont
  {Vines}, \citenamefont {Limtrakul}, \citenamefont {Truhlar},\ and\
  \citenamefont {Illas}}]{JaLuKo14}%
  \BibitemOpen
  \bibfield  {author} {\bibinfo {author} {\bibfnamefont {P.}~\bibnamefont
  {Janthon}}, \bibinfo {author} {\bibfnamefont {S.}~\bibnamefont {Luo}},
  \bibinfo {author} {\bibfnamefont {S.~M.}\ \bibnamefont {Kozlov}}, \bibinfo
  {author} {\bibfnamefont {F.}~\bibnamefont {Vines}}, \bibinfo {author}
  {\bibfnamefont {J.}~\bibnamefont {Limtrakul}}, \bibinfo {author}
  {\bibfnamefont {D.~G.}\ \bibnamefont {Truhlar}},\ and\ \bibinfo {author}
  {\bibfnamefont {F.}~\bibnamefont {Illas}},\ }\bibfield  {title} {\bibinfo
  {title} {{Bulk Properties of Transition Metals: A Challenge for the Design of
  Universal Density Functionals}},\ }\href@noop {} {\bibfield  {journal}
  {\bibinfo  {journal} {Journal of Chemical Theory and Computations}\ }\textbf
  {\bibinfo {volume} {10}},\ \bibinfo {pages} {3832} (\bibinfo {year}
  {2014})}\BibitemShut {NoStop}%
\bibitem [{\citenamefont {Gao}\ \emph {et~al.}(2016)\citenamefont {Gao},
  \citenamefont {Abtew}, \citenamefont {Cai}, \citenamefont {Sun},
  \citenamefont {Zhang},\ and\ \citenamefont {Zhang}}]{GaAbCa16}%
  \BibitemOpen
  \bibfield  {author} {\bibinfo {author} {\bibfnamefont {W.}~\bibnamefont
  {Gao}}, \bibinfo {author} {\bibfnamefont {T.~A.}\ \bibnamefont {Abtew}},
  \bibinfo {author} {\bibfnamefont {T.}~\bibnamefont {Cai}}, \bibinfo {author}
  {\bibfnamefont {Y.-Y.}\ \bibnamefont {Sun}}, \bibinfo {author} {\bibfnamefont
  {S.}~\bibnamefont {Zhang}},\ and\ \bibinfo {author} {\bibfnamefont
  {P.}~\bibnamefont {Zhang}},\ }\bibfield  {title} {\bibinfo {title} {{On the
  applicability of hybrid functionals for predicting fundamental properties of
  metals}},\ }\href@noop {} {\bibfield  {journal} {\bibinfo  {journal} {Solid
  State Communications}\ }\textbf {\bibinfo {volume} {234-235}},\ \bibinfo
  {pages} {10} (\bibinfo {year} {2016})}\BibitemShut {NoStop}%
\bibitem [{\citenamefont {Sharada}\ \emph {et~al.}(2019)\citenamefont
  {Sharada}, \citenamefont {Karlsson}, \citenamefont {Maimaiti}, \citenamefont
  {Voss},\ and\ \citenamefont {Bligaard}}]{ShKaMa19}%
  \BibitemOpen
  \bibfield  {author} {\bibinfo {author} {\bibfnamefont {S.~M.}\ \bibnamefont
  {Sharada}}, \bibinfo {author} {\bibfnamefont {R.~K.~B.}\ \bibnamefont
  {Karlsson}}, \bibinfo {author} {\bibfnamefont {Y.}~\bibnamefont {Maimaiti}},
  \bibinfo {author} {\bibfnamefont {J.}~\bibnamefont {Voss}},\ and\ \bibinfo
  {author} {\bibfnamefont {T.}~\bibnamefont {Bligaard}},\ }\bibfield  {title}
  {\bibinfo {title} {{Adsorption on transition metal surfaces: Transferability
  and accuracy of DFT using the ADs41 dataset}},\ }\href@noop {} {\bibfield
  {journal} {\bibinfo  {journal} {Phys. Rev B}\ }\textbf {\bibinfo {volume}
  {100}},\ \bibinfo {pages} {035439} (\bibinfo {year} {2019})}\BibitemShut
  {NoStop}%
\bibitem [{\citenamefont {Perrichon}\ \emph {et~al.}(2020)\citenamefont
  {Perrichon}, \citenamefont {Granhed}, \citenamefont {Romanelli},
  \citenamefont {Piovano}, \citenamefont {Lindman}, \citenamefont {Hyldgaard},
  \citenamefont {Wahnstr{\"o}m},\ and\ \citenamefont {Karlsson}}]{PeGrRo20}%
  \BibitemOpen
  \bibfield  {author} {\bibinfo {author} {\bibfnamefont {A.}~\bibnamefont
  {Perrichon}}, \bibinfo {author} {\bibfnamefont {E.~J.}\ \bibnamefont
  {Granhed}}, \bibinfo {author} {\bibfnamefont {G.}~\bibnamefont {Romanelli}},
  \bibinfo {author} {\bibfnamefont {A.}~\bibnamefont {Piovano}}, \bibinfo
  {author} {\bibfnamefont {A.}~\bibnamefont {Lindman}}, \bibinfo {author}
  {\bibfnamefont {P.}~\bibnamefont {Hyldgaard}}, \bibinfo {author}
  {\bibfnamefont {G.}~\bibnamefont {Wahnstr{\"o}m}},\ and\ \bibinfo {author}
  {\bibfnamefont {M.}~\bibnamefont {Karlsson}},\ }\bibfield  {title} {\bibinfo
  {title} {{Unraveling the Ground-State Structure of BaZrO$_3$ by Neutron
  Scattering Experiments and First-Principles Calculations}},\ }\href@noop {}
  {\bibfield  {journal} {\bibinfo  {journal} {Chem. Mater.}\ }\textbf {\bibinfo
  {volume} {32}},\ \bibinfo {pages} {2824} (\bibinfo {year}
  {2020})}\BibitemShut {NoStop}%
\bibitem [{\citenamefont {Momma}\ and\ \citenamefont {Izumi}(2011)}]{VESTA3}%
  \BibitemOpen
  \bibfield  {author} {\bibinfo {author} {\bibfnamefont {K.}~\bibnamefont
  {Momma}}\ and\ \bibinfo {author} {\bibfnamefont {F.}~\bibnamefont {Izumi}},\
  }\bibfield  {title} {\bibinfo {title} {{VESTA 3 for three-dimensional
  visualization of crystal, volumetric and morphology data}},\ }\href@noop {}
  {\bibfield  {journal} {\bibinfo  {journal} {J. Appl. Cryst.}\ }\textbf
  {\bibinfo {volume} {44}},\ \bibinfo {pages} {1272} (\bibinfo {year}
  {2011})}\BibitemShut {NoStop}%
\bibitem [{\citenamefont {Myshlyavtsev}\ and\ \citenamefont
  {Stishenko}(2015)}]{MYSHLYAVTSEV201551}%
  \BibitemOpen
  \bibfield  {author} {\bibinfo {author} {\bibfnamefont {A.~V.}\ \bibnamefont
  {Myshlyavtsev}}\ and\ \bibinfo {author} {\bibfnamefont {P.~V.}\ \bibnamefont
  {Stishenko}},\ }\bibfield  {title} {\bibinfo {title} {{Potential of lateral
  interactions of {CO} on {P}t (111) fitted to recent STM images}},\
  }\href@noop {} {\bibfield  {journal} {\bibinfo  {journal} {Surface Science}\
  }\textbf {\bibinfo {volume} {642}},\ \bibinfo {pages} {51} (\bibinfo {year}
  {2015})}\BibitemShut {NoStop}%
\bibitem [{\citenamefont {Schimka}\ \emph {et~al.}(2010)\citenamefont
  {Schimka}, \citenamefont {Harl}, \citenamefont {Stroppa}, \citenamefont
  {Gr\"{u}neis}, \citenamefont {Marsman}, \citenamefont {Mittendorfer},\ and\
  \citenamefont {Kresse}}]{rpa:ads}%
  \BibitemOpen
  \bibfield  {author} {\bibinfo {author} {\bibfnamefont {L.}~\bibnamefont
  {Schimka}}, \bibinfo {author} {\bibfnamefont {J.}~\bibnamefont {Harl}},
  \bibinfo {author} {\bibfnamefont {A.}~\bibnamefont {Stroppa}}, \bibinfo
  {author} {\bibfnamefont {A.}~\bibnamefont {Gr\"{u}neis}}, \bibinfo {author}
  {\bibfnamefont {M.}~\bibnamefont {Marsman}}, \bibinfo {author} {\bibfnamefont
  {F.}~\bibnamefont {Mittendorfer}},\ and\ \bibinfo {author} {\bibfnamefont
  {G.}~\bibnamefont {Kresse}},\ }\bibfield  {title} {\bibinfo {title} {Accurate
  surface and adsorption energies from many-body perturbation theory},\
  }\href@noop {} {\bibfield  {journal} {\bibinfo  {journal} {Nat. Mater.}\
  }\textbf {\bibinfo {volume} {9}},\ \bibinfo {pages} {741} (\bibinfo {year}
  {2010})}\BibitemShut {NoStop}%
\bibitem [{\citenamefont {Grinberg}\ \emph {et~al.}(2002)\citenamefont
  {Grinberg}, \citenamefont {Yourdshahyan},\ and\ \citenamefont
  {Rappe}}]{grinberg2002co}%
  \BibitemOpen
  \bibfield  {author} {\bibinfo {author} {\bibfnamefont {I.}~\bibnamefont
  {Grinberg}}, \bibinfo {author} {\bibfnamefont {Y.}~\bibnamefont
  {Yourdshahyan}},\ and\ \bibinfo {author} {\bibfnamefont {A.~M.}\ \bibnamefont
  {Rappe}},\ }\bibfield  {title} {\bibinfo {title} {{CO on Pt (111) puzzle: A
  possible solution}},\ }\href@noop {} {\bibfield  {journal} {\bibinfo
  {journal} {J. Chem. Phys.}\ }\textbf {\bibinfo {volume} {117}},\ \bibinfo
  {pages} {2264} (\bibinfo {year} {2002})}\BibitemShut {NoStop}%
\bibitem [{\citenamefont {Olsen}\ \emph {et~al.}(2003)\citenamefont {Olsen},
  \citenamefont {Philipsen},\ and\ \citenamefont {Baerends}}]{olsen2003co}%
  \BibitemOpen
  \bibfield  {author} {\bibinfo {author} {\bibfnamefont {R.}~\bibnamefont
  {Olsen}}, \bibinfo {author} {\bibfnamefont {P.}~\bibnamefont {Philipsen}},\
  and\ \bibinfo {author} {\bibfnamefont {E.}~\bibnamefont {Baerends}},\
  }\bibfield  {title} {\bibinfo {title} {{CO} on {P}t (111): A puzzle
  revisited},\ }\href@noop {} {\bibfield  {journal} {\bibinfo  {journal} {J.
  Chem. Phys.}\ }\textbf {\bibinfo {volume} {119}},\ \bibinfo {pages} {4522}
  (\bibinfo {year} {2003})}\BibitemShut {NoStop}%
\bibitem [{\citenamefont {Stroppa}\ \emph {et~al.}(2007)\citenamefont
  {Stroppa}, \citenamefont {Termentzidis}, \citenamefont {Paier}, \citenamefont
  {Kresse},\ and\ \citenamefont {Hafner}}]{stroppa2007co}%
  \BibitemOpen
  \bibfield  {author} {\bibinfo {author} {\bibfnamefont {A.}~\bibnamefont
  {Stroppa}}, \bibinfo {author} {\bibfnamefont {K.}~\bibnamefont
  {Termentzidis}}, \bibinfo {author} {\bibfnamefont {J.}~\bibnamefont {Paier}},
  \bibinfo {author} {\bibfnamefont {G.}~\bibnamefont {Kresse}},\ and\ \bibinfo
  {author} {\bibfnamefont {J.}~\bibnamefont {Hafner}},\ }\bibfield  {title}
  {\bibinfo {title} {{CO adsorption on metal surfaces: A hybrid functional
  study with plane-wave basis set}},\ }\href@noop {} {\bibfield  {journal}
  {\bibinfo  {journal} {Phys. Rev. B}\ }\textbf {\bibinfo {volume} {76}},\
  \bibinfo {pages} {195440} (\bibinfo {year} {2007})}\BibitemShut {NoStop}%
\bibitem [{\citenamefont {Alaei}\ \emph {et~al.}(2008)\citenamefont {Alaei},
  \citenamefont {Akbarzadeh}, \citenamefont {Gholizadeh},\ and\ \citenamefont
  {de~Gironcoli}}]{alaei2008c}%
  \BibitemOpen
  \bibfield  {author} {\bibinfo {author} {\bibfnamefont {M.}~\bibnamefont
  {Alaei}}, \bibinfo {author} {\bibfnamefont {H.}~\bibnamefont {Akbarzadeh}},
  \bibinfo {author} {\bibfnamefont {H.}~\bibnamefont {Gholizadeh}},\ and\
  \bibinfo {author} {\bibfnamefont {S.}~\bibnamefont {de~Gironcoli}},\
  }\bibfield  {title} {\bibinfo {title} {{CO}/{P}t (111): {GGA} density
  functional study of site preference for adsorption},\ }\href@noop {}
  {\bibfield  {journal} {\bibinfo  {journal} {Phys. Rev. B}\ }\textbf {\bibinfo
  {volume} {77}},\ \bibinfo {pages} {085414} (\bibinfo {year}
  {2008})}\BibitemShut {NoStop}%
\bibitem [{\citenamefont {Stroppa}\ and\ \citenamefont
  {Kresse}(2008)}]{KresseNJP}%
  \BibitemOpen
  \bibfield  {author} {\bibinfo {author} {\bibfnamefont {A.}~\bibnamefont
  {Stroppa}}\ and\ \bibinfo {author} {\bibfnamefont {G.}~\bibnamefont
  {Kresse}},\ }\bibfield  {title} {\bibinfo {title} {The shortcomings of
  semi-local and hybrid functionals: what we can learn from surface science
  studies},\ }\href@noop {} {\bibfield  {journal} {\bibinfo  {journal} {New J.
  Phys.}\ }\textbf {\bibinfo {volume} {10}},\ \bibinfo {pages} {063020}
  (\bibinfo {year} {2008})}\BibitemShut {NoStop}%
\bibitem [{\citenamefont {Lazi{\'c}}\ \emph {et~al.}(2010)\citenamefont
  {Lazi{\'c}}, \citenamefont {Alaei}, \citenamefont {Atodiresei}, \citenamefont
  {Caciuc}, \citenamefont {Brako},\ and\ \citenamefont
  {Bl{\"u}gel}}]{lazic10p045401}%
  \BibitemOpen
  \bibfield  {author} {\bibinfo {author} {\bibfnamefont {P.}~\bibnamefont
  {Lazi{\'c}}}, \bibinfo {author} {\bibfnamefont {M.}~\bibnamefont {Alaei}},
  \bibinfo {author} {\bibfnamefont {N.}~\bibnamefont {Atodiresei}}, \bibinfo
  {author} {\bibfnamefont {V.}~\bibnamefont {Caciuc}}, \bibinfo {author}
  {\bibfnamefont {R.}~\bibnamefont {Brako}},\ and\ \bibinfo {author}
  {\bibfnamefont {S.}~\bibnamefont {Bl{\"u}gel}},\ }\bibfield  {title}
  {\bibinfo {title} {Density functional theory with nonlocal correlation: {A}
  key to the solution of the {CO} adsorption puzzle},\ }\href@noop {}
  {\bibfield  {journal} {\bibinfo  {journal} {Phys. Rev. B}\ }\textbf {\bibinfo
  {volume} {81}},\ \bibinfo {pages} {045401} (\bibinfo {year}
  {2010})}\BibitemShut {NoStop}%
\bibitem [{\citenamefont {Janthon}\ \emph {et~al.}(2017)\citenamefont
  {Janthon}, \citenamefont {Vines}, \citenamefont {Sirijaraensre},
  \citenamefont {Limtrakul},\ and\ \citenamefont {Illas}}]{janthon2017adding}%
  \BibitemOpen
  \bibfield  {author} {\bibinfo {author} {\bibfnamefont {P.}~\bibnamefont
  {Janthon}}, \bibinfo {author} {\bibfnamefont {F.}~\bibnamefont {Vines}},
  \bibinfo {author} {\bibfnamefont {J.}~\bibnamefont {Sirijaraensre}}, \bibinfo
  {author} {\bibfnamefont {J.}~\bibnamefont {Limtrakul}},\ and\ \bibinfo
  {author} {\bibfnamefont {F.}~\bibnamefont {Illas}},\ }\bibfield  {title}
  {\bibinfo {title} {Adding pieces to the {CO}/{P}t (111) puzzle: the role of
  dispersion},\ }\href@noop {} {\bibfield  {journal} {\bibinfo  {journal} {J.
  Phys. Chem. C}\ }\textbf {\bibinfo {volume} {121}},\ \bibinfo {pages} {3970}
  (\bibinfo {year} {2017})}\BibitemShut {NoStop}%
\bibitem [{\citenamefont {Hammer}\ and\ \citenamefont
  {N{\o}rskov}(1995{\natexlab{a}})}]{HaNo95}%
  \BibitemOpen
  \bibfield  {author} {\bibinfo {author} {\bibfnamefont {B.}~\bibnamefont
  {Hammer}}\ and\ \bibinfo {author} {\bibfnamefont {J.~K.}\ \bibnamefont
  {N{\o}rskov}},\ }\bibfield  {title} {\bibinfo {title} {Electronic factors
  determining the reactivity of metal surfaces},\ }\href@noop {} {\bibfield
  {journal} {\bibinfo  {journal} {Surface Science}\ }\textbf {\bibinfo {volume}
  {343}},\ \bibinfo {pages} {211} (\bibinfo {year}
  {1995}{\natexlab{a}})}\BibitemShut {NoStop}%
\bibitem [{\citenamefont {Hammer}\ and\ \citenamefont
  {N{\o}rskov}(1995{\natexlab{b}})}]{Noblest}%
  \BibitemOpen
  \bibfield  {author} {\bibinfo {author} {\bibfnamefont {B.}~\bibnamefont
  {Hammer}}\ and\ \bibinfo {author} {\bibfnamefont {J.~K.}\ \bibnamefont
  {N{\o}rskov}},\ }\bibfield  {title} {\bibinfo {title} {Why gold is the
  noblest of all the metals},\ }\href@noop {} {\bibfield  {journal} {\bibinfo
  {journal} {Nature}\ }\textbf {\bibinfo {volume} {376}},\ \bibinfo {pages}
  {238} (\bibinfo {year} {1995}{\natexlab{b}})}\BibitemShut {NoStop}%
\bibitem [{\citenamefont {Blyholder}(1964)}]{Blyholder}%
  \BibitemOpen
  \bibfield  {author} {\bibinfo {author} {\bibfnamefont {G.~J.}\ \bibnamefont
  {Blyholder}},\ }\bibfield  {title} {\bibinfo {title} {Performance of a
  non-local {van der Waals} density functional on the dissociation of {H}$_2$
  on metal surfaces},\ }\href@noop {} {\bibfield  {journal} {\bibinfo
  {journal} {J. Phys. Chem.}\ }\textbf {\bibinfo {volume} {68}},\ \bibinfo
  {pages} {2772} (\bibinfo {year} {1964})}\BibitemShut {NoStop}%
\bibitem [{Note6()}]{Note6}%
  \BibitemOpen
  \bibinfo {note} {The functional choice also has an indirect effect on the
  adsorption description and the competition between sites. This is because the
  XC functional choice sets the lattice constant, which, in turn, adjusts the
  electronic structure.}\BibitemShut {Stop}%
\bibitem [{\citenamefont {Togo}\ and\ \citenamefont {Tanaka}(2015)}]{Togo2015}%
  \BibitemOpen
  \bibfield  {author} {\bibinfo {author} {\bibfnamefont {A.}~\bibnamefont
  {Togo}}\ and\ \bibinfo {author} {\bibfnamefont {I.}~\bibnamefont {Tanaka}},\
  }\bibfield  {title} {\bibinfo {title} {{First principles phonon calculations
  in materials science}},\ }\href
  {https://doi.org/10.1016/j.scriptamat.2015.07.021} {\bibfield  {journal}
  {\bibinfo  {journal} {Scr. Mater.}\ }\textbf {\bibinfo {volume} {108}},\
  \bibinfo {pages} {1} (\bibinfo {year} {2015})}\BibitemShut {NoStop}%
\bibitem [{\citenamefont {Kruse}\ \emph {et~al.}(2019)\citenamefont {Kruse},
  \citenamefont {Banas},\ and\ \citenamefont {Sponer}}]{KrBaSp2019}%
  \BibitemOpen
  \bibfield  {author} {\bibinfo {author} {\bibfnamefont {H.}~\bibnamefont
  {Kruse}}, \bibinfo {author} {\bibfnamefont {P.}~\bibnamefont {Banas}},\ and\
  \bibinfo {author} {\bibfnamefont {J.}~\bibnamefont {Sponer}},\ }\bibfield
  {title} {\bibinfo {title} {{Investigations of Stacked DNA Base-Pair Steps:
  Highly Accurate Stacking Interaction Energies, Energy Decomposition, and
  Many-Body Stacking Effects}},\ }\href@noop {} {\bibfield  {journal} {\bibinfo
   {journal} {J. Chem. Theory Comput.}\ }\textbf {\bibinfo {volume} {15}},\
  \bibinfo {pages} {95} (\bibinfo {year} {2019})}\BibitemShut {NoStop}%
\bibitem [{\citenamefont {Riplinger}\ \emph {et~al.}(2016)\citenamefont
  {Riplinger}, \citenamefont {Pinski}, \citenamefont {Becker}, \citenamefont
  {Valeev},\ and\ \citenamefont {Neese}}]{RiPiBe2016}%
  \BibitemOpen
  \bibfield  {author} {\bibinfo {author} {\bibfnamefont {C.}~\bibnamefont
  {Riplinger}}, \bibinfo {author} {\bibfnamefont {P.}~\bibnamefont {Pinski}},
  \bibinfo {author} {\bibfnamefont {U.}~\bibnamefont {Becker}}, \bibinfo
  {author} {\bibfnamefont {E.~F.}\ \bibnamefont {Valeev}},\ and\ \bibinfo
  {author} {\bibfnamefont {F.}~\bibnamefont {Neese}},\ }\bibfield  {title}
  {\bibinfo {title} {{Sparse Maps -- A systematic Infrastructure for
  Reduced-Scaling Electronic Structure Methods. II. Linear-Scaling Domain Based
  Pair Natural Orbital Coupled Cluster Theory}},\ }\href@noop {} {\bibfield
  {journal} {\bibinfo  {journal} {J. Chem. Phys.}\ }\textbf {\bibinfo {volume}
  {144}},\ \bibinfo {pages} {024109} (\bibinfo {year} {2016})}\BibitemShut
  {NoStop}%
\bibitem [{\citenamefont {Kim}\ \emph {et~al.}(2020{\natexlab{b}})\citenamefont
  {Kim}, \citenamefont {Gould}, \citenamefont {Rocca},\ and\ \citenamefont
  {Lebeque}}]{KiGoRo2020}%
  \BibitemOpen
  \bibfield  {author} {\bibinfo {author} {\bibfnamefont {M.}~\bibnamefont
  {Kim}}, \bibinfo {author} {\bibfnamefont {T.}~\bibnamefont {Gould}}, \bibinfo
  {author} {\bibfnamefont {D.}~\bibnamefont {Rocca}},\ and\ \bibinfo {author}
  {\bibfnamefont {S.}~\bibnamefont {Lebeque}},\ }\bibfield  {title} {\bibinfo
  {title} {{Establishing the accuracy of density functional approaches for the
  description of noncovalent interactions in biomolecules}},\ }\href@noop {}
  {\bibfield  {journal} {\bibinfo  {journal} {Phys. Chem. Chem. Phys.}\
  }\textbf {\bibinfo {volume} {22}},\ \bibinfo {pages} {21685} (\bibinfo {year}
  {2020}{\natexlab{b}})}\BibitemShut {NoStop}%
\bibitem [{\citenamefont {Toyoda}\ \emph
  {et~al.}(2009{\natexlab{a}})\citenamefont {Toyoda}, \citenamefont {Nakano},
  \citenamefont {Hamada}, \citenamefont {Lee}, \citenamefont {Yanagisawa},\
  and\ \citenamefont {Morikawa}}]{toyoda09p2912}%
  \BibitemOpen
  \bibfield  {author} {\bibinfo {author} {\bibfnamefont {K.}~\bibnamefont
  {Toyoda}}, \bibinfo {author} {\bibfnamefont {Y.}~\bibnamefont {Nakano}},
  \bibinfo {author} {\bibfnamefont {I.}~\bibnamefont {Hamada}}, \bibinfo
  {author} {\bibfnamefont {K.}~\bibnamefont {Lee}}, \bibinfo {author}
  {\bibfnamefont {S.}~\bibnamefont {Yanagisawa}},\ and\ \bibinfo {author}
  {\bibfnamefont {Y.}~\bibnamefont {Morikawa}},\ }\bibfield  {title} {\bibinfo
  {title} {First-principles study of benzene on noble metal surfaces:
  {A}dsorption states and vacuum level shifts},\ }\href@noop {} {\bibfield
  {journal} {\bibinfo  {journal} {Surf. Sci.}\ }\textbf {\bibinfo {volume}
  {603}},\ \bibinfo {pages} {2912} (\bibinfo {year}
  {2009}{\natexlab{a}})}\BibitemShut {NoStop}%
\bibitem [{\citenamefont {Toyoda}\ \emph
  {et~al.}(2009{\natexlab{b}})\citenamefont {Toyoda}, \citenamefont {Nakano},
  \citenamefont {Hamada}, \citenamefont {Lee}, \citenamefont {Yanagisawa},\
  and\ \citenamefont {Morikawa}}]{toyoda09p78}%
  \BibitemOpen
  \bibfield  {author} {\bibinfo {author} {\bibfnamefont {K.}~\bibnamefont
  {Toyoda}}, \bibinfo {author} {\bibfnamefont {Y.}~\bibnamefont {Nakano}},
  \bibinfo {author} {\bibfnamefont {I.}~\bibnamefont {Hamada}}, \bibinfo
  {author} {\bibfnamefont {K.}~\bibnamefont {Lee}}, \bibinfo {author}
  {\bibfnamefont {S.}~\bibnamefont {Yanagisawa}},\ and\ \bibinfo {author}
  {\bibfnamefont {Y.}~\bibnamefont {Morikawa}},\ }\bibfield  {title} {\bibinfo
  {title} {First-principles study of the pentacene/{C}u(111) interface:
  {A}dsorption states and vacuum level shifts},\ }\href@noop {} {\bibfield
  {journal} {\bibinfo  {journal} {J. Electron Spectrosc. Relat. Phenom.}\
  }\textbf {\bibinfo {volume} {174}},\ \bibinfo {pages} {78} (\bibinfo {year}
  {2009}{\natexlab{b}})}\BibitemShut {NoStop}%
\bibitem [{\citenamefont {Berland}\ \emph {et~al.}(2011)\citenamefont
  {Berland}, \citenamefont {Chakarova-K{\"a}ck}, \citenamefont {Cooper},
  \citenamefont {Langreth},\ and\ \citenamefont
  {Schr{\"o}der}}]{berland11p135001}%
  \BibitemOpen
  \bibfield  {author} {\bibinfo {author} {\bibfnamefont {K.}~\bibnamefont
  {Berland}}, \bibinfo {author} {\bibfnamefont {S.~D.}\ \bibnamefont
  {Chakarova-K{\"a}ck}}, \bibinfo {author} {\bibfnamefont {V.~R.}\ \bibnamefont
  {Cooper}}, \bibinfo {author} {\bibfnamefont {D.~C.}\ \bibnamefont
  {Langreth}},\ and\ \bibinfo {author} {\bibfnamefont {E.}~\bibnamefont
  {Schr{\"o}der}},\ }\bibfield  {title} {\bibinfo {title} {A van der {W}aals
  density functional study of adenine on graphene: Single-molecular adsorption
  and overlayer binding},\ }\href@noop {} {\bibfield  {journal} {\bibinfo
  {journal} {J. Phys.: Condens. Matter}\ }\textbf {\bibinfo {volume} {23}},\
  \bibinfo {pages} {135001} (\bibinfo {year} {2011})}\BibitemShut {NoStop}%
\bibitem [{\citenamefont {Cafferty}\ \emph {et~al.}(2013)\citenamefont
  {Cafferty}, \citenamefont {Gallego}, \citenamefont {Chen}, \citenamefont
  {Farley}, \citenamefont {Eritja},\ and\ \citenamefont {Hud}}]{Cafferty2013}%
  \BibitemOpen
  \bibfield  {author} {\bibinfo {author} {\bibfnamefont {B.~J.}\ \bibnamefont
  {Cafferty}}, \bibinfo {author} {\bibfnamefont {I.}~\bibnamefont {Gallego}},
  \bibinfo {author} {\bibfnamefont {M.~C.}\ \bibnamefont {Chen}}, \bibinfo
  {author} {\bibfnamefont {K.~I.}\ \bibnamefont {Farley}}, \bibinfo {author}
  {\bibfnamefont {R.}~\bibnamefont {Eritja}},\ and\ \bibinfo {author}
  {\bibfnamefont {N.~V.}\ \bibnamefont {Hud}},\ }\bibfield  {title} {\bibinfo
  {title} {{Efficient Self-Assembly in Water of Long Noncovalent Polymers by
  Nucleobase Analogues}},\ }\href {https://doi.org/10.1021/ja312155v}
  {\bibfield  {journal} {\bibinfo  {journal} {J. Am. Chem. Soc.}\ }\textbf
  {\bibinfo {volume} {135}},\ \bibinfo {pages} {2447} (\bibinfo {year}
  {2013})}\BibitemShut {NoStop}%
\bibitem [{\citenamefont {Smith}\ \emph {et~al.}(2018)\citenamefont {Smith},
  \citenamefont {Fraccia}, \citenamefont {Todisco}, \citenamefont {Zanchetta},
  \citenamefont {Zhu}, \citenamefont {Hayden}, \citenamefont {Bellini},\ and\
  \citenamefont {Clark}}]{Smith2018}%
  \BibitemOpen
  \bibfield  {author} {\bibinfo {author} {\bibfnamefont {G.~P.}\ \bibnamefont
  {Smith}}, \bibinfo {author} {\bibfnamefont {T.~P.}\ \bibnamefont {Fraccia}},
  \bibinfo {author} {\bibfnamefont {M.}~\bibnamefont {Todisco}}, \bibinfo
  {author} {\bibfnamefont {G.}~\bibnamefont {Zanchetta}}, \bibinfo {author}
  {\bibfnamefont {C.}~\bibnamefont {Zhu}}, \bibinfo {author} {\bibfnamefont
  {E.}~\bibnamefont {Hayden}}, \bibinfo {author} {\bibfnamefont
  {T.}~\bibnamefont {Bellini}},\ and\ \bibinfo {author} {\bibfnamefont {N.~A.}\
  \bibnamefont {Clark}},\ }\bibfield  {title} {\bibinfo {title} {{Backbone-Free
  Duplex-Stacked Monomer Nucleic Acids Exhibiting Watson–Crick
  Selectivity}},\ }\href {https://doi.org/10.1073/pnas.1721369115} {\bibfield
  {journal} {\bibinfo  {journal} {Proc. Natl. Acad. Sci U.S.A.}\ }\textbf {\bibinfo {volume} {115}},\
  \bibinfo {pages} {E7658} (\bibinfo {year} {2018})}\BibitemShut {NoStop}%
\bibitem [{\citenamefont {Hamann}(2013)}]{ONCV}%
  \BibitemOpen
  \bibfield  {author} {\bibinfo {author} {\bibfnamefont {D.~R.}\ \bibnamefont
  {Hamann}},\ }\bibfield  {title} {\bibinfo {title} {Optimized norm-conserving
  {Vanderbilt} pseudopotentials},\ }\href@noop {} {\bibfield  {journal}
  {\bibinfo  {journal} {Phys. Rev. B}\ }\textbf {\bibinfo {volume} {88}},\
  \bibinfo {pages} {085117} (\bibinfo {year} {2013})}\BibitemShut {NoStop}%
\bibitem [{\citenamefont {Schlipf}\ and\ \citenamefont {Gygi}(2015)}]{sg15}%
  \BibitemOpen
  \bibfield  {author} {\bibinfo {author} {\bibfnamefont {M.}~\bibnamefont
  {Schlipf}}\ and\ \bibinfo {author} {\bibfnamefont {F.}~\bibnamefont {Gygi}},\
  }\bibfield  {title} {\bibinfo {title} {Optimization algorithm for the
  generation of {ONCV} pseudopotentials},\ }\href@noop {} {\bibfield  {journal}
  {\bibinfo  {journal} {Comput. Phys. Commun.}\ }\textbf {\bibinfo {volume}
  {196}},\ \bibinfo {pages} {36} (\bibinfo {year} {2015})}\BibitemShut
  {NoStop}%
\bibitem [{\citenamefont {Garrity}\ \emph {et~al.}(2014)\citenamefont
  {Garrity}, \citenamefont {Bennett}, \citenamefont {Rabe},\ and\ \citenamefont
  {Vanderbilt}}]{GBRV}%
  \BibitemOpen
  \bibfield  {author} {\bibinfo {author} {\bibfnamefont {K.~F.}\ \bibnamefont
  {Garrity}}, \bibinfo {author} {\bibfnamefont {J.~W.}\ \bibnamefont
  {Bennett}}, \bibinfo {author} {\bibfnamefont {K.~M.}\ \bibnamefont {Rabe}},\
  and\ \bibinfo {author} {\bibfnamefont {D.}~\bibnamefont {Vanderbilt}},\
  }\bibfield  {title} {\bibinfo {title} {Pseudopotentials for high-throughput
  {DFT} calculations},\ }\href@noop {} {\bibfield  {journal} {\bibinfo
  {journal} {Comput. Mater. Sci.}\ }\textbf {\bibinfo {volume} {81}},\ \bibinfo
  {pages} {446} (\bibinfo {year} {2014})}\BibitemShut {NoStop}%
\bibitem [{\citenamefont {Gonze}\ \emph {et~al.}(2005)\citenamefont {Gonze},
  \citenamefont {Rignanese}, \citenamefont {Verstraete}, \citenamefont
  {Beuken}, \citenamefont {Pouillon}, \citenamefont {Caracas}, \citenamefont
  {Jollet}, \citenamefont {Torrent}, \citenamefont {Zerah}, \citenamefont
  {Mikami}, \citenamefont {Ghosez}, \citenamefont {Veithen}, \citenamefont
  {Raty}, \citenamefont {Olevanov}, \citenamefont {Bruneval}, \citenamefont
  {Reining}, \citenamefont {Godby}, \citenamefont {Onida}, \citenamefont
  {Hamann},\ and\ \citenamefont {Allen}}]{abinit05}%
  \BibitemOpen
  \bibfield  {author} {\bibinfo {author} {\bibfnamefont {X.}~\bibnamefont
  {Gonze}}, \bibinfo {author} {\bibfnamefont {M.}~\bibnamefont {Rignanese}},
  \bibinfo {author} {\bibfnamefont {M.}~\bibnamefont {Verstraete}}, \bibinfo
  {author} {\bibfnamefont {J.~M.}\ \bibnamefont {Beuken}}, \bibinfo {author}
  {\bibfnamefont {Y.}~\bibnamefont {Pouillon}}, \bibinfo {author}
  {\bibfnamefont {R.}~\bibnamefont {Caracas}}, \bibinfo {author} {\bibfnamefont
  {F.}~\bibnamefont {Jollet}}, \bibinfo {author} {\bibfnamefont
  {M.}~\bibnamefont {Torrent}}, \bibinfo {author} {\bibfnamefont
  {G.}~\bibnamefont {Zerah}}, \bibinfo {author} {\bibfnamefont
  {M.}~\bibnamefont {Mikami}}, \bibinfo {author} {\bibfnamefont
  {P.}~\bibnamefont {Ghosez}}, \bibinfo {author} {\bibfnamefont
  {M.}~\bibnamefont {Veithen}}, \bibinfo {author} {\bibfnamefont {J.~Y.}\
  \bibnamefont {Raty}}, \bibinfo {author} {\bibfnamefont {V.}~\bibnamefont
  {Olevanov}}, \bibinfo {author} {\bibfnamefont {F.}~\bibnamefont {Bruneval}},
  \bibinfo {author} {\bibfnamefont {L.}~\bibnamefont {Reining}}, \bibinfo
  {author} {\bibfnamefont {R.}~\bibnamefont {Godby}}, \bibinfo {author}
  {\bibfnamefont {G.}~\bibnamefont {Onida}}, \bibinfo {author} {\bibfnamefont
  {D.~R.}\ \bibnamefont {Hamann}},\ and\ \bibinfo {author} {\bibfnamefont
  {D.~C.}\ \bibnamefont {Allen}},\ }\bibfield  {title} {\bibinfo {title} {A
  brief introduction to the abinit software package},\ }\href@noop {}
  {\bibfield  {journal} {\bibinfo  {journal} {Z. Kristallogr.}\ }\textbf
  {\bibinfo {volume} {220}},\ \bibinfo {pages} {558} (\bibinfo {year}
  {2005})}\BibitemShut {NoStop}%
\bibitem [{\citenamefont {Dabo}\ \emph {et~al.}(2008)\citenamefont {Dabo},
  \citenamefont {Kozinsky}, \citenamefont {Singh-Miller},\ and\ \citenamefont
  {Marzari}}]{NicolaENVprb}%
  \BibitemOpen
  \bibfield  {author} {\bibinfo {author} {\bibfnamefont {I.}~\bibnamefont
  {Dabo}}, \bibinfo {author} {\bibfnamefont {B.}~\bibnamefont {Kozinsky}},
  \bibinfo {author} {\bibfnamefont {N.~E.}\ \bibnamefont {Singh-Miller}},\ and\
  \bibinfo {author} {\bibfnamefont {N.}~\bibnamefont {Marzari}},\ }\bibfield
  {title} {\bibinfo {title} {Electrostatics in periodic boundary conditions and
  real-space corrections},\ }\href@noop {} {\bibfield  {journal} {\bibinfo
  {journal} {Phys. Rev B}\ }\textbf {\bibinfo {volume} {77}},\ \bibinfo {pages}
  {115139} (\bibinfo {year} {2008})}\BibitemShut {NoStop}%
\bibitem [{\citenamefont {Makov}\ and\ \citenamefont {Payne}(1995)}]{Makov}%
  \BibitemOpen
  \bibfield  {author} {\bibinfo {author} {\bibfnamefont {G.}~\bibnamefont
  {Makov}}\ and\ \bibinfo {author} {\bibfnamefont {M.~C.}\ \bibnamefont
  {Payne}},\ }\bibfield  {title} {\bibinfo {title} {Periodic boundary
  conditions in ab initio calculations},\ }\href@noop {} {\bibfield  {journal}
  {\bibinfo  {journal} {Phys. Rev. B}\ }\textbf {\bibinfo {volume} {51}},\
  \bibinfo {pages} {4014} (\bibinfo {year} {1995})}\BibitemShut {NoStop}%
\bibitem [{\citenamefont {Tao}\ \emph {et~al.}(2018)\citenamefont {Tao},
  \citenamefont {Jiao}, \citenamefont {Mo}, \citenamefont {Yang}, \citenamefont
  {Zhu}, \citenamefont {Hyldgaard},\ and\ \citenamefont
  {Perdew}}]{NanovdWScale}%
  \BibitemOpen
  \bibfield  {author} {\bibinfo {author} {\bibfnamefont {J.}~\bibnamefont
  {Tao}}, \bibinfo {author} {\bibfnamefont {Y.}~\bibnamefont {Jiao}}, \bibinfo
  {author} {\bibfnamefont {Y.}~\bibnamefont {Mo}}, \bibinfo {author}
  {\bibfnamefont {Z.-H.}\ \bibnamefont {Yang}}, \bibinfo {author}
  {\bibfnamefont {J.-X.}\ \bibnamefont {Zhu}}, \bibinfo {author} {\bibfnamefont
  {P.}~\bibnamefont {Hyldgaard}},\ and\ \bibinfo {author} {\bibfnamefont
  {J.~P.}\ \bibnamefont {Perdew}},\ }\bibfield  {title} {\bibinfo {title}
  {First-principles study of the binding energy between nanostructures and its
  scaling with system size},\ }\href@noop {} {\bibfield  {journal} {\bibinfo
  {journal} {Phys. Rev. B}\ }\textbf {\bibinfo {volume} {97}},\ \bibinfo
  {pages} {155143} (\bibinfo {year} {2018})}\BibitemShut {NoStop}%
\bibitem [{\citenamefont {Wittea}\ \emph {et~al.}(2019)\citenamefont {Wittea},
  \citenamefont {Neaton},\ and\ \citenamefont {Head-Gordon}}]{NeatonLimit}%
  \BibitemOpen
  \bibfield  {author} {\bibinfo {author} {\bibfnamefont {J.}~\bibnamefont
  {Wittea}}, \bibinfo {author} {\bibfnamefont {J.~B.}\ \bibnamefont {Neaton}},\
  and\ \bibinfo {author} {\bibfnamefont {M.}~\bibnamefont {Head-Gordon}},\
  }\bibfield  {title} {\bibinfo {title} {Push it to the limit: comparing
  periodic and local approaches to density functional theory for intermolecular
  interactions},\ }\href@noop {} {\bibfield  {journal} {\bibinfo  {journal}
  {Mol. Phys.}\ }\textbf {\bibinfo {volume} {117}},\ \bibinfo {pages} {1298}
  (\bibinfo {year} {2019})}\BibitemShut {NoStop}%
\bibitem [{\citenamefont {Cartz}\ \emph {et~al.}(1979)\citenamefont {Cartz},
  \citenamefont {Srinivasa}, \citenamefont {Riedner}, \citenamefont
  {Jorgonsen},\ and\ \citenamefont {Workton}}]{Worlton1979}%
  \BibitemOpen
  \bibfield  {author} {\bibinfo {author} {\bibfnamefont {L.}~\bibnamefont
  {Cartz}}, \bibinfo {author} {\bibfnamefont {S.~R.}\ \bibnamefont
  {Srinivasa}}, \bibinfo {author} {\bibfnamefont {R.~J.}\ \bibnamefont
  {Riedner}}, \bibinfo {author} {\bibfnamefont {J.~D.}\ \bibnamefont
  {Jorgonsen}},\ and\ \bibinfo {author} {\bibfnamefont {T.~G.}\ \bibnamefont
  {Workton}},\ }\bibfield  {title} {\bibinfo {title} {Effects of pressure
  onbonding in black phosphorus},\ }\href@noop {} {\bibfield  {journal}
  {\bibinfo  {journal} {J. Chem. Phys.}\ }\textbf {\bibinfo {volume} {71}},\
  \bibinfo {pages} {1718} (\bibinfo {year} {1979})}\BibitemShut {NoStop}%
\bibitem [{\citenamefont {Racioppi}\ \emph {et~al.}(2022)\citenamefont
  {Racioppi}, \citenamefont {Lolur}, \citenamefont {Hyldgaard},\ and\
  \citenamefont {Rahm}}]{ChiDFT22}%
  \BibitemOpen
  \bibfield  {author} {\bibinfo {author} {\bibfnamefont {S.}~\bibnamefont
  {Racioppi}}, \bibinfo {author} {\bibfnamefont {P.}~\bibnamefont {Lolur}},
  \bibinfo {author} {\bibfnamefont {P.}~\bibnamefont {Hyldgaard}},\ and\
  \bibinfo {author} {\bibfnamefont {M.}~\bibnamefont {Rahm}},\ }\bibfield
  {title} {\bibinfo {title} {{ A Density Functional Theory for the Average
  Electron Energy}},\ }\href@noop {} {\bibfield  {journal} {\bibinfo  {journal}
  {Submitted, https://doi.org/10.26434/chemrxiv-2022-22s75-v2}\ } (\bibinfo
  {year} {2022})}\BibitemShut {NoStop}%
\end{thebibliography}
%

\cleardoublepage
\FloatBarrier\clearpage
\pagebreak
\widetext
\begin{center}
\textbf{\large Supplementary Materials for:\\
A second-generation range-separated hybrid van der Waals density functional}
\end{center}

	This supplementary-information (SI) document contain Fig.\ S 1 and Fig.\ S 2  and Tables
	S I though S XXII that substantiate the discussion and presentation in
	the main text of our paper.

\maketitle

\FloatBarrier\clearpage
\widetext

\setcounter{equation}{0}
\setcounter{section}{0}
\setcounter{subsection}{0}
\setcounter{figure}{0}
\setcounter{table}{0}
\setcounter{page}{1}
\makeatletter
\renewcommand{\theequation}{S \arabic{equation}}
\renewcommand{\thetable}{S \Roman{table}}
\renewcommand{\thefigure}{S \arabic{figure}}
\renewcommand{\bibnumfmt}[1]{[#1]}


\begin{table*}[!htbp]
	\caption{XC functionals: abbreviations, code-nature or QE-input specifications, and literature overview. Where relevant, 
	we also list the Fock-exchange mixing that we used in the hybrid benchmarking. All benchmark results that (in the following 
	tables and in the main text discussion) are marked `(OB)' are taken from the orbital-based-DFT assessment summarized in 
	Ref.\ \onlinecite{gmtkn55}. All other benchmarks are provided here, using the stated QE `input$\_$dft' 
	specification (and the benchmarking strategy and setup that are defined and discussed in appendices A and B).
	}
	\label{tab:XCfunctionalNames}
	\begin{ruledtabular}
	\begin{tabular}{lcccl}
		Abbreviation & (DFT-type) XC name & Code/QE input & Mixing & Litterature \\
		\hline
		PBE         &  PBE & pbe &  - &  Refs.\ \onlinecite{pebuer96,pebuwa96} \\
		revPBE+D3   & revPBE+D3 & revpbe/Grimme-D3 &  -  &      Refs.\ \onlinecite{pebuer96,grimme3} \\
		SCAN+D3(OB) & (Orbital-based) SCAN+D3 & (From GMTKN55 paper) &  - &      Refs.\ \onlinecite{SCAN,grimme3,gmtkn55}\\
		HSE+D3      & HSE+D3 & hse/Grimme-D3 &  0.25 &      Refs.\ \onlinecite{HSE03,HSE06,grimme3} \\
		HSE+D3(OB)  & (Orbital-based) HSE+D3 & (From GMTKN55 paper) & 0.25 &      Refs.\ \onlinecite{HSE03,HSE06,grimme3,gmtkn55} \\
		B3LYP+D3(OB)  & (Orbital-based) B3LYP+D3 & (From GMTKN55 paper) & 0.25 &      Refs.\ \onlinecite{BeckeIII,LYP,grimme3,gmtkn55} \\
		\hline
		vdW-DF1     & vdW-DF & vdw-df &     - &   Refs.\ \onlinecite{Dion,thonhauser} \\
		vdW-DF2     & vdW-DF2 & vdw-df2 &    - &   Ref.\ \onlinecite{Dion,lee10p081101}\\
		rVV10       & revised VV10 & rvv10 &   - &      Refs.\ \onlinecite{vv10,Sabatini2013p041108} \\
		\hline
		C09 & vdW-DF-C09 & vdw-df-c09  & - &  Refs.\ \onlinecite{Dion,cooper10p161104} \\
		OB86 & vdW-optB86r & vdw-df-ob86 & - & Refs.\ \onlinecite{Dion,vdwsolids} \\
		DF3-opt1  & vdW-DF3-opt1 & vdw-df3-opt1 &   - &   Refs.\ \onlinecite{Dion,ChBeTh20} \\
		DF3-opt2  & vdW-DF3-opt2 & vdw-df3-opt2 & -  &   Refs.\ \onlinecite{Dion,ChBeTh20} \\
		OBK8 & vdW-optB88 & vdw-df-obk8 & - &  Refs.\ \onlinecite{Dion,thonhauser,optx}\\
		\hline
		CX & vdW-DF-cx & vdw-df-cx   & - &  Refs.\ \onlinecite{Dion,behy14} \\
		B86R & rev-vdW-DF2 & vdw-df2-b86r  & - &  Refs.\ \onlinecite{Dion,lee10p081101,hamada14} \\
		\hline
		CX0 & vdW-DF-cx+0 & vdw-df-cx0   & 0.25 & \onlinecite{Dion,behy14,DFcx02017} \\
		CX0P & Zero-param.\ vdW-DF-cx+0 & vdw-df-cx0p & 0.20 &  Ref.\ \onlinecite{behy14,DFcx02017,JiScHy18b}\\
		DF2-BR0 & rev-vdW-DF2+0& vdw-df2-br0 & 0.25 & Refs.\ \onlinecite{lee10p081101,hamada14,DFcx02017} and this work.\\
		\hline
		AHCX & vdW-DF-ahcx & vdw-df-ahcx & 0.20 & Refs. \onlinecite{Dion,behy14,DefineAHCX} \\
		AHCX$_{25}$ & vdW-DF-ahcx  & vdw-df-ahcx & 0.25 & Refs.\ \onlinecite{Dion,behy14,DefineAHCX} \\
		DF2-AH & vdW-DF2-ah & vdw-df2-ah & 0.20   & Refs. \onlinecite{lee10p081101,DefineAHCX} \\
		AHBR$_{20}$ & vdW-DF2-ahbr & vdw-df2-ahbr & 0.20 & Refs.\ \onlinecite{lee10p081101,hamada14,DefineAHCX} and this work. \\
		AHBR & vdW-DF2-ahbr & vdw-df2-ahbr & 0.25 & Refs.\ \onlinecite{lee10p081101,hamada14,DefineAHCX} and this work. \\
	\end{tabular}
	
	\end{ruledtabular}
\end{table*}

\begin{figure}[!htbp]
\centering
        \includegraphics[width=0.85\columnwidth]{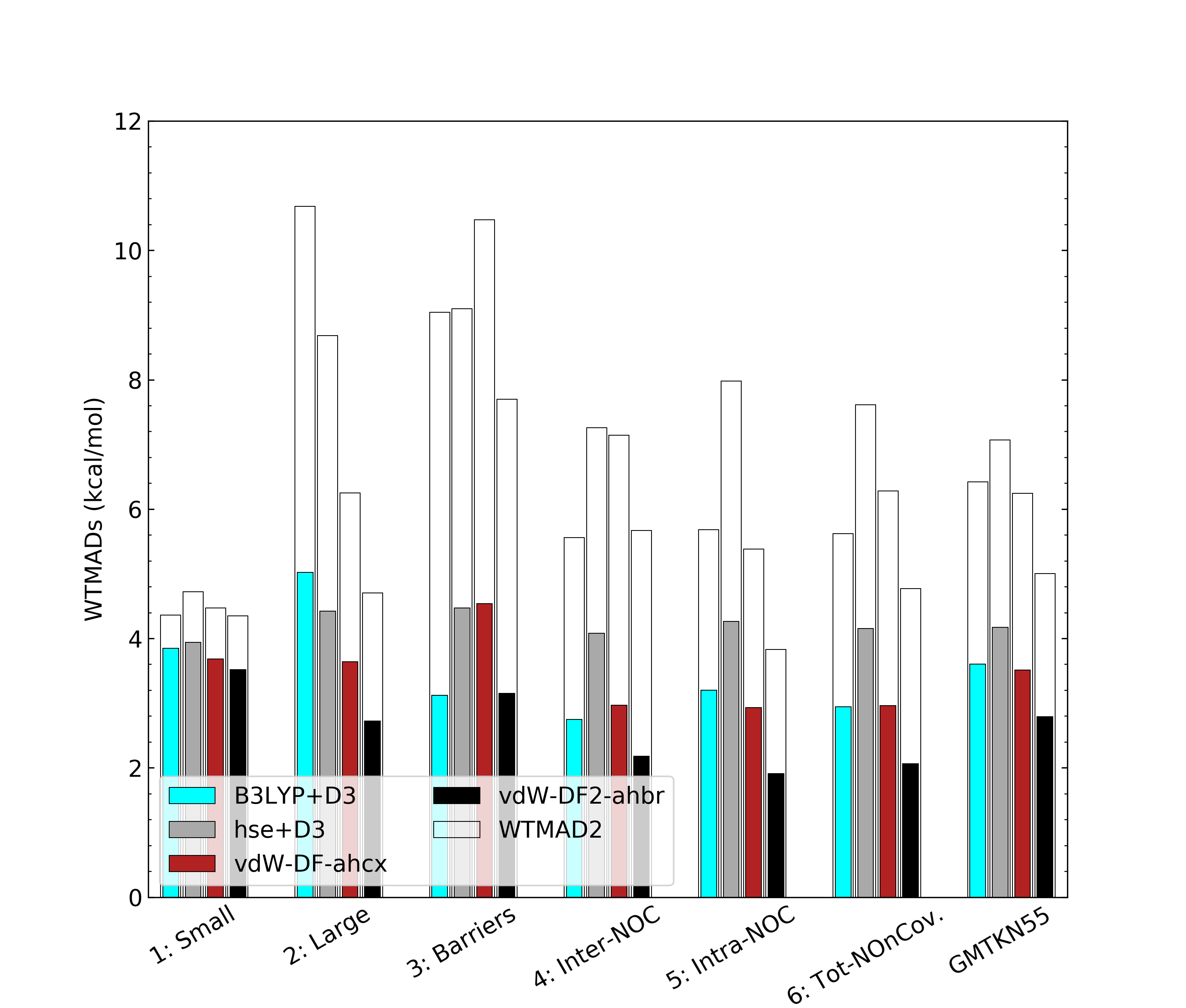}
        \caption{Full GMTKN55 benchmark-suite \cite{gmtkn55} performance comparisons among 
	dispersion-corrected B3LYP+D3, HSE+D3, and RSH vdW-DFs (vdW-DF-ahcx or AHCX and 
	vdW-DF2-ahbr or AHBR). The labels summarize the nature of the 6 benchmark groups, i.e., 
	they identify the type of molecular property that the specific GMTKN55 group primarily assess. 
	Solid (open) bars reflect weighted mean-absolute deviation WTMAD1
	(WTMAD2) measures defined (and suggested ifor molecular benchmarking) 
	in Ref.\ \onlinecite{gmtkn55}, from where we also include 
	the B3LYP+D3 assessment.
\label{fig:GMTKN55compReleasesWTMAD}
}
\end{figure}

\begin{figure}[!hbp]
\centering
\includegraphics[width=0.85\columnwidth]{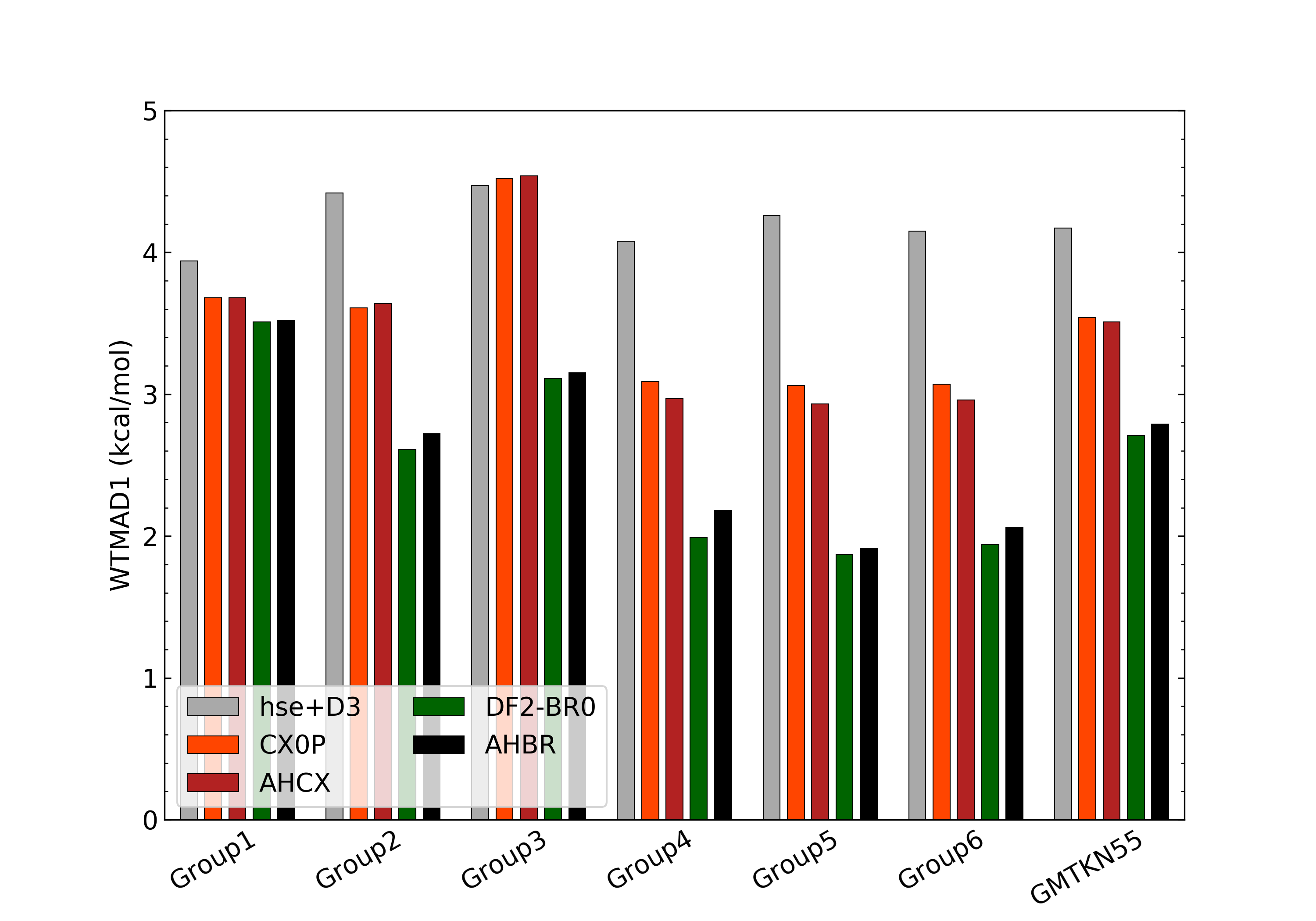}
\caption{Full GMTKN55 benchmark-suite performance comparison among unscreened hybrid 
	vdW-DFs (CX0P, DF2-BR0), RSH vdW-DFs (AHCX and AHBR), and dispersion-corrected HSE.
	See Ref.\ \onlinecite{gmtkn55} (Fig.\ S 1, above) for
        a full description (brief summary) of the 6 GMTKN55 benchmark groups.
\label{fig:GMTKN55compReleases}
}
\end{figure}

\FloatBarrier\clearpage

\subsection{Planewave benchmarking on GMTKN55}

\begin{table*}[!htbp]
	\caption{Comparison of functional performance of regular (density explicit) vdW-DFs and associated unscreened hybrid and RSH vdW-DFs, as asserted by our planewave DFT benchmarking across the full GMTKN55 suite \cite{gmtkn55} on broad molecular properties; The benchmark groups 1-6 are defined and introduced in Ref.\ \onlinecite{gmtkn55}. We report computed values (in kcal/mol)
	for the	weighted-mean-absolute-deviation measure `WTMAD1' that is introduced and discussed
	in Ref.\ \onlinecite{gmtkn55}. For comparison, we also report our performance assessment
	for PBE, for the best-performing \cite{gmtkn55} dispersion-corrected GGA (revPBE+D3), of 
	a meta-GGA (SCAN+D3), and of the PBE-based hybrid (HSE-D3).  We have used
	the electrostatic-environment handling to accurately characterize the performance
	on the G21EA, WATER21, and AHB21 benchmark sets, in spite of the presence of
	negative ions and radicals. Entries marked `OB' are literature orbital-based 
	DFT results \cite{gmtkn55}, that we include to illustrate that there is some
	but also not an excessive dependence of the code nature (and basis-set
	choice) in this broad molecular benchmarking. 
	}
	\label{tab:benchmark_summary}
	\begin{ruledtabular}
	\begin{tabular}{lccccccc}
		XC functional & Group 1 & Group 2 & Group 3 & Group 4 & Group 5 & Group 6 (4\&5) & GMTKN55  \\
		\hline
		PBE                      & 5.27 &      7.37 &      7.09 &     10.29 &     11.90 &     10.98 &     8.03 \\
		revPBE+D3                & 4.91 &      5.55 &      6.30 &      3.41 &      4.27 &      3.78 &     4.76 \\
		SCAN+D3(OB)              & 4.81 &      4.55 &      6.29 &      4.40 &      3.61 &      4.06 &     4.67 \\
		HSE+D3                   & 3.94 &      4.42 &      4.47 &      4.08 &      4.26 &      4.15 &     4.17 \\
		HSE+D3(OB)               & 3.56 &      4.21 &      4.14 &      4.26 &      3.73 &      4.03 &     3.92 \\
		B3LYP+D3(OB)             & 3.85 &      5.02 &      3.12 &      2.75 &      3.20 &      2.94 &     3.60 \\
		\hline
		vdW-DF1                  & 6.99 &      7.83 &      3.48 &      4.31 &      6.04 &      5.05 &     5.94 \\
		vdW-DF2                  & 8.59 &      9.52 &      3.53 &      3.17 &      4.27 &      3.64 &     6.21 \\
		rVV10                    & 5.15 &      6.17 &      6.20 &      4.95 &      5.07 &      5.00 &     5.39 \\
		\hline
		C09 (vdW-DF-c09)         & 5.20 &      5.53 &      8.44 &      4.29 &      6.10 &      5.07 &     5.62 \\
		OB86 (vdW-DF-optB86r)    & 4.74 &      4.65 &      7.18 &      3.81 &      4.63 &      4.16 &     4.81 \\
		DF3-opt1             & 5.31 &      5.22 &      8.90 &      5.40 &      6.13 &      5.72 &     5.91 \\
		DF3-opt2             & 4.87 &      4.66 &      7.67 &      4.69 &      4.60 &      4.65 &     5.11 \\
		OBK8 (vdW-DF-optB88)     & 5.03 &      4.77 &      6.23 &      3.48 &      4.56 &      3.94 &     4.73 \\
		\hline
		CX (vdW-DF-cx)           & 4.95 &      4.99 &      7.56 &      3.18 &      4.04 &      3.55 &     4.75 \\
		B86R (rev-vdW-DF2)       & 4.95 &      4.82 &      6.99 &      3.22 &      3.77 &      3.45 &     4.62 \\
		\hline
		CX0 (vdW-DF-cx0,0.25)    & 3.55 &      3.40 &      3.88 &      3.38 &      3.04 &      3.24 &     3.45 \\
		CX0P (vdW-DF-cx0p,0.20)  & 3.68 &      3.61 &      4.52 &      3.09 &      3.06 &      3.07 &     3.54 \\
		DF2-BR0 (B86R+0,0.25)    & 3.51 &      2.61 &      3.11 &      1.99 &      1.87 &      1.94 &     2.71 \\
		\hline
		AHCX (vdW-DF-ahcx,0.20)  & 3.68 &      3.64 &      4.54 &      2.97 &      2.93 &      2.96 &     3.51 \\
		AHCX$_{25}$ (0.25)         & 3.55 &      3.43 &      3.89 &      3.25 &      2.86 &      3.09 &     3.40 \\
		DF2-AH (vdW-DF2-ah,0.20)   & 6.90 &      7.27 &      3.12 &      2.41 &      3.92 &      3.05 &     5.01 \\
		AHBR$_{20}$ (0.20)         & 3.67 &      3.02 &      3.82 &      2.22 &      2.17 &      2.20 &     3.02 \\
		AHBR (vdW-DF2-ahbr,0.25) & 3.52 &      2.72 &      3.15 &      2.18 &      1.91 &      2.06 &     2.79 \\
	\end{tabular}
	
	\end{ruledtabular}
\end{table*}

\begin{table*}
	\caption{Comparison of functional performance of regular vdW-DFs and associated 
	unscreened hybrid and RSH vdW-DFs across the full GMTKN55 suite \cite{gmtkn55}; 
	Same underlying benchmark data as in Table \ref{tab:benchmark_summary} but values
	(in kcal/mol) are here reported for a different  weighted-mean-absolute-deviation
	measure, WTMAD2, that is also introduced and discussed in Ref.\ \onlinecite{gmtkn55}. 
	\label{tab:benchmark_summary2}
	}
	\begin{tabular}{lccccccc}
		\hline \hline
		XC functional & Group 1 & Group 2 & Group 3 & Group 4 & Group 5 & Group 6 (4\&5) & GMTKN55  \\
		\hline
		PBE                      & 6.23 &     15.72 &     16.36 &     16.61 &     20.42 &     18.48 &    13.91 \\
		revPBE+D3                & 5.82 &     10.30 &     14.88 &      6.36 &      8.08 &      7.20 &     8.26 \\
		SCAN+D3(OB)              & 5.31 &      7.86 &     14.94 &      8.50 &      6.61 &      7.58 &     7.86 \\
		HSE+D3                   & 4.73 &      8.68 &      9.10 &      7.26 &      7.98 &      7.61 &     7.07 \\
		HSE+D3(OB)               & 4.14 &      8.67 &      9.01 &      7.79 &      7.07 &      7.44 &     6.80 \\
		B3LYP+D3(OB)             & 4.36 &     10.28 &      9.04 &      5.56 &      5.68 &      5.62 &     6.42 \\
		\hline
		vdW-DF1                  & 8.11 &     14.25 &     10.54 &     10.38 &     11.17 &     10.77 &    10.47 \\
		vdW-DF2                  &10.05 &     17.24 &     11.38 &      6.71 &      7.07 &      6.88 &    10.13 \\
		rVV10                    & 6.04 &     11.89 &     15.52 &     10.24 &      9.99 &     10.12 &     9.82 \\
		\hline
		C09 (vdW-DF-c09)         & 6.61 &     10.82 &     20.10 &      8.05 &     11.39 &      9.69 &    10.25 \\
		OB86 (vdW-DF-optB86r)    & 5.93 &      8.91 &     17.80 &      7.38 &      8.47 &      7.92 &     8.73 \\
		DF3-opt1             & 6.98 &      9.91 &     21.35 &     10.12 &     12.43 &     11.25 &    11.00 \\
		DF3-opt2             & 6.23 &      9.16 &     19.12 &      9.34 &      9.01 &      9.18 &     9.53 \\
		OBK8 (vdW-DF-optB88)     & 6.03 &      8.72 &     16.14 &      6.22 &      8.26 &      7.22 &     8.24 \\
		\hline
		CX (vdW-DF-cx)           & 6.19 &      9.60 &     18.23 &      7.83 &      7.37 &      7.60 &     8.85 \\
		B86R (rev-vdW-DF2)       & 6.08 &      9.15 &     17.57 &      6.41 &      7.18 &      6.79 &     8.34 \\
		\hline
		CX0 (CX+0,0.25)          & 4.44 &      5.64 &      8.74 &      7.80 &      5.57 &      6.71 &     6.09 \\
		CX0P (vdW-DF-cx0p,0.20)  & 4.46 &      6.24 &     10.40 &      7.38 &      5.58 &      6.50 &     6.32 \\ 
		DF2-BR0 (B86R+0,0.25)    & 4.34 &      4.49 &      7.56 &      4.78 &      3.76 &      4.28 &     4.76 \\
		\hline
		AHCX (vdW-DF-ahcx,0.20)  & 4.47 &      6.25 &     10.47 &      7.14 &      5.38 &      6.28 &     6.24 \\
		AHCX$_{25}$ (0.25)         & 4.45 &      5.64 &      8.81 &      7.68 &      5.27 &      6.50 &     6.01 \\
		DF2-AH (vdW-DF2-ah,0.20)   & 8.55 &     13.39 &      8.02 &      4.79 &      6.52 &      5.63 &     8.11 \\
		AHBR$_{20}$ (0.20)         & 4.39 &      5.32 &      9.46 &      5.04 &      4.26 &      4.65 &     5.30 \\
		AHBR (vdW-DF2-ahbr,0.25) & 4.35 &      4.70 &      7.70 &      5.67 &      3.83 &      4.77 &     5.00 \\
		\hline
	\end{tabular}
\end{table*}

\begin{table*}
	\caption{Comparison of functional performance of regular vdW-DFs and associated 
	unscreened hybrid and RSH vdW-DFs across the full GMTKN55 suite \cite{gmtkn55}; 
	Same underlying benchmark data as in Table \ref{tab:benchmark_summary} but values
	(in kcal/mol) are here reported for so-called total weighted-mean-absolute-deviation
	measure, TMAD, that simply averages MAD values over the number of bencmarks in each group \cite{Jana20,DefineAHCX}.
	For Group 4 we also include the TMAD estimates (marked with an asteriks *)
	that results when ignoring the impact of WATER27 \cite{Jana20}.
	\label{tab:benchmark_summary3}
	}
	\begin{tabular}{lccccccc}
		\hline \hline
		XC functional & Group 1 & Group 2 & Group 3 & Group 4* & Group 4 & Group 5 & Group 6 (4\&5) \\
		\hline
		PBE                      & 5.44 &      6.92 &      6.41 &  1.37 &  1.55 &      2.28 &      1.86 \\
		revPBE+D3                & 5.20 &      5.74 &      5.68 &  0.58 &  0.75 &      0.75 &      0.75 \\
		HSE+D3                   & 4.40 &      3.58 &      3.64 &  0.69 &  1.10 &      0.72 &      0.94 \\
		\hline
		vdW-DF1                  & 8.02 &     11.69 &      2.91 &  0.93 &  1.50 &      1.36 &      1.44 \\
		vdW-DF2                  &10.28 &     14.87 &      2.91 &  0.69 &  0.77 &      1.20 &      0.96 \\
		rVV10                    & 5.34 &      6.37 &      5.55 &  0.74 &  1.61 &      0.90 &      1.31 \\
		\hline
		C09 (vdW-DF-c09)         & 5.72 &      6.17 &      7.81 &  0.66 &  1.27 &      0.94 &      1.13 \\
		OB86 (vdW-DF-optB86r)    & 4.99 &      5.01 &      6.60 &  0.55 &  0.97 &      0.72 &      0.86 \\
		DF3-opt1             & 5.91 &      5.53 &      8.22 &  0.88 &  1.95 &      0.89 &      1.49 \\
		DF3-opt2             & 5.25 &      4.96 &      7.06 &  0.69 &  1.47 &      0.75 &      1.16 \\
		OBK8 (vdW-DF-optB88)     & 5.28 &      5.64 &      5.65 &  0.52 &  0.92 &      0.81 &      0.87 \\
		\hline
		CX (vdW-DF-cx)           & 5.36 &      5.42 &      6.98 &  0.47 &  0.67 &      0.63 &      0.65 \\
		B86R (rev-vdW-DF2)       & 5.24 &      5.34 &      6.41 &  0.46 &  0.84 &      0.65 &      0.76 \\
		\hline
		CX0 (CX+0,0.25)          & 4.18 &      3.16 &      3.17 &  0.50 &  0.71 &      0.50 &      0.62 \\
		CX0P (vdW-DF-cx0p,0.20)  & 4.08 &      3.40 &      3.84 &  0.45 &  0.65 &      0.49 &      0.58 \\
		DF2-BR0 (B86R+0,0.25)    & 4.36 &      3.05 &      2.46 &  0.34 &  0.52 &      0.31 &      0.43 \\
		\hline
		AHCX (vdW-DF-ahcx,0.20)  & 4.09 &      3.43 &      3.85 &  0.44 &  0.62 &      0.45 &      0.55 \\
		AHCX$_{25}$ (0.25)       & 4.17 &      3.18 &      3.16 &  0.49 &  0.68 &      0.46 &      0.58 \\
		DF2-AH (vdW-DF2-ah,0.20) & 9.31 &     11.77 &      2.59 &  0.53 &  0.56 &      1.02 &      0.75  \\
		AHBR$_{20}$ (0.20)       & 4.19 &      3.38 &      3.18 &  0.35 &  0.56 &      0.39 &      0.49 \\
		AHBR (vdW-DF2-ahbr,0.25) & 4.33 &      3.15 &      2.48 &  0.36 &  0.54 &      0.34 &      0.45 \\
		\hline
	\end{tabular}
\end{table*}

\begin{table*}
	\caption{Functional performance of vdW-DFs and associated 
	unscreened hybrid and RSH vdW-DFs for individual benchmark sets of the
	GMTKN55 group 3: Barriers heights \cite{gmtkn55}. The nature and importance of these
	benchmark sets are discussed in the main text. All entries are MAD values in kcal/mol.
	\label{tab:barrier_mad}
	}
	\begin{tabular}{lccccccc}
		\hline \hline
		XC functional & BH76 & BHPERI & BHDIV10 & INV24 & BHROT27 & PX13 & WCPT18  \\
		\hline
		PBE                      & 8.46 & 3.85 & 8.10 & 3.07 & 0.54 &12.12 & 8.71 \\
		PBE (OB)                 & 9.15 & 3.95 & 8.23 & 2.67 & 0.47 &11.54 & 8.61 \\
		revPBE+D3                & 7.38 & 5.74 & 7.47 & 2.20 & 0.48 & 9.29 & 7.22 \\
		revPBE+D3(OB)            & 8.32 & 6.29 & 7.83 & 2.18 & 0.37 & 8.75 & 7.22 \\
		SCAN+D3(OB)              & 7.77 & 5.50 & 6.62 & 1.16 & 0.84 & 8.34 & 6.22 \\
		HSE+D3                   & 4.21 & 2.83 & 4.75 & 1.13 & 0.64 & 7.38 & 4.57 \\
		HSE+D3(OB)               & 4.58 & 2.52 & 4.56 & 1.16 & 0.61 & 6.02 & 4.01 \\
		B3LYP+D3(OB)             & 5.70 & 1.18 & 3.22 & 1.05 & 0.41 & 4.33 & 2.27 \\
		\hline
		vdW-DF1                  & 6.33 & 2.37 & 4.35 & 2.07 & 0.44 & 2.36 & 2.45 \\
		vdW-DF2                  & 6.90 & 3.08 & 4.67 & 2.32 & 0.49 & 1.14 & 1.75 \\
		rVV10                    & 8.26 & 4.57 & 6.82 & 2.48 & 0.51 & 9.23 & 6.96 \\
		\hline
		C09 (vdW-DF-c09)         & 9.86 & 8.68 & 9.06 & 1.95 & 0.49 &14.36 &10.30 \\
		OB86 (vdW-DF-optB86r)    & 9.11 & 6.91 & 7.69 & 1.93 & 0.45 &11.63 & 8.48 \\
		DF3-opt1             &10.60 & 8.83 & 9.46 & 2.10 & 0.53 &15.18 &10.82 \\ 
		DF3-opt2             & 9.83 & 7.39 & 8.11 & 2.01 & 0.47 &12.43 & 9.19 \\
		OBK8 (vdW-DF-optB88)     & 8.60 & 5.64 & 6.56 & 1.91 & 0.46 & 9.58 & 6.79 \\
		\hline
		CX (vdW-DF-cx)           & 9.15 & 7.20 & 8.19 & 1.96 & 0.45 &12.80 & 9.14 \\
		B86R (rev-vdW-DF2)       & 9.22 & 6.08 & 7.46 & 2.06 & 0.45 &11.36 & 8.27 \\
		\hline
		CX0 (CX+0,0.25)          & 4.19 & 3.38 & 3.57 & 0.97 & 0.55 & 6.04 & 3.46 \\
		CX0P (vdW-DF-cx0p,0.20)  & 5.09 & 4.07 & 4.46 & 0.83 & 0.53 & 7.35 & 4.57 \\
		DF2-BR0 (B86R+0,0.25)    & 4.05 & 1.81 & 2.77 & 0.90 & 0.51 & 4.58 & 2.56 \\
		\hline
		AHCX (vdW-DF-ahcx,0.20)  & 5.15 & 4.05 & 4.51 & 0.84 & 0.54 & 7.30 & 4.54 \\
		AHCX$_{25}$ (0.25)         & 4.26 & 3.33 & 3.62 & 0.95 & 0.57 & 5.98 & 3.43 \\
		DF2-AH (vdW-DF2-ah,0.20) & 4.14 & 3.48 & 4.00 & 1.08 & 0.41 & 2.53 & 2.50 \\
		AHBR$_{20}$ (0.20)         & 5.08 & 2.65 & 3.74 & 0.84 & 0.50 & 5.88 & 3.58 \\
		AHBR (vdW-DF2-ahbr,0.25) & 4.14 & 1.84 & 2.84 & 0.89 & 0.53 & 4.56 & 2.54 \\
		\hline
	\end{tabular}
\end{table*}

\begin{table*}
	\caption{Functional performance of regular vdW-DFs and associated 
	unscreened hybrid and RSH vdW-DFs for individual benchmark sets of the
	GMTKN55 group 1: small-system properties \cite{gmtkn55}. This table 
	reports on benchmarks containing charged systems. The calculations of negatively charged ions in
	the G21EA set are dramatically affected by self-interaction errors \cite{BurkeSIE}
	and requires the here-discussed electrostatic environment assessment procedure
	for a meaningful planewave assessment, see appendix and 
	Refs.\ \onlinecite{BurkeSIE,DefineAHCX}. All entries are MAD values in kcal/mol.
	\label{tab:smallcharged_mad}
	}
	\begin{tabular}{lcccccccc}
		\hline \hline
		XC functional & G21EA & G21IP & DIPCS10 & PA26 & SIE4x4 & ALK8 & RC21 & BH76RC  \\
		\hline
		PBE                      & 3.07 & 4.45 & 5.12 & 1.84 &21.61 & 2.67 & 4.49 & 3.37 \\
		PBE (OB)                 & 3.43 & 3.85 & 4.51 & 1.97 &23.44 & 2.78 & 5.48 & 4.09 \\ 
		revPBE+D3                & 2.83 & 4.72 & 4.89 & 4.41 &21.67 & 3.70 & 4.02 & 2.87 \\
		revPBE+D3(OB)            & 2.75 & 4.20 & 4.81 & 4.73 &23.43 & 3.61 & 4.85 & 2.76 \\ 
		SCAN+D3(OB)              & 3.64 & 4.69 & 4.92 & 3.18 &17.99 & 3.45 & 6.69 & 3.38 \\
		HSE+D3                   & 3.40 & 4.31 & 3.19 & 2.66 &13.58 & 4.67 & 4.77 & 1.79 \\
		HSE+D3(OB)               & 2.68 & 3.70 & 3.05 & 2.65 &14.49 & 4.67 & 4.75 & 2.35 \\
		B3LYP+D3(OB)             & 1.91 & 3.55 & 4.73 & 2.87 &18.06 & 2.48 & 2.44 & 2.25 \\
		\hline
		vdW-DF1                  & 5.68 & 5.19 & 9.11 & 5.06 &22.31 & 5.93 & 3.23 & 4.27 \\
		vdW-DF2                  & 9.66 & 8.11 &16.38 & 4.46 &21.73 & 7.60 & 5.55 & 4.97 \\
		rVV10                    & 3.39 & 4.77 & 4.69 & 2.29 &21.03 & 1.93 & 3.33 & 2.85 \\
		\hline
		C09 (vdW-DF-c09)         & 2.79 & 3.96 & 5.99 & 1.64 &23.93 & 2.13 & 7.82 & 3.45 \\
		OB86 (vdW-DF-optB86r)    & 3.36 & 3.82 & 4.83 & 2.11 &23.56 & 2.29 & 5.71 & 3.29 \\
		DF3-opt1             & 3.96 & 3.65 & 4.43 & 1.54 &23.96 & 1.79 & 8.05 & 3.58 \\
		DF3-opt2             & 4.72 & 3.85 & 5.42 & 1.66 &23.66 & 2.53 & 6.10 & 3.36 \\
		OBK8 (vdW-DF-optB88)     & 4.69 & 4.20 & 5.42 & 2.97 &23.00 & 3.02 & 3.92 & 3.32 \\
		\hline
		CX (vdW-DF-cx)           & 2.80 & 3.91 & 5.93 & 1.93 &23.80 & 2.59 & 6.53 & 3.31 \\
		B86R (rev-vdW-DF2)       & 4.50 & 3.82 & 5.30 & 1.83 &23.52 & 3.38 & 4.94 & 3.29 \\
		\hline
		CX0 (CX+0,0.25)          & 2.36 & 3.47 & 3.72 & 2.98 &15.20 & 2.01 & 4.62 & 2.32 \\
		CX0P (vdW-DF-cx0p,0.20)  & 2.10 & 3.45 & 3.68 & 2.68 &16.89 & 2.10 & 4.90 & 2.27 \\
		DF2-BR0 (B86R+0,0.25)    & 2.54 & 3.85 & 5.98 & 2.61 &14.94 & 3.18 & 2.99 & 2.53 \\
		\hline
		AHCX (vdW-DF-ahcx,0.20)  & 2.17 & 3.51 & 3.92 & 2.74 &17.00 & 2.02 & 4.85 & 2.30 \\
		AHCX$_{25}$ (0.25)       & 2.45 & 3.52 & 3.89 & 3.05 &15.34 & 1.88 & 4.54 & 2.34 \\
		DF2-AH (vdW-DF2-ah,0.20) & 6.52 & 7.63 &15.48 & 5.02 &15.26 & 6.39 & 4.80 & 4.97 \\
		AHBR$_{20}$ (0.20)       & 2.48 & 3.66 & 5.04 & 2.32 &16.75 & 3.11 & 3.17 & 2.42 \\
		AHBR (vdW-DF2-ahbr,0.25) & 2.32 & 3.75 & 5.62 & 2.67 &15.09 & 3.05 & 2.94 & 2.55 \\
		\hline
	\end{tabular}
\end{table*}

\begin{table*}
	\caption{Functional performance of regular vdW-DFs and associated 
	unscreened hybrid and RSH vdW-DFs for individual benchmark sets of the
	GMTKN55 group 1: small-system properties \cite{gmtkn55}. This table focuses on
        all-neutral systems. We abbreviate GMTKN55 benchmark ALKBDE10 as `ALKB', YBDE18 as `YBDE',
	HEAVYSB11 as 'HSB', and TAUT15 as 'TAUT'. All entries are MAD values in kcal/mol.
	\label{tab:smallneutral_mad}
	}
	\begin{tabular}{lcccccccccc}
		\hline \hline
		XC functional & W4-11 & ALKB & YBDE & AL2X6 & HSB & NBPRC & G2RC & FH51 & TAUT & DC13 \\
		\hline
		PBE                      &  7.44 & 4.98 & 6.39 & 4.05 & 3.71 & 2.88 & 5.85 & 3.34 & 2.05 & 10.67 \\
		PBE (OB)                 & 14.96 & 6.21 & 5.91 & 4.26 & 4.58 & 2.82 & 6.29 & 3.40 & 1.81 & 10.31 \\
		revPBE+D3                &  5.88 & 5.15 & 5.91 & 1.91 & 3.40 & 2.03 & 5.60 & 3.26 & 1.94 &  9.38 \\
		revPBE+D3(OB)            &  7.57 & 5.16 & 4.41 & 2.07 & 2.72 & 1.98 & 6.16 & 3.34 & 1.55 &  8.87 \\
		SCAN+D3(OB)              &  4.08 &19.27 & 3.12 & 2.13 & 6.64 & 2.51 & 6.39 & 2.75 & 1.74 &  7.29 \\
		HSE+D3                   &  6.77 & 5.78 & 3.35 & 1.22 & 2.36 & 2.68 & 6.48 & 2.63 & 1.36 &  8.24 \\
		HSE+D3(OB)               &  3.56 & 5.57 & 1.41 & 0.93 & 1.14 & 2.57 & 6.21 & 2.47 & 1.19 &  7.09 \\
		B3LYP+D3(OB)             &  3.40 & 4.39 & 4.72 & 2.71 & 3.30 & 2.00 & 2.73 & 2.61 & 1.16 & 10.14 \\
		\hline
		vdW-DF1                  & 12.98 & 4.15 &13.28 & 6.61 & 8.36 & 6.84 & 7.10 & 5.57 & 1.60 & 17.06 \\
		vdW-DF2                  & 18.69 & 4.02 &16.04 & 7.65 & 9.44 & 8.74 & 9.43 & 6.63 & 1.78 & 24.21 \\
		rVV10                    &  6.93 & 4.57 & 8.34 & 2.04 & 4.25 & 2.96 & 4.14 & 3.59 & 1.94 & 13.13 \\
		\hline
		C09 (vdW-DF-c09)         & 11.14 & 7.03 & 4.53 & 1.59 & 2.76 & 2.26 & 7.09 & 3.50 & 1.93 &  9.41 \\
		OB86 (vdW-DF-optB86r)    &  6.57 & 5.72 & 5.44 & 1.00 & 2.53 & 1.57 & 6.16 & 3.32 & 1.80 &  6.87 \\
		DF3-opt1             & 14.36 & 8.38 & 4.76 & 1.58 & 3.23 & 2.22 & 6.93 & 3.50 & 2.03 &  8.51 \\
		DF3-opt2             &  8.86 & 6.53 & 5.31 & 0.99 & 2.55 & 1.28 & 5.91 & 3.29 & 1.86 &  6.65 \\
		OBK8 (vdW-DF-optB88)     &  4.40 & 5.08 & 7.56 & 1.65 & 3.57 & 2.36 & 5.41 & 3.74 & 1.71 &  9.10 \\
		\hline
		CX (vdW-DF-cx)           &  8.55 & 6.29 & 5.15 & 1.47 & 2.80 & 1.61 & 6.77 & 3.35 & 1.83 &  7.88 \\
		B86R (rev-vdW-DF2)       &  6.97 & 5.88 & 5.96 & 2.35 & 2.81 & 1.82 & 5.58 & 3.12 & 1.81 &  7.26 \\
		\hline
		CX0 (CX+0,0.25)          &  7.56 & 5.60 & 3.90 & 0.76 & 2.66 & 1.27 & 4.53 & 2.48 & 1.07 &  8.79 \\
		CX0P (vdW-DF-cx0p,0.20)  &  5.01 & 5.27 & 3.80 & 0.81 & 2.27 & 1.24 & 4.76 & 2.49 & 1.21 &  8.49 \\
		DF2-BR0 (B86R+0,0.25)    &  9.49 & 5.44 & 5.71 & 1.61 & 4.05 & 1.15 & 3.30 & 1.96 & 1.04 &  6.18 \\
		\hline
		AHCX (vdW-DF-ahcx,0.20)  &  4.99 & 5.16 & 3.92 & 0.76 & 2.28 & 1.16 & 4.71 & 2.47 & 1.23 &  8.35 \\
		AHCX$_{25}$ (0.25)       &  7.56 & 5.41 & 4.04 & 0.70 & 2.66 & 1.15 & 4.44 & 2.44 & 1.09 &  8.63 \\
		DF2-AH (vdW-DF2-ah,0.20) & 26.00 & 7.92 &14.97 & 5.83 & 9.71 & 7.17 & 5.94 & 4.40 & 1.39 & 18.17 \\
		AHBR$_{20}$ (0.20)       &  6.55 & 4.76 & 5.52 & 1.86 & 3.50 & 1.26 & 3.52 & 2.16 & 1.20 &  6.07 \\
		AHBR (vdW-DF2-ahbr,0.25) &  9.50 & 5.29 & 5.81 & 1.73 & 4.02 & 1.23 & 3.30 & 1.98 & 1.06 &  6.18 \\ 
		\hline
	\end{tabular}
\end{table*}

\begin{table*}
	\caption{Functional performance of regular vdW-DFs and associated 
	unscreened hybrid and RSH vdW-DFs for individual benchmark sets of the
	GMTKN55 group 2: large-system isomerizations \cite{gmtkn55}. All entries
	are MAD values in kcal/mol.
	\label{tab:largeiso_mad}
	}
	\begin{tabular}{lccccccccc}
		\hline \hline
		XC functional & MB16-43 & DARC & RSE43 & BSR36 & CDIE20 & ISO34 & ISOL24 & C60ISO & PArel \\
		\hline
		PBE                      & 22.60 &  7.13 & 2.54 & 7.65 & 1.78 & 1.73 & 6.88 &10.06 & 1.93 \\ 
		PBE (OB)                 & 22.78 &  6.94 & 3.10 & 7.67 & 1.81 & 1.80 & 6.83 &11.06 & 1.81 \\
		revPBE+D3                & 25.28 &  4.28 & 1.93 & 1.70 & 1.54 & 1.49 & 4.82 & 8.93 & 1.69 \\
		revPBE+D3(OB)            & 27.11 &  3.71 & 2.31 & 1.80 & 1.50 & 1.50 & 4.56 & 9.82 & 1.53 \\
		SCAN+D3(OB)              & 17.77 &  2.01 & 1.29 & 1.28 & 1.45 & 1.30 & 3.23 & 6.01 & 1.50 \\
		HSE+D3                   & 15.48 &  2.65 & 1.25 & 3.83 & 1.30 & 1.42 & 2.42 & 2.51 & 1.34 \\
		HSE+D3(OB)               & 14.27 &  2.11 & 1.49 & 3.83 & 1.32 & 1.34 & 2.64 & 2.43 & 1.16 \\
		B3LYP+D3(OB)             & 24.84 &  8.03 & 1.72 & 3.35 & 1.00 & 1.78 & 5.80 & 2.22 & 1.18 \\
		\hline
		vdW-DF1                  & 57.84 & 15.99 & 1.22 & 4.10 & 1.24 & 2.85 & 9.49 &10.73 & 1.76 \\
		vdW-DF2                  & 75.22 & 22.24 & 1.13 & 4.94 & 1.05 & 3.97 &12.69 &10.43 & 2.17 \\
		rVV10                    & 21.11 &  8.75 & 1.77 & 3.46 & 1.45 & 1.94 & 6.81 &10.55 & 1.51 \\
		\hline
		C09 (vdW-DF-c09)         & 27.04 &  3.90 & 2.31 & 2.44 & 1.29 & 1.59 & 2.86 &12.17 & 1.99 \\
		OB86 (vdW-DF-optB86r)    & 21.38 &  1.66 & 2.05 & 1.05 & 1.26 & 1.32 & 2.88 &11.84 & 1.70 \\
		DF3-opt1             & 25.00 &  2.53 & 2.54 & 0.53 & 1.30 & 1.46 & 2.81 &11.59 & 2.04 \\
		DF3-opt2             & 20.28 &  1.66 & 2.17 & 1.36 & 1.17 & 1.27 & 2.90 &12.09 & 1.77 \\
		OBK8 (vdW-DF-optB88)     & 23.46 &  5.18 & 1.73 & 0.27 & 1.17 & 1.68 & 4.61 &11.34 & 1.42 \\
		\hline
		CX (vdW-DF-cx)           & 24.27 &  1.71 & 2.21 & 1.27 & 1.34 & 1.49 & 2.65 &12.01 & 1.86 \\
		B86R (rev-vdW-DF2)       & 22.37 &  2.97 & 2.14 & 0.64 & 1.24 & 1.35 & 3.68 &11.95 & 1.68 \\
		\hline
		CX0 (CX+0,0.25)          & 14.09 &  5.14 & 0.73 & 0.80 & 0.80 & 1.30 & 2.32 & 2.16 & 1.07 \\
		CX0P (vdW-DF-cx0p,0.20)  & 15.31 &  4.34 & 0.98 & 0.88 & 0.90 & 1.33 & 2.28 & 3.40 & 1.18 \\
		DF2-BR0 (B86R+0,0.25)    & 18.88 &  1.16 & 0.72 & 0.30 & 0.74 & 1.02 & 1.59 & 2.15 & 0.92 \\
		\hline
		AHCX (vdW-DF-ahcx,0.20)  & 15.41 &  4.00 & 1.01 & 0.74 & 0.96 & 1.32 & 2.25 & 3.99 & 1.19 \\
		AHCX$_{25}$ (0.25)       & 14.15 &  4.71 & 0.75 & 0.62 & 0.87 & 1.30 & 2.28 & 2.83 & 1.08 \\
		DF2-AH (vdW-DF2-ah,0.20) & 67.26 & 15.70 & 0.82 & 4.38 & 0.76 & 3.10 & 9.34 & 2.80 & 1.74 \\
		AHBR$_{20}$ (0.20)       & 18.97 &  1.27 & 0.98 & 0.29 & 0.88 & 1.05 & 2.10 & 3.88 & 1.04 \\
		AHBR (vdW-DF2-ahbr,0.25) & 18.95 &  1.06 & 0.74 & 0.35 & 0.80 & 1.02 & 1.82 & 2.72 & 0.94 \\
		\hline
	\end{tabular}
\end{table*}

\begin{table*}
	\caption{Functional performance of regular vdW-DFs and associated 
	unscreened hybrid and RSH vdW-DFs for individual benchmark sets of the
	GMTKN55 group 4: intermolecular noncovalent (NOC) interactions \cite{gmtkn55}.
	We report MAD values in kcal/mol, abbreviating benchmark name HEAVY28
	as 'HEAVY', PNICO23 as `PNICO', CARBHB12 as 'CARBH', and WATER27 as 'WATER'.
	\label{tab:interNOC}
	}
	\begin{tabular}{lcccccccccccc}
		\hline \hline
		XC functional & RG18 & ADIM6 & S22 & S66 & HEAVY & CARBH & PNICO & HAL59 & CHB6 & IL16 & AHB21 & WATER  \\
		\hline
		PBE                      & 0.28 & 3.36 & 2.56 & 2.13 & 0.50 & 1.05 & 0.85 & 1.41 & 0.75 & 1.34 & 0.85 &  3.47 \\
		PBE (OB)                 & 0.28 & 3.38 & 2.55 & 2.11 & 0.47 & 1.09 & 0.82 & 1.30 & 0.79 & 1.49 & 0.83 &  2.83 \\
		revPBE+D3                & 0.08 & 0.12 & 0.36 & 0.25 & 0.30 & 1.10 & 0.84 & 0.82 & 0.88 & 0.60 & 1.01 &  2.63 \\
		revPBE+D3(OB)            & 0.09 & 0.25 & 0.43 & 0.28 & 0.29 & 1.10 & 0.88 & 0.72 & 0.90 & 0.77 & 1.04 &  3.51 \\
		SCAN+D3(OB)              & 0.18 & 0.12 & 0.47 & 0.43 & 0.27 & 1.38 & 1.08 & 1.03 & 0.45 & 0.98 & 1.67 & 10.15 \\
		HSE+D3                   & 0.11 & 0.25 & 0.54 & 0.39 & 0.38 & 1.41 & 0.86 & 0.64 & 1.24 & 0.38 & 1.32 &  5.73 \\ 
		HSE+D3(OB)               & 0.13 & 0.14 & 0.52 & 0.39 & 0.44 & 1.47 & 0.97 & 0.73 & 1.29 & 0.31 & 1.32 &  6.29 \\ 
		B3LYP+D3(OB)             & 0.13 & 0.11 & 0.31 & 0.26 & 0.34 & 0.88 & 0.48 & 0.57 & 1.41 & 0.76 & 0.33 &  4.07 \\ 
		\hline
		vdW-DF1                  & 0.59 & 0.23 & 1.26 & 0.69 & 0.31 & 0.44 & 0.76 & 0.50 & 1.00 & 2.61 & 1.79 &  7.80 \\ 
		vdW-DF2                  & 0.35 & 0.51 & 0.69 & 0.32 & 0.14 & 0.46 & 0.39 & 0.69 & 1.13 & 1.82 & 1.05 &  1.75 \\
		rVV10                    & 0.16 & 0.28 & 0.42 & 0.43 & 0.40 & 1.59 & 0.91 & 1.47 & 0.36 & 0.82 & 1.28 & 11.23 \\ 
		\hline
		C09 (vdW-DF-c09)         & 0.12 & 0.51 & 0.43 & 0.38 & 0.18 & 1.14 & 0.95 & 1.22 & 0.27 & 1.22 & 0.86 &  7.95 \\ 
		OB86 (vdW-DF-optB86r)    & 0.22 & 0.75 & 0.30 & 0.35 & 0.14 & 0.92 & 0.68 & 1.05 & 0.33 & 0.62 & 0.67 &  5.66 \\
		DF3-opt1             & 0.11 & 0.42 & 0.49 & 0.48 & 0.22 & 1.62 & 1.31 & 1.50 & 0.13 & 1.82 & 1.58 & 13.66 \\
		DF3-opt2             & 0.23 & 0.59 & 0.32 & 0.41 & 0.21 & 1.32 & 0.99 & 1.32 & 0.27 & 0.94 & 1.04 & 10.04 \\
		OBK8 (vdW-DF-optB88)     & 0.08 & 0.82 & 0.30 & 0.36 & 0.13 & 0.79 & 0.52 & 1.00 & 0.55 & 0.50 & 0.62 &  5.37 \\ 
		\hline
		CX (vdW-DF-cx)           & 0.37 & 0.05 & 0.36 & 0.28 & 0.24 & 0.76 & 0.66 & 0.94 & 0.50 & 0.37 & 0.65 &  2.88 \\ 
		B86R (rev-vdW-DF2)       & 0.07 & 0.21 & 0.45 & 0.36 & 0.22 & 0.85 & 0.56 & 1.00 & 0.30 & 0.37 & 0.64 &  5.10 \\ 
		\hline
		CX0 (CX+0,0.25)          & 0.47 & 0.78 & 0.41 & 0.42 & 0.26 & 0.58 & 0.41 & 0.58 & 0.71 & 0.32 & 0.60 &  2.94 \\ 
		CX0P (vdW-DF-cx0p,0.20)  & 0.45 & 0.61 & 0.30 & 0.30 & 0.25 & 0.59 & 0.44 & 0.62 & 0.56 & 0.32 & 0.50 &  2.82 \\
		DF2-BR0 (B86R+0,0.25)    & 0.06 & 0.04 & 0.27 & 0.25 & 0.35 & 0.48 & 0.24 & 0.57 & 0.52 & 0.49 & 0.46 &  2.50 \\
		\hline
		AHCX (vdW-DF-ahcx,0.20)  & 0.43 & 0.56 & 0.26 & 0.27 & 0.25 & 0.60 & 0.44 & 0.63 & 0.53 & 0.33 & 0.50 &  2.71 \\ 
		AHCX$_{25}$ (0.25)       & 0.47 & 0.71 & 0.36 & 0.37 & 0.26 & 0.58 & 0.40 & 0.59 & 0.67 & 0.32 & 0.61 &  2.79 \\ 
		DF2-AH (vdW-DF2-ah,0.20) & 0.26 & 0.53 & 0.43 & 0.19 & 0.14 & 0.37 & 0.43 & 0.39 & 0.80 & 1.76 & 0.61 &  0.77 \\ 
		AHBR$_{20}$ (0.20)       & 0.07 & 0.09 & 0.31 & 0.28 & 0.32 & 0.56 & 0.29 & 0.63 & 0.41 & 0.44 & 0.42 &  2.95 \\  
		AHBR (vdW-DF2-ahbr,0.25) & 0.20 & 0.06 & 0.29 & 0.26 & 0.35 & 0.50 & 0.25 & 0.58 & 0.51 & 0.49 & 0.48 &  2.52 \\ 
		\hline
	\end{tabular}
\end{table*}

\begin{table*}
	\caption{Functional performance of regular vdW-DFs and associated 
	unscreened hybrid and RSH vdW-DFs for individual benchmark sets of the
	GMTKN55 group 5: intramolecular NOC interactions \cite{gmtkn55}.
	We report MAD values in kcal/mol, abbreviating benchmark name Amino20x4
	as `Amino', PCON21 as 'PCONF', and BUT14DIOL as 'B14D'.
	\label{tab:intraNOC}
	}
	\begin{tabular}{lccccccccc}
		\hline \hline
		XC functional & IDISP & ICONF & ACONF & Amino & PCONF & MCONF & SCONF & UPU23 & B14D  \\
		\hline
		PBE                     & 10.91 & 0.43 & 0.61 & 0.53 & 3.59 & 1.83 & 0.37 & 2.02 & 0.25 \\ 
		PBE (OB)                & 10.78 & 0.43 & 0.61 & 0.51 & 3.48 & 1.80 & 0.35 & 1.99 & 0.26 \\
		revPBE+D3               &  3.25 & 0.31 & 0.05 & 0.35 & 1.01 & 0.44 & 0.51 & 0.60 & 0.26 \\
		revPBE+D3(OB)           &  3.14 & 0.32 & 0.09 & 0.37 & 0.87 & 0.44 & 0.51 & 0.47 & 0.31 \\
		SCAN+D3(OB)             &  2.15 & 0.31 & 0.13 & 0.22 & 0.47 & 0.46 & 0.66 & 0.39 & 0.40 \\
		HSE+D3                  &  2.96 & 0.29 & 0.19 & 0.29 & 1.34 & 0.31 & 0.21 & 0.70 & 0.20 \\
		HSE+D3(OB)              &  2.55 & 0.30 & 0.14 & 0.27 & 1.05 & 0.27 & 0.26 & 0.55 & 0.26 \\
		B3LYP+D3(OB)            &  3.57 & 0.29 & 0.05 & 0.21 & 0.53 & 0.22 & 0.30 & 0.61 & 0.31 \\
		\hline
		vdW-DF1                 &  7.56 & 0.51 & 0.38 & 0.53 & 0.60 & 0.58 & 1.04 & 0.55 & 0.50 \\
		vdW-DF2                 &  7.89 & 0.52 & 0.12 & 0.38 & 0.40 & 0.48 & 0.52 & 0.53 & 0.13 \\
		rVV10                   &  3.91 & 0.33 & 0.12 & 0.33 & 0.73 & 0.41 & 1.08 & 0.43 & 0.75 \\
		\hline
		C09 (vdW-DF-c09)        &  3.31 & 0.24 & 0.14 & 0.35 & 0.94 & 0.74 & 1.39 & 0.64 & 0.72 \\
		OB86 (vdW-DF-optB86r)   &  2.57 & 0.22 & 0.07 & 0.25 & 0.77 & 0.57 & 0.87 & 0.66 & 0.51 \\
		DF3-opt1            &  2.79 & 0.26 & 0.19 & 0.42 & 0.90 & 0.57 & 1.58 & 0.37 & 0.94 \\
		DF3-opt2            &  2.89 & 0.20 & 0.11 & 0.27 & 0.69 & 0.43 & 1.06 & 0.41 & 0.68 \\
		OBK8 (vdW-DF-optB88)    &  3.54 & 0.24 & 0.11 & 0.23 & 0.75 & 0.53 & 0.75 & 0.64 & 0.50 \\
		\hline
		CX (vdW-DF-cx)          &  2.27 & 0.26 & 0.11 & 0.25 & 0.75 & 0.39 & 0.81 & 0.47 & 0.37 \\
		B86R (rev-vdW-DF2)      &  2.69 & 0.24 & 0.05 & 0.22 & 0.68 & 0.33 & 0.76 & 0.37 & 0.48 \\
		\hline
		CX0 (CX+0,0.25)         &  1.94 & 0.21 & 0.05 & 0.23 & 0.40 & 0.55 & 0.26 & 0.65 & 0.20 \\
		CX0P (vdW-DF-cx0p,0.20) &  1.79 & 0.20 & 0.06 & 0.22 & 0.42 & 0.50 & 0.35 & 0.60 & 0.23 \\
		DF2-BR0 (B86R+0,0.25)   &  1.26 & 0.22 & 0.09 & 0.18 & 0.22 & 0.17 & 0.13 & 0.36 & 0.20 \\
		\hline
		AHCX (vdW-DF-ahcx,0.20) &  1.61 & 0.21 & 0.08 & 0.22 & 0.40 & 0.46 & 0.32 & 0.59 & 0.22 \\
		AHCX$_{25}$ (0.25)        &  1.72 & 0.21 & 0.07 & 0.22 & 0.36 & 0.50 & 0.22 & 0.63 & 0.19 \\
		DF2-AH (vdW-DF2-ah,0.20)&  6.28 & 0.41 & 0.14 & 0.37 & 0.33 & 0.32 & 0.68 & 0.51 & 0.14 \\
		AHBR$_{20}$ (0.20)        &  1.75 & 0.21 & 0.10 & 0.18 & 0.32 & 0.18 & 0.22 & 0.34 & 0.24 \\
		AHBR (vdW-DF2-ahbr,0.25)&  1.50 & 0.23 & 0.11 & 0.19 & 0.24 & 0.16 & 0.11 & 0.36 & 0.19 \\
		\hline
	\end{tabular}
\end{table*}

\FloatBarrier\clearpage

\subsection{Parameters for the analytical-exchange hole modeling and convergence tests}

\begin{table*}[!htbp]
	\caption{\label{tab:paramHmore} Parameters in the rational function defining $\mathcal{H}(s)$ in the
	HJS AH model \cite{HJS08,DefineAHCX} in its description for the exchange functionals PBEx, PBEsolx, 
	rPW86 (exchange in vdW-DF2), cx13 (exchange in vdW-DF-cx), and B86R (exchange in vdW-DF2-b86r).
	The determination of the PBEx, PBSEsolx, rPW86, and LV-rPW86 (or cx-13) parameters are given
	in Ref.\ \onlinecite{DefineAHCX}, and repeated for the reader's convenience; The last column
	gives parameters that we provide for the AH modeling of B86R and hence for the AHBR specification.
	}
	\begin{tabular*}{0.9\textwidth}{@{\extracolsep{\fill}}lrrrrr}
		\hline \hline
		&     PBEx &  PBEsolx &    rPW86 &  LV-rPW86 &        B86R \\
		\hline
		$a_2$ &      0.0154999  &      0.0045881  &      0.0000006  &      0.0024387  &      0.0045620  \\
		$a_3$ &     -0.0361006  &     -0.0085784  &      0.0402647  &     -0.0041526  &     -0.0087000  \\
		$a_4$ &      0.0379567  &      0.0072956  &     -0.0353219  &      0.0025826  &      0.0073696  \\
		$a_5$ &     -0.0186715  &     -0.0032019  &      0.0116112  &      0.0000012  &     -0.0030244  \\
		$a_6$ &      0.0017426  &      0.0006049  &     -0.0001555  &     -0.0007582  &      0.0003868  \\
		$a_7$ &      0.0019076  &      0.0000216  &      0.0000504  &      0.0002764  &      0.0000944  \\
		$b_1$ &     -2.7062566  &     -2.1449453  &     -1.8779594  &     -2.2030319  &     -2.2089330  \\
		$b_2$ &      3.3316842  &      2.0901104  &      1.5198811  &      2.1759315  &      2.1968353  \\
		$b_3$ &     -2.3871819  &     -1.1935421  &     -0.5383109  &     -1.2997841  &     -1.2662249  \\
		$b_4$ &      1.1197810  &      0.4476392  &      0.1352399  &      0.5347267  &      0.4689964  \\
		$b_5$ &     -0.3606638  &     -0.1172367  &     -0.0428465  &     -0.1588798  &     -0.1165714  \\
		$b_6$ &      0.0841990  &      0.0231625  &      0.0117903  &      0.0367329  &      0.0207188  \\
		$b_7$ &     -0.0114719  &     -0.0035278  &      0.0033791  &     -0.0077318  &     -0.0029772  \\
		$b_8$ &      0.0016928  &      0.0005399  &     -0.0000493  &      0.0012667  &      0.0005982  \\
		$b_9$ &      0.0015054  &      0.0000158  &      0.0000071  &      0.0000008  &      0.0000047  \\
		\hline \hline
	\end{tabular*}
\end{table*}

\begin{table}[!htbp]
	\caption{\label{tab:vdWDFasymImpact}
	Scope of impact by spurious intercell vdW-attraction on the GMTKN55 benchmarks sets on the MAD
	values (in kcal/mol) that characterize CX performance on individual-GMTKN55 benchmarks sets,
	given our choice of a 10 {\AA} vacuum padding. The characterization rests on
	self-consistent electron-density variations (computed for the roughly 2450 different 
	GMTKN55 systems \cite{gmtkn55} in a pilot study using the \textsc{AbInit} PPs \cite{abinit05} at 
	80 Ry wavefunction-energy cutoff) and the extraction of effective per-unit-cell 
	$C_6^{\rm mol}$ asymptotic-vdW interaction coefficients \cite{Dion,kleis08p205422,NanovdWScale},
	see Appendix A.  We list assessments of impact (in kcal/mol) for the benchmarks that we find are most 
	susceptible to offsets from such spurious vdW coupling in our planewave setup. 
	}
	\begin{ruledtabular}
	\begin{tabular}{lc}
		Benchmark &   Spurious-vdW impact \\
		\hline
		G21EA    &      0.010  \\
		ALK8     &      0.002  \\
		ALKBDE10 &      0.002  \\
		WATER27  &      0.001  \\ 
		MB16-43  &      0.001  \\
	\end{tabular}
	\end{ruledtabular}
\end{table}

\FloatBarrier\clearpage

\subsection{Bulk-system performance}

\begin{table*}[!htbp]
	\caption{\label{tab:bulklattice} Comparison of computed bulk lattice constants $a$ and 
	experiment values, back-corrected to zero point energy and thermal effects \cite{Tran19}. All entries in {\AA}.
	}
	\begin{ruledtabular}
\begin{tabular}{l|ccc|c|ccc|c}
                            & CX                          & AHCX                        & AHCX$_{0.25}$              & DF2-AH                        & b86R                        & AHBR$_{0.20}$               & AHBR                      & Exper.*                        \\ 
\hline                                                                                                                                             
\textbf{Ag}                 & $4.065$                     & $4.078$                     & $4.082$                    & $4.287$                       & $4.104$                     & $4.115$                     & $4.118$                   & $\mathbf{4.070}$            \\ 
\hline                                                                                                                                             
\textbf{Au}                 & $4.101$                     & $4.098$                     & $4.097$                    & $4.305$                       & $4.134$                     & $4.127$                     & $4.126$                   & $\mathbf{4.067}$            \\ 
\hline                                                                                                                                             
\textbf{Al}                 & $4.041$                     & $4.040$                     & $4.039$                    & $4.044$                       & $4.030$                     & $4.032$                     & $4.033$                   & $\mathbf{4.022}$            \\ 
\hline                                                                                                                                             
\textbf{C}                  & $3.561$                     & $3.545$                     & $3.541$                    & $3.573$                       & $3.565$                     & $3.548$                     & $3.544$                   &  $\mathbf{3.553}$            \\ 
\hline                                                                                                                                             
\textbf{Cu}                 & $3.576$                     & $3.587$                     & $3.592$                    & $3.736$                       & $3.602$                     & $3.613$                     & $3.617$                   & $\mathbf{3.599}$            \\ 
\hline                                                                                                                                            
\textbf{GaAs}                & $5.705$                     & $5.640$                     & $5.628$                    & $5.758$                       & $5.733$                     & $5.661$                     & $5.644$                   & $\mathbf{5.638}$            \\ 
\hline                                                                                                                                             
\textbf{LiF}                & $4.052$                     & $4.012$                     & $4.002$                    & $4.026$                       & $4.036$                     & $4.004$                     & $3.996$                   & $\mathbf{3.972}$            \\ 
\hline                                                                                                                                             
\textbf{MgO}                & $4.243$                     & $4.205$                     & $4.197$                    & $4.222$                       & $4.225$                     & $4.194$                     & $4.187$                   & $\mathbf{4.189}$            \\ 
\hline                                                                                                                                             
\textbf{NaCl}                & $5.661$                     & $5.623$                     & $5.612$                    & $5.635$                       & $5.626$                     & $5.603$                     & $5.597$                   & $\mathbf{5.569}$            \\ 
\hline                                                                                                                                             
\textbf{SiC}                & $4.374$                     & $4.353$                     & $4.348$                    & $4.377$                       & $4.377$                     & $4.356$                     & $4.351$                   & $\mathbf{4.346}$            \\ 
\hline                                                                                                                                            
\textbf{Pt}                 & $3.929$                     & $3.910$                     & $3.906$                    & $4.057$                       & $3.952$                     & $3.929$                     & $3.925$                   & $\mathbf{3.917}$            \\ 
\hline                                                                                                                                             
\textbf{Rh}                 & $3.786$                     & $3.760$                     & $3.754$                    & $3.875$                       & $3.806$                     & $3.776$                     & $3.770$                   & $\mathbf{3.786}$            \\ 
\hline                                                                                                                                             
\textbf{Si}                 & $5.462$                     & $5.441$                     & $5.435$                    & $5.476$                       & $5.465$                     & $5.444$                     & $5.439$                   & $\mathbf{5.411}$            \\ 
\end{tabular}                                                                                                                                   \end{ruledtabular}
\end{table*}
                                                                                                                                                   
\begin{table*}
	\caption{\label{tab:bulkcohesive} Comparison of computed bulk cohesive energies $E_{\rm coh}$ and 
	experiment values, back-corrected to zero point energy and thermal effects \cite{Tran19}. All entries in eV.
	}
	\begin{ruledtabular}
\begin{tabular}{l|ccc|c|ccc|c}
                            & CX                          & AHCX                        & AHCX$_{0.25}$              & DF2-AH                        & b86R                        & AHBR$_{0.20}$               & AHBR                      & Exper.*                        \\ 
\hline                                                                                                                                             
\textbf{Ag}                 & $2.955$                     & $2.774$                     & $2.737$                    & $2.100$                       & $2.779$                     & $2.592$                     & $2.549$                   & $\mathbf{2.964}$            \\ 
\hline                                                                                                                                             
\textbf{Au}                 & $3.634$                     & $3.440$                     & $3.398$                    & $2.469$                       & $3.402$                     & $3.205$                     & $3.158$                   & $\mathbf{3.835}$            \\ 
\hline                                                                                                                                             
\textbf{Al}                 & $3.642$                     & $3.430$                     & $3.421$                    & $2.516$                       & $3.439$                     & $3.251$                     & $3.200$                   & $\mathbf{3.431}$            \\ 
\hline                                                                                                                                             
\textbf{C}                  & $7.891$                     & $7.565$                     & $7.553$                    & $6.570$                       & $7.777$                     & $7.414$                     & $7.424$                   & $\mathbf{7.452}$            \\ 
\hline                                                                                                                                             
\textbf{Cu}                 & $3.781$                     & $3.348$                     & $3.264$                    & $2.551$                       & $3.582$                     & $3.160$                     & $3.064$                   & $\mathbf{3.513}$            \\ 
\hline                                                                                                                                             
\textbf{GaAs}                & $3.358$                     & $3.317$                     & $3.321$                    & $2.739$                       & $3.242$                     & $3.190$                     & $3.179$                   & $\mathbf{3.337}$            \\ 
\hline                                                                                                                                            
\textbf{LiF}                & $4.405$                     & $4.399$                     & $4.374$                    & $4.501$                       & $4.553$                     & $4.418$                     & $4.382$                   & $\mathbf{4.457}$            \\ 
\hline                                                                                                                                            
\textbf{MgO}                & $5.110$                     & $5.057$                     & $5.105$                    & $4.889$                       & $5.247$                     & $5.065$                     & $5.020$                   & $\mathbf{5.203}$            \\ 
\hline                                                                                                                                            
\textbf{NaCl}                & $3.225$                     & $3.258$                     & $3.245$                    & $3.182$                       & $3.230$                     & $3.191$                     & $3.179$                   & $\mathbf{3.337}$            \\ 
\hline                                                                                                                                            
\textbf{SiC}                & $6.590$                     & $6.406$                     & $6.393$                    & $5.698$                       & $6.514$                     & $6.301$                     & $6.296$                   & $\mathbf{6.478}$            \\ 
\hline                                                                                                                                             
\textbf{Pt}                 & $6.226$                     & $5.524$                     & $5.259$                    & $3.941$                       & $5.999$                     & $5.131$                     & $4.930$                   & $\mathbf{5.866}$            \\ 
\hline                                                                                                                                             
\textbf{Rh}                 & $6.367$                     & $5.244$                     & $4.972$                    & $3.956$                       & $6.389$                     & $5.164$                     & $4.856$                   & $\mathbf{5.783}$            \\ 
\hline                                                                                                                                             
\textbf{Si}                 & $4.758$                     & $4.664$                     & $4.624$                    & $4.155$                       & $4.679$                     & $4.563$                     & $4.531$                   & $\mathbf{4.685}$            \\ 
\end{tabular}                                                                                                                                   \end{ruledtabular}
\end{table*}
                                                                                                                                                   
\begin{table*}
	\caption{\label{tab:bulkmodulus} Comparison of computed bulk-modulus constants $B_0$ and 
	experiment values, back-corrected to zero point energy and thermal effects \cite{Tran19}. All entries in GPa.
	}
	\begin{ruledtabular}
\begin{tabular}{l|ccc|c|ccc|c}                                                                                                                        
                            & CX                          & AHCX                        & AHCX$_{0.25}$              & DF2-AH                        & b86R                        & AHBR$_{0.20}$               & AHBR                      & Exper.*                        \\ 
\hline                                                                                                                                             
\textbf{Ag}                 & $115.3$                     & $104.8$                     & $104.0$                    & $63.9$                        & $102.4$                     & $95.2$                      & $94.6$                    & $\mathbf{105.7}$            \\ 
\hline                                                                                                                                             
\textbf{Au}                 & $170.5$                     & $167.8$                     & $166.8$                    & $94.2$                        & $153.4$                     & $152.0$                     & $151.3$                   & $\mathbf{182.0}$            \\ 
\hline                                                                                                                                             
\textbf{Al}                 & $78.2$                      & $82.4$                      & $82.5$                     & $74.6$                        & $78.7$                      & $81.9$                      & $81.8$                    & $\mathbf{72.2}$             \\ 
\hline                                                                                                                                             
\textbf{C}                  & $439.8$                     & $466.2$                     & $472.6$                    & $434.8$                       & $434.1$                     & $461.9$                     & $468.5$                   & $\mathbf{454.7}$            \\ 
\hline                                                                                                                                             
\textbf{Cu}                 & $163.3$                     & $148.3$                     & $146.0$                    & $91.1$                        & $151.3$                     & $141.4$                     & $136.0$                   & $\mathbf{144.3}$            \\ 
\hline                                                                                                                                             
\textbf{GaAs}                & $64.8$                      & $73.8$                      & $76.7$                     & $59.7$                        & $61.2$                      & $71.6$                      & $74.2$                    & $\mathbf{76.7}$             \\ 
\hline                                                                                                                                            
\textbf{LiF}                & $68.3$                      & $74.5$                      & $75.9$                     & $75.7$                        & $69.7$                      & $74.9$                      & $76.0$                    & $\mathbf{76.3}$             \\ 
\hline                                                                                                                                            
\textbf{MgO}                & $153.3$                     & $168.2$                     & $171.8$                    & $167.4$                       & $158.3$                     & $171.7$                     & $174.9$                   & $\mathbf{169.8}$            \\ 
\hline                                                                                                                                            
\textbf{NaCl}                & $24.9$                      & $26.1$                      & $26.5$                     & $27.3$                        & $26.2$                      & $26.9$                      & $27.0$                    & $\mathbf{27.6}$             \\ 
\hline                                                                                                                                            
\textbf{SiC}                & $215.0$                     & $228.5$                     & $231.8$                    & $214.5$                       & $212.6$                     & $226.7$                     & $230.1$                   & $\mathbf{229.1}$            \\ 
\hline                                                                                                                                            
\textbf{Pt}                 & $284.0$                     & $297.7$                     & $298.1$                    & $187.9$                       & $264.0$                     & $277.9$                     & $279.3$                   & $\mathbf{285.5}$            \\ 
\hline                                                                                                                                             
\textbf{Rh}                 & $295.8$                     & $312.7$                     & $317.3$                    & $225.2$                       & $275.7$                     & $298.0$                     & $303.3$                   & $\mathbf{277.1}$            \\ 
\hline                                                                                                                                             
\textbf{Si}                 & $90.1$                      & $96.5$                      & $98.3$                     & $90.7$                        & $89.1$                      & $95.9$                      & $97.5$                    & $\mathbf{101.3}$            \\ 
\end{tabular}
\end{ruledtabular}
\end{table*}

\FloatBarrier

\subsection{Testing on DNA base-pair assembly}
\FloatBarrier
\begin{table*}[!h]
	\caption{\label{tab:DNAstep} Comparison of vdW-DF tool-chain performance for DNA assembly: Stepping
	energies defined by stacking two Watson-Crick (WC) base pairs, as discussed in the main text.
	A superscript `GBRV' identifies more electron-sparse calculations (performed with ultrasoft 
	PPs \cite{GBRV} at 50 Ry wavefunction-energy cutoff) than supplement our standard ONCV-SG15 PP/160 Ry characterizations.
	The coupled-cluster reference energies (as well as the atomic configurations) for 10 stacked base-pair 
	combinations are taken from Ref.\ \onlinecite{KrBaSp2019}; They are computed in DLPNO-CCSD(T), 
	Ref.\ \onlinecite{RiPiBe2016}.  All entries are in kcal/mol.
	}
\begin{ruledtabular}
\begin{tabular}{l|cc|ccc|ccc|c}
	& B3LYP+D3(BJ)$^a$ & CX$^{\rm GBRV}$ & CX & AHCX & AHCX$_{0.25}$ & B86R   & AHBR$_{0.20}$ & AHBR & DLPNO-CCSD(T)\\
\hline
	ApA  & -13.72 & -14.04 & -14.33 & -15.93 & -16.33 & -12.34 & -12.64 & -12.73 &  -12.95 \\
	ApC  & -12.86 & -13.16 & -13.45 & -14.95 & -15.34 & -11.49 & -11.72 & -11.79 &  -11.88 \\
	ApG  & -13.47 & -13.75 & -14.03 & -15.60 & -16.01 & -12.19 & -12.44 & -12.52 &  -12.53 \\
	ApT  & -12.05 & -12.30 & -12.64 & -14.10 & -14.49 & -10.63 & -10.89 & -10.98 &  -11.00 \\
	CpC  & -11.38 & -11.86 & -12.14 & -13.63 & -14.01 & -10.24 & -10.43 & -10.50 &  -10.53 \\
	CpG  & -17.02 & -17.12 & -17.42 & -19.31 & -19.78 & -15.92 & -16.36 & -16.49 &  -16.23 \\
	GpC  & -14.97 & -15.13 & -15.48 & -17.06 & -17.47 & -13.54 & -13.81 & -13.89 &  -13.94 \\
	TpA  & -13.70 & -13.94 & -14.25 & -15.86 & -16.27 & -12.46 & -12.82 & -12.93 &  -12.92 \\
	TpC  & -12.51 & -13.05 & -13.36 & -14.79 & -15.16 & -11.32 & -11.52 & -11.59 &  -11.58 \\
	TpG  & -14.99 & -15.10 & -15.41 & -17.15 & -17.59 & -13.76 & -14.17 & -14.30 &  -14.17 \\
\hline
	MD   & -      & -1.17 & -1.48 & -3.06 & -3.47 &  0.38 & 0.09  & 0.00 & - \\
	MAD  & 0.89   &  1.17 &  1.48 &  3.06 &  3.47 &  0.38 & 0.12  & 0.08 & - \\
\hline
$^a$Ref.\ \onlinecite{KrBaSp2019}.
\end{tabular}
\end{ruledtabular}
\end{table*}

\begin{table*}
	\caption{\label{tab:DNAbpair} Comparison of vdW-DF tool-chain performance for descriptions
	of DNA assembly by base-pair stacking, here focusing on molecular-pairing contributions, 
	$\Delta E_{\rm B-pair}'$, for which there are also couple-cluster reference results \cite{KrBaSp2019}. All entries are in kcal/mol. 
	}
\begin{ruledtabular}
\begin{tabular}{l|c|ccc|ccc|c}
	& CX$^{GBRV}$ & CX & AHCX & AHCX$_{0.25}$ & B86R   & AHBR$_{0.20}$ & AHBR & DLPNO-CCSD(T)$^{a}$\\
\hline
	ApA  & -14.75 & -15.02 & -16.51 & -16.89 & -12.29 & -12.50 & -12.57 &  -13.16 \\
	ApC  & -14.77 & -15.04 & -16.46 & -16.83 & -12.39 & -12.53 & -12.59 &  -13.24 \\
	ApG  & -15.37 & -15.63 & -17.14 & -17.52 & -13.02 & -13.19 & -13.26 &  -13.67 \\
	ApT  & -13.07 & -13.29 & -14.76 & -15.13 & -10.76 & -10.96 & -11.02 &  -11.62 \\
	CpC  & -15.41 & -15.67 & -17.06 & -17.41 & -12.92 & -13.00 & -13.04 &  -13.50 \\
	CpG  & -19.09 & -19.38 & -21.32 & -21.80 & -17.17 & -17.60 & -17.73 &  -18.11 \\
	GpC  & -17.34 & -17.64 & -19.20 & -19.59 & -14.99 & -15.17 & -15.24 &  -15.77 \\
	TpA  & -14.76 & -14.98 & -16.59 & -16.98 & -12.60 & -12.88 & -12.97 &  -13.58 \\
	TpC  & -14.74 & -14.99 & -16.35 & -16.70 & -12.18 & -12.30 & -12.34 &  -12.90 \\
	TpG  & -16.90 & -17.16 & -18.89 & -19.32 & -14.82 & -15.16 & -15.27 &  -15.73 \\
\hline
	MD   &  -1.49 & -1.75  & -3.30  & -3.69 &  0.82  &  0.60  & 0.52 & - \\
	MAD  &   1.49 &  1.75  &  3.30  &  3.69 &  0.82  &  0.60  & 0.52 & - \\
\hline
$^a$Ref.\ \onlinecite{KrBaSp2019}.
\end{tabular}
\end{ruledtabular}
\end{table*}

\FloatBarrier\clearpage

\end{document}